\documentclass[10pt,a4paper]{article}
\usepackage{jheppub}

\usepackage{placeins}
\usepackage{epsfig}
\usepackage{latexsym}
\usepackage{amsfonts}
\usepackage[bbgreekl]{mathbbol}
\usepackage{mathtools}
\usepackage{amsthm}
\usepackage{amsbsy}
\usepackage{multirow}
\usepackage{slashed}
\usepackage{float}
\usepackage{esint}
\usepackage{epsfig}
\usepackage{lscape}
\usepackage{amsmath,latexsym,amssymb}
\usepackage{graphicx}
\usepackage{psfrag}
\usepackage[latin1]{inputenc}
\usepackage[T1]{fontenc}
\usepackage{subfigure}
\usepackage{nicefrac}
\usepackage[stable]{footmisc}
\usepackage{rotating}
\usepackage{afterpage}
\usepackage{bbm}

\def\NO{\nonumber}

\newcommand{\be}{\begin{equation}}
\newcommand{\ee}{\end{equation}}

\def\bea{\begin{eqnarray}}
\def\eea{\end{eqnarray}}

\def\beqx{\begin{displaymath}}
\def\eeqx{\end{displaymath}}

\newcommand{\bmat}{\left(\begin{array}}
\newcommand{\emat}{\end{array}\right)}

\newcommand{\ca}{{\cal A}}
\newcommand{\cd}{{\cal D}}
\newcommand{\ce}{{\cal E}}
\newcommand{\ck}{{\cal K}}

\newcommand{\cm}{{\cal M}}
\newcommand{\cl}{{\cal L}}
\newcommand{\cf}{{\cal F}}
\newcommand{\cg}{{\cal G}}

\newcommand{\co}{{\cal O}}
\newcommand{\cp}{{\cal P}}
\newcommand{\cs}{{\cal S}}
\newcommand{\cq}{{\cal Q}}
\newcommand{\ct}{{\cal T}}

\newtheorem{lemma}{Lemma}[section]

\def\a{\alpha}
\def\b{\beta}

\def\d{\delta}
\def\e{\epsilon}
\def\f{\phi}
\def\g{\gamma}

\def\j{\psi}
\def\k{\kappa}
\def\l{\lambda}
\def\m{\mu}
\def\n{\nu}

    \def\om{\omega}
\def\p{\pi}

    \def\th{\theta}
\def\r{\rho}
\def\s{\sigma}

\def\x{\xi}
\def\z{\zeta}
\def\D{\Delta}
\def\F{\Phi}

\def\L{\Lambda}

    \def\Om{\Omega}
\def\P{\Pi}

\def\S{\Sigma}

\def\ve{\varepsilon}
\def\vr{\varrho}

\def\vth{\vartheta}

\def\vf{\varphi}

\def\bs{\bbalpha}
\def\bs{\bbbeta}
\def\bs{\bbgamma}
\def\bs{\bbdelta}
\def\bs{\bbepsilon}
\def\bs{\bbeta}
\def\bs{\bbsigma}

\def\ca{{\cal A}}
\def\cb{{\cal B}}
\def\cc{{\cal C}}
\def\cd{{\cal D}}
\def\ce{{\cal E}}
\def\cf{{\cal F}}
\def\cg{{\cal G}}
\def\ch{{\cal H}}

\def\ck{{\cal K}}
\def\cl{{\cal L}}
\def\cm{{\cal M}}

\def\co{{\cal O}}
\def\cp{{\cal P}}
\def\cq{{\cal Q}}
\def\car{{\cal R}}
\def\cs{{\cal S}}
\def\ct{{\cal T}}

\def\mf#1{\ensuremath{\mathfrak{#1}}} 
\def\bb#1{\ensuremath{\mathbb{#1}}} 

\def\bo{{\raise-.3ex\hbox{\large$\Box$}}}               
\def\pa{\partial}                                       
\def\face{{\raise.2ex\hbox{$\displaystyle \bigodot$}\mskip-2.2mu \llap {$\ddot
        \smile$}}}                                   
\def\>{\rangle}                                      
\def\<{\langle}                                      

\newcommand{\sub}[1]{{}_{(#1)}{}}    					 
\def\wt#1{\widetilde{#1}}                            
\def\Hat#1{\widehat{#1}}                             
\def\lbar#1{\ensuremath{\overline{#1}}}              
\def\abs#1{\left| #1\right|}                         
\def\leftrightarrowfill{$\mathsurround=0pt \mathord\leftarrow \mkern-6mu
        \cleaders\hbox{$\mkern-2mu \mathord- \mkern-2mu$}\hfill
        \mkern-6mu \mathord\rightarrow$}        
\def\dvec#1{\vbox{\ialign{##\crcr
        \leftrightarrowfill\crcr\noalign{\kern-1pt\nointerlineskip}
        $\hfil\displaystyle{#1}\hfil$\crcr}}}           

\def\-{\hphantom{-}}

\title{Lifshitz holography: The whole shebang}
\author[a]{Wissam Chemissany} 
\author[b]{and Ioannis Papadimitriou}

\affiliation[a]{Department of Physics and SITP, Stanford University, Stanford, California 94305 USA}
\affiliation[b]{Instituto de F\'isica Te\'orica UAM/CSIC, Universidad Aut\'onoma de Madrid, Madrid 28049, Spain}

\emailAdd{wissam@stanford.edu} 
\emailAdd{ioannis.papadimitriou@csic.es}

\abstract{We provide a general algorithm for constructing the holographic dictionary for any asymptotically locally Lifshitz background, with or without hyperscaling violation, and for any values of the dynamical exponents $z$ and $\th$, as well as the vector hyperscaling violating exponent \cite{Gouteraux:2012yr}, that are compatible with the null energy condition. The analysis is carried out for a very general bottom up model of gravity coupled to a massive vector field and a dilaton with arbitrary scalar couplings. The solution of the radial Hamilton-Jacobi equation is obtained recursively in the form of a graded expansion in eigenfunctions of two commuting operators \cite{Chemissany:2014xpa}, which are the appropriate generalization of the dilatation operator for non scale invariant and Lorentz violating boundary conditions. The Fefferman-Graham expansions, the sources and 1-point functions of the dual operators, the Ward identities, as well as the local counterterms required for holographic renormalization all follow from this asymptotic solution of the radial Hamilton-Jacobi equation. We also find a family of exact backgrounds with $z>1$ and $\th>0$ corresponding to a marginal deformation shifting the vector hyperscaling violating parameter and we present an example where the conformal anomaly contains the only $z=2$ conformal invariant in $d=2$ with four spatial derivatives.    }


\preprint{IFT UAM/CSIC-14-073}

\begin{document}  \maketitle

\section{Introduction}
\label{intro}

The use of holographic techniques in order to gain insight into the strongly coupled dynamics of condensed matter systems has attracted considerable interest in the last few years. Gravity duals to quantum critical points exhibiting Lifshitz \cite{Koroteev:2007yp,Kachru:2008yh,Taylor:2008tg} or Schr\"{o}dinger \cite{Balasubramanian:2008dm,Son:2008ye} symmetry have been put forward and studied extensively. More recently, scaling geometries where translations in the radial coordinate is not an isometry but only a conformal isometry have been proposed as gravity duals to non-relativistic systems exhibiting hyperscaling violation \cite{Charmousis:2010zz,Huijse:2011ef,Iizuka:2011hg,Ogawa:2011bz,Shaghoulian:2011aa,Dong:2012se,Iizuka:2012iv,Gouteraux:2012yr}. Hyperscaling violating Lifshitz (hvLf) geometries are characterized by two dynamical exponents, the Lorentz violating exponent $z$ and the hyperscaling violating parameter $\th$, and take the form   
\be\label{thetaz}
ds_{d+2}^2 =\ell^2  u^{-\frac{2(d-\th)}{d}} \left( - u^{-2(z-1)} dt^2 + du^2 + dx_a^2 \right),
\ee
where $d$ is the number of spatial dimensions, $a=1,\ldots, d$, and $\ell$ is the Lifshitz radius. This metric is invariant under time and spatial translations, as well as spatial rotations, but under the anisotropic scaling transformation 
\be
\label{scaletrans} 
x_{a} \to \l x_{a},\quad 
t \to \l^{z} t,\quad
u \to \l u,
\ee
it transforms homogeneously according to 
\be
ds_{d+2}^{2} \to \l^{2\th/d} ds_{d+2}^{2}.
\ee
Hence, (\ref{scaletrans}) is only a {\em conformal} isometry of (\ref{thetaz}) unless $\th=0$, which corresponds to the scale invariant Lifshitz (Lif) geometry. For $z=1$ the metric (\ref{thetaz}) coincides with the (non-compact part of the) near horizon geometry of relativistic D$p$ branes \cite{Itzhaki:1998dd,Boonstra:1998mp,Gherghetta:2001iv,Kanitscheider:2008kd,Wiseman:2008qa}, with the hyperscaling violating exponent $\th$ given by 
\be\label{theta-p}
\th=\frac{(p-3)^2}{p-5},\quad d=p.
\ee    
This special case not only provides insight into the physics described by hyperscaling violating backgrounds, but also is an important guide in developing the holographic dictionary for such backgrounds.   

As for D$p$ branes, the holographic relation between the energy scale of the dual field theory and the radial coordinate $u$ can be unambiguously identified through a supergravity probe calculation \cite{Peet:1998wn,Dong2012}. This determines that the ultraviolet (UV) of the dual theory is located at $u=0$, independently of the value of $\theta$, in agreement with the relativistic case $z=1$ \cite{Itzhaki:1998dd,Boonstra:1998mp,Gherghetta:2001iv,Kanitscheider:2008kd,Wiseman:2008qa}. It follows that the proper identification of the boundary of the geometry (\ref{thetaz}) through a conformal compactification requires a Weyl transformation to the ``dual frame'' \cite{Duff:1994fg,Kanitscheider:2008kd}, where the metric becomes Lifshitz, thus providing an unambiguous definition of the boundary. In the conformal case, $\th=0$, such a potential ambiguity does not arise since no field redefinition (including Weyl frame transformations) change the asymptotic behavior of the metric. Given that the curvature invariants scale with $u$ as
\be
R\propto u^{-\frac{2\th}{d}},\quad R_{\m\n}R^{\m\n}\propto R_{\m\n\r\s}R^{\m\n\r\s}\propto u^{-\frac{4\th}{d}},
\ee
one might be tempted to conclude that e.g. for $\th>0$ there is a curvature singularity as we approach the UV at $u=0$. However, given that geometries of the form (\ref{thetaz}) with $\th\neq 0$ generically require the presence of a linear dilaton that tends to $\pm\infty$ as $u\to 0$, such statements are not well defined since we can tune the curvature singularity at will by changing Weyl frame. In particular, in the dual frame the curvature singularity is completely absorbed in the dilaton. Since this is the proper holographic frame in the case $\th\neq 0$, there are no restrictions on $\th$ imposed by requiring absence of curvature singularities in the UV. In the IR one can apply the criterion of  \cite{Gubser:2000nd}, which again provides an unambiguous statement about curvature singularities in the presence of scalars. 

Restrictions on $\th$ and $z$ do arise, however, from the null energy condition (NEC)
\be
T_{\mu\nu}k^\mu k^\nu\geq0,
\ee
where $k^\m$ is an arbitrary null vector field, i.e. $k^\mu k_\mu = 0$. The NEC leads to the two constraints
\be
(d-\theta)(d(z-1)-\theta) \geq 0,\quad 
(d-\theta+z)(z-1) \geq 0.
\ee
Including the relativistic case, $z=1$, the solutions of these constraints are:
\begin{align}
	\label{ranges}
	\begin{tabular}{|c|cc|}
		\hline
		I  & $z\leq 0$        & $\th \geq d$ \\\hline
		II & $0 <z \leq 1$ & $\th\geq d+z$\\\hline
		IIIa & \multirow{2}{*}{$1\leq z\leq 2$} & $\th \leq d(z-1)$ \\
		IIIb &          & $d\leq \th\leq d+z$\\ \hline 
		IVa & \multirow{2}{*}{$2<z\leq \frac{2d}{d-1}$} & $\th\leq d$ \\
		IVb &         & $d(z-1)\leq\th\leq d+z$ \\\hline
		V & $z>\frac{2d}{d-1}$ & $\th\leq d$ \\\hline
	\end{tabular}
\end{align}
For $\th=0$ all cases except I and II admit solutions, which leads to the condition $z\geq 1$. A comparison with the relativistic case is instructive. 
From (\ref{theta-p}) follows that for $p\leq 4$ we have $\th\leq 0$, corresponding to case IIIa. For $p=5$ (\ref{theta-p}) is ill defined but it can be understood as the limit $\th\to -\infty$ or $\th\to+\infty$, corresponding respectively to cases IIIa and II. Finally, $p=6$ gives $\th=9> d+z=7$ and so it belongs to case II. It is well known that there are no well defined Fefferman-Graham asymptotic expansions in the case of D$6$ branes \cite{Kanitscheider:2008kd}, which reflects the fact that there is no decoupling limit \cite{Itzhaki:1998dd}. A general criterion for the existence of well defined asymptotic expansions is the volume divergence of the on-shell action. For the metric (\ref{thetaz}) in the Einstein frame we get
\be
S\sim \int \frac{d u}{u^{d+z+1-\th}}, 
\ee    
which diverges as $u\to 0$ provided
\be
\th\leq d+z.
\ee
This criterion is independent of the choice of Weyl frame. It follows that all cases except I and II admit well defined asymptotic expansions. Asymptotic expansions, therefore, exist for $z>1$, but not for $z<1$, and so we will mostly focus on the case $z>1$ in the following. 

For an extensive list of references on non-relativistic  backgrounds, their hyperscaling violating versions and possible string theory embeddings we refer the reader to the following recent papers and references therein \cite{Gath:2012pg,Gouteraux:2012yr,Bueno:2012vx,Narayan:2012wn}. The body of literature most relevant to us here, however, concerns earlier work on holographic renormalization and the holographic dictionary for asymptotically Lifshitz spacetimes \cite{Ross:2009ar,Horava:2009vy,Ross:2011gu,Mann:2011hg,Baggio:2011cp,Baggio:2011ha,Griffin:2011xs,Griffin:2012qx}.
These papers focus mainly on the Einstein-Proca theory, i.e. 
gravity coupled to a massive vector field, mostly without any scalars and only with conformal (Lifshitz) boundary conditions. Moreover, the emphasis is often on the physically interesting but rather special case $d=z=2$. Our aim here is to extend these analyses to the case of general hvLf boundary conditions.   

Besides the aforementioned studies on the first principles construction of the holographic dictionary for asymptotically Lifshitz backgrounds of the Einstein-Proca theory, there are few examples where the non-relativistic dictionary has been inferred from a related relativistic dictionary for asymptotically AdS backgrounds. In \cite{Chemissany:2011mb} a 
4-dimensional model that admits $z=2$ Lifshitz backgrounds was obtained by a dimensional reduction of an axion-dilaton system in 5 dimensions that can be embedded in Type IIB supergravity. In particular, the $z=2$ Lifshitz backgrounds are obtained from the reduction of 5-dimensional Schr\"odinger solutions of the axion-dilaton theory with $z=0$, which are asymptotically AdS$_5$. This connection was utilized in \cite{Chemissany:2012du} in order to deduce the holographic dictionary for the Lifshitz backgrounds from the dictionary for asymptotically locally AdS solutions of the axion-dilaton theory developed in \cite{Papadimitriou:2011qb}. The same model was revisited in \cite{Christensen:2013lma,Christensen:2013rfa} using the vielbein formalism and a connection between the structure of the sources and Newton-Cartan geometry on the boundary was proposed. Another way to relate the Lifshitz and AdS boundary conditions is a scaling limit where $z\to\infty$. The resulting asymptotic geometry is AdS$_{2} \times \bb R^{d-1}$. This limit, however, is not very useful in practice because the holographic dictionary for the limiting spacetime is not fully understood -- due to the non-compact $\bb R^{d-1}$ directions and the well-known subtleties associated with AdS$_2$ holography. Finally, one can study Lifshitz backgrounds with dynamical exponent infinitesimally close to the relativistic value, i.e. $z = 1+\e,$ where $\e$ is small  
\cite{Korovin:2013bua,Korovin:2013nha}. This corresponds to deforming the relativistic CFT with an irrelevant operator and so the analysis must be done with a UV cut-off. 
 
The main goal of the present paper is a systematic derivation of the holographic dictionary for general asymptotically Lif and hvLf backgrounds, for generic values of the dynamical exponents $z$ and $\th$. In particular, the aim here is not a detailed discussion of the physics of a specific model, but rather the construction of a general algorithm from which the physics can be systematically extracted for {\em any} model that admits Lif and hvLf backgrounds. Moreover, throughout this paper we adopt the point of view that the field theory exhibiting Lifshitz or hyperscaling violating Lifshitz symmetry is at the UV -- {\em not } in the IR -- since the physics of Lif or hvLf geometries in the IR can be simply extracted by studying the corresponding UV theory. The IR physics of a geometry which, for example, starts as AdS in the UV and runs to hvLf in the IR (or at some intermediate energy scale) can be studied using standard well known tools for asymptotically locally AdS holography. There is no need for new machinery in that case. Here we are therefore concerned exclusively with backgrounds which are asymptotically locally Lif or hvLf {\em in the UV}. For $\th>d+z$ such backgrounds will generically require a different UV completion, but we will not be concerned with this case here.         

Our algorithm for constructing the holographic dictionary hinges upon a certain asymptotic solution of the radial Hamilton-Jacobi (HJ) equation \cite{Boer:1999xf,Martelli:2002sp, Papadimitriou:2004ap,Papadimitriou:2010as}, subject to asymptotically Lif or hvLf boundary conditions. This asymptotic solution of the radial HJ equation not only provides the necessary local boundary counterterms to render the on-shell action finite, but also is required in order for the variational problem to be well defined both for asymptotically locally AdS \cite{Papadimitriou:2005ii} and asymptotically non AdS \cite{Papadimitriou:2010as} backgrounds. Moreover, the procedure of holographic renormalization based on such an asymptotic solution of the HJ equation is completely equivalent to the traditional method based on asymptotic solutions of the equations of motion \cite{Henningson:1998gx,deHaro:2000xn,Skenderis:2002wp}. However, there are two crucial differences between our use of the radial HJ equation and the way it is used in most of the literature. Firstly, we do not need to make an ansatz for the solution of the HJ solution. Finding the correct ansatz becomes increasingly difficult in the presence of matter fields and especially when non AdS boundary conditions are imposed. Moreover, the number of equations obtained by inserting an ansatz into the HJ equation is in general greater than the number of unknown parameters of the ansatz and so the system is overdetermined. Instead, the way we solve the HJ equation is by setting up a recursion procedure based on the covariant expansion of the HJ solution in eigenfunctions of a suitable operator.  For scale invariant boundary conditions this operator is usually the relativistic \cite{Papadimitriou:2004ap} or non-relativistic \cite{Ross:2011gu,Griffin:2011xs} dilatation operator. For more general boundary conditions, such as non-conformal branes or hvLf backgrounds, a generalized dilatation operator is required, such as the one discussed in \cite{Papadimitriou:2011qb} for relativistic non scale invariant boundary conditions. One of the main results of the present paper is the identification of a suitable set of commuting operators that lead to a recursive solution of the HJ equation with Lif or hvLf boundary conditions \cite{Chemissany:2014xpa}. A second point where our approach differs from other approaches to the holographic dictionary is that at no point do we use the general second order equations of motion. In particular, the asymptotic Fefferman-Graham expansions are obtained by integrating the first order flow equations corresponding to the asymptotic solution of the HJ equation. In this way there is no need for making an ansatz for the asymptotic solutions of the equations of motion -- the asymptotic form is determined algorithmically by integrating order by order the flow equations. This is particularly useful in the case of non AdS boundary conditions where the form of the asymptotic expansions is a priori unknown and may even contain multiple scales \cite{Papadimitriou:2011qb}.   

The paper is organized as follows. In Section \ref{model} we present a general bottom up model that admits both
Lif and hvLf backgrounds and we formulate its dynamics in the radial Hamiltonian formalism, which we use later in order to develop the  holographic dictionary. Section \ref{bgrnds} concerns exclusively homogeneous but anisotropic background solutions of the model presented in Section \ref{model}. Both Lif and hvLf backgrounds are discussed in detail and the holographic dictionary for the minisuperspace of homogeneous asymptotically Lif and hvLf backgrounds is obtained. This serves as a self contained warm up for the derivation of the general dictionary for asymptotically {\em locally} Lif and hvLf backgrounds that will follow, but also it provides a general description of anisotropic holographic renormalization group (RG) flows. In Section \ref{algorithm} we discuss the boundary conditions corresponding to asymptotically locally Lif and hvLf backgrounds and we present a general algorithm for solving the radial HJ equation iteratively for such backgrounds. This is achieved by covariantly expanding the solution of the HJ equation in simultaneous eigenfunctions of two commuting operators, which as we show are the appropriate generalization of the dilatation operator for anisotropic and non scale invariant boundary conditions. The full holographic dictionary, i.e. the Fefferman-Graham asymptotic expansions, the identification of the sources and 1-point functions of the dual operators, the holographic Ward identities and the conformal anomalies, as well as the covariant boundary counterterms that render the on-shell action finite all follow directly from general asymptotic solution of the HJ equation as is discussed in Section \ref{ward}. Finally, a number of examples are worked out in Section \ref{examples}, and a few technical results are presented in the appendices.

\section{The model and radial Hamiltonian formalism}
\label{model}

The minimal field content that supports Lifshitz solutions is a massive vector field, or a massless vector field and a scalar, coupled to Einstein-Hilbert gravity. A more general model that includes both these cases and  supports in addition hyperscaling violating solutions is the action introduced in \cite{Gouteraux:2012yr}, namely
\be\label{action-0}
S=\frac{1}{2\k^2}\int_\cm d^{d+2}x\sqrt{-g}\left(R[g]-\a \pa_\m\f\pa^\m\f-Z(\f)F^2-W(\phi)A^{2}-V(\f)\right)+S_{GH},
\ee
where $\k^2=8\p G_{d+2}$ is the gravitational constant in $d+2$ dimensions and $S_{GH}$ is the Gibbons-Hawking term 
\be
S_{GH}=\frac{1}{2\k^2}\int_{\pa\cm} d^{d+1}x\sqrt{-\g}2K.
\ee
The functions $Z(\f)$, $W(\f)$ and $V(\f)$ are arbitrary, subject only to the condition that the equations of motion admit the desired asymptotic solutions. We will derive these conditions in detail in the subsequent analysis. Moreover, the parameter $\a>0$  
can be removed by a rescaling of the scalar field, but we keep it to facilitate direct comparison with the existing literature, where different conventions are used. Finally, we do not include Chern-Simons terms  here in order to keep the spacetime dimension arbitrary throughout most of our analysis. Such terms can be  incorporated in the analysis though, once a choice of spacetime dimension has been made.

We want to generalize the action (\ref{action-0}) in two crucial ways, however. Firstly, in order to consistently describe this theory in a Hamiltonian language we need to maintain the $U(1)$ gauge invariance in the presence of a mass term for the vector field. This can be done straightforwardly by introducing a St\"uckelberg field $\om$ and replacing 
\be
A_\m\to B_\m=A_\m-\pa_\m\om,
\ee
so that the $U(1)$ gauge transformation 
\be
A_\m\to A_\m+\pa_\m\L,\quad \om\to \om+\L,
\ee
leaves $B_\m$ invariant. As it turns out, the preservation of the $U(1)$ gauge invariance has important implications for the holographic dictionary.

Secondly, in order to be able to develop the holographic dictionary for asymptotically Lifshitz and hyperscaling violating Lifshitz backgrounds simultaneously, it is necessary to go to a generic Weyl frame by means of the 
Weyl transformation
\be
g\to e^{2\x\f}g,
\ee
of the action (\ref{action-0}), with $\x$ an arbitrary parameter. As we shall see later, $\x$ is related to the hyperscaling violation exponent $\th$ in the Einstein frame. With these generalizations, the model we will study is
defined by the action  
\be\label{action}
S_\x=\frac{1}{2\k^2}\int_\cm d^{d+2}x\sqrt{-g}e^{d\x \f}\left(R[g]
-\a_\x \pa_\m\f\pa^\m\f-Z_\x(\f)F^2-W_\x(\phi)B^{2}-V_\x(\f)\right)
+S^\x_{GH},
\ee 
where
\be
\a_\x=\a-d(d+1)\x^2,\quad
Z_\x(\f)=e^{-2\x\f}Z(\f),\quad
W_\x(\f)=W(\f),\quad
V_\x(\f)=e^{2\x\f}V(\f),
\ee
and the Gibbons-Hawking term now takes the form
\be
S_{GH}^\x=\frac{1}{2\k^2}\int_{\pa\cm} d^{d+1}x\sqrt{-\g}2e^{d\x\f}K.
\ee
The equations of motion following from this action are
\begin{align}
\label{eoms}
\begin{aligned}
& R_{\m\n}-\frac12 R g_{\m\n}=d\x\nabla_\m\nabla_\n\f+\left(d^2\x^2+\a_\x\right)\pa_\m\f\pa_\n\f
+2Z_\x(\f)F_{\m\r}F_\n{}^\r+ W_\x(\phi) B_{\mu}B_{\nu}\\
&\phantom{more}-g_{\m\n}\left(\left(d^2\x^2+\frac{\a_\x}{2}\right)\pa_\r\f\pa^\r\f+d\x\square_g\f
+\frac12 Z_\x(\f)F^2+\frac{1}{2}W_\x(\phi) B^{2}+\frac12 V_\x(\f)\right),\\
&\nabla_\m\left(e^{d\x\f}Z_\x(\f)F^\m{}_\n\right)=\frac{1}{2}e^{d\x\f}W_\x(\phi) B_{\nu},\\
&\nabla_\n\left(e^{d\x\f}W_\x(\f)B^\n\right)=0,\\
&2\a\square_g\f+2d\x\a\pa_\m\f\pa^\m\f-e^{-2\x\f}Z'(\f)F^2-W'(\f)B^2-e^{2\x\f}V'(\f)=0.
\end{aligned}
\end{align}
We will not need these equations in the subsequent analysis, except for demonstrating that the first order equations we will derive for background homogeneous solutions solve these equations.

\begin{flushleft}
{\bf Radial Hamiltonian formalism}
\end{flushleft}

The starting point for the derivation of the holographic dictionary for the action (\ref{action}) is a radial Hamiltonian description of the dynamics, where the radial coordinate is interpreted as the Hamiltonian `time'. We start by the standard ADM decomposition of the metric \cite{ADM} as 
\be\label{ADM-metric}
ds^2=(N^2+N_iN^i)dr^2+2N_idr dx^i+\g_{ij}(r,x)dx^idx^j,
\ee
where $N$ and $N_i$ are respectively the shift and lapse functions, and $\g_{ij}$ is the induced metric on the radial slices $\S_r$. In terms of these variables the action (\ref{action}) can be written as a radial integral over the Lagrangian
\bea\label{lagrangian}
L&=&\frac{1}{2\k^2}\int d^{d+1}x\sqrt{-\g}Ne^{d\x\f}\left(\left(1+\frac{d^2\x^2}{\a_\x}\right) K^2-K^{ij}K_{ij} -\frac{\a_\x}{N^2}\left(\dot\f-N^i\pa_i\f-\frac{d\x}{\a_\x}N K\right)^2
\right.\NO\\
&&\left.-\frac{2}{N^2}Z_\x(\f)(F_{ri}-N^kF_{ki})(F_{r}{}^i-N^lF_{l}{}^i)
-\frac{1}{N^2}W_\x(\f)\left(A_r-N^iA_i-\dot\om+N^i\pa_i\om\right)^2\right.\NO\\
&&\left.+R[\g]-\a_\x\pa_i\f\pa^i\f-Z_\x(\f)F_{ij}F^{ij}-W_\x(\f)B_iB^i-V_\x(\f)\rule{0cm}{0.6cm}\right),
\eea
where the extrinsic curvature $K_{ij}$ is given by
\be
K_{ij}=\frac{1}{2N}\left(\dot\g_{ij}-D_iN_j-D_jN_i\right),
\ee
and $D_i$ denotes the covariant derivative with respect to the induced metric $\g_{ij}$. Moreover, we will use the notation $K=\g^{ij}K_{ij}$ to denote the trace of the extrinsic curvature. Since no radial derivatives of $N$, $N_i$ or $A_r$ appear in this Lagrangian, the corresponding canonical momenta vanish identically and these fields play 
the role of Lagrange multipliers, imposing the usual first class constraints which we will derive shortly. The canonical momenta for the rest of the fields are 
\bea\label{momenta}
&&\p^{ij}=\frac{\d L}{\d\dot\g_{ij}}=\frac{1}{2\k^2}\sqrt{-\g}e^{d\x\f}\left(
K\g^{ij}-K^{ij}+\frac{d\x}{N}\g^{ij}\left(\dot\f-N^k\pa_k\f\right)\right),\NO\\
&&\p^{i}=\frac{\d L}{\d\dot A_{i}}=-\frac{1}{2\k^2}\sqrt{-\g}e^{d\x\f}Z_\x(\f)
\frac{4}{N}\g^{ij}\left(F_{rj}-N^kF_{kj}\right),\NO\\
&&\p_\f=\frac{\d L}{\d\dot\f}=\frac{1}{2\k^2}\sqrt{-\g}e^{d\x\f}
\left(2d\x K-\frac{2\a_\x}{N}(\dot\f-N^i\pa_i\f)\right),\NO\\
&&\p_\om=\frac{\d L}{\d\dot\om}=-\frac{1}{2\k^2}\sqrt{-\g}e^{d\x\f}W_\x(\f)\frac2N
\left(\dot\om-N ^i\pa_i\om-A_r+N^iA_i\right).
\eea
These relations can be  inverted to obtain the generalized velocities in terms of the canonical momenta 
\bea
&&\dot\g_{ij}=-\frac{4\k^2}{\sqrt{-\g}}e^{-d\x\f}N\left(\p_{ij}-\frac{\a_\x+d^2\x^2}{d\a}\p\g_{ij}-\frac{\x}{2\a}\p_\f\g_{ij}\right)+D_iN_j+D_jN_i,\NO\\
&&\dot A_i=-\frac{\k^2}{2}\frac{1}{\sqrt{-\g}}e^{-d\x\f}Z_\x^{-1}(\f)N\p_i+\pa_iA_r+N^kF_{ki},\NO\\
&&\dot\f=-\frac{\k^2}{\a}\frac{1}{\sqrt{-\g}}e^{-d\x\f}N(\p_\f-2\x\p)+N^i\pa_i\f,\NO\\
&&\dot\om=-\frac{\k^2}{\sqrt{-\g}}e^{-d\x\f}W^{-1}_\x(\f)N\p_\om
+N^i\pa_i\om+A_r-N^iA_i.\label{inverted-momenta}
\eea
The Hamiltonian is then  obtained as the Legendre transform of the Lagrangian, namely 
\be
H=\int d^{d+1}x\left(\dot\g_{ij}\p^{ij}+\dot A_i\p^i+\dot\f\p_\f+\dot\om\p_\om\right)-L
=\int d^{d+1}x\left(N\ch+N_i\ch^i+A_r\cf\right),
\ee
where the local densities $\ch$, $\ch^i$ and $\cf$ are given by
\begin{align}
\label{constraints}
\boxed{
\begin{aligned}
&\ch=-\frac{\k^2}{\sqrt{-\g}}e^{-d\x\f}\left(2\p^{ij}\p_{ij}-\frac2d\p^2+\frac{1}{2\a}\left(\p_\f-2\x\p\right)^2
+\frac14Z^{-1}_\x(\f)\p^i\p_i+\frac12W^{-1}_\x(\f)\p_\om^2\right)\\
&\rule{0.8cm}{0cm}+\frac{\sqrt{-\g}}{2\k^2}e^{d\x\f}\left(-R[\g]+\a_\x\pa^i\f\pa_i\f+Z_\x(\f)F^{ij}F_{ij}+W_\x(\f)B^iB_i+V_\x(\f)\right),\\
&\ch^i=-2D_j\p^{ji}+F^i{}_j\p^j+\p_\f\pa^i\f-B^i\p_\om,\\
& \cf=-D_i\p^i+\p_\om.
\end{aligned}}
\end{align}
These three quantities appear in the Hamiltonian as coefficients of the three Lagrange multipliers $N$, $N_i$, and $A_r$ respectively, and so the corresponding Hamilton equations yield the three constraints 
\be\label{constraints0}
\ch=\ch^i=\cf=0.
\ee
These first class constraints reflect the full diffeomorphism and $U(1)$ gauge invariance of the action (\ref{action}). In particular, this would not have been the case had we not used the St\"uckelberg mechanism to preserve the $U(1)$ symmetry in the presence of a mass for the vector field. This plays a critical role in our construction of the holographic dictionary. 

The constraints (\ref{constraints0}) are the basis of the radial Hamilton-Jacobi formulation of the model (\ref{action}). The key new ingredient provided by the Hamilton-Jacobi formalism is the alternative expression for the canonical momenta as gradients of a functional $\cs[\g,A,\f,\om]$ of the induced fields, namely 
\be\label{HJ-momenta}
\p^{ij}=\frac{\d \cs}{\d\g_{ij}},\quad \p^i=\frac{\d\cs}{\d A_i},\quad \p_\f=\frac{\d\cs}{\d\f},\quad \p_\om=\frac{\d\cs}{\d\om}.
\ee
Inserting these expressions for the momenta in the constraints (\ref{constraints}) leads to a set of functional partial differential equations for $\cs[\g,A,\f,\om]$, which is often known as Hamilton's principal function. A fundamental property of the Hamilton-Jacobi approach to the dynamical problem is that the Hamilton-Jacobi equations, i.e. the constraints (\ref{constraints0}), together with the relations (\ref{HJ-momenta}) expressing the momenta as gradients of a `potential' $\cs[\g,A,\f,\om]$, provide a full description of the dynamics. In particular, there is no need to consider the second order equations of motion (\ref{eoms}). By constructing suitable solutions of the Hamilton-Jacobi equations, therefore, we can provide a complete description of the classical dynamical problem, and hence of the holographic dictionary. 

Our main objective in the subsequent analysis will therefore be to develop a systematic algorithm for solving the Hamilton-Jacobi equations (\ref{constraints0}), subject to the desired boundary conditions. In fact, we only need to focus on the Hamiltonian constraint $\ch=0$, as the other two can be  satisfied by construction. In particular, the momentum constraint $\ch^i=0$ simply requires the functional $\cs$ to be invariant with respect to diffeomorphisms on the radial slices $\S_r$, while the constraint $\cf=0$ imposes $U(1)$ invariance, i.e. it simply requires that $\cs$ depends on $A_i$ and $\om$ only through the gauge-invariant filed $B_i$. Provided then we look for ${\rm Diff}_{\S_r}$--invariant solutions $\cs[\g,B,\f]$, the only equation we need to solve is the Hamiltonian constraint $\ch=0$. Of course, the other two constraints will also play a crucial role in the construction of the holographic dictionary, giving rise to certain Ward identities.   

Given a solution $\cs[\g,B,\f]$ of the Hamilton-Jacobi equations, the radial trajectories of the induced fields 
can be obtained by integrating the first order equations 
(\ref{inverted-momenta}), where the canonical momenta are expressed as gradients of the given solution of the Hamilton-Jacobi equations as in (\ref{HJ-momenta}). With the gauge choice 
\be\label{gauge-fixing}
N=1,\quad N_i=0,\quad A_r=0,
\ee
which we will adopt from now on, these first order equations take the form
\begin{align}
\label{flow-eqs}
\boxed{
\begin{aligned}
&\dot\g_{ij}=-\frac{4\k^2}{\sqrt{-\g}}e^{-d\x\f}\left(\left(\g_{ik}\g_{jl}-\frac{\a_\x+d^2\x^2}{d\a}\g_{ij}\g_{kl}\right)\frac{\d}{\d\g_{kl}}-\frac{\x}{2\a}\g_{ij}\frac{\d}{\d\f}\right)\cs,\\
&\dot A_i=-\frac{\k^2}{2}\frac{1}{\sqrt{-\g}}e^{-d\x\f}Z_\x^{-1}(\f)
\g_{ij}\frac{\d}{\d A_j}\cs,\\
&\dot\f=-\frac{\k^2}{\a}\frac{1}{\sqrt{-\g}}e^{-d\x\f}
\left(\frac{\d}{\d\f}-2\x\g_{ij}\frac{\d}{\d\g_{ij}}\right)\cs,\\
&\dot\om=-\frac{\k^2}{\sqrt{-\g}}e^{-d\x\f}W^{-1}_\x(\f)\frac{\d}{\d\om}\cs.
\end{aligned}}
\end{align}
We will use these first order equations in two different but complementary ways. Firstly, making an ansatz for a class of background solutions, these first order equations become analogous to first order BPS equations, while Hamilton's principal function $\cs$ plays the role of a fake superpotential \cite{Freedman:2003ax}. We will discuss this in detail in Section \ref{bgrnds}. 

The second major application of these equations will be to obtain the asymptotic Fefferman-Graham expansions of the fields, and as a result the holographic dictionary, from the general asymptotic solution of the Hamilton-Jacobi equation subject to specified boundary conditions. The systematic construction of this general asymptotic solution of the Hamilton-Jacobi equation is the subject of Section \ref{algorithm}. As we shall see, the general asymptotic solution contains a number of undetermined integration functions. In the Hamilton-Jacobi language these are the `initial' momenta contained in a complete integral of the Hamilton-Jacobi equation, while in the holographic context they correspond to the renormalized momenta. Via the flow equations (\ref{flow-eqs}) these undetermined functions give rise to the normalizable modes in the Fefferman-Graham expansions of the fields. The non-normalizable modes, on the other hand, appear as the integration functions of the first order flow equations themselves. The Hamilton-Jacobi formalism, therefore, provides a natural qualitative division of the asymptotic data into two classes, data arising from the integration of the Hamilton-Jacobi equation and data arising from the integration of the first order flow equations. This division in most cases coincides with the separation of the asymptotic data into sources and 1-point functions in the holographic context, but there are exceptions to this rule. An obvious exception is the case of scalars or vector fields with two normalizable modes. More generally, the symplectic form on the space of asymptotic solutions, parameterized by the modes arising from the integration of the Hamilton-Jacobi equation and the first order flow equations, will not be diagonal. The way to identify the sources and 1-point functions out of these asymptotic data in such cases is to diagonalize the symplectic form \cite{Papadimitriou:2010as}.

\section{Holography for homogeneous anisotropic backgrounds}
\label{bgrnds}

As a prelude to the general analysis of asymptotically locally Lif and hvLf backgrounds, and in order to outline several of the key steps of our method, it is very instructive to start by discussing the Hamiltonian formalism and the holographic dictionary within the minisuperspace of homogeneous, yet anisotropic, background solutions of the equations of motion.

In particular, in this section we will consider solutions described by the ansatz
\be\label{bkgrd-ansatz}
ds^2=dr^2-e^{2f(r)}dt^2+e^{2h(r)}\d_{ab}dx^adx^b,\quad A=a(r)dt,\quad \f=\f(r),\quad \om=\om(r), 
\ee
where $a,b=1,\ldots,d$. Inserting this ansatz in the equations of motion (\ref{eoms}) gives the set of equations
\bea
&&2d\dot f\dot h+d(d-1)\dot h^2=\a_\x \dot\f^2-2d\x (\dot f+d\dot h)\dot\f-2Z_\x e^{-2f}\dot a^2+e^{-2 f}W_{\xi}a^2-V_\x,\NO\\
&&\ddot f+(\dot f+d \dot h) \dot f=-d\x \dot f\dot\f+\frac{2(d-1)}{d}Z_\x e^{-2f}\dot a^2+e^{-2 f}W_{\xi}a^2-\frac1d V_\x-\x\left(\ddot\f+(\dot f+d\dot h)\dot\f+d\x\dot\f^2\right),\NO\\
&&\ddot h+(\dot f+d \dot h) \dot h=-d\x \dot h\dot\f-\frac1d \left(2Z_\x e^{-2f} \dot a^2+V_\x+d\x\left(\ddot\f+(\dot f+d\dot h)\dot\f+d\x\dot\f^2\right)\right),\NO\\
&&\pa_r\left(e^{d\x\f-2f}Z_\x\dot a\right)+(\dot f+d\dot h)e^{d\x\f-2f}Z_\x\dot a=\frac{1}{2}e^{d\xi \f} W_{\xi}e^{-2f}a,\NO\\
&&2\a\left(\ddot\f+(\dot f+d\dot h)\dot\f +d\x\dot\f^2\right)+2Z'e^{-2\x\f-2f} \dot a^2+e^{-2f}W' a^2-e^{2\x\f}V'=0,\NO\\
&&\dot \om=0.\label{Gauss-Codazzi-simplified}
\eea
These equations, except the first and the last, are the equations of motion following from the effective point particle Lagrangian
\bea\label{pp-lagrangian}
L_{eff}&=&\frac{1}{2\k^2}e^{f+dh+d\x\f}\left(\left(1+\frac{d^2\x^2}{\a_\x}\right)(\dot f+d\dot h)^2-(\dot f^2+d\dot h^2)-\a_\x\left(\dot \f-\frac{d\x}{\a_\x}(\dot f+d\dot h)\right)^2\right.\NO\\
&&\left.\rule{0cm}{0.6cm}\rule{2.8cm}{0cm}+2Z_\x(\f)e^{-2f}\dot a^2-W_\x(\f)\dot\om^2+W_\x(\f)e^{-2f}a^2-V_\x(\f)\right),
\eea
which is obtained by inserting the ansatz (\ref{bkgrd-ansatz}) in (\ref{lagrangian}) and setting the Lagrange multipliers to the values in (\ref{gauge-fixing}). The first equation in (\ref{Gauss-Codazzi-simplified}) is the energy conservation equation, while the last one is related to the conserved quantity 
\be
e^{f+dh+d\x\f}W_\x(\f)\dot\om.
\ee
The values of these conserved quantities are zero in the gravitational context, which can be derived by keeping the Lagrange multipliers $N$ and $A_r$ in the effective point particle Lagrangian.   

The canonical momenta following from the Lagrangian (\ref{pp-lagrangian}) are
\bea\label{pp-momenta}
&&\p_f=\frac{1}{\k^2}e^{f+dh+d\x\f}d\left(\dot h+\x\dot\f\right),\quad \p_h=\frac{1}{\k^2}e^{f+dh+d\x\f}d\left(\dot f+(d-1)
\dot h+d\x\dot\f\right),\\ &&\p_a=\frac{2}{\k^2}e^{-f+dh+d\x\f}Z_\x\dot a,\quad \p_{\f}=-\frac{\a_\x}{\k^2}e^{f+dh+d\x\f}\left(\dot\f-\frac{d\x}{\a_\x}(\dot f+d\dot h)\right),\quad
\p_\om=-\frac{1}{\k^2}e^{f+dh+d\x\f}W_\x\dot\om,\NO
\eea
and the corresponding Hamiltonian is
\bea\label{pp-hamiltonian}
H_{eff}&=&\frac{\k^2}{2}e^{-f-dh-d\x\f}\left(\frac{1}{d}\p_f\left(2\p_h-(d-1)\p_f\right)-\frac{1}{\a}\left(\p_\f-\x(\p_f+\p_h)\right)^2+\frac12Z_\x^{-1}e^{2f}\p_a^2-W_\x^{-1}\p_\om^2\right)\NO\\
&&+\frac{1}{2\k^2}e^{f+dh+d\x\f}\left(-W_\x(\f)e^{-2f}a^2+V_\x(\f)\right).
\eea
This Hamiltonian is conserved, but invariance under radial reparameterizations -- which would be manifest in (\ref{pp-lagrangian}) had we not gauge-fixed the einbein -- requires that it is in fact zero. The Hamilton-Jacobi equation therefore is 
\be\label{pp-HJ}
H_{eff}=0,
\ee
with the canonical momenta expressed as gradients of a function $\cs_{eff}(f,h,a,\f,\om)$ of the generalized coordinates so that (\ref{pp-HJ}) becomes a partial differential equation (PDE) for the function $\cs_{eff}(f,h,a,\f,\om)$.

\subsection{Hamiltonian algorithm for the holographic dictionary}
\label{procedure}

The full holographic dictionary for the backgrounds (\ref{bkgrd-ansatz}) can be constructed from suitable solutions $\cs_{eff}(f,h,a,\f,\om)$ of the HJ equation (\ref{pp-HJ}), without ever using the second order equations (\ref{Gauss-Codazzi-simplified}). To this end it is very important to understand the relation between solutions of the HJ equation and solutions of the equations of motion. In particular, the most general solution of the  equations of motion can be obtained from a {\em complete integral} of the Hamilton-Jacobi equation, i.e. a solution $\cs_{eff}(f,h,a,\f,\om;\lbar\p_{f},\lbar\p_{h},\lbar\p_{a},\lbar\p_{\f},\lbar\p_{\om})$ that contains as many integration constants as generalized coordinates. These integration constants will eventually be identified with the renormalized momenta, i.e. the renormalized 1-point functions \cite{Papadimitriou:2010as}. Such a complete integral is clearly not the most general solution of the HJ equation, but it is all that is needed in order to describe the general solution of the equations of motion. However, the solutions of the HJ equation generically contain branch cuts in field space, and so a given complete integral may not cover the entire solution space, but rather a subset. A {\em discrete} set of complete integrals is sufficient to cover the entire space of solutions of the second order equations of motion.  

There are two types of solutions of the HJ equations we will need: 
\begin{itemize}

\item {\em Exact solutions of the HJ equation}
  
These are special but exact solutions of the HJ equations that can be understood as `fake superpotentials' \cite{Freedman:2003ax}. Typically they are obtained by finding suitable ans\"atze that render the HJ equation tractable. Moreover, any discrete branch of the HJ equation is acceptable.\footnote{In the familiar case of Poincar\'e domain walls this branch cut ambiguity is related to the two discrete choices for the coefficient of the quadratic term in the superpotential. One choice describes RG flows due to a deformation by a relevant operator while the other choice corresponds to an RG flow due to a vacuum expectation value \cite{Papadimitriou:2004rz}.}
The corresponding exact backgrounds that solve the  equations of motion are obtained by integrating the flow equations (\ref{flow-eqs}). Such solutions may or may not contain any integration parameters and they are generically interpreted as RG flows of the dual theory.

\item {\em An asymptotic complete integral of the HJ equation}

This type of solution is the main tool in the construction of the holographic map. It is only required to be an {\em asymptotic} solution of the HJ equation,
in the sense explained in Fig. \ref{fig1}, but must contain all integration constants required of a complete integral. In order to include these integration constants the asymptotic solution must be obtained up to and including the finite terms in $\cs_{eff}(f,h,a,\f,\om;\lbar\p_{f},\lbar\p_{h},\lbar\p_{a},\lbar\p_{\f},\lbar\p_{\om})$. These finite terms are exactly the terms that are not completely determined in the asymptotic solution and so are parameterized in terms of a number of undetermined integration constants. 
Moreover, the condition that the solution must be valid in the asymptotic region $\ca$ in configuration space requires that a particular branch of the Hamilton-Jacobi solution be chosen. In the Poincar\'e domain wall example this is the well known fact that only a superpotential with a quadratic term that corresponds to a deformation can be used to construct the holographic dictionary \cite{Papadimitriou:2004rz}. Constructing such an asymptotic complete integral and deriving the holographic map for asymptotically Lifshitz and hvLf backgrounds is the main purpose of this paper. We now describe this construction within the  minisuperspace (\ref{bkgrd-ansatz}) of homogeneous backgrounds, postponing the general case for Section \ref{algorithm}.      
\begin{figure}
\begin{center}
\scalebox{0.3}{\rotatebox{90}{\includegraphics[trim= 4.5cm 2.5cm 8.0cm 3.5cm,clip,width=\textwidth,keepaspectratio]{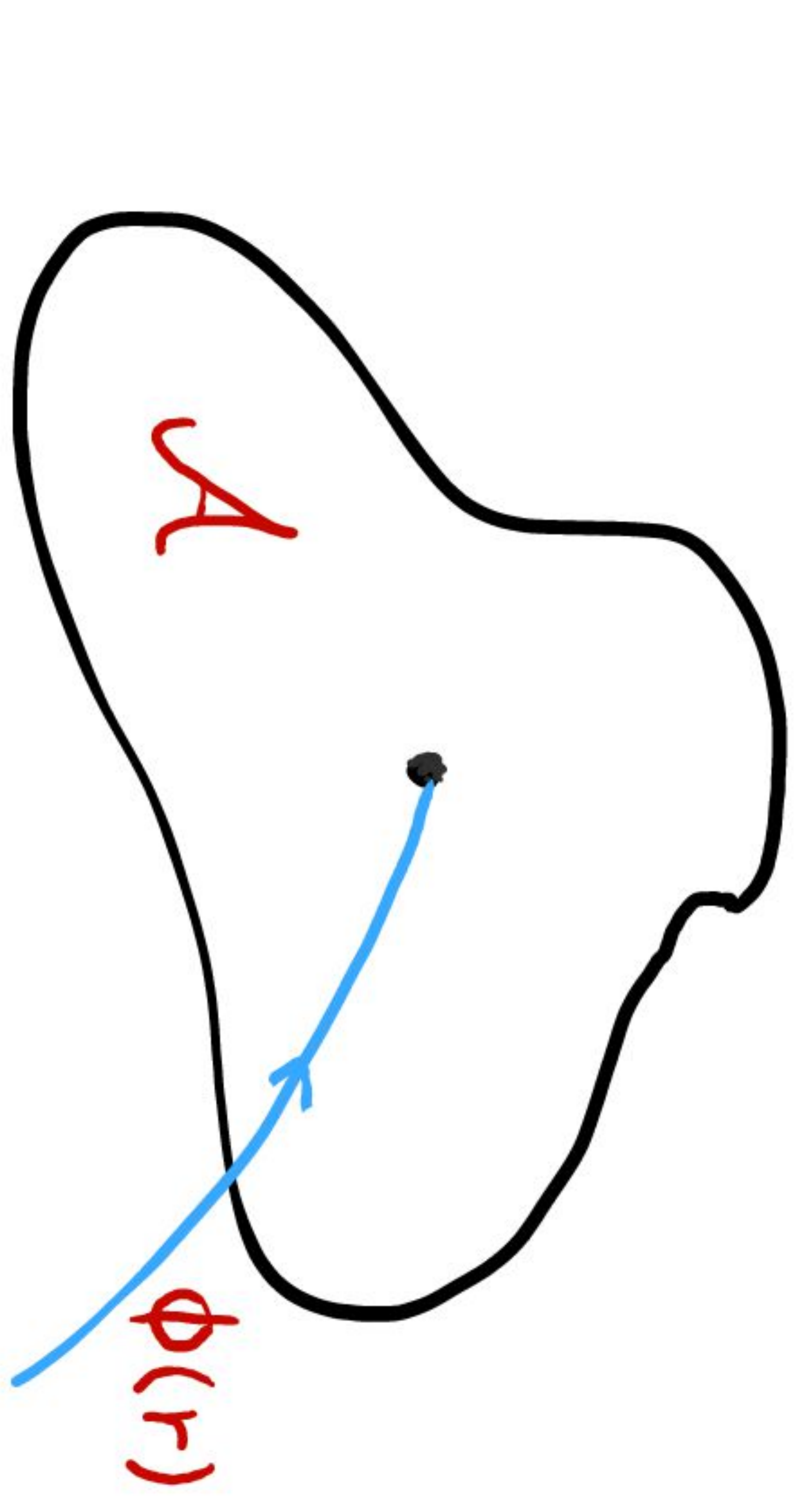}}}
\end{center}
\caption{The asymptotic form of the fields, collectively denoted by $\f(r)$ in this plot, as the radial coordinate $r$ tends to the UV defines a region $\ca$ in configuration space $\cc$, namely $\ca_{\ve_{r_o}} := \{\f(r)\in\cc\; |\; |\f(r)-\f(\infty)|\leq \ve_{r_o}, \, \forall\, r>r_o \}$, where $r_o$ is a radial UV cut-off and $\ve_{r_o}>0$ is arbitrarily small. This in turn defines the concept of an asymptotic solution of the Hamilton-Jacobi equation as a solution valid in the region $\ca$ in configuration space for any arbitrarily small $\ve_{r_o}>0$.}
\label{fig1}
\end{figure}

\end{itemize}

\begin{flushleft}
{\bf Asymptotic complete integral and the Fefferman-Graham expansions}
\end{flushleft}

Although we are focusing on homogeneous solutions for now, the asymptotic complete integral we want to construct must still correspond to the zero-derivative asymptotic solution of the HJ equation in the full theory, even when the fields have arbitrary spacetime dependent sources. Since for a renormalizable holographic dual the divergent part of the on-shell action must be local in these sources, as well as diffeomorphism and gauge invariant, it follows that   
the most general form of the divergent part of the HJ solution in the full theory must be of the form
\be\label{div-HJ}
\cs=\frac{1}{\k^2}\int d^{d+1}x\sqrt{-\g}U(\f,B^iB_i),
\ee
for some `superpotential' $U$. This restriction, however, does not apply to the finite part of the asymptotic complete integral, for which there is no requirement of locality. This observation is crucial in order to obtain the full complete integral with the correct number of integration constants, which clearly cannot be obtained from the superpotential $U$ that contains up to two integration constants. However, once the divergent part is determined, the finite part can be  obtained in terms of a number of undetermined integration constants, as we will show shortly. 

The form (\ref{div-HJ}) of the divergent part of the general asymptotic HJ solution implies that the divergent part of the complete integral $\cs_{eff}$ we are interested in for the homogeneous backgrounds takes the form
\be\label{pp-HJ-ansatz}
\cs_{eff}=\frac{1}{\k^2}e^{f+dh}U(\f,-e^{-2f}a^2).
\ee
Defining $X:=\f$, $Y:=- e^{-2f}a^2$, and inserting this point particle HJ function in the Hamiltonian leads to the following PDE for the superpotential $U(X,Y)$:
\begin{align}
\label{master}
\boxed{
\begin{aligned}
&\frac{1}{\a}\left(U_X-\x(d+1)U+2\x YU_Y\right)^2
-\frac{1}{d}(U-2Y U_Y)^2-\left(U+2YU_Y\right)\left(U-2YU_Y\right)\\
&+2Z_\x^{-1}Y U_Y^2= e^{2d\x X}\left(W_\x Y+V_\x\right),
\end{aligned}}
\end{align}
where the subscripts $X$ and $Y$ denote partial derivatives w.r.t. the corresponding variable. The superpotential equation (\ref{master}) significantly simplifies the problem of determining the divergent part of the general asymptotic complete integral, since we have to solve a PDE in only two variables, but can also be used to obtain exact solutions. 

Identifying the canonical momenta (\ref{pp-momenta}) with the gradients of (\ref{pp-HJ}) and the ansatz (\ref{pp-HJ-ansatz}) leads to the first order flow equations  
\bea
&&\dot f=2e^{-d\x X}\left(YU_Y
+\left(\frac{\a_\x}{2d\a}U+\frac{\x}{2\a}U_X
-\frac{\a_\x+d^2\x^2}{d\a}YU_Y\right)\right),\NO\\
&&\dot h=2e^{-d\x X}\left(\frac{\a_\x}{2d\a}U+\frac{\x}{2\a}U_X
-\frac{\a_\x+d^2\x^2}{d\a}YU_Y\right),\NO\\
&&\dot a=-e^{-d\x X}Z_\x^{-1}(X)U_Ya,\label{pp-flow}
\eea
and 
\bea
&&\dot X=-\frac{1}{\a}e^{-d\x X}
\left(U_X-(d+1)\x U+2\x YU_Y\right),\NO\\
&&\dot Y=-4e^{-d\x X}Y\left(\frac{\a_\x}{2d\a}U+\frac{\x}{2\a}U_X+\frac{(d-1)\a+d\x^2}{d\a}YU_Y+\frac12Z_\x^{-1}(X)U_Y\right).\label{pp-flow-XY}
\eea 
Given any solution of the superpotential equation (\ref{master}), asymptotic or exact, the flow equations(\ref{pp-flow-XY}) can be integrated to obtain the trajectories of $X$ and $Y$. Inserting those in turn in (\ref{pp-flow}), $f$, $h$ and $a$ can be determined as well. As we stressed earlier, solutions obtained in this way automatically satisfy the second order equations of motion (\ref{Gauss-Codazzi-simplified}). 

A last point we must address is the finite part of the asymptotic complete integral, which as we explained cannot be assumed to be of the form (\ref{pp-HJ-ansatz}). To this end let us consider a solution $\cs_o$ of the HJ equation, which without loss of generality can be taken to be of the form (\ref{pp-HJ-ansatz}). We then seek to determine the possible infinitesimal deformations of this solution, which should give us the full set of integration constants that parameterize a complete integral. Inserting 
\be
\cs=\cs_o+\d\cs,
\ee   
in (\ref{pp-HJ}) and keeping terms up to linear order in $\d\cs$ gives the linear PDE
\bea
&&\left[\left(U-\frac{d-1}{d}(U-2YU_Y)+\frac{\x}{\a}\left(U_X-\x(d+1)U+2\x YU_Y\right)\right)\frac{\pa}{\pa f}\right.\NO\\
&&\left.+
\left(\frac{1}{d}(U-2YU_Y)+\frac{\x}{\a}\left(U_X-\x(d+1)U+2\x YU_Y\right)\right)\frac{\pa}{\pa h}\right.\NO\\
&&\left.-\frac{1}{\a}\left(U_X-\x(d+1)U+2\x YU_Y\right)\frac{\pa}{\pa X}-Z_\x^{-1}U_Ya\frac{\pa}{\pa a}\right]\d\cs=0.
\eea
Comparing this with the flow equations (\ref{pp-flow}) and (\ref{pp-flow-XY}) we see that this equation can be written in the form 
\be
\left(\dot f\frac{\pa}{\pa f}+\dot h\frac{\pa}{\pa h}+\dot \f\frac{\pa}{\pa \f}+\dot a\frac{\pa}{\pa a}\right)\d\cs=0\quad \Leftrightarrow\quad \pa_r\d\cs=0,
\ee
which shows that only the finite part of the solution $\cs_o$ can be deformed. To determine the complete set of deformations it suffices to consider this equation in the leading asymptotic limit as $r\to\infty$ so that the radial derivative is replaced by the dilatation operator $\d_D$ \cite{Papadimitriou:2004ap}:
\be\label{modes}\boxed{
\d_D\d\cs=0.}
\ee  
The characteristic surfaces of this linear first order PDE
determine the deformation parameters of the solution $\cs_o$, which correspond to the full set of normalizable modes. 

Various solutions of the superpotential equation (\ref{master}) will be discussed in detail in Section \ref{sup}, including the derivation of the general asymptotic complete integral for Lif and hvLf backgrounds.

\subsection{Lif solutions}
\label{Lifshitz-solution}

In order for the equations (\ref{Gauss-Codazzi-simplified}) to admit Lifshitz solutions, the potentials in the action (\ref{action}) must be of the form
\be \label{exp-potentials}
V_{\xi}=V_o e^{2  (\r +\xi ) \f},\qquad  Z_{\xi}=Z_o e^{-2  (\xi +\n )\f},\qquad W_{\xi}=W_o e^{2 \s \f},
\ee
at least asymptotically, where the various constants are constrained in a way we will specify momentarily. In this section we will assume that this is the exact form of the potentials, but more general potentials will be considered later on. 
 
The Lifshitz solutions take the form
\be\label{Lif-sol} 
ds^{2}=d r^2-e^{2 z r} dt^{2}+e^{2r} d\vec{x}^2,\quad A=\frac{\cq}{\e Z_o} e^{\e r} dt,\quad \f=\m r,\quad \om=const.,
\ee
where the various parameters are related as follows: 
\begin{align}
\label{Lif-parameters}
\boxed{
\begin{aligned}
& \rho=-\xi,\quad \n=-\x+\frac{\e-z}{\m},\quad \s=\frac{z-\e}{\m},\\
&\e=\frac{(\a_\x+d^2\x^2)\m^2-d\m\x+z(z-1)}{z-1},\quad \cq^2=\frac12Z_o(z-1)\e,\\ 
& W_o= 2 Z_o \e (d + z + d \m\x  - \e),\quad 
V_o=-d (1 + \m \x) (d + z + d \m \x) - (z - 1) \e.
\end{aligned} }
\end{align}
Note that a possible additive constant in the scalar field has been absorbed in the Lifshitz radius $\ell$, which we set to 1. These solutions are related in the Einstein frame to the hvLf solutions of \cite{Gouteraux:2012yr}. We will discuss the connection of these solutions to hvLf solutions shortly. Moreover, various limits of these solutions deserve special attention.

\begin{flushleft}
{\bf Special limits}
\end{flushleft}

\begin{itemize}
\item[i)]  $W_o=0,\, \cq\neq 0$: 

This case is interesting because it corresponds to a massless $U(1)$ gauge field, and so the Action becomes the Einstein-Maxwell-Dilaton (EMD) action. The values of the parameters in this case simplify as follows: 
\bea 
&&
\r=-\x,\quad \n=(d-1)\x+\frac{d}{\m},\quad \e=d+z+d\x\m,\quad \cq^2=\frac12 Z_o(z-1)(d+z+d\x\m),\NO\\
&&
(\a_\x+d^2\x^2)\m^2-d\x z\m-d(z-1)=0,\quad V_o=-(d+z+d\x\m)(d+z-1+d\x\m).
\eea
In the Einstein frame this case corresponds to hvLf solutions with 
\bea
&& \th \leq d(z-1),\quad {\rm or}\quad \th\geq d,\quad {\rm for }\quad 1<z\leq 2,\NO\\
&& \th \leq d,\quad {\rm or}\quad \th\geq d(z-1),\quad {\rm for }\quad z> 2,
\eea
which are compatible with the NEC solutions III-V provided also $\th<d+z$. Setting $\x=0$ in these solutions we recover the anisotropic solutions obtained in \cite{Taylor:2008tg}. Note that necessarily $\m\neq 0$ in this case, and so a running scalar is required to support these solutions. The limiting case $\th=d+z$ leads to $\cq=0$ and was discussed in \cite{Gath:2012pg}. However there are more solutions
with $\cq=0$ which we discuss now. 

\item[ii)]  $W_o=0,\, \cq=0$: 

This case also corresponds to a massless $U(1)$ gauge field but now the gauge field in not switched on in the background. The values of the parameters in this case are: 
\bea \label{Q=0-Lif}
&&
\r=-\x,\quad \n=-\x-\frac{z}{\m},\quad \s=\frac{z}{\m},\quad \e=0,\NO\\
&&
(\a_\x+d^2\x^2)\m^2-d\m\x+z(z-1)=0,\quad V_o=-d (1 + \m \x) (d + z + d \m \x).
\eea
As we shall see, these solutions in the Einstein frame are hvLf solutions with 
\be\label{range-th}
\th \leq \frac d2 -\frac d2\sqrt{1+\frac{z(z-1)}{d}}, \quad {\rm or}\quad \th \geq \frac d2 +\frac d2\sqrt{1+\frac{z(z-1)}{d}}.
\ee
These solutions include the zero vector field solution with $\th=d+z$ discussed in \cite{Gath:2012pg}, but the fact that any $\th$ in the range (\ref{range-th}) leads to a solution with $W_o=0$ and $\cq=0$ was missed in \cite{Gath:2012pg} because only the case $\e=d+z+d\m\x$ was considered there.

\item[iii)] $\m=0$: 

This is another important special case, where  non-relativistic conformal invariance is recovered at least asymptotically. The parameters of the solution now take the simpler form
\bea 
   && \rho=-\xi,\quad \n=-\frac{(d+z-1)\x}{z-1},\quad \s=\frac{d\x}{z-1},\quad \e=z,\NO\\
   &&\cq^2=\frac12Z_o z (z-1),\quad W_o= 2d z  Z_o,\quad V_o=-(d  (d + z ) +z(z - 1)).
\eea
The scalar can be set identically to zero in this case, so that the action (\ref{action}) reduces to Einstein-Proca theory \cite{Kachru:2008yh}. The scalar is not identically zero necessarily in this case, however, and so it is important to keep $\x$ as a parameter. Firstly, when we generalize these solutions to inhomogeneous solutions with dependence on the transverse coordinates we will see that there can be non-zero subleading terms in the scalar. Moreover, if the potentials (\ref{exp-potentials}) are suitably modified at subleading orders, then the scalar can acquire not trivial radial dependence. Both cases of constant scalar and and non-constant scalar with $\m=0$ will be studied in detail in Section \ref{examples}.

\item[iv)] D$p$ branes in the dual frame:

Finally, it is useful as a reference to obtain the relativistic D$p$ brane solutions by setting $z=1$ in (\ref{Lif-sol}). The resulting family of solutions with parameters 
\bea
&&d=p,\quad z=1,\quad \a_\x=\frac{4(p-1)(4-p)}{(7-p)^2},\quad 
\x=\frac{2(p-3)}{p(7-p)},\quad \m=\frac{(7-p)(p-3)}{2(5-p)},
\eea
corresponds to D$p$ branes in the dual frame \cite{Itzhaki:1998dd,Boonstra:1998mp}.

\end{itemize}

\subsection{hvLf solutions}

By means of the coordinate transformation 
\be
u=\frac{|\th|}{d}r^{\frac{d}{\th}},\quad \th\neq0, 
\ee
and a suitable rescaling of the time and spatial coordinates, the hvLf metric (\ref{thetaz}) takes the form
\be\label{hvLf}
ds^2=dr^2-r^{2\n_z}dt^2+r^{2\n_1}d\vec{x}^2,
\ee
where 
\be
\n_z=1-\frac{dz}{\th},\quad \n_1=1-\frac{d}{\th}.
\ee
Note that in this coordinate system the UV is located at $r\to\infty$ for $\th < 0$ and at $r=0$ for $\th> 0$. Inserting this ansatz in the equations of motion (\ref{Gauss-Codazzi-simplified}), together with the homogeneous ansatz
\be
A=\frac{\cq}{\e Z_o} r^{\e}dt,\quad \f=\m\log r,\quad \om=const.,
\ee 
for the rest of the fields, we find that such solutions exist provided 
\begin{align}
\label{hvLf-parameters}
\boxed{
\begin{aligned} 
   & \mu  (\xi +\rho )=-1,\quad \n=-\xi-\frac{\n_{z}-\e}{\mu },\quad \s=\frac{\n_{z}-\e-1}{\mu },\quad \cq^2=\frac{1}{2}Z_o(\n_{z}-\n_{1})\e,\\
   &\e=\frac{ \left(\alpha_\xi+d^2 \xi ^2 \right)\mu ^2-d \xi (\n_{1}+1) \mu  -\n_{1} (d+\n_{z}-1) +\n_{z}(\n_{z}-1)}{\n_{z}-\n_{1}},\\ 
   &W_o= 2 \e Z_o (d (\n_{1}+\mu  \xi )+\n_{z}-1-\e),\\
   & V_o=\e (\n_{1}-\n_{z})-d (\n_{1}+\mu  \xi ) (d (\n_{1}+\mu  \xi )+\n_{z}-1).
\end{aligned}}
\end{align}
As for the Lifshitz solutions, the additive constant in the scalar field has been absorbed into the Lifshitz radius, which we set to 1. Note that these solutions do not exist for $\m=0$, and so they always require a running dilaton. Moreover, the parameter $\x$ in these solutions is somewhat redundant as we can always set it to zero by a redefinition of $\th$. For $d=2$ and $\x=0$ they reduce to the solutions discussed in Section 3.2.2 of \cite{Gouteraux:2012yr}. Note in particular that the independent metric and gauge field hyperscaling violating parameters discussed in \cite{Gouteraux:2012yr} are related to our parameters $\th$ and $\m$ respectively.

\begin{flushleft}
{\bf Special limits}
\end{flushleft}

\begin{itemize}
\item[i)]  $W_o=0,\, \cq\neq 0$:

As for the Lifshitz solutions, there are two cases with massless vector. Namely $\cq\neq 0$ and $\cq=0$. In the former case the hvLf solutions of the EMD model satisfy the following conditions:  
\bea
&&(\x+\r)\m=-1,\quad \n=(d-1)\x+\frac{d\n_1-1}{\m},\quad \e=d (\n_{1}+\mu  \xi )+\n_{z}-1,\NO\\
&& (\a_\x+d^2\x^2)\m^2-(1+\n_z)d\x\m-d\n_1(\n_z-\n_1+1)=0,\\
&&\cq^2=\frac{Z_o}{2} (\n_z-\n_1)(\n_z+d\n_1-1+d\x\m),\,
V_o=-\left(\n_z+(d-1)\n_1+d\x\m\right)\left(\n_z+d\n_1-1+d\x\m\right).\NO 
\eea
These solutions are related to the finite charge density solutions in \cite{Charmousis2010a}. Note that, as for the Lifshitz solutions, there is a limiting case of this class of solutions that has $\cq=0$ and $\n_z+d\n_1-1+d\x\m=0$. For $\x=0$ this is the corresponding Lifshitz solution we discussed above but now in the Einstein frame, and it is also the $\cq=0$ solution discussed in \cite{Gath:2012pg}. However, as in the Lifshitz case, there are more solutions with $\cq=0$.

\item[ii)]  $W_o=0,\, \cq= 0$:

The class of hvLf solutions with $\cq=0$ corresponds to the parameter space
\bea
&&(\x+\r)\m=-1,\quad \n=-\x-\frac{\n_z}{\m},\quad \e=0,\NO\\
&&  \left(\alpha_\xi+d^2 \xi ^2 \right)\mu ^2-d \xi (\n_{1}+1) \mu  -\n_{1} (d+\n_{z}-1) +\n_{z}(\n_{z}-1)=0,\\
&&
V_o=-\left(\n_z+(d-1)\n_1+d\x\m\right)\left(\n_z+d\n_1-1+d\x\m\right).\NO 
\eea
Setting $\x=0$ in these solutions we reproduce the Einstein frame version of the Lifshitz solutions (\ref{Q=0-Lif}) with $\th$ in the range (\ref{range-th}).

\item[iii)]  D$p$ branes in the Einstein frame:

Finally, from the relativistic limit $z=1$ of the hvLf solutions (\ref{hvLf-parameters}) we recover the Einstein frame version of the D$p$ brane solutions with parameters 
\bea
&& d=p,\quad z=1,\quad \x=0,\quad \th=\frac{(p-3)^2}{(p-5)},
\quad \m=\frac{p(7-p)}{2(p-3)}, \quad \a=\frac{4(9-p)}{p(7-p)^2}.
\eea

\end{itemize}

\subsection{Weyl transforming hvLf solutions to Lif solutions}

As we have already mentioned, hvLf and Lif solutions are conformally related. This is immediately obvious from the metric (\ref{thetaz}), but it is useful to see how all the parameters of the solutions transform under the relevant conformal transformation, and in particular to clarify the role of the Weyl frame parameter $\x$.  

Starting with the hvLf (\ref{hvLf}) metric and introducing the new coordinates 
\be
r=e^{-\frac{\th}{d}\bar r},
\quad t=\frac{\abs\th}{d}\bar t,\quad 
x^a=\frac{\abs\th}{d}\bar x^a,
\ee
we obtain 
\be
ds^2=\left(\frac{\th}{d}\right)^2e^{\frac{-2\th\bar r}{d}}\left(d\bar r^2
-e^{2z\bar r}d\bar t^2+e^{2\bar r}d\vec{\bar x}^2\right),
\ee
while the scalar is given by
\be
\f=\m_h\log r=-\frac{\th}{d}\m_h\bar r\equiv \m_L\bar r.
\ee
Note that the UV is located at $\bar r\to\infty$ for all values of $\th\neq 0$. It follows that the hvLf metric (\ref{hvLf}) can be written as 
\be\label{Lifshitz2hvLf}
g_h=e^{-\frac{2\th}{d\m_L}\f}g_L,
\ee 
where $\m_L=-\th\m_h/d$ and $g_L$ is a Lifshitz metric with radius $\ell_L=|\th|\ell_h/d$. 

We now observe that if a metric $g_o$ solves the equations of motion (\ref{eoms}) with $\x=0$, then $g=e^{-2\x\f}g_o$ solves the equations of motion with non-zero $\x$. In particular, let $g_{L}=e^{-2\x_L\f}g_o$ be a Lifshitz metric and $g_{h}=e^{-2\x_h\f}g_o$ a hvLf one with hyperscaling violating parameter $\th$ that solve the equations of motion corresponding respectively to $\x=\x_L$ and $\x=\x_h$. The two metrics are therefore related as
\be
g_h=e^{2(\x_L-\x_h)\f}g_L.
\ee
Comparing this with (\ref{Lifshitz2hvLf}), we arrive at the following mapping of the parameters of the dual frame Lifshitz background corresponding to a given hvLf background:
\begin{align}
\label{}
\boxed{
\begin{aligned}
& \ell_L=\frac{|\th|}{d}\ell_h,\quad \m_L=-\frac{\th}{d}\m_h,\quad \e_L=-\frac{\th}{d}\e_h,\quad \cq_L=-\left(\frac{\th}{d}\right)^2\cq_h\\
&\x_L-\x_h=-\frac{\th}{d\m_L}=\frac{1}{\m_h},\quad \a_{\xi{}_L}=\a_{\xi{}_h}-d (d+1) \left(\x_L^2-\x_h^2\right).
\end{aligned}}
\end{align}
In practice we are interested mostly in the case $\x_h=0$, so that the hvLf metric solves the equations of motion in the Einstein frame.

This relation between Lifshitz and hvLf solutions can be utilized in order to transform such hvLf backgrounds into Lifshitz backgrounds in a different Weyl frame. This is exactly analogous to the way D$p$ branes with $p\leq 4$, were studied in \cite{Kanitscheider:2008kd} by going to a Weyl frame where the geometry is asymptotically locally AdS. The method we develop in the following in order to systematically construct the holographic dictionary is directly applicable to Lif backgrounds in any Weyl frame and to hvLf backgrounds in the Einstein frame with $\th<0$. This restriction for hvLf in the Einstein frame is related to the fact in the coordinate system (\ref{hvLf}) the UV is located at $r=0$ for $\th>0$. However, for any $\th$, we can work in the dual frame where the hvLf backgrounds become Lifshitz. We will therefore work entirely in the dual frame from now on and consider Lifshitz asymptotics only. In this way we are able to develop the holographic dictionary for both Lif and hvLf with any $\th$ simultaneously. This is the reason for allowing for a non-zero Weyl parameter $\x$ throughout our analysis.
It is useful to keep in mind that the combination of parameters $d\m\x$ from now on can be understood as 
\be
d\m\x=-\th,
\ee
where $\th$ is the hyperscaling violating parameter of the corresponding hvLf background in the Einstein frame.

\subsection{Lif boundary conditions as a second class constraint}

From the solutions (\ref{Lif-sol}) follows that Lifshitz boundary conditions amount to the asymptotic relations 
\be\label{pp-asymptotics}
\dot f\sim z, \quad \dot h\sim 1,\quad \dot a\sim \e a,\quad \dot X\sim \m,\quad \dot Y \sim 2(\e-z)Y.
\ee 
Recall that $X:=\f$ and $Y:=B_iB^i=-e^{-2f}a^2$. 
Inserting these asymptotic expressions in the flow equations (\ref{pp-flow-XY}) and (\ref{pp-flow}), one finds that the resulting set of linear PDEs for $U(X,Y)$ admit an asymptotic solution for the superpotential $U(X,Y)$ provided asymptotically 
\be\label{Yo}
Y\sim Y_o(X):=-\frac{z-1}{2\e}Z_\x^{-1}(X),\qquad 0< |Y_o^{-1}(Y-Y_o)|<<1.
\ee
The corresponding superpotential $U(X,Y)$ takes the form
\be\label{asymptotic-superpotential}\boxed{
U(X,Y)=e^{d\x X}\left(d+z+d\m\x-1-\e Z_\x(X)(Y-Y_o)+\co(Y-Y_o)^2\right).}
\ee

It is important to pause for a moment and clarify the significance of these asymptotic conditions since they play a key role in the construction of the holographic dictionary for anisotropic backgrounds and throughout the subsequent analysis. Using the definition of the variable $Y$ we can express the time component of the vector field as   
\be\label{pp-constraint-asymptotic}
a= \sqrt{\frac{z-1}{2\e}}Z_\x^{-1/2}(\f)e^f\sqrt{1+Y_o^{-1}(Y-Y_o)}.
\ee 
This expression can be seen as a change of variables in configuration space (a special canonical transformation), trading the variable $a$ in favor of $Y-Y_o$, without any physical significance. The non-trivial condition, however, comes from demanding Lif asymptotics, i.e. that asymptotically $Y-Y_o\to 0$. The reason why this is particularly significant is that 
setting 
\be\label{pp-constraint}
a= \sqrt{\frac{z-1}{2\e}}Z_\x^{-1/2}(\f)e^f,
\ee
is not compatible with any integral of motion of the equations (\ref{Gauss-Codazzi-simplified}) and so 
amounts to a {\em second class constraint}. Another way this constraint can be deduced is the fact there is no superpotential $U(X)$ -- crucially without any dependence on $Y$ -- that leads to the asymptotics (\ref{pp-asymptotics}) via the flow equations (\ref{pp-flow}) and  (\ref{pp-flow-XY}). In Appendix \ref{constrained-systems} we show how such a constrained system can be described in a Hamiltonian language, either by solving explicitly the constraint at the start, or by using Dirac's algorithm for constrained systems. As long as we keep at least the linear term in $Y-Y_o$ in (\ref{asymptotic-superpotential}), which corresponds to a deviation from the constraint surface (\ref{pp-constraint}), the standard Hamiltonian analysis applies, however. Demanding that a Taylor expansion in $Y-Y_o$ be compatible with the dynamics is equivalent to requiring that (\ref{pp-constraint}) be a {\em consistent truncation} of the theory. In other words, we are asking that the effective potential\footnote{The effective potential for $Y-Y_o$ is not the superpotential $U(X,Y)$ which does have a linear term in $Y-Y_o$. This is rather counterintuitive if one compares the mode $Y-Y_o$ with an elementary scalar field. In that case an extremum of the potential is also an extremum of the superpotential. However, this analogy has its limitations because the mode $Y-Y_o$ is a composite field.} for the fluctuation $Y-Y_o$ has no linear term and that the quadratic term (mass) is such that $Y-Y_o$ sources a relevant operator. As we will see shortly, this leads to further conditions for the potentials parameterizing the Lagrangian (\ref{pp-lagrangian}), besides the leading asymptotic form (\ref{exp-potentials}).

\subsection{Fefferman-Graham expansions and anisotropic RG flows from a superpotential}
\label{sup}

In the previous subsection we determined that imposing Lifshitz asymptotics requires the superpotential $U(X,Y)$ to have the asymptotic form (\ref{asymptotic-superpotential}). In order to obtain asymptotically Lif backgrounds that correspond to deformations of the `ground states' (\ref{Lif-sol}), such as anisotropic renormalization group (RG) flows, we need an exact solution of the superpotential equation (\ref{master}) that satisfies the asymptotic condition (\ref{asymptotic-superpotential}). In this subsection we 
make use of various ans\"atze to simplify the superpotential equation and we present a class of exact solutions corresponding to a certain marginal deformation of the backgrounds (\ref{Lif-sol}). We also obtain the general solution to the superpotential equation (\ref{master}) with the asymptotic condition (\ref{asymptotic-superpotential}) in the form of a Taylor expansion in $Y-Y_o$, which can be used to determine the general asymptotic complete integral and the Fefferman-Graham expansions. It is worth pointing out that a solution $U(X,Y)$ of (\ref{master}) cannot be polynomial in $Y$ for the physical range of the various parameters. Combined with the asymptotic condition (\ref{asymptotic-superpotential}), this implies that any superpotential can be expressed as a non-truncating Taylor series in $Y-Y_o$, although there can be non-analytic terms starting at the normalizable order.

\begin{flushleft}
{\bf Superpotential I:}
\end{flushleft}

An important special case of the Lagrangian (\ref{pp-lagrangian}) occurs when the potentials are exactly -- not merely asymptotically -- exponentials as in (\ref{exp-potentials}), i.e. 
\be 
V_{\xi}=V_o,\qquad  Z_{\xi}=Z_o e^{-2  (\xi +\n )X},\qquad W_{\xi}=W=W_o e^{-2  (\xi +\n ) X},
\ee
with the various parameters satisfying the relations (\ref{Lif-parameters}). Since this holds asymptotically anyway, this example captures the essential physics for general asymptotically Lif and hvLf backgrounds. 

The superpotential equation (\ref{master}) in this case can be reduced to an ordinary differential equation (ODE) via the ansatz 
\be\label{ansatzI} 
U(X,Y)=e^{d\x X}w\left(Y Z_\x(X)\right), 
\ee  
for some function $w(y)$ of $y\equiv Y Z_\x(X)$. Inserting the ansatz (\ref{ansatzI}) into (\ref{master}) we get a first order ODE
for $w(y)$:
\be\label{ansatzI-eq} \boxed{
(\a_2 y +\a_1)yw'^2+\b yww'+\g w^2=\d y+\ve, }
\ee
where
\be
\a_2=4\left(d\n^2+(d-1)\a\right),\,
\a_1=2 d\a,\,
\b=4\left(d\n\x+\a\right),\,
\g=d\x^2-(d+1)\a,\,
\d=\frac{d\a W_o}{Z_o},\,
\ve=d\a V_o.
\ee
The asymptotic condition (\ref{asymptotic-superpotential}) determines that $w(y)$ must satisfy 
\be\label{ansatzI-bc} 
w(y_o)=d(1+\m\x)+z-1,\qquad w'(y_o)=-\e,\qquad y_o:= Y_o(X) Z_\x(X)= -\frac{z-1}{2\e}. 
\ee

Equation (\ref{ansatzI-eq}) can be transformed into an Abel equation of the first kind \cite{Polyanin1995}, which is in general non-integrable. For special ranges of the parameters it admits analytic solutions of the form $w = \sqrt{a+b y}$, which are special cases of the more general class of solutions derived from superpotential II below. For generic values of the parameters, however, we can obtain the solution to (\ref{ansatzI-eq}) subject to the initial conditions (\ref{ansatzI-bc}) in the form of a Taylor expansion around $y_o$, including potential non-analytic terms at normalizable order. In particular, for generic values of the parameters the solution of (\ref{ansatzI-eq}) subject to the initial conditions (\ref{ansatzI-bc}) takes the form\footnote{For $\D_+= (d+z+d\m\x)/2$ the Breitenlohner-Freedman (BF) bound \cite{Breitenlohner1982a,Breitenlohner1982c} is saturated and various logarithmic terms appear starting at order $(y-y_o)^2\log|y-y_o|$.}
\be\label{supI-sol}
w(y)=\sum_{n=0}^\infty\left( w_n (y-y_o)^n+|y-y_o|^{\frac{d+z+d\m\x}{d+z+d\m\x-\D_+}}\wt w_n (y-y_o)^n\right),\quad |y-y_o|<<1,
\ee 
where $w_0$ and $w_1$ are determined by (\ref{ansatzI-bc}), $\wt w_0$ is an integration constant, and the scaling dimension $\D_+> (d+z+d\m\x)/2$ is defined via the asymptotic behavior 
\be
y-y_o\sim e^{-(d+z+d\m\x-\D_+)r}. 
\ee
When $d+z+d\m\x-\D_+=0$ we have instead $y-y_o\sim r^{-1}$
and so $y-y_o$ is the source of a marginally relevant operator in this case. As we shall see, the value of $\D_+$ is related to $w_2$, which is determined by the quadratic equation
\be 
4(\a_2y_o+\a_1)y_o w_2^2+\left((4\a_2+3\b)y_ow_1+2\g w_o\right)w_2+(\a_2+\b+\g)w_1^2=0. 
\ee 
The two roots of this equation are 
\be\label{supI-roots}
w_2^\pm=\frac{\e^2}{(z-1)w_o}\left(\frac12(z-1)-w_o+\frac{d(z-1)\left(4\e-w_o-1\mp\sqrt{\cd}\right)}{4\left(\left(1-\frac{d\x^2}{\a}\right)w_o-d\right)}\right),
\ee
where
\be
\cd=\left(4\e-w_o-1+\frac{2(w_o+1-z)}{z-1}\right)^2-\frac{4w_o(w_o+1-z)}{z-1}\left(\frac{2}{d}\left(1-\frac{d\x^2}{\a}\right)(2\e-w_o-1)+\frac{z}{z-1}\right).
\ee
For $\x=\m=0$ these roots reduce to
\be\label{supI-roots-simple}
w_2^{\pm}=-\frac{z^2\left(d(d+z)-2(z-1)(2d-1)\pm d\sqrt{(d+z)^2-8(z-1)(d-z)}\right)}{4(z-1)(d+z-1)}.
\ee
The coefficients $w_n$ with 
\be
2< n < n_c:=\frac{d+z+d\m\x}{d+z+d\m\x-\D_+},
\ee
can be obtained recursively from the linear equations
\bea\label{supI-recursion}
&&\left[4(n+2)(\a_2 y_o+\a_1)y_o w_2+\left(2(n+2)\a_2+(n+3)\b \right)y_o w_1+2\g w_o\right]w_{n+2}=\NO\\
&&-(\a_2 y_o+\a_1)y_o\sum_{k=2}^n(k+1)(n-k+3)w_{k+1}w_{n-k+3}\NO\\
&&-\sum_{k=1}^n(k+1)\left[(n-k+2)(2\a_2 y_o+\a_1)+\b y_o\right]w_{k+1}w_{n-k+2}\NO\\
&&-\sum_{k=0}^n\left[(k+1)((n-k+1)\a_2+\b)+\g\right]w_{k+1}w_{n-k+1},\quad n\geq 1.
\eea
These are all the terms that are needed to determine the asymptotic solutions of the fields via the flow equations, since the terms $w_n$ with $n> n_c$, 
as well as the terms $\wt w_n$ with $n\geq 1$, are subleading relative to the normalizable modes.  When $\D_+=d+z+d\m\x$, however, the mode $y-y_o\sim 1/r$ goes to zero only logarithmically and $n_c\to\infty$, which means that all terms in the solution (\ref{supI-sol}) must be kept in this case to obtain the correct asymptotic solution of the HJ equation. This is reminiscent of what happens in the case of Improved Holographic QCD \cite{Gursoy:2007cb,Gursoy:2007er,Papadimitriou:2011qb} and it is important in order to correctly renormalize the often studied Einstein-Proca theory for $d=z=2$ when the marginally relevant deformation $y-y_o$ is turned on as in e.g. \cite{Cheng:2009df}. All terms must also be determined in order to obtain an exact background solution. Backgrounds with $\wt w_n=0$ can be obtained through the recursion relations (\ref{supI-recursion}) applied to any $n>2$.

These results are in agreement with those of 
\cite{Ross:2009ar,Baggio:2011cp,Gouteraux:2012yr}, which were obtained through an analysis of the linearized fluctuations of the equations of motion. Here we have derived these in a simpler way using only the superpotential equation (\ref{master}). There is no need for studying linearized perturbations of the equations of motion (except for computing 2-point functions, of course), or indeed using the second order equations, since the full asymptotic expansions can be obtained from an asymptotic complete integral of the HJ equation.

Inserting the solution (\ref{supI-sol}) in the flow equations (\ref{pp-flow}) and (\ref{pp-flow-XY}) leads to the first order equations
\bea\label{supI-flow-eqs}
\dot X&=&\m-\frac{\e}{\a}Z_\x\left(\x+2\n\left(1+\frac{z-1}{\e^2}w_2\right)\right)(Y-Y_o)+\co(Y-Y_o)^2,\NO\\
\dot Y-\dot Y_o &=&-\left(w_o+1-2\e+2z-\D_+\right)(Y-Y_o)+\co(Y-Y_o)^2,\NO\\
\dot f&=&z-\frac{z-1}{2\a_1}\left(\b+2\g-\a_1+\left(\b-2\a_1\right)\frac{(z-1)}{\e^2}w_2\right)Y_o^{-1}(Y-Y_o)+\co(Y-Y_o)^2,\NO\\
\dot h &=& 1-\frac{z-1}{2\a_1}\left(\b+2\g+\a_1+\b\frac{(z-1)}{\e^2}w_2\right)Y_o^{-1}(Y-Y_o)+\co(Y-Y_o)^2,\NO\\
\dot a&=&\e a\left(1+\frac{z-1}{\e^2}w_2Y_o^{-1}(Y-Y_o)+\co(Y-Y_o)^2\right),
\eea
where\footnote{Note that the reason why $\D_+$ appears in the leading asymptotic form of $Y-Y_o$ is that in order to determine the most general asymptotic expansion we must pick the exponent corresponding to the most dominant solution asymptotically. This is an explicit example where a choice between different discrete branches of the HJ solution needs to be made. }
\be\label{dimensions}\boxed{
\D_{\pm}=\frac12\left(w_o+1-2\e+\frac{(z-1)\left(1+\frac{d\n\x}{\a}\right)}{\left(1-\frac{d\x^2}{\a}\right)w_o-d}\left(2\e\pm\sqrt{\cd}\right)\right).}
\ee
Note that for $\x=\m=0$ we have $\e=z$ and so
\be\label{dimensions-simple}
\D_{\pm}=\frac12\left(d+z\pm\sqrt{(d+z)^2-8(z-1)(d-z)}\right).
\ee
These first order equations can be  integrated to obtain the full set of asymptotic expansions, including the normalizable and non-normalizable modes. In particular, the non-normalizable modes appear as integration constants of these first order equations. Namely, the leading asymptotic form of the fields takes the form
\be
X\sim \m r+ \f_o,\quad Y-Y_o= c_3 e^{-\left(w_o+1-2\e+2z-\D_+\right)r},\quad f\sim z r+ c_4,\quad h\sim r + c_5,\quad a\sim a_o e^{\e r}, 
\ee
where $\f_o$, $c_3$, $c_4$, $c_5$ and $a_o$ are integration constants, and we have kept the notation of \cite{Ross:2009ar} to facilitate the comparison of the modes. However, $a_o$ is fixed by the boundary condition (\ref{pp-constraint}) in terms of the other parameters as 
\be 
a_o= \sqrt{\frac{z-1}{2\e}}Z_o^{-1/2}e^{c_4+(\n+\x)\f_o}.
\ee
It corresponds to a source of a marginal operator with respect to Lifshitz boundary conditions, which do not want to turn on. Moreover, if $\D_+\geq d+z+d\m\x$, then the mode $c_3$ must also be set to zero since otherwise $Y_o^{-1}(Y-Y_o)$ is not vanish asymptotically and the Taylor expansion in $Y-Y_o$ breaks down. In terms of the dual theory, in that case $c_3$ sources a marginal or irrelevant operator relative to the Lifshitz theory. Finally, this asymptotic form of the scalar is valid assuming $\m\neq 0$. If $\m=0$ then one has to look at subleading terms of the potential, and in particular at the mass term, to determine the asymptotic form of the scalar. 

To determine the normalizable modes we need to consider the most general deformations of the solution (\ref{supI-sol}) of the HJ equation, as was discussed in Section \ref{procedure}. We showed that this can be done by finding the characteristics of the linear PDE defined by the dilatation operator. The dilatation operator itself is obtained from the asymptotic form of the non-normalizable modes through     
\bea
\pa_r &=& \dot f\frac{\pa}{\pa f}+\dot h\frac{\pa}{\pa h}+\dot \f\frac{\pa}{\pa \f}+(\dot Y-\dot Y_o) \frac{\pa}{\pa (Y-Y_o)}\NO\\
&\sim & z\frac{\pa}{\pa f}+\frac{\pa}{\pa h}+\m\frac{\pa}{\pa \f}-\left(w_o+1-2\e+2z-\D_+\right)(Y-Y_o) \frac{\pa}{\pa (Y-Y_o)} =: \d_D.
\eea
Note that the Lifshitz boundary condition has changed the form of the dilatation operator, replacing the derivative with respect to $a$ with a derivative with respect to $Y-Y_o$. This reflects the fact that Lifshitz boundary conditions fix the mode $a_o$ and so we cannot consider variations with respect to $a_o$ without changing the variational problem. To determine the normalizable modes, therefore, we need to find the characteristics of the linear PDE
\be
\left(z\frac{\pa}{\pa f}+\frac{\pa}{\pa h}+\m\frac{\pa}{\pa \f}-\left(w_o+1-2\e+2z-\D_+\right)(Y-Y_o) \frac{\pa}{\pa (Y-Y_o)}\right)\d\cs=0.
\ee
Assuming $\m\neq 0$, a convenient basis for the three independent characteristics is 
\be
\cc_1= \m f-z\f,\quad \cc_2=\m h-\f,\quad \cc_3=\m\log|Y-Y_o|+\left(w_o+1-2\e+2z-\D_+\right)\f,
\ee
and so the most general\footnote{The most general deformation, of course, corresponds to adding arbitrary functions of these characteristics. However, we are only interested in a complete integral and for this it suffices to consider constant coefficients multiplying a given function of the characteristics.} deformation of the solution (\ref{supI-sol}) of the HJ equation can be written in the form
\be\label{supI-deformations}
\d\cs = q_1 e^{f-z\f/\m}+q_2 e^{h-\f/\m}+ q_3 (Y-Y_o) e^{(w_o+1-2\e+2z-\D_+) \f/\m},
\ee
where $q_i$ are the normalizable modes.\footnote{This is a special case of $\Hat\cs_{reg}$ in (\ref{Sreg}) in Section \ref{ward} for general asymptotically hvLf backgrounds.} Note that the parameter $\wt w_0$ in (\ref{supI-sol}) can be expressed in terms of these deformation parameters. The fact that there are only three independent normalizable modes, while there are apparently four sources is due to the fact that we consider homogeneous solutions. A fourth deformation of the HJ solution is the energy, but such a deformation is not allowed in a model that comes from gravity since the Hamiltonian vanishes due to diffeomorphism invariance. The source conjugate to the energy is the radial cutoff $r_o$, which can be used to eliminate one of the sources for homogeneous solutions. We choose to eliminate $\f_o$. From (\ref{supI-deformations}) we find that the symplectic form on the space of asymptotic solutions \cite{Papadimitriou:2010as} takes the form
\bea
\Om &=& \d q_1\wedge \d c_4+\d q_2\wedge\d c_5+ \d q_3\wedge \d c_3. 
\eea
As we shall see in Section \ref{ward}, the modes $q_1$, $q_2$ and $q_3$ are related respectively to the energy density, spatial stress tensor and scalar operator dual to $Y-Y_o$ \cite{Ross:2009ar}.
Finally, from the momenta (\ref{pp-momenta}) we see that the deformations (\ref{supI-deformations}) will modify the flow equations (\ref{supI-flow-eqs}) according to 
\bea
\d\dot f &\sim & \k^2 e^{-f-dh-d\x\f}\left(\frac1d\left(\frac{\pa}{\pa h}-(d-1)\frac{\pa}{\pa f}\right)+\frac{\x}{\a}\left(\frac{\pa}{\pa\f}-\x\left(\frac{\pa}{\pa f}+\frac{\pa}{\pa h}\right)\right)\right)\d\cs\NO\\
& \sim &\k^2 e^{-f-dh-d\x\f}\left(-\left(\frac{d-1}{d}+\frac{\x^2}{\a}\right)q_1+\left(\frac1d-\frac{\x^2}{\a}\right)q_2\right),\NO\\\NO\\
\d\dot h &\sim &  \k^2 e^{-f-dh-d\x\f}\left(\frac1d\frac{\pa}{\pa f}+\frac{\x}{\a}\left(\frac{\pa}{\pa\f}-\x\left(\frac{\pa}{\pa f}+\frac{\pa}{\pa h}\right)\right)\right)\d\cs\NO\\
& \sim &\k^2 e^{-f-dh-d\x\f}\left(\left(\frac{1}{d}-\frac{\x^2}{\a}\right)q_1-\frac{\x^2}{\a}q_2\right),\NO\\\NO\\
\d\dot \f &\sim &  \k^2 e^{-f-dh-d\x\f}\left(-\frac{\x}{\a}\left(\frac{\pa}{\pa\f}-\x\left(\frac{\pa}{\pa f}+\frac{\pa}{\pa h}\right)\right)\right)\d\cs\NO\\
& \sim &\k^2 e^{-f-dh-d\x\f}\frac{\x^2}{\a}\left(q_1+q_2\right),\NO\\\NO\\
\d\dot a &\sim & 0,
\eea
where we have used the fact that the sources $a_o$ and $\f_o$ have been fixed. Since these terms correspond to the normalizable modes in the asymptotic expansions, the latter are only needed up to this order.

\begin{flushleft}
{\bf Superpotential II:}
\end{flushleft}

We now consider an ansatz that allows us to separate variables in the superpotential equation (\ref{master}), and as a result, to obtain exact hvLf solutions that correspond to marginal deformations of the backgrounds (\ref{Lif-sol}). Inserting the ansatz  
\be\label{ansatzII} 
U(X,Y)=\ve_0 e^{d\x X}\sqrt{\ve_1e^{2\x X}u^2(X)+\ve_2v^2(X)Y}, 
\ee
where $\ve_{0,1,2}=\pm 1$ are independent signs, in the superpotential equation (\ref{master}) leads to the three equations
\begin{align}
\label{reduced-HJ}
\begin{aligned}
& v'^2=\a \ve_2 W(X)\geq 0,\\
& 2vv'uu'-u^2\left(2\a v^2+ v'^2\right)=\a \ve_1 v^2\left(V(X)-\frac{v^2}{2}\ve_2Z^{-1}(X)\right),\\
&u^2\left( u'^2-\frac{(d+1)\a}{d}u^2\right)=\a \ve_1 u^2 V(X).
\end{aligned} 
\end{align}
The first and second equations can be integrated directly to obtain
\begin{align}
\label{}
\begin{aligned}
& v=\pm\sqrt{\a} \int^X dX'\sqrt{\ve_2 W(X')},\\
& \ve_1u^2=\left\{\begin{matrix}
\a v \vth^{-1}\int^X \vth\left(V-\frac12 v^2\ve_2Z^{-1}\right)v'^{-1}, & v'\neq 0,\\
-\frac12\left(V-\frac12 v^2\ve_2Z^{-1}\right), & v'=0,\\
\end{matrix}\right.
\end{aligned} 
\end{align}
where 
\be
\vartheta(X)\equiv e^{-2\a\int^X\frac{v}{v'}}.
\ee
However, $u$ must also satisfy the last equation in (\ref{reduced-HJ}), which leads to a constraint relating $V(X)$, $Z(X)$ and $W(X)$. Any solution of these equations is a solution to the original superpotential equation (\ref{master}), but in order for this superpotential to correspond to Lif or hvLf solutions the asymptotic conditions (\ref{asymptotic-superpotential}) must also be satisfied. 

Expanding the ansatz (\ref{ansatzII}) around the asymptotic curve $(X,Y_o(X))$ we obtain
\bea
&&U(X,Y_o(X))\sim \ve_0 e^{d\x X}\sqrt{\ve_1e^{2\x X}u^2(X)+\ve_2v^2(X)Y_o(X)},\NO\\
&&U_X(X,Y_o(X))\sim d\x U(X,Y_o(X))+\frac{\ve_0e^{d\x X}\left(\ve_1\x e^{2\x X}u^2+\ve_1e^{2\x X}uu'+\ve_2vv'Y_o(X)\right)}{\sqrt{\ve_1e^{2\x X}u^2(X)+\ve_2v^2(X)Y_o(X)}},\NO\\
&&U_Y(X,Y_o(X))\sim \frac{\ve_0\ve_2e^{d\x X}v^2}{2\sqrt{\ve_1e^{2\x X}u^2(X)+\ve_2v^2(X)Y_o(X)}}.
\eea
Comparing these with the asymptotic conditions following from  (\ref{asymptotic-superpotential}) determines 
\bea\label{uv-asymptotics}
&&\ve_0={\rm sgn}\{d(1+\m\x)+z-1\},\NO\\
&&u^2(X)\sim \ve_1d(1+\m\x)\left(d(1+\m\x)+z-1\right)e^{-2\x X},\NO\\
&&v^2(X)\sim -2\ve_2\e Z_o\left(d(1+\m\x)+z-1\right)e^{-2(\n+\x)X}.
\eea
Inserting the asymptotic condition for $v^2$ in the first equation in (\ref{reduced-HJ}) leads to a constraint on the parameters of the solutions, namely
\be\label{supII-constraint}
(\n+\x)^2=-\a\left(\frac{d+z+d\m\x-\e}{d+z+d\m\x-1}\right).
\ee

Before determining the possible solutions of this constraint, it is instructive to derive it in an alternative way. Inserting the ansatz (\ref{ansatzII}) in the flow equations (\ref{pp-flow-XY}) (\ref{pp-flow})
we obtain 
\bea
&&\pa_\r X=\ve_1uu'+\ve_2vv'\wt Y,\NO\\
&&\pa_\r{\wt Y}=2\wt Y\left(\a (\ve_1u^2+\ve_2v^2\wt Y)+\ve_1\left(u'-\frac{v'}{v}u\right)^2\right),\NO\\
&&\pa_\r f=-\left(\x\left(\ve_1uu'+\ve_2vv'\wt Y\right)+\a\left(\frac1d \ve_1u^2+\ve_2v^2\wt Y\right)\right),\NO\\
&&\pa_\r h=-\left(\x\left(\ve_1uu'+\ve_2vv'\wt Y\right)+\frac{\a}{d} \ve_1u^2\right),\NO\\
&&\pa_\r a=\ve_1a\left(\left(u'-\frac{v'}{v}u\right)^2+\frac{d-1}{d}\a u^2\right),
\eea
where $\wt Y\equiv e^{-2\x X}Y$ and the radial coordinate $\r$ is defined by 
\be
-\ve_0\a e^{-\x X}\sqrt{\ve_1u^2+\ve_2v^2\wt{Y}}\pa_r \equiv \pa_\r.
\ee
Combining the first two flow equations leads to a first order equation for $\wt Y$ as a function of $X$: 
\be\label{eq1} 
(\ve_1uu'+\ve_2vv'\wt Y)\wt Y'(X)=2\wt Y\left(\a (\ve_1u^2+\ve_2v^2\wt Y)+\ve_1\left(u'-\frac{v'}{v}u\right)^2\right). 
\ee 
This is an Abel equation of the second kind \cite{Polyanin1995}, which is in general non-integrable but there are known integrable classes. In particular, this equation can be  solved for the $u$ and $v$ in (\ref{uv-asymptotics}). The solution is 
\be\label{Y-sol}
Y=\frac{d-\th}{4\e}Z_\x^{-1}(X)\left(1\pm\sqrt{1+c\; e^{-2(\a+\n(\n+\x))X}}\right),
\ee
where $c$ is an integration constant. Since $d-\th\neq -(z-1)$ (otherwise $u$ and $v$ vanish identically), the only way this solution can be compatible with the asymptotic condition (\ref{Yo}) is that the parameters of the solution satisfy $\a+\n(\n+\x)=0$ and the integration constant is chosen appropriately so that $Y=Y_o(X)$ identically. It can be  checked that this condition on the parameters is precisely the constraint (\ref{supII-constraint}). It is also the condition for the dimension $\D_+$ in (\ref{dimensions}) to be equal to $d+z-\th$ and therefore, the operator dual to the deformation $Y-Y_o$
is a marginal operator. Indeed, (\ref{Y-sol}) can be written as 
\be
Y= \wt c\; Y_o,
\ee
for an arbitrary constant $\wt c$ and so $Y-Y_o=(\wt c-1) Y_o$. The boundary condition (\ref{Yo}), however requires that we turn off the source for this operator and so we must set $\wt c=1$. With the source for $Y-Y_o$
set to zero the corresponding background solutions are identical to the backgrounds (\ref{Lif-sol}), but for the specific set of parameters that satisfy (\ref{supII-constraint}). However, turning on a source  for $Y-Y_o$ in this case leads to a marginal deformation of the dual theory, which can be seen as a shift in the exponent $\e$. 

The parameter space allowed by the marginality condition (\ref{supII-constraint}) turns out to be rather restricted, but non-empty. One can show that there is no solution with $\m=0$ and finite $\x$, or with $\x=0$. Solving the constraint for $\a\m^2$ in terms of $\m\x$, $d$ and $z$ we get 
\be\label{exact-sol} 
\a\m^2=\frac{z-1}{2}\left(\pm\sqrt{(1+\m\x-z)^2-4d\m\x(1+\m\x)}+\frac{(1+\m\x)}{z-1}(z-1+2d\m\x)-z\right)\geq 0. 
\ee
Recall that $-d\m\x$ is the hyperscaling violating exponent $\th$ in the Einstein frame, while $\a\m^2\geq 0$ is related to the independent vector hyperscaling violating parameter discussed in \cite{Gouteraux:2012yr}.
\begin{itemize}

\item[i)] {$z>1$:}

For $z>1$ we must choose the plus sign in (\ref{exact-sol}). The quantity inside the bracket then is positive provided either of the following two conditions holds:
\bea\label{conditions}
&(a) & (1+\m\x)\left(d(1+\m\x)+z-1\right)\leq 0,\NO\\
&(b) & \frac{2d-z-2\sqrt{d^2-dz+dz^2}}{4d-1} \leq 1+\m\x < -\frac{z-1}{d}, \quad z\leq \frac{2d-1}{d-1}.
\eea
The first condition requires  
\be
-\frac{z-1}{d} < 1+\m\x \leq  0 \Leftrightarrow 
d\leq \th <d+z-1,
\ee
which is compatible with the NEC provided 
\be
1<z \leq \frac{2d}{d-1}.
\ee
In terms of $\th$ these solutions can be summarized as follows:
\begin{align}
\label{exact-sol-z>1}
\boxed{
\begin{aligned}
(a) &&& 1<z \leq \frac{2d}{d-1}, && d\leq \th <d+z-1, \\
 &&&   && W_o\leq 0,\quad \e \geq d+z-\th,\quad  \ve_0=1,\quad \ve_1=\ve_2=-1,\\&&& &&\\
(b) &&& 1<z \leq \frac{2d-1}{d-1}, && d+z-1<\th\leq \frac{d(2d+z-1+2\sqrt{d^2-dz+dz^2})}{4d-1},\\
 &&&   && W_o>0,\quad \e < d+z-\th,\quad  \ve_0=-1,\quad \ve_1=\ve_2=1. 
\end{aligned}}
\end{align}
These are solutions of type IIIb or IVb in terms of the classification (\ref{ranges}). The case $\th =d+z-1$ ($\n=0$) corresponds to the trivial solution $u=v=0$. The case $\th=d$ corresponding to $\e=z$ and $W_o=0$ is obtained as the scaling limit $\m\to 0$ keeping $\m\x=-1$ fixed.

\item[ii)] {$z<1$:}

For $z<1$ the minus sign in (\ref{exact-sol}) must be chosen. The RHS of (\ref{exact-sol}) is then positive provided
\be
d+z-1\leq \th\leq d,
\ee
which violates the NEC except for the limiting case $\th=d$ as above, but now with $z\leq 0$.

\end{itemize}

\begin{flushleft}
{\bf Superpotential III:}
\end{flushleft}

As a final example, we consider the Taylor expansion of the general superpotential $U(X,Y)$, without any simplifying assumptions for the potentials of the Lagrangian except for the asymptotic conditions (\ref{exp-potentials}). However, as we already anticipated, additional consistency conditions will arise by requiring that a Taylor expansion in $Y-Y_o$ be consistent with the asymptotic expansion, as required by the Lifshitz boundary conditions. The analysis here is  a straightforward generalization of the analysis for superpotential I above.   

We start by expanding the superpotential $U(X,Y)$ in a Taylor series in $Y-Y_o$ as 
\bea
&&U=U_0+U_1(Y-Y_o)+U_2(Y-Y_o)^2+\co(Y-Y_o)^3,\NO\\
&&U_X=\left(U'_0-Y'_oU_1\right)+\left(U_1'-2Y_o'U_2\right)(Y-Y_o)+\co(Y-Y_o)^2,\NO\\
&&U_Y=U_1+2U_2(Y-Y_o)+\co(Y-Y_o)^2.
\eea
In order to simplify the subsequent formulas we reparameterize the coefficients $U_m(X)$ as 
\be
U_m(X)=e^{(d+1)\x X}Y_o^{-m}(X)u_m(X).
\ee   
Clearly, this expansion is well defined only if 
\be
\left|\frac{U_m(Y-Y_o)}{U_{m-1}}\right|=\left|\frac{u_m(Y-Y_o)}{Y_o u_{m-1}}\right|<< 1,\quad \forall\; m\geq 1.
\ee
In fact, there are three distinct requirements this superpotential must fulfill in general:

\begin{itemize}

\item[i)] {\em Asymptotic conditions}

The asymptotic form (\ref{asymptotic-superpotential}) of the superpotential determines the asymptotic behavior of the coefficients $u_0(\f)$ and $u_1(\f)$ to be
\bea\label{u-asymptotics}
&&u_0(\f)\sim \left(z-1+d(1+\m\x)\right)e^{-\x\f},\NO\\
&&u_1(\f)\sim \frac12 (z-1)e^{-\x\f}.
\eea 
More generally,
\be
u_n(\f)\sim (-1)^n\left(\frac{z-1}{2\e}\right)^ne^{-\x \f}w_n,
\ee
where $w_n$ are the coefficients of the Taylor expansion (\ref{supI-sol}).

\item[ii)] {\em Hamilton-Jacobi equation}

Inserting the formal Taylor expansion in the superpotential equation (\ref{master}) leads to a set of equations for the coefficients $u_m(\f)$. The first three orders in $Y-Y_o$ give respectively
\bea
&&\frac{1}{2\a}\left(u_0'+\frac{Z'}{Z}u_1\right)^2+\left(-\frac{2\e}{z-1}+\frac{2(d-1)}{d}\right)u_1^2
+\frac2d u_0u_1-\frac{d+1}{2d}u_0^2=\frac12 V-\frac{z-1}{4\e}WZ^{-1},\label{0th-order-0} \NO \\ \\
&&\left[\frac{2}{\a}\left(u_0'+\frac{Z'}{Z}u_1\right)\frac{Z'}{Z}-\frac{8\e}{z-1}u_1+\frac{4}{d}\left(u_0+2(d-1)u_1\right)\right]u_2\NO\\
&&+\frac{1}{\a}\left(u_0'+\frac{Z'}{Z}u_1\right)\left(u_1'+\frac{Z'}{Z}u_1\right)+\left(\frac{2(2d-1)}{d}-\frac{2\e}{z-1}\right)u_1^2-\frac{d-1}{d}u_0u_1=-\frac{z-1}{4\e}WZ^{-1}.\label{0th-order-1}\NO \\ \\
&&\left[\frac{2}{\a}\left(u_0'+\frac{Z'}{Z}u_1\right)\frac{Z'}{Z}-\frac{8\e}{z-1}u_1+\frac{4}{d}\left(u_0+2(d-1)u_1\right)\right]3u_3\NO\\
&&+\frac{2}{\a}\left(u_0'+\frac{Z'}{Z}u_1\right)u_2'+16\left(\frac{1}{4\a}\left(\frac{Z'}{Z}\right)^2+\frac{d-1}{d}-\frac{\e}{z-1}\right)u_2^2\NO\\
&&+\left(\frac1\a\left(u_0'+\frac{Z'}{Z}u_1+u_1'+\frac{Z'}{Z}u_1\right)\frac{Z'}{Z}-\frac{d-3}{2d}u_0+\left(\frac{8(d-1)+3}{d}-\frac{4\e}{z-1}\right)u_1\right)4u_2\NO\\
&&+\frac1\a\left(u_1'+\frac{Z'}{Z}u_1\right)^2+\left(3-\frac1d\right)u_1^2=0.\label{0th-order-2}
\eea
Note that these equations alone do not completely determine the functions $u_n(\f)$ in the Taylor expansion of the superpotential.

\item[iii)] {\em Consistency of the Taylor expansion}

A final condition on the functions $u_n(\f)$ is imposed by requiring that the Taylor expansion is consistent with the asymptotic expansion. To derive this consistency condition we need to write the flow equations (\ref{pp-flow}) and (\ref{pp-flow-XY}) in terms of the functions $u_n(\f)$, namely
\bea
\dot X&=&-\frac1\a e^{\x X}\left(u_0'+\frac{Z'}{Z}u_1\right)
-\frac1\a e^{\x X}\left(u'_1+(u_1+2u_2)\frac{Z'}{Z}\right)Y_o^{-1}(Y-Y_o)+\co(Y-Y_o)^2,\NO\\
\dot Y-\dot Y_o&=&-\frac12 Y_o e^{\x X}\left(\frac{2}{\a}\left(u_0'+\frac{Z'}{Z}u_1\right)\frac{Z'}{Z}-\frac{8\e}{z-1}u_1+\frac{4}{d}\left(u_0+2(d-1)u_1\right)\right)\NO\\
&&+e^{\x X}\left(-\frac2d\left(u_0+u_1+4(d-1)(u_1+u_2)\right)+\frac{4\e}{z-1}(u_1+2u_2)\right.\NO\\
&&\left.-\frac{2\x}{\a}\left(u_0'+\frac{Z'}{Z}u_1\right)-\frac1\a\left(u_1'+\frac{Z'}{Z}(u_1+2u_2)\right)\frac{Z'}{Z}\right)(Y-Y_o)+\co(Y-Y_o)^2,\label{supIII-flow-XY}
\eea
and
\bea
\dot f&=&e^{\x X}\left(\frac1d\left(u_0+2(d-1)u_1\right)+\frac{\x}{\a}\left(u'_0+\frac{Z'}{Z}u_1\right)\right)\NO\\
&&+e^{\x X}\left(\frac{1}{d}\left((2d-1)u_1+4(d-1)u_2\right)+\frac{\x}{\a}\left(u'_1+\frac{Z'}{Z}(u_1+2u_2)\right)\right)Y_o^{-1}(Y-Y_o)\NO\\
&&+\co(Y-Y_o)^2,\NO\\
\dot h&=&e^{\x X}\left(\frac1d(u_0-2u_1)+\frac{\x}{\a}\left(u_0'+\frac{Z'}{Z}u_1\right)\right)\NO\\
&&-e^{\x X}\left(\frac1d(u_1+4u_2)-\frac{\x}{\a}\left(u_1'+\frac{Z'}{Z}(u_1+2u_2)\right)\right)Y_o^{-1}(Y-Y_o)+\co(Y-Y_o)^2,\NO\\
\frac{\dot a}{a}&=&\frac{2\e}{z-1}e^{\x X}\left(u_1+2u_2Y_o^{-1}(Y-Y_o)+\co(Y-Y_o)^2\right),\label{supIII-flow}
\eea
The consistency condition comes from the inhomogeneous term in the flow equation for $Y-Y_o$, which must vanish identically in order for the Taylor expansion to be well defined. Note that if the inhomogeneous term is not zero then $Y_o^{-1}(Y-Y_o)$ does not vanish asymptotically. This condition holds automatically for the asymptotic form (\ref{u-asymptotics}) of $u_0$ and $u_1$ and the leading form of $Z$ in (\ref{exp-potentials}), but it imposes a non-trivial condition on the subleading terms of $u_0$ and $u_1$ (or of $Z$ if one views this as an equation for $Z$.)

\end{itemize}

These three conditions on the superpotential completely determine the coefficients $u_n(\f)$ in the Taylor expansion. Notice that the inhomogeneous term in the $Y-Y_o$ flow equation is identical to the coefficient of $u_2$ and $u_3$ in (\ref{0th-order-1}) and (\ref{0th-order-2}) respectively. Since this term must vanish, $u_2$ is eliminated from (\ref{0th-order-1}) and $u_3$ from (\ref{0th-order-2}). Equations (\ref{0th-order-0}) and (\ref{0th-order-1}) then become two equations for $u_0(\f)$ and $u_1(\f)$, while (\ref{0th-order-2}) becomes a Riccati equation for $u_2(\f)$. Higher order terms are determined by first order linear equations that are derived from higher orders in $Y-Y_o$ of the HJ equation. Since $u_0(\f)$ and $u_1(\f)$ must also satisfy the constraint coming from the consistency of the Taylor expansion, there are three equations for these two functions, and hence there is an implicit constraint on the three potentials $V$, $W$ and $Z$. The three equations are\footnote{Later on we will impose one more condition on the functions $u_0(\f)$ and $u_1(\f)$, namely (\ref{u-condition}), so that there is effectively only one arbitrarily specifiable function. This condition, however, is only necessary for our algorithm to apply in its simplest form and it can in principle be relaxed.} 
\begin{align}
\label{potentials}
\boxed{
\begin{aligned}
&V=\frac{1}{\a}\left(u_0'+\frac{Z'}{Z}u_1\right)(u_0'-2u_1')-\frac{4\e}{z-1}u_1^2
-\frac1d(u_0-2u_1)\left((d+1)u_0-2u_1\right),\\
&W=-\frac{4\e}{z-1}Z\left(\frac{1}{\a}\left(u_0'+\frac{Z'}{Z}u_1\right)\left(u_1'+\frac{Z'}{Z}u_1\right)+\left(\frac{2(2d-1)}{d}-\frac{2\e}{z-1}\right)u_1^2-\frac{d-1}{d}u_0u_1\right),\\
&\frac{2}{\a}\left(u_0'+\frac{Z'}{Z}u_1\right)\frac{Z'}{Z}-\frac{8\e}{z-1}u_1+\frac{4}{d}\left(u_0+2(d-1)u_1\right)=0.
\end{aligned} }
\end{align}
However, in a bottom up approach the potentials $V$, $W$ and $Z$ are a priori unspecified and so we can in fact {\em define} the potentials in terms of the two functions $u_0(\f)$ and $u_1(\f)$ of the superpotential, which are only subject to the asymptotic conditions (\ref{u-asymptotics}). Given these functions, the Riccati equation (\ref{0th-order-2}) can be solved for $u_2$ and the higher order coefficients $u_n$ are determined by solving the linear equations coming from the higher order terms in the Taylor expansion of the HJ equation. The leading asymptotic form of these will be identical to the one obtained from the superpotential I above, but they can potentially differ at subleading orders due to the choice of subleading terms in  $u_0(\f)$ and $u_1(\f)$. Finally, the Fefferman-Graham asymptotic expansions are obtained by integrating the flow equations (\ref{supIII-flow-XY}) and (\ref{supIII-flow}). Note that since the leading asymptotic form of these expansions is the same as for the superpotential I above, the non-normalizable modes remain the same as in that case. Moreover, since the form of dilatation operator is determined by the non-normalizable modes, it follows that the analysis of the finite part of the asymptotic complete integral, and hence the normalizable modes, are again the same as in the superpotential I case. The only exception occurs in the case $\m=0$, where the subleading terms in $u_0(\f)$ and $u_1(\f)$ determine the asymptotic form of the scalar. But the corresponding normalizable and non-normalizable modes can be  determined by the same procedure in that case too.

\section{Recursive solution of the HJ equation for asymptotically locally Lif backgrounds}
\label{algorithm}

In the previous section we considered exclusively homogeneous backgrounds, for which we obtained the general asymptotic solution of the Hamilton-Jacobi equation, the Fefferman-Graham expansions, as well as the non-normalizable and normalizable modes corresponding respectively to the sources and 1-point functions of the dual operators. We now extend this analysis to incorporate sources with arbitrary spatial and time dependence. Note that the solution of the HJ equation we obtained in Section \ref{bgrnds} is still relevant in the presence of arbitrary spacetime-dependent sources, since it appears as the leading zero derivative solution of the HJ equation. What we will be mainly concerned with in this section, therefore, is the systematic construction of the subleading terms in the HJ solution that contain transverse derivatives.

\subsection{Locally Lif boundary conditions} 
\label{bcs}

Before we address the derivative terms in the solution of the HJ equation, however, we need to identify the most general spacetime-dependent sources allowed by Lifshitz boundary conditions. To this end we consider again the most general diffeomorphism and gauge invariant solution of the general HJ equation (\ref{constraints0}), containing no transverse derivatives. As we have argued in the previous section this takes the form 
\be\label{HJ-zero-order-solution}
\cs\sub{0}=\frac{1}{\k^2}\int d^{d+1}x\sqrt{-\g}U(\f,B_iB^i),
\ee
where $U(X,Y)$ is some superpotential. Note that $U(1)$ gauge invariance dictates that it is $B_iB^i$ that should appear in the superpotential and not $A_iA^i$, and so $\cs\sub{0}$ in fact contains transverse derivatives, but in a rather trivial way.    

The relation between the superpotential $U(X,Y)$ and the asymptotic form of the fields is provided by the flow equations (\ref{flow-eqs}), which now become 
\bea
&&\dot\g_{ij}=4e^{-d\x X}\left(U_YB_iB_j
+\left(\frac{\a_\x}{2d\a}U+\frac{\x}{2\a}U_X
-\frac{\a_\x+d^2\x^2}{d\a}YU_Y\right)\g_{ij}\right),\NO\\
&&\dot A_i=-e^{-d\x X}Z_\x^{-1}(X)U_YB_i,\NO\\
&&\dot\f=-\frac{1}{\a}e^{-d\x X}
\left(U_X-(d+1)\x U+2\x YU_Y\right),\NO\\
&&\dot\om=-2e^{-d\x\f}W_\x^{-1}(\f)D_i\left(U_Y B^i\right).\label{leading-flow-eqs}
\eea  
In order to accommodate anisotropic solutions we parameterize the induced fields on the radial slice 
$\S_r$ in terms of fields compatible with the anisotropy. In particular, we decompose the induced metric $\g_{ij}$ and vector field $A_i$ as\footnote{This is merely a field redefinition, as is the parameterization of the metric in terms of vielbeins in \cite{Ross:2011gu}, since the spin connection is not treated as an independent field. We thank Simon Ross for useful comments on this.} 
\bea\label{decomposition}
&&\g_{ij}dx^idx^j=-(n^2-n_a n^a)dt^2+2n_adtdx^a+\s_{ab}dx^a dx^b,\NO\\
&&A_idx^i=a dt+ A_a dx^a,\quad B_idx^i=b dt+ B_a dx^a,\quad b=a-\pa_t\om,\quad B_a=A_a-\pa_a\om,
\eea
where the indices $a,b$ run from 1 to $d$ and $\s_{ab}(r,t,x)$, $n_a(r,t,x)$, $n(r,t,x)$, $a(r,t,x)$ and $A_a(r,t,x)$ 
are the fields in terms of which we will parameterize the dynamics. In terms of the anisotropic fields the flow equations (\ref{leading-flow-eqs}) take the form
\bea
&&\pa_r n^2=4e^{-d\x\f}\left(-U_Y(b-n^aB_a)^2
+\left(\frac{\a_\x}{2d\a}U+\frac{\x}{2\a}U_X
-\frac{\a_\x+d^2\x^2}{d\a}YU_Y\right)n^2\right),\NO\\
&&\dot n_a=4e^{-d\x\f}\left(U_YbB_a
+\left(\frac{\a_\x}{2d\a}U+\frac{\x}{2\a}U_X
-\frac{\a_\x+d^2\x^2}{d\a}YU_Y\right)n_a\right),\NO\\
&&\dot\s_{ab}=4e^{-d\x\f}\left(U_YB_aB_b
+\left(\frac{\a_\x}{2d\a}U+\frac{\x}{2\a}U_X
-\frac{\a_\x+d^2\x^2}{d\a}YU_Y\right)\s_{ab}\right),\NO\\
&&\dot a=-e^{-d\x\f}Z_\x^{-1}(\f)U_Yb,\NO\\
&&\dot A_a=-e^{-d\x\f}Z_\x^{-1}(\f)U_YB_a,\NO\\
&&\dot \f=-\frac{1}{\a}e^{-d\x\f}
\left(U_X-(d+1)\x U+2\x YU_Y\right),\NO\\
&&\dot\om \sim 0,\label{leading-flow-eqs-anisotropic}
\eea 
where we have used the leading asymptotic form of the flow equation for the St\"uckelberg field. 
 
The Lifshitz metric (\ref{Lif-sol}) implies that the most general asymptotic form of the fields $n$ and $n_a$ compatible with locally Lif asymptotics is
\be\label{asymp-anisotropic}
n\sim e^{r z} n_{(0)}(t,x),\quad
n_{a}\sim e^{r(z+1-\b)} n_{(0)a}(t,x),\quad
\s_{ab}\sim e^{2 r} g_{(0)ab}(t,x),
\ee
where $n_{(0)}(t,x)$, $n_{(0)a}(t,x)$, and $g_{(0)ab}(t,x)$ are arbitrary functions of the transverse coordinates and the constant $\b$ is to be determined. Since $\g_{tt}=-n^{2}+n_{a}n^{a}$, requiring that $n_{a}n^{a}$ is at most divergent as $n^{2}$ imposes the restriction
\be
\b \geq 0.
\ee
Inserting the asymptotic behaviors (\ref{asymp-anisotropic}) in the flow equations (\ref{leading-flow-eqs-anisotropic}) 
leads to a set of asymptotic conditions on the superpotential, namely  
\bea
&&4e^{-d\x X}\left(\frac{\a_\x}{2d\a}U+\frac{\x}{2\a}U_X
-\frac{\a_\x+d^2\x^2}{d\a}YU_Y\right)\sim 2,\label{asymptotic-conditions-lifshitz-1}\\
&&|U_Y B_aB_b|<< e^{d\x X}|\s_{ab}|,\label{asymptotic-conditions-lifshitz-2}\\
&&-4e^{-d\x X}U_Y(b-n^aB_a)^2\sim 2(z-1)n^2,\label{asymptotic-conditions-lifshitz-3}\\
&&4e^{-d\x X}U_YbB_a\sim (z-1-\b)n_a.\label{asymptotic-conditions-lifshitz-4}
\eea
Using the inverse metric
\be
\g^{-1}=\left(\begin{matrix}
-\frac{1}{n^2} & \frac{n^a}{n^2} \\
\frac{n^a}{n^2} & \s^{ab}-\frac{n^an^b}{n^2}
\end{matrix}\right),
\ee 
(\ref{asymptotic-conditions-lifshitz-3}) implies
\be\label{U-1}
YU_Y=U_YB_iB^i=U_YB^aB_a-U_Y\frac{(b-n^aB_a)^2}{n^2}\sim
U_YB^aB_a+\frac12(z-1)e^{d\x X}\sim \frac12(z-1)e^{d\x X}, 
\ee
where we have used  (\ref{asymptotic-conditions-lifshitz-2}) in the last step. Inserting this in (\ref{asymptotic-conditions-lifshitz-1}) gives 
\be\label{U-2}
\a_\x U+d\x U_X\sim \left(d\a+(z-1)(\a_\x+d^2\x^2)\right) e^{d\x X}.
\ee  
Moreover, using the leading form of the flow equation for $\om$ to replace $\dot a $ and $\dot A_a$ with $\dot b $ and $\dot B_a$ respectively in the vector flow equations, we see that the latter require that the time component, $b$, and the spatial component, $B_a$, behave in the same way asymptotically, which we parameterize as
\be
b\sim b\sub{0}(t,x)e^{\e r},\quad B_a\sim B\sub{0}_a(t,x)e^{\e r},
\ee
where $b\sub{0}(t,x)$ and $B\sub{0}_a(t,x)$ are arbitrary functions of the transverse coordinates and the exponent $\e$ is as yet unspecified. Using this asymptotic form of $B_a$ in the vector flow equation together with (\ref{U-1}) we find 
\be\label{U-3}
Y\sim -\frac{z-1}{2\e}Z_\x^{-1}(X)=:Y_o(X),
\ee
which is the asymptotic constraint (\ref{Yo}) we found for the homogeneous solutions. Moreover,
\be
n^aB_a\sim n\sub{0}^aB\sub{0}_ae^{(z-1-\b+\e)r},
\ee
and so
\be
Y=B_aB^a-\frac{(b-n^aB_a)^2}{n^2}\sim -\frac{(b-n^aB_a)^2}{n^2}\sim Y\sub{0}(t,x) e^{\d r},
\ee
where, assuming $B\sub{0}_a\neq 0$,
\be
\d=\left\{\begin{array}{lll}
2(\e-z), &\quad z-1-\b < 0, &\quad Y\sub{0}=-b\sub{0}^2/n\sub{0}^2,\\
2(\e-1-\b), &\quad z-1-\b>0, &\quad Y\sub{0}=-(n\sub{0}^aB\sub{0}_a)^2/n\sub{0}^2,\\
2(\e-z), &\quad z-1-\b=0, &\quad Y\sub{0}=-(b\sub{0}- n\sub{0}^aB\sub{0}_a)^2/n\sub{0}^2.
\end{array}\right.
\ee
However, (\ref{U-1}) implies that, if $B\sub{0}_a\neq 0$, in order to satisfy (\ref{asymptotic-conditions-lifshitz-2}) we must demand that
\be\label{delta}
\d>2\e-2,
\ee
which requires that either $z<1$ or $\b<0$. The latter contradicts the above asymptotic conditions and so it is not an acceptable solution. Moreover, we have argued that $z<1$ corresponds to the solutions I and II of the NEC in (\ref{ranges}) and since $\th\geq d+z$ in those cases, there are no well-defined asymptotic expansions. A possible exception is the marginal case $\th=d+z$ with $0\leq z <1$, but we will not consider this here. The  only alternative, therefore, is to require
\be\label{Ba=0}
B\sub{0}_a(t,x)=0, 
\ee 
in which case 
\be
\d=2(\e-z),\quad Y\sub{0}=-b\sub{0}^2/n\sub{0}^2.
\ee
Note that the inequality (\ref{delta}) need not hold in this case since (\ref{asymptotic-conditions-lifshitz-2}) is automatically satisfied. Moreover, (\ref{asymptotic-conditions-lifshitz-4}) determines
\be
(z-1-\b)n\sub{0}_a=0,
\ee
in this case, which can be solved by either 
setting $\b=z-1$ and leaving $n\sub{0}_a(t,x)$  arbitrary, or by setting $n\sub{0}_a(t,x)=0$ in which case $\b$ does not arise at all. Since we want to keep all possible sources compatible with Lif asymptotics, we set 
\be 
\b=z-1, 
\ee 
and keep $n\sub{0}_a(t,x)$ unconstrained. 

To summarize, from this asymptotic analysis we have determined that locally Lifshitz boundary conditions amount to the gauge-invariant asymptotic constraint
\be\label{Lifshitz-constraint}\boxed{
B_i\sim B_{oi}=\sqrt{-Y_o(X)} \;\bb n_i,}
\ee
where $\bb n_i=(n,0)$ is the unit normal to the constant time surfaces and $Y_o(X)$ is defined in (\ref{U-3}). This is a covariant way of writing the scalar constraint (\ref{U-3}) and the spatial vector constraint (\ref{Ba=0}). This covariant form of the asymptotic constraint allows us to obtain the corresponding asymptotic form of the covariant momenta 
\bea\label{asymptotic-Lifshitz-momenta}
&&K_{ij}\sim \g_{ij}-2\e Z_\x(\f)B_iB_j,\NO\\
&&\p^{ij}\sim \frac{1}{2\k^2}\sqrt{-\g}e^{d\x\f}\left(\left(d+d\m\x+z-1\right)\g^{ij}+2\e Z_\x(\f)B^iB^j\right),\NO\\
&&\p^i\sim -\frac{1}{2\k^2}\sqrt{-\g}e^{d\x\f}Z_\x(\f)4\e B^i,\NO\\
&&\p_\f\sim \frac{1}{2\k^2}\sqrt{-\g}e^{d\x\f}\left(2d\x\left(d+z\right)-2\a_\x\m\right),
\eea
which can be integrated to obtain the leading asymptotic from of the zero order solution of the Hamilton-Jacobi equation:
\be\label{asymptotic-sol}\boxed{
\cs\sub{0}\sim\frac{1}{\k^2}\int_{\S_r} d^{d+1}x\sqrt{-\g}e^{d\x\f}\left(d(1+\m\x)+\frac12(z-1)-\e Z_\x(\f)B_iB^i\right).}
\ee
The asymptotic form of the momentum conjugate to the St\"uckelberg field $\om$ following from this HJ solution is 
\be
\p_\om\sim \frac{\d \cs\sub{0}}{\d\om}\sim -\frac{2}{\k^2}\e\sqrt{-\g}D_i\left(e^{d\x\f}Z_\x(\f)B^i\right),
\ee
which as we shall see shortly is subleading relative to the rest of the momenta in a precise sense that we will specify. In terms of the superpotential, the asymptotic conditions (\ref{asymptotic-Lifshitz-momenta}) imply the following conditions on the superpotential $U(X,Y)$ and its first derivatives: 
\begin{align}
\label{U-asymptotics}
\boxed{
\begin{aligned}
& U(X,Y_o(X))\sim e^{d\x X}\left(d(1+\m\x)+z-1\right),\\
& U_Y(X,Y_o(X))\sim -\e e^{d\x X}Z_\x(X),\\
& U_X(X,Y_o(X))\sim e^{d\x X}\left(-\m\a_\x+d\x(d+z)\right).
\end{aligned}}
\end{align}
Inserting these in the superpotential equation (\ref{master}) one recovers the relations (\ref{Lif-parameters}) between the various parameters. As we have seen from the homogeneous solutions in Section \ref{bgrnds}, there are additional constraints on the superpotential at subleading orders, coming from the consistency of the Taylor expansion in $B_i-B_{oi}$. Moreover, there are more sources appearing at subleading order due to the constraint (\ref{Lifshitz-constraint}). We will revisit these points later on, when we develop the recursive algorithm for determining the subleading terms of the HJ solution and when discussing the general Fefferman-Graham expansions.

\subsection{Graded expansion in eigenfunctions of the derivative and gradation operators}

A solution of the HJ equation of the form (\ref{HJ-zero-order-solution}) captures all zero derivative terms. However, the general asymptotic solution of the HJ equation with spacetime-dependent sources contains asymptotically subleading terms with transverse derivatives acting on the induced fields.   
In order to account for these terms in a systematic way, and to consistently impose Lif boundary conditions, we are going to seek a solution in the form of a covariant expansion in eigenfunctions of a suitable functional operator. This is analogous to the expansion in the dilatation operator for asymptotically locally AdS spaces introduced in \cite{Papadimitriou:2004ap} or its generalization to asymptotically non AdS -- but relativistic --  backgrounds in \cite{Papadimitriou:2011qb}. The anisotropy introduced by the Lif boundary conditions, however, necessitates some generalization of the formalism. The dilatation operator method has been extended to Lifshitz backgrounds without a linear dilaton in the vielbein formalism \cite{Ross:2011gu} and in Lifshitz gravity \cite{Griffin:2011xs}. However, the expansion we develop is both fully covariant and applicable in the presence of a linear dilaton, which is necessary in order to accommodate hvLf backgrounds.

The leading order solution of the Hamilton-Jacobi equation in this covariant expansion is of the form (\ref{HJ-zero-order-solution}). Since the superpotential $U(\f,B^2)$ depends on the choice of the potentials $V(\f)$, $Z(\f)$ and $W(\f)$ in the Lagrangian, which we want to keep as general as possible at this stage, we demand that (\ref{HJ-zero-order-solution}) be an eigenfunction of the functional operator we expand in for any choice of $U(\f,B^2)$. There are two operators that satisfy this criterion, namely
\be\label{operators}\boxed{
\Hat\d:= \int d^{d+1}x\left(2\g_{ij}\frac{\d}{\d\g_{ij}}+B_i\frac{\d}{\d B_i}\right),\quad \d_B:=\int d^{d+1}x\left(2Y^{-1}B_iB_j\frac{\d}{\d\g_{ij}}+B_i\frac{\d}{\d B_i}\right),}
\ee
for which it is easy to check that
\be
\Hat\d \cs\sub{0}= (d+1) \cs\sub{0},\quad \d_B \cs\sub{0}= \cs\sub{0},
\ee
and so $\cs\sub{0}$ is an eigenfunction of both $\Hat\d$ and $\d_B$, with respective eigenvalues $d+1$ and $1$, for any $U(\f,B^2)$. Crucially, these operators commute 
\be
[\Hat\d,\d_B]=0,
\ee
which means that if $\cs\sub{2k}$ is an eigenfunction of $\Hat\d$, then so is $\d_B\cs\sub{2k}$ with the same eigenvalue. This allows us to expand $\cs$ covariantly in a {\em double} expansion. 

In order to construct the covariant expansion, we need to understand the structure of the eigenfunctions of $\Hat\d$ and $\d_B$. As we have argued, any function of $B^2$ (and trivially of $\f$) is automatically an eigenfunction of both operators. It therefore remains to understand how these operators act on terms with transverse derivatives, $\pa_i$. From the structure of the Hamiltonian constraint follows that any derivative expansion of the Hamilton-Jacobi functional will contain only even number of derivatives. Covariance then requires that for every pair of derivatives there is either an inverse metric, $\g^{ij}$, or a factor of $B^iB^j$ with which the two derivatives are contracted. A simple counting exercise then shows that $\Hat\d$ counts the number of derivatives. Namely, any functional $\cs\sub{2k}$ containing $2k$ derivatives is an eigenfunction of $\Hat\d$ with eigenvalue $d+1-2k$, where $d+1$ is the contribution of the volume element. 

The eigenvalues of the operator $\d_B$ follow from the observation that it satisfies 
\be
\d_B\s^{ij}=0,
\ee  
where 
\be\label{projector}
\s^i_j:=\d^i_j-Y^{-1}B^iB_j,
\ee
is a projection operator: 
\be
\s^i_k\s^k_j=\s^i_j.
\ee
This implies that an eigenfunction $\cs\sub{2k}$ of $\Hat\d$ with $2k$ derivatives can be split in a sum of up to $k+1$ terms containing $0, 1, \ldots, k$ powers of $\s^{ij}$. This can be achieved systematically as follows. Terms in which all $2k$ derivatives are contracted with $B^i$ are eigenfunctions of $\d_B$ with eigenvalue $1-2k$, since every factor of $B^i$ contributes $-1$ to the eigenvalue and the $1$ comes from the volume element. Next, we consider terms where $2k-2$ derivatives are contracted with $B^i$ and $2$ derivatives are contracted with $\g^{ij}$. Such terms are not eigenfunctions of $\d_B$ but they can be written as a sum of two eigenfunctions of $\d_B$ with eigenvalues $1-2(k-1)$ and $1-2k$ by writing 
\be
\g^{ij}=\s^{ij} + Y^{-1}B^iB^j.
\ee
This process can be repeated for all terms with $2k$ derivatives in order to split $\cs\sub{2k}$ into a sum of eigenfunctions of $\d_B$ with eigenvalues $1-2\ell$, $\ell=0,1,\ldots,k$.  

This analysis shows that we can formally expand the solution of the Hamilton-Jacobi equation covariantly in a graded expansion in eigenfunctions of both $\Hat\d$ and $\d_B$, namely
\be\label{covariant-expansion-s}
\cs=\sum_{k=0}^\infty\cs\sub{2k}=\sum_{k=0}^\infty\sum_{\ell=0}^k\cs\sub{2k,2\ell},
\ee
where 
\be
\Hat\d\cs\sub{2k,2\ell}=(d+1-2k)\cs\sub{2k,2\ell},\quad  
\d_B\cs\sub{2k,2\ell}=(1-2\ell)\cs\sub{2k,2\ell},
\ee
and $\cs\sub{0,0}=\cs\sub{0}$ is given by (\ref{HJ-zero-order-solution}). We will refer to the operator $\Hat \d$ as the `derivative operator' since it counts transverse derivatives, while $\d_B$ we will call the `gradation operator'. It should be stressed, however, that there is an inherent assumption of locality for these expansions in local eigenfunctions of the operators $\Hat\d$ and $\d_B$ to be meaningful. This assumption is of course not valid for the finite part of the solution of the HJ equation, i.e. the renormalized 
on-shell action. However, this is of no concern right now. Our strategy is to develop a recursive algorithm that determines iteratively increasingly asymptotically subleading terms in the solution of the HJ equation {\em assuming locality}. This recursive procedure breaks down exactly at the order where the finite contribution to the solution occurs. This finite part is required in order for the asymptotic solution of the HJ equation to qualify as a complete integral, and it is necessary for the derivation of the Fefferman-Graham expansions and the identification of the normalizable modes. As in the case of homogeneous solutions in Section \ref{bgrnds}, the finite non-local part must be addressed separately, and it will be the main subject of Section \ref{ward}.

\vspace{0.2in}
\begin{flushleft}
{\bf Expansion of the canonical momenta}
\end{flushleft}
Since the canonical momenta are related to the solution of the Hamilton-Jacobi equation via (\ref{HJ-momenta}), one might expect that the momenta defined via
\be
\p\sub{2k,2\ell}^{ij}=\frac{\d\cs\sub{2k,2\ell}}{\d\g_{ij}},\quad
\p\sub{2k,2\ell}^i=\frac{\d\cs\sub{2k,2\ell}}{\d A_i},\quad
\p_\f\sub{2k,2\ell}=\frac{\d\cs\sub{2k,2\ell}}{\d\f},\quad
\p_\om\sub{2k,2\ell}=\frac{\d\cs\sub{2k,2\ell}}{\d\om},
\ee  
are also eigenfunctions of $\Hat\d$ and $\d_B$. This is in fact not true, and it should be emphasized that the subscripts in the momenta do not denote their eigenvalues under $\Hat\d$ and $\d_B$, since they are not eigenfunctions. The subscripts on the momenta instead indicate that they are gradients of the corresponding eigenfunctions $\cs\sub{2k,2\ell}$. The action of  $\Hat\d$ and $\d_B$ on these momenta can be  obtained using the commutation relations
\begin{align}
\begin{aligned}
&&&\left[\Hat\d,\frac{\d}{\d\g_{ij}}\right]=-2\frac{\d}{\d\g_{ij}}, &&\quad  \left[\d_B,\frac{\d}{\d\g_{ij}}\right]=-2Y^{-2}B^iB^jB_kB_l\frac{\d}{\d\g_{kl}},\\ 
&&&\left[\Hat\d,\frac{\d}{\d B_i}\right]=-\frac{\d}{\d B_i},&&\quad  \left[\d_B,\frac{\d}{\d B_i}\right]=-\frac{\d}{\d B_i}-4Y^{-1}\s^i_kB_l\frac{\d}{\d\g_{kl}},  \\ 
&&&\left[\Hat\d,\frac{\d}{\d \f}\right]=0,  &&\quad  \left[\d_B,\frac{\d}{\d \f}\right]=0,\\ 
&&&\left[\Hat\d,\frac{\d}{\d \om}\right]=-D_i\frac{\d}{\d B_i},&& \quad 
\left[\d_B,\frac{\d}{\d \om}\right]=-D_i\left(\frac{\d}{\d B_i}+4Y^{-1}\s^i_kB_l\frac{\d}{\d\g_{kl}}\right).
\end{aligned}
\end{align}
The results are summarized in Table \ref{operator-momenta}.
\begin{table}
\begin{center}
\begin{tabular}{|l|l|l|}
\hline \hline &&\\
 &$\Hat\d$ &  $\d_B$  \\\hline &&\\
 
 $\p\sub{2k,2\ell}^{ij}$ & $(d-1-2k)\p\sub{2k,2\ell}^{ij}$ & $(1-2\ell)\p\sub{2k,2\ell}^{ij}-2Y^{-2}B^iB^jB_kB_l\p\sub{2k,2\ell}^{kl}$\\&&\\
 
 $\p\sub{2k,2\ell}$ & $(d+1-2k)\p\sub{2k,2\ell}$ & $(1-2\ell)\p\sub{2k,2\ell}$\\&&\\
 
 $B_kB_l\p\sub{2k,2\ell}^{kl}$& $(d+1-2k)B_kB_l\p\sub{2k,2\ell}^{kl}$ & $(1-2\ell)B_kB_l\p\sub{2k,2\ell}^{kl}$\\&&\\
 
 $\p\sub{2k,2\ell}^{i}$ & $(d-2k)\p\sub{2k,2\ell}^{i}$ & $-2\ell\p\sub{2k,2\ell}^{i}-4Y^{-1}\s^i_kB_l\p\sub{2k,2\ell}^{kl}$ \\&&\\
 
 $B_k\p\sub{2k,2\ell}^{k}$ & $(d+1-2k)B_k\p\sub{2k,2\ell}^{k}$ & $(1-2\ell)B_k\p\sub{2k,2\ell}^{k}$ \\&&\\
 
 $\p_\f\sub{2k,2\ell}$ & $(d+1-2k)\p_\f\sub{2k,2\ell}$ & $(1-2\ell)\p_\f\sub{2k,2\ell}$ \\&&\\
 
 $\p_\om\sub{2k,2\ell}$ & $(d+1-2k)\p_\om\sub{2k,2\ell}-D_i\p\sub{2k,2\ell}^{i}$ & $(1-2\ell)\p_\om\sub{2k,2\ell}$ \\
 && $\phantom{space}-D_i\left(\p\sub{2k,2\ell}^{i}+4Y^{-1}\s^i_kB_l\p\sub{2k,2\ell}^{kl}\right)$\\
&&\\\hline
\end{tabular}
\end{center}
\caption{Action of the operators $\Hat\d$ and $\d_B$ on the canonical momenta.}
\label{operator-momenta}
\end{table}
From the expressions in Table \ref{operator-momenta}  the complete set of linearly independent eigenfunctions of both $\Hat\d$ and $\d_B$ that are linear in the canonical momenta can be constructed. These eigenfunctions are listed in Table \ref{momentum-eigenfunctions}, along with their eigenvalues under $\Hat\d$ and $\d_B$.    
\begin{table}
\begin{center}
\begin{tabular}{|l|c|c|}
\hline \hline &&\\
 &$\Hat\d$ &  $\d_B$  \\\hline &&\\
 
 $\s^i_k\s^j_l\p\sub{2k,2\ell}^{kl}$ & $d-1-2k$ & $1-2\ell$\\&&\\
 
 $\s^i_k B_l\p\sub{2k,2\ell}^{kl}$ & $d-2k$ & $2-2\ell$\\&&\\
 
 $B_kB_l\p\sub{2k,2\ell}^{kl}$& $d+1-2k$ & $1-2\ell$\\&&\\
 
 $\p\sub{2k,2\ell}$& $d+1-2k$ & $1-2\ell$\\&&\\
 
 $\cp\sub{2k,2\ell}^{i}:=\s^i_k\left(\p\sub{2k,2\ell}^{k}+2Y^{-1}B_l\p\sub{2k,2\ell}^{kl}\right)$ & $d-2k$ & $-2\ell$\\&&\\
 
 $B_k\p\sub{2k,2\ell}^{k}$ & $d+1-2k$ & $1-2\ell$\\&&\\
 
 $\p_\f\sub{2k,2\ell}$ & $d+1-2k$ & $1-2\ell$ \\&&\\
  
  $\p_\om\sub{2k,2\ell}-D_i\p\sub{2k,2\ell}^i$ & $d+1-2k$ & $1-2\ell$ \\
&&\\\hline
\end{tabular}
\end{center}
\caption{The complete set of simultaneous eigenfunctions of $\Hat\d$ and $\d_B$ linear in the canonical momenta, along with their eigenvalues. }
\label{momentum-eigenfunctions}
\end{table}
The eigenfunctions in Table \ref{momentum-eigenfunctions} in turn allow us to decompose any quantity that involves the canonical momenta in terms of these eigenfunctions. For example, the metric and vector momenta can be decomposed in terms of eigenfunctions of $\Hat\d$ and $\d_B$ as follows: 
\bea\label{mom-dec}
\p\sub{2k,2\ell}^{ij}&=&\left(\s^i_k+Y^{-1}B^iB_k\right)\left(\s^j_l+Y^{-1}B^jB_l\right)\p\sub{2k,2\ell}^{kl}\NO\\
&=&\s^i_k\s^j_l\p\sub{2k,2\ell}^{kl}+Y^{-1}\left(\s^i_kB^j+\s^j_kB^i\right)B_l\p\sub{2k,2\ell}^{kl}+Y^{-2}B^iB^jB_kB_l\p\sub{2k,2\ell}^{kl},\NO\\\NO\\
\p\sub{2k,2\ell}^{i}&=&\left(\s^i_k+Y^{-1}B^iB_k\right)\p\sub{2k,2\ell}^{k}\NO\\
&=&\cp\sub{2k,2\ell}^i-2Y^{-1}\s^i_kB_l\p\sub{2k,2\ell}^{kl}+Y^{-1}B^iB_k\p\sub{2k,2\ell}^{k},\NO\\\NO\\
D_i\p\sub{2k,2\ell}^i&=&D_i\cp\sub{2k,2\ell}^i-2D_i\left(Y^{-1}\s^i_kB_l\p\sub{2k,2\ell}^{kl}\right)+D_i\left(Y^{-1}B^iB_k\p\sub{2k,2\ell}^{k}\right),
\eea
where the quantity $\cp\sub{2k,2\ell}^i$ is defined in Table \ref{momentum-eigenfunctions}. For future reference we decompose all scalar quantities that are quadratic in the canonical momenta in terms of the eigenfunctions of these operators in Table \ref{quadratic-momenta}. We will need these eigenfunctions in the next subsection in order to analyze the Hamiltonian constraint and to develop the recursion algorithm.
\begin{table}
\begin{center}
\begin{tabular}{|l|c|c|}
\hline \hline &&\\
 &$\Hat\d$ &  $\d_B$  \\\hline &&\\
 
 $\frac{1}{\sqrt{-\g}}\s^i_k\s^j_l\p\sub{2k,2\ell}^{kl}\p\sub{2k',2\ell'}_{ij}$ & $d+1-2k-2k'$ & $1-2\ell-2\ell'$\\&&\\
 
 $\frac{1}{\sqrt{-\g}}Y^{-2}B_iB_jB_kB_l\p\sub{2k,2\ell}^{ij}\p\sub{2k',2\ell'}^{ij}$ & $d+1-2k-2k'$ & $1-2\ell-2\ell'$\\&&\\
 
 $\frac{1}{\sqrt{-\g}}\s_{ij}B_k\p\sub{2k,2\ell}^{ik}B_l\p\sub{2k',2\ell'}^{jl}$ & $d+1-2k-2k'$ & $3-2\ell-2\ell'$\\&&\\
 
 $\frac{1}{\sqrt{-\g}}\cp\sub{2k,2\ell}_kB_l\p\sub{2k',2\ell'}^{kl}$ & $d+1-2k-2k'$ & $1-2\ell-2\ell'$\\&&\\
 
 $\frac{1}{\sqrt{-\g}}\cp\sub{2k,2\ell}^i\cp\sub{2k',2\ell'}_i$ & $d+1-2k-2k'$ & $-1-2\ell-2\ell'$\\&&\\
 
  $\frac{1}{\sqrt{-\g}}Y^{-1}B_iB_j\p\sub{2k,2\ell}^{i}\p\sub{2k',2\ell'}^{j}$ & $d+1-2k-2k'$ & $1-2\ell-2\ell'$\\&&\\
 
 $\frac{1}{\sqrt{-\g}}\p_\f\sub{2k,2\ell}\p_\f\sub{2k',2\ell'}$ & $d+1-2k-2k'$ & $1-2\ell-2\ell'$\\&&\\
 
 $\frac{1}{\sqrt{-\g}}D_i\cp\sub{2k,2\ell}^iD_j\cp\sub{2k',2\ell'}^j$ & $d-1-2k-2k'$ & $-1-2\ell-2\ell'$\\&&\\
 
 $\frac{1}{\sqrt{-\g}}D_i\cp\sub{2k,2\ell}^iD_j\left(Y^{-1}\s^j_kB_l\p\sub{2k',2\ell'}^{kl}\right)$ & $d-1-2k-2k'$ & $1-2\ell-2\ell'$\\&&\\
 
 $\frac{1}{\sqrt{-\g}}D_i\cp\sub{2k,2\ell}^iD_j\left(Y^{-1}B^jB_k\p\sub{2k',2\ell'}^{k}\right)$ & $d-1-2k-2k'$ & $-1-2\ell-2\ell'$\\&&\\
 
 $\frac{1}{\sqrt{-\g}}D_i\left(Y^{-1}\s^i_kB_l\p\sub{2k,2\ell}^{kl}\right)D_j\left(Y^{-1}\s^j_pB_q\p\sub{2k',2\ell'}^{pq}\right)$ & $d-1-2k-2k'$ & $3-2\ell-2\ell'$\\&&\\
 
 $\frac{1}{\sqrt{-\g}}D_i\left(Y^{-1}\s^i_kB_l\p\sub{2k,2\ell}^{kl}\right)D_j\left(Y^{-1}B^jB_p\p\sub{2k',2\ell'}^{p}\right)$ & $d-1-2k-2k'$ & $1-2\ell-2\ell'$\\&&\\
 
 $\frac{1}{\sqrt{-\g}}D_i\left(Y^{-1}B^iB_k\p\sub{2k,2\ell}^{k}\right)D_j\left(Y^{-1}B^jB_l\p\sub{2k',2\ell'}^{l}\right)$ & $d-1-2k-2k'$ & $-1-2\ell-2\ell'$\\
 
&&\\\hline
\end{tabular}
\end{center}
\caption{The complete set of simultaneous scalar eigenfunctions of $\Hat\d$ and $\d_B$ that are quadratic in the canonical momenta, along with their eigenvalues. }
\label{quadratic-momenta}
\end{table}

\begin{flushleft}
{\bf Expansion of the first class constraints}
\end{flushleft}
In order to develop a recursive algorithm for solving the Hamilton-Jacobi equations in terms of eigenfunctions of the derivative and gradation operators we must expand the first class constraints (\ref{constraints}) in eigenfunctions of these operators. The momentum and $U(1)$ gauge constraints are linear in the momenta and so they can be decomposed in eigenfunctions of $\Hat\d$ and $\d_B$ using the eigenfunctions in Table \ref{momentum-eigenfunctions}. The Hamiltonian constraint, however, is quadratic in the momenta and the eigenfunctions in Table \ref{quadratic-momenta} are required instead. Let us consider each constraint in turn.

\begin{flushleft}
{\em $U(1)$ constraint:}
\end{flushleft}
The $U(1)$ constraint  
\be
\p_\om-D_i\p^i=0, 
\ee
can be immediately decomposed in eigenfunctions of $\Hat\d$ and $\d_B$ using the last eigenfunction in Table \ref{momentum-eigenfunctions}. Namely,
\be
\sum_{k,\ell}\left(\p_\om\sub{2k,2\ell}-D_i\p\sub{2k,2\ell}^i\right)=0,
\ee
and hence 
\be
\p_\om\sub{2k,2\ell}=D_i\p\sub{2k,2\ell}^i,\quad \forall k,\ell. 
\ee

\begin{flushleft}
{\em Momentum constraint:}
\end{flushleft}
Using the $U(1)$ constraint we can write the momentum constraint in the form
\be
-2D_j\p^{ji}+F^i{}_j\p^j+\p_\f\pa^i\f-B^iD_j\p^j=0,
\ee
which can be expanded in eigenfunctions of $\Hat\d$ so that for all $k$
\be
-2D_j\p\sub{2k}^{ji}+F^i{}_j\p\sub{2k}^j+\p_\f\sub{2k}\pa^i\f-B^iD_j\p\sub{2k}^j=0.
\ee
Using the decomposition of the momenta in eigenfunctions of both the derivative and gradation operators in (\ref{mom-dec}), this can be in turn written as 
\bea
&&-2\sum_{\ell=0}^kD_j\left(\s^i_k\s^j_l\p\sub{2k,2\ell}^{kl}+Y^{-1}\left(\s^i_kB^j+\s^j_kB^i\right)B_l\p\sub{2k,2\ell}^{kl}+Y^{-2}B^iB^jB_kB_l\p\sub{2k,2\ell}^{kl}\right)\NO\\
&&+\left(\s^i_p+Y^{-1}B^iB_p\right)F^p{}_j\sum_{\ell=0}^k\left(\cp\sub{2k,2\ell}^j-2Y^{-1}\s^j_kB_l\p\sub{2k,2\ell}^{kl}+Y^{-1}B^jB_k\p\sub{2k,2\ell}^{k}\right)\NO\\
&&+\left(\s^i_p+Y^{-1}B^iB_p\right)\pa^p\f\sum_{\ell=0}^k\p_\f\sub{2k,2\ell}\NO\\
&&-B^i\sum_{\ell=0}^kD_j\left(\cp\sub{2k,2\ell}^j-2Y^{-1}\s^j_kB_l\p\sub{2k,2\ell}^{kl}+Y^{-1}B^jB_k\p\sub{2k,2\ell}^{k}\right)=0.
\eea
Matching terms of equal eigenvalues under $\d_B$ we obtain the two conditions
\begin{align}
\begin{aligned}
&\s^i_pF^p{}_j\left(\cp\sub{2k,2\ell}^j+Y^{-1}B^jB_k\p\sub{2k,2\ell}^{k}-2Y^{-1}\s^j_kB_l\p\sub{2k,2\ell+2}^{kl}\right)=0,\\&\\
&-2D_j\left(Y^{-2}B^iB^jB_kB_l\p\sub{2k,2\ell}^{kl}\right)
+Y^{-1}B^iB_pF^p{}_j\left(\cp\sub{2k,2\ell}^j+Y^{-1}B^jB_k\p\sub{2k,2\ell}^{k}\right)\\
&+Y^{-1}B^iB_k\pa^k\f\p_\f\sub{2k,2\ell}
-B^iD_j\left(\cp\sub{2k,2\ell}^j+Y^{-1}B^jB_k\p\sub{2k,2\ell}^{k}\right)\\
&-2D_j\left(\s^i_k\s^j_l\p\sub{2k,2\ell+2}^{kl}+Y^{-1}\left(\s^i_kB^j+\s^j_kB^i\right)B_l\p\sub{2k,2\ell+2}^{kl}\right)+\s^i_k\pa^k\f\p_\f\sub{2k,2\ell+2}\\
&-2Y^{-2}B^iB_pF^p{}_j\s^j_kB_l\p\sub{2k,2\ell+2}^{kl}+2B^iD_j\left(Y^{-1}\s^j_kB_l\p\sub{2k,2\ell+2}^{kl}\right)=0,
\end{aligned} 
\end{align}
for all $0\leq\ell\leq k$. In particular, we note the special cases
\begin{align}
\begin{aligned}
&\s^i_pF^p{}_j\left(\cp\sub{2k,2k}^j+Y^{-1}B^jB_k\p\sub{2k,2k}^{k}\right)=0,\\&\\
&\s^i_pF^p{}_j\s^j_kB_l\p\sub{2k,0}^{kl}=0,\\&\\
&-2D_j\left(Y^{-2}B^iB^jB_kB_l\p\sub{2k,2k}^{kl}\right)
+Y^{-1}B^iB_pF^p{}_j\left(\cp\sub{2k,2k}^j+Y^{-1}B^jB_k\p\sub{2k,2k}^{k}\right)\\
&+Y^{-1}B^iB_k\pa^k\f\p_\f\sub{2k,2k}
-B^iD_j\left(\cp\sub{2k,2k}^j+Y^{-1}B^jB_k\p\sub{2k,2k}^{k}\right)=0,\\&\\
&-2D_j\left(\s^i_k\s^j_l\p\sub{2k,0}^{kl}+Y^{-1}\left(\s^i_kB^j+\s^j_kB^i\right)B_l\p\sub{2k,0}^{kl}\right)+\s^i_k\pa^k\f\p_\f\sub{2k,0}\\
&-2Y^{-2}B^iB_pF^p{}_j\s^j_kB_l\p\sub{2k,0}^{kl}+2B^iD_j\left(Y^{-1}\s^j_kB_l\p\sub{2k,0}^{kl}\right)=0.
\end{aligned} 
\end{align}

\begin{flushleft}
{\em Hamiltonian constraint:}
\end{flushleft}
The Hamiltonian constraint in (\ref{constraints}) is quadratic in the canonical momenta and it is the dynamical equation that determines the Hamilton-Jacobi function $\cs$. In particular, using the decomposition of the momenta in terms of the eigenfunctions of $\Hat\d$ and $\d_B$, we will turn the Hamiltonian constraint into a tower of linear equations for $\cs\sub{2k,2\ell}$, which can be solved iteratively. 

Expanding the Hamiltonian constraint in eigenfunctions of $\Hat\d$ and isolating terms with the same eigenvalue we obtain for $k>0$  
\bea\label{recursion-relations}
&&\frac{2\k^2}{\sqrt{-\g}}e^{-d\x\f}\left\{2\left(\g_{ik}\g_{jl}-\frac1d\g_{ij}\g_{kl}\right)\p\sub{0}^{ij}\p\sub{2k}^{kl}\right.\NO\\
&&\left.\phantom{moremorem}
+\frac{1}{2\a}\left(\p_\f\sub{0}-2\x\p\sub{0}\right)\left(\p_\f\sub{2k}-2\x\p\sub{2k}\right)+\frac14Z^{-1}_\x\p\sub{0}^i\p\sub{2k}_i\right\}=\car\sub{2k},
\eea
where 
\bea\label{hsource}
\car\sub{2k}&=&\frac{\sqrt{-\g}}{2\k^2}e^{d\x\f}\left(-R[\g]+\a_\x\pa^i\f\pa_i\f+Z_\x(\f)F^{ij}F_{ij}\right)\d_{k,1}\NO\\
&&-\frac12\frac{\k^2}{\sqrt{-\g}}e^{-d\x\f}\sum_{m=0}^{k-1}W^{-1}_\x(\f)\p_\om\sub{2m}\p_\om\sub{2k-2m-2}\NO\\
&&-\frac{\k^2}{\sqrt{-\g}}e^{-d\x\f}\sum_{m=1}^{k-1}\left\{2\left(\g_{ik}\g_{jl}-\frac1d\g_{ij}\g_{kl}\right)\p\sub{2m}^{ij}\p\sub{2k-2m}^{kl}
+\frac14Z^{-1}_\x(\f)\p\sub{2m}^i\p\sub{2k-2m}_i\right.\NO\\
&&\left.+\frac{1}{2\a}\left(\p_\f\sub{2m}-2\x\p\sub{2m}\right)\left(\p_\f\sub{2k-2m}-2\x\p\sub{2k-2m}\right)\right\}.
\eea
We have written these constraints in the form of inhomogeneous linear equations for $\cs\sub{2k}$ by collecting all momenta coming from $\cs\sub{2k}$ on the LHS and grouping terms that originate in $\cs\sub{2k'}$ with $k'<k$ in the inhomogeneous term $\car\sub{2k}$. There is an exception to this, however, because as we have seen above the $\Hat\d$ eigenvalue of $\frac{1}{\sqrt{-\g}}\p_\om\sub{2k}\p_\om\sub{0}$ is $d-1-2k$ instead of $d+1-2k$, and therefore, this term must be included in the source $\car\sub{2k+2}$. Inserting the the zero order momenta 
\be
\p\sub{0}^{ij}=\frac{1}{\k^2} \sqrt{-\g}\left(\frac12 \g^{ij}U-U_Y B^iB^j\right),\quad
\p\sub{0}^i=\frac{1}{\k^2} \sqrt{-\g}2 U_Y B^i,\quad
\p_\f\sub{0}=\frac{1}{\k^2} \sqrt{-\g} U_X,
\ee  
in these recursion relations we obtain 
\begin{align}
\begin{aligned}
&e^{-d\x\f}\left\{\frac1\a\left(U_X-(d+1)\x U+2\x Y U_Y\right)\p_\f\sub{2k}-4U_YB_iB_j\p\sub{2k}^{ij}\right.\\
&\left.
-\frac{2}{d\a}\left(\a_\x U-2(\a_\x+d^2\x^2)Y U_Y+d\x U_X\right)\p\sub{2k}+Z_\x^{-1} U_Y B_i\p\sub{2k}^i\right\}=\car\sub{2k},\quad k>0.
\end{aligned}
\end{align}
Finally, using Tables \ref{momentum-eigenfunctions} and \ref{quadratic-momenta} these recursion relations can be expanded in eigenfunctions of $\d_B$ as  
\begin{align}
\label{recursion-relations-U-momentum}
\boxed{
\begin{aligned}
&e^{-d\x\f}\left\{\frac1\a\left(U_X-(d+1)\x U+2\x Y U_Y\right)\p_\f\sub{2k,2\ell}-4U_YB_iB_j\p\sub{2k,2\ell}^{ij}\right.\\
&\left.
-\frac{2}{d\a}\left(\a_\x U-2(\a_\x+d^2\x^2)Y U_Y+d\x U_X\right)\p\sub{2k,2\ell}+Z_\x^{-1} U_Y B_i\p\sub{2k,2\ell}^i\right\}=\car\sub{2k,2\ell},
\end{aligned}}
\end{align}
for all $k>0$ and $0\leq\ell\leq k$. These recursion relations are the basis of our algorithm for systematically solving the Hamilton-Jacobi equation. We now explain how this can be achieved.

\begin{flushleft}
{\bf Recursion relations}
\end{flushleft}
We now turn to the question of how the recursion relations (\ref{recursion-relations-U-momentum}) can be utilized in order to determine the terms $\cs\sub{2k,2\ell}$ of the Hamilton-Jacobi functional. A number of useful results that we will need in this section is presented in Appendix \ref{functional-operators}. In particular, in the appendix we define the unintegrated versions of the functional operators $\Hat\d$ and $\d_B$, namely,
\be 
\Hat d:= \left(2\g_{ij}\frac{\d}{\d\g_{ij}}+B_i\frac{\d}{\d B_i}\right),\quad d_B:=\left(2Y^{-1}B_iB_j\frac{\d}{\d\g_{ij}}+B_i\frac{\d}{\d B_i}\right). 
\ee
Using these unintegrated operators we can rewrite 
(\ref{recursion-relations-U-momentum})  in the form
\begin{align}
\label{recursion-d}
\begin{aligned}
&e^{-d\x\f}\left\{\frac1\a\left(U_X-(d+1)\x U+2\x Y U_Y\right)\p_\f\sub{2k,2\ell}\right.\\
&\left.+\left(
(2Y+Z_\x^{-1})U_Y +\frac{1}{d\a}\left(\a_\x U-2(\a_\x+d^2\x^2)Y U_Y+d\x U_X\right)\right)B_i\p\sub{2k,2\ell}^i\right.\\
&\left.
-\frac{1}{d\a}\left(\a_\x U-2(\a_\x+d^2\x^2)Y U_Y+d\x U_X\right)\Hat d\cs\sub{2k,2\ell}-2YU_Yd_B\cs\sub{2k,2\ell}\right\}=\car\sub{2k,2\ell}.
\end{aligned}
\end{align}
This form of the recursion relations allows us to utilize the fact that $\cs\sub{2k,2\ell}$ is a simultaneous eigenfunction of both $\Hat\d$ and $\d_B$. Some attention is required, however, in understanding the structure of various total derivative terms. Writing 
\be
\cs\sub{2k,2\ell}=\int d^{d+1}x\cl\sub{2k,2\ell},
\ee
and using the results of Appendix \ref{functional-operators}, we have 
\bea
&&\Hat d\cs\sub{2k,2\ell}=(d+1-2k)\cl\sub{2k,2\ell}+\pa_i\Hat u\sub{2k,2\ell}^i,\NO\\
&&d_B\cs\sub{2k,2\ell}=(1-2\ell)\cl\sub{2k,2\ell}+\pa_i\left(u_B\sub{2k,2\ell}^i+v_B\sub{2k,2\ell}^i\right),
\eea 
as well as 
\bea
&&\Hat \d\cl\sub{2k,2\ell}=(d+1-2k)\cl\sub{2k,2\ell},\NO\\
&&\d_B\cl\sub{2k,2\ell}=(1-2\ell)\cl\sub{2k,2\ell}+\pa_iv_B\sub{2k,2\ell}^i,
\eea 
where we have invoked Lemma \ref{functional-lemma-1} to deduce that $\cl\sub{2k,2\ell}$ is an eigenfunction of $\Hat \d$, without any total derivative term. Combining these relations one can show that the operators $\Hat\d$ and $\d_B$ act on the total derivative terms as follows:
\bea
&&\Hat\d \Hat u\sub{2k,2\ell}^i=(d+1-2k)\Hat u\sub{2k,2\ell}^i,\NO\\
&&\Hat\d\left(u_B\sub{2k,2\ell}^i+v_B\sub{2k,2\ell}^i\right)=(d+1-2k)\left(u_B\sub{2k,2\ell}^i+v_B\sub{2k,2\ell}^i\right),\NO\\
&&\d_B\Hat u\sub{2k,2\ell}^i=(1-2\ell)\Hat u\sub{2k,2\ell}^i-(d+1-2k)v_B\sub{2k,2\ell}^i,\NO\\
&&\d_B\left(u_B\sub{2k,2\ell}^i+v_B\sub{2k,2\ell}^i\right)=(1-2\ell)u_B\sub{2k,2\ell}^i.
\eea
However, $\cl\sub{2k,2\ell}$ is only defined up to a total derivative and so we are free to define
\be
\wt \cl\sub{2k,2\ell}:=\cl\sub{2k,2\ell}+\frac{1}{1-2\ell}\pa_i\left(u_B\sub{2k,2\ell}^i+v_B\sub{2k,2\ell}^i\right).
\ee
Using the action of $\Hat\d$ and $\d_B$ on the total derivative terms we now find 
\bea
&&\Hat d\cs\sub{2k,2\ell}=(d+1-2k)\wt\cl\sub{2k,2\ell}+\pa_i\Hat{\wt u}\sub{2k,2\ell}^i,\NO\\
&&d_B\cs\sub{2k,2\ell}=(1-2\ell)\wt\cl\sub{2k,2\ell},
\eea 
where
\be
\Hat{\wt u}\sub{2k,2\ell}^i=\Hat u\sub{2k,2\ell}^i-\left(\frac{d+1-2k}{1-2\ell}\right)\left(u_B\sub{2k,2\ell}^i+v_B\sub{2k,2\ell}^i\right),
\ee
and it satisfies
\be
\Hat\d\Hat{\wt u}\sub{2k,2\ell}^i=(d+1-2k)\Hat{\wt u}\sub{2k,2\ell}^i,\quad \d_B\Hat{\wt u}\sub{2k,2\ell}^i=(1-2\ell)\Hat{\wt u}\sub{2k,2\ell}^i.
\ee
More generally we define 
\be
\cl^\l\sub{2k,2\ell}:=\wt \cl\sub{2k,2\ell}+\frac{1-\l}{d+1-2k}\pa_i\Hat{\wt u}\sub{2k,2\ell}^i,
\ee
where $\l$ is an arbitrary parameter, so that 
\bea
&&\Hat d\cs\sub{2k,2\ell}=(d+1-2k)\cl^\l\sub{2k,2\ell}+\l\pa_i\Hat{\wt u}\sub{2k,2\ell}^i,\NO\\
&&d_B\cs\sub{2k,2\ell}=(1-2\ell)\cl^\l\sub{2k,2\ell}+(\l-1)\left(\frac{1-2\ell}{d+1-2k}\right)\pa_i\Hat{\wt u}\sub{2k,2\ell}^i.
\eea 

Inserting these expression in the recursion relation (\ref{recursion-d}) we obtain
\bea\label{recursion-relations-U}
&&\car\sub{2k,2\ell}=e^{-d\x\f}\left\{\rule{0.cm}{0.6cm}\frac1\a\left(U_X-(d+1)\x U+2\x Y U_Y\right)\p_\f\sub{2k,2\ell}\right.\NO\\
&&\left.+\left(
(2Y+Z_\x^{-1})U_Y +\frac{1}{d\a}\left(\a_\x U-2(\a_\x+d^2\x^2)Y U_Y+d\x U_X\right)\right)B_i\p\sub{2k,2\ell}^i\right.\NO\\
&&\left.
-\left(\frac{1}{d\a}\left(\a_\x U-2(\a_\x+d^2\x^2)Y U_Y+d\x U_X\right)(d+1-2k)+2YU_Y(1-2\ell)\right)\cl\sub{2k,2\ell}\right.\\
&&\left.-\left(\frac{1}{d\a}\left(\a_\x U-2(\a_\x+d^2\x^2)Y U_Y+d\x U_X\right)(d+1-2k)\l+2YU_Y(\l-1)(1-2\ell)\right)\frac{\pa_i\Hat{\wt u}\sub{2k,2\ell}^i}{d+1-2k}\right\},\NO
\eea
where we have dropped the superscript $\l$ in $\cl^\l\sub{2k,2\ell}$. Provided the ratio of the functions $YU_Y$ and $\left(\a_\x U-2(\a_\x+d^2\x^2)Y U_Y+d\x U_X\right)$ is constant, a suitable choice of the parameter $\l$ eliminates the total derivative term. However, we will keep the total derivative term for the time being and proceed with solving these recursive equations. On the way we will determine the minimal condition the superpotential $U(X,Y)$ must satisfy so that this total derivative term can be eliminated.

\subsection{Taylor expansion in the Lifshitz constraint}

The expansion of the HJ functional in eigenfunctions of the commuting operators $\Hat\d$ and $\d_B$ and the corresponding recursion relations  (\ref{recursion-relations-U}) are not specific to Lif boundary conditions. In order to incorporate these we must impose the asymptotic constraint (\ref{Lifshitz-constraint}). This means that, in addition to the expansion in eigenfunctions of $\Hat\d$ and $\d_B$, the solution of the HJ equation must take the form of a Taylor expansion in $B_i-B_{oi}$. In particular, these two expansions must be consistent with each other, and so each term $\cs\sub{2k,2\ell}$ in the graded covariant expansion must admit a Taylor expansion in $B_i-B_{oi}$. This Taylor expansion, except from imposing Lif boundary conditions, will allows us to eliminate the functional derivative with respect to $B_i$ in the recursion relations (\ref{recursion-relations-U}), leading to tractable linear functional differential equations in one variable.    

The Taylor expansion in $B_i-B_{oi}$ for the zero order solution $\cs\sub{0}$ can be immediately obtained from the Taylor expansion of the superpotential $U(X,Y)$ in $Y-Y_o$ in Section \ref{bgrnds}, using the identity
\be
Y-Y_o=2B_o^i(B_i-B_{oi})+(B^i-B_{o}^i)(B_i-B_{oi}).
\ee
More generally we expand $\cl\sub{2k,2\ell}$ in a functional Taylor expansion in $B_i-B_{oi}$ as
\bea\label{constraint-exp}
\cl\sub{2k,2\ell}[\g(x),B(x),\f(x)]&=&\cl^0_{(2k,2\ell)}[\g(x),\f(x)]\\
&&+\int d^{d+1}x'(B_i(x')-B_{oi}(x'))\cl^{1i}_{(2k,2\ell)}[\g(x),\f(x);x']+\co\left(B-B_o\right)^2.\NO
\eea
However, since the operators $\Hat\d$ and $\d_B$ depend on $B_i$ as well, they must also be Taylor expanded. Considering $\Hat\d$ first, we evaluate  
\be
\Hat d\cs\sub{2k,2\ell}=2\g_{ij}\frac{\d}{\d\g_{ij}}\cs^0\sub{2k,2\ell}+\Hat d\cs^1\sub{2k,2\ell}+\co(B-B_o),
\ee
where
\bea
\Hat d\cs^1\sub{2k,2\ell}&=&\Hat d(x)\int d^{d+1}x'\int d^{d+1}x''(B_i(x'')-B_{oi}(x''))\cl^{1i}_{(2k,2\ell)}[\g(x'),\f(x');x'']\NO\\
&=&\int d^{d+1}x'B_{oi}(x)\cl^{1i}_{(2k,2\ell)}[\g(x'),\f(x');x]\NO\\
&&-\int d^{d+1}x'B_{oi}(x)\cl^{1i}_{(2k,2\ell)}[\g(x'),\f(x');x]+\co(B-B_o)\NO\\
&=&\co(B-B_o),
\eea
and we have made use of the identity (\ref{b-derivative-gamma}) in the third line. An analogous result holds for $d_B$. This leads to the following identities  
\bea\label{reduced-operator-action}
&&\Hat d^0\cs^0_{(2k,2\ell)}=(d+1-2k)\cl^0_{(2k,2\ell)}+\l\pa_i\Hat{\wt u}_{(2k,2\ell)}^{0i},\NO\\
&&d^0_B\cs^0_{(2k,2\ell)}=(1-2\ell)\cl^0_{(2k,2\ell)}+(\l-1)\left(\frac{1-2\ell}{d+1-2k}\right)\pa_i\Hat{\wt u}_{(2k,2\ell)}^{0i},
\eea 
where the operators
\be
\Hat d^0:=2\g_{ij}\frac{\d}{\d\g_{ij}},\quad d^0_B:=2Y_o^{-1}B_{oi}B_{oj}\frac{\d}{\d\g_{ij}},
\ee
are respectively the pullbacks of the operators $\Hat d$ and $d_B$ on the constrained submanifold $B_i=B_{oi}$. Note that since $B_{oi}\propto \bb n_i$, the unit normal to the constant time slices, it follows that the pullback of the gradation operator, $\d^0_B$, counts {\em time} derivatives. Moreover, the pullback of the projection operator (\ref{projector}) becomes the spatial metric (see Table \ref{anisotropic-geometry})
\be
\bs_{ij}=\g_{ij}-Y_o^{-1}B_{oi}B_{oj}=\g_{ij}+\bb n_i\bb n_j.
\ee
The covariant expansion in simultaneous eigenfunctions of $\Hat\d^0$ and $\d^0_B$, therefore, is a derivative expansion with the number of derivatives given by the eigenvalue of $\Hat\d^0$ and graded according to the number of time derivatives, counted by the eigenvalue of $\d^0_B$.

\begin{flushleft}
{\bf Taylor expansion of the HJ equation}
\end{flushleft}
The HJ equation for the zero order solution $\cl\sub{0}$
is the superpotential equation (\ref{master}). Since $\cl\sub{0}$ depends on $B_i$ only though $Y=B_iB^i$ the Taylor expansion of the superpotential equation in $B_i-B_{oi}$ is equivalent to the Taylor expansion in $Y-Y_o$ we discussed in the superpotential III part of Section \ref{bgrnds}. All the results there carry over, except that the flow equations must be generalized to account for components that were identically zero for homogeneous backgrounds. For now, we only need equations (\ref{0th-order-0}), (\ref{0th-order-1}) and (\ref{0th-order-2}), which follow from the Taylor expansion of the superpotential.    

The HJ equations for $\cl\sub{2k,2\ell}$ with $k>0$  are the recursion relations (\ref{recursion-relations-U}). 
Inserting the expansion (\ref{constraint-exp}) and using the identity (\ref{b-derivative-phi}) the first two orders in $B_i-B_{oi}$ give the following two equations:
\bea\label{kth-order-0-prelim}
&&\hspace{-0.8cm}\co(1):\NO\\\NO\\
&&e^{-\x\f}\car^0_{(2k,2\ell)}[\g(x),\f(x)]=\frac1\a\left(u_0'+\frac{Z'}{Z}u_1\right)\frac{\d}{\d\f}\int d^{d+1}x'\cl^0_{(2k,2\ell)}[\g(x'),\f(x')]\NO\\
&&+\left(\frac{1}{2\a}\left(u_0'+\frac{Z'}{Z}u_1\right)\frac{Z'}{Z}+\frac{1}{d}(u_0+2(d-1)u_1)-\frac{2\e}{z-1}u_1\right)B_{oi}\int d^{d+1}x'\cl^{1i}_{(2k,2\ell)}[\g(x'),\f(x');x]\NO\\
&&-\left[(d+1-2k)\left(\frac{\x}{\a}\left(u_0'+\frac{Z'}{Z}u_1\right)+\frac{1}{d}(u_0-2u_1)\right)+2(1-2\ell)u_1\right]\cl^0_{(2k,2\ell)}[\g(x),\f(x)]\NO\\
&&-\left[(d+1-2k)\l\left(\frac{\x}{\a}\left(u_0'+\frac{Z'}{Z}u_1\right)+\frac{1}{d}(u_0-2u_1)\right)+2(\l-1)(1-2\ell)u_1\right]\frac{\pa_i\Hat{\wt u}^{0i}_{(2k,2\ell)}}{d+1-2k},
\eea
\bea\label{kth-order-1-prelim}
&&\hspace{-0.2cm}\co(B-B_o):\NO\\\NO\\
&&e^{-\x\f}\left.\frac{\d\car_{(2k,2\ell)}[\g(x),\f(x)]}{\d(B_i(x')-B_{oi}(x'))}\right|_{B=B_o}-\left[\frac1\a\left(u_1'+(u_1+2u_2)\frac{Z'}{Z}\right)\p_{\f(2k,2\ell)}^0+4(u_1+2u_2)\bb n_k\bb n_l\p_{(2k,2\ell)}^{0kl}\right.\NO\\
&&\left.-\left(\frac{\x}{\a}\left(u_1'+(u_1+2u_2)\frac{Z'}{Z}\right)-\frac1d(u_1+4u_2)\right)2\p^0_{(2k,2\ell)}\right]2Y_o^{-1}B_o^i\d^{(d+1)}(x-x')\NO\\
&&=\frac1\a\left(u_0'+\frac{Z'}{Z}u_1\right)\frac{\d \p^{0i}_{(2k,2\ell)}(x')}{\d\f(x)}\NO\\
&&+\left(\frac{\x}{\a}\left(u_0'+\frac{Z'}{Z}u_1\right)+\frac{1}{d}(u_0+2(d-1)u_1)-\frac{2\e}{z-1}u_1\right)\p^{0i}_{(2k,2\ell)}\d^{(d+1)}(x-x')\NO\\
&&-\left(\frac{1}{2\a}\left(u_1'+(u_1+2u_2)\frac{Z'}{Z}\right)\frac{Z'}{Z}-\frac{1}{d}(u_1+4u_2)+2(u_1+2u_2)-\frac{4\e}{z-1}u_2\right)2\bb n^i\bb n_j\p^{0j}_{(2k,2\ell)}\d^{(d+1)}(x-x')\NO\\
&&-\left[(d+1-2k)\left(\frac{\x}{\a}\left(u_0'+\frac{Z'}{Z}u_1\right)+\frac{1}{d}(u_0-2u_1)\right)+2(1-2\ell)u_1\right]\cl_{(2k,2\ell)}^{1i}[\g(x),\f(x);x']\NO\\
&&-\left[(d+1-2k)\l\left(\frac{\x}{\a}\left(u_0'+\frac{Z'}{Z}u_1\right)+\frac{1}{d}(u_0-2u_1)\right)+2(\l-1)(1-2\ell)u_1\right]\frac{\pa_j\Hat{\wt u}^{1ij}_{(2k,2\ell)}(x,x')}{d+1-2k}\\
&&+\left(\frac{1}{2\a}\left(u_0'+\frac{Z'}{Z}u_1\right)\frac{Z'}{Z}+\frac{1}{d}(u_0+2(d-1)u_1)-\frac{2\e}{z-1}u_1\right)2B_{oj}\int d^{d+1}y\cl^{2ij}_{(2k,2\ell)}[\g(y),\f(y);x,x'].\NO
\eea
where
\be
\p^{0ij}_{(2k,2\ell)}:=\frac{\d\cs^0_{(2k,2\ell)}}{\d \g_{ij}},\quad \p^{0}_{\f(2k,2\ell)}:=\frac{\d\cs^0_{(2k,2\ell)}}{\d \f},
\ee
and
\be\label{vector-momentum}
\p^{0i}_{(2k,2\ell)}:=\frac{\d\cs^1_{(2k,2\ell)}}{\d B_i}=
\int d^{d+1}x'\cl^{1i}_{(2k,2\ell)}[\g(x'),\f(x');x].
\ee
It must be stressed that with this definition of $\p^{0ij}_{(2k,2\ell)}$ and $\p^{0}_{\f(2k,2\ell)}$ these quantities {\em are not} the $\co(B-B_o)^0$ terms in the Taylor expansion of the corresponding momenta. In fact, using (\ref{b-derivative-gamma}) and (\ref{b-derivative-phi}) we find
\begin{align}
\label{zero-order-momenta}
\boxed{
\begin{aligned}
&\left.\p^{ij}_{(2k,2\ell)}\right|_{\co(B-B_o)^0}=\p^{0ij}_{(2k,2\ell)}-\frac12Y_o^{-1}B_o^iB_o^jB_{ok}\p^{0k}_{(2k,2\ell)},\\
&\left.\p_{\f(2k,2\ell)}\right|_{\co(B-B_o)^0}=\p^{0}_{\f(2k,2\ell)}+\frac12\frac{Z'_\x}{Z_\x}B_{ok}\p^{0k}_{(2k,2\ell)}.
\end{aligned}}
\end{align}

We will not present the equations for $\co(B-B_o)^2$ and higher here, but note that provided $B_i-B_{oi}$ sources a relevant operator, there is always some order at which the Taylor expansion can be truncated since higher order terms are subleading relative to the normalizable modes. At which order the Taylor expansion can be truncated depends on the leading asymptotic behavior of $B-B_o$, which was discussed in Section \ref{bgrnds}. Moreover, we can identify some generic features that apply to the higher order equations as well. Firstly, recall that the Taylor expansion in $B-B_o$ is well defined provided the inhomogeneous term in the flow equation for $Y-Y_o$ in (\ref{supIII-flow-XY}) vanishes. As in the expansion of the superpotential $U(X,Y)$, this constraint appears as the coefficient of $\cl_{(2k,2\ell)}^{1i}$ in (\ref{kth-order-0-prelim}) and of $\cl_{(2k,2\ell)}^{2ij}$ in (\ref{kth-order-1-prelim}). It follows that (\ref{kth-order-0-prelim}) is a decoupled equation for $\cl_{(2k,2\ell)}^{0}$ and (\ref{kth-order-1-prelim}) is a decoupled equation for $\cl_{(2k,2\ell)}^{1i}$. It is easy to show that this continues to hold in higher order equations so that the $\co(B-B_o)^m$ equation determines $\cl_{(2k,2\ell)}^{mi_1i_2\ldots i_m}$.

Another generic feature of these equations is the structure of the total derivative terms. In particular, the relative coefficient of the two total derivative terms remains the same for any order. It follows that  imposing a single condition on the functions $u_0(\f)$ and $u_1(\f)$, in addition to the three equations (\ref{potentials}), ensures that the total derivative terms can be eliminated from all equations at any order. Namely, if 
\be
\frac{\x}{\a}\left(u_0'+\frac{Z'}{Z}u_1\right)+\frac{1}{d}(u_0-2u_1)= c u_1,
\ee  
holds for some constant $c$, then the total derivative terms can be eliminated by setting 
\be
\l=\frac{2(1-2\ell)}{(d+1-2k)c+2(1-2\ell)}.
\ee 
The constant $c$ cannot take any value, however, since the asymptotic conditions (\ref{u-asymptotics}) require that
\be
c=\frac{2}{z-1}.
\ee
We will therefore restrict our attention to theories that satisfy 
\be\label{u-condition}\boxed{
\frac{\x}{\a}\left(u_0'+\frac{Z'}{Z}u_1\right)+\frac{1}{d}(u_0-2u_1)= \frac{2}{z-1} u_1,}
\ee  
in addition to (\ref{potentials}). Using the third equation in (\ref{potentials}), this condition (\ref{u-condition}) can alternatively be written as 
\be
\left(\frac{\e-z}{z-1}\right)\a u_1=\frac14 \frac{Z'_\x}{Z_\x}\left(u_0'+\frac{Z'}{Z}u_1\right).
\ee
Imposing this relation between $u_1$ and $u_0$ implies that the functions $V(\f)$, $W(\f)$ and $Z(\f)$ are all parameterized in terms of one arbitrary function through (\ref{potentials}). Note however, that (\ref{u-condition}) is automatically satisfied by the asymptotic form (\ref{u-asymptotics}) of the functions $u_1$ and $u_0$ and so it imposes no additional constraint on the parameters of generic Lif solutions. It only constrains the structure of the subleading terms in $u_1$ and $u_0$ and in this sense it is a mild restriction. However, we believe that imposing this restriction is not essential in order to solve the equations (\ref{kth-order-0-prelim}) and (\ref{kth-order-1-prelim}), but we have found no alternative way to solve them in the generic case. Of course, in special cases one can use an ansatz to solve these equations, but besides being very inefficient, this approach cannot be applied to the general case. 

Incorporating the conditions (\ref{u-condition}) and (\ref{potentials}) and decomposing the $\co(B-B_o)$ equation in spacelike and timelike parts the recursion equations for the first two orders in $B_i-B_{oi}$ become    
\bea \label{kth-order-0}
&&\hspace{-0.5cm}\rule{13.5cm}{.02cm}\NO\\
&&\hspace{-0.4cm}\co(1):\NO\\\NO\\
&&\frac1\a\left(u_0'+\frac{Z'}{Z}u_1\right)\frac{\d}{\d\f}\int d^{d+1}x'\cl^0_{(2k,2\ell)}-\frac{2u_1}{z-1}\cc_{k,\ell}\cl^0_{(2k,2\ell)}=e^{-\x\f}\car^0_{(2k,2\ell)},\\\NO\\
&&\hspace{-0.4cm}\co(B-B_o)\,\,\mbox{\em spacelike}:\NO\\\NO\\
&&\frac1\a\left(u_0'+\frac{Z'}{Z}u_1\right)\frac{\d }{\d\f(x)}\int d^{d+1}x''Z_\x^{-\frac12}(\f(x')) \bs^i_j(x') \cl_{(2k,2\ell)}^{1j}[\g(x''),\f(x'');x']\NO\\
&&-\frac{2u_1}{z-1}\cc_{k,\ell}Z_\x^{-\frac12}\bs^i_j\cl_{(2k,2\ell)}^{1j}[\g(x),\f(x);x']=e^{-\x\f}Z_\x^{-\frac12}\bs^i_j\car^{1j}_{(2k,2\ell)}[\g(x),\f(x);x'], \label{kth-order-1-space} \\ \NO\\
&&\hspace{-0.4cm}\co(B-B_o)\,\,\mbox{\em timelike},\quad u_0'+\frac{Z'}{Z}u_1\neq 0:\NO\\\NO\\
&&\frac1\a\left(u_0'+\frac{Z'}{Z}u_1\right)\frac{\d}{\d\f(x)}\int d^{d+1}x''\Om(\f(x')) B_{oj}(x') \cl_{(2k,2\ell)}^{1j}[\g(x''),\f(x'');x']\NO\\
&&-\frac{2u_1}{z-1}\cc_{k,\ell}\Om B_{oj}\cl_{(2k,2\ell)}^{1j}[\g(x),\f(x);x']=e^{-\x\f}\Om B_{oj}\Hat\car^{1j}_{(2k,2\ell)}[\g(x),\f(x);x'],\label{kth-order-1-time}\\\NO\\
&&\hspace{-0.4cm}\co(B-B_o)\,\,\mbox{\em timelike},\quad u_0'+\frac{Z'}{Z}u_1= 0:\NO\\\NO\\
&&\left(\frac{1}{\a}\left(u_1'+(u_1+2u_2)\frac{Z'}{Z}\right)\frac{Z'}{Z}-\frac{2}{d}(u_1+4u_2)+4(u_1+2u_2)-\frac{8\e}{z-1}u_2-\frac{2u_1}{z-1}\cc_{k,\ell}\right)\times\NO\\
&&B_{oj} \cl_{(2k,2\ell)}^{1j}[\g(x),\f(x);x']=e^{-\x\f} B_{oj}\Hat\car^{1j}_{(2k,2\ell)}[\g(x),\f(x);x'],\label{kth-order-1-time-degenerate}\\
&&\hspace{-0.5cm}\rule{13.5cm}{.02cm}\NO
\eea 
where we have defined the constants
\be
\cc_{k,\ell}:=d+1-2k+(z-1)(1-2\ell),
\ee
and the function  
\be\label{Omega} 
\Om(\f):=\exp\left(2\a\int d\f \frac{\frac{1}{2\a}\left(u_1'+(u_1+2u_2)\frac{Z'}{Z}\right)\frac{Z'}{Z}-\frac{1}{d}(u_1+4u_2)+2(u_1+2u_2)-\frac{4\e}{z-1}u_2}{u_0'+\frac{Z'}{Z}u_1}\right), 
\ee
which is defined provided $u_0'+\frac{Z'}{Z}u_1\neq 0$. Moreover, the source $B_{oj}\Hat\car^{1j}_{(2k,2\ell)}$ in the last two equations is given by
\bea\label{t-source}
&&B_{oj}\Hat\car^{1j}_{(2k,2\ell)}[\g(x),\f(x);x']:=B_{oj}\car^{1j}_{(2k,2\ell)}[\g(x),\f(x);x']\NO\\
&&-2e^{\x\f}\left[\frac1\a\left(u_1'+(u_1+2u_2)\frac{Z'}{Z}\right)\p_{\f(2k,2\ell)}^0+4(u_1+2u_2)\bb n_k\bb n_l\p_{(2k,2\ell)}^{0kl}\right.\NO\\
&&\left.\hspace{3cm}-\left(\frac{\x}{\a}\left(u_1'+(u_1+2u_2)\frac{Z'}{Z}\right)-\frac1d(u_1+4u_2)\right)2\p^0_{(2k,2\ell)}\right]\d^{(d+1)}(x-x').
\eea

\subsection{Solving the recursion equations}

In Section \ref{bgrnds} we determined the $k=0$ solution of the HJ equation as a Taylor expansion in $B_i-B_{oi}$ through the superpotential equation (\ref{master}). Given this $k=0$ solution, in the previous subsection we derived the equations that, at each order in $k>0$ and $\ell$, determine the first two orders in the Taylor expansion in $B_i-B_{oi}$, namely $\co(1)$ and $\co(B-B_o)$. These equations provide a recursive algorithm that allows us to obtain the solution of the HJ equation at order $k+1$ from the solution at order $k$. Namely, given the solution of the HJ at order $k$, the corresponding canonical momenta determine the inhomogeneous term in the linear equations for the order $k+1$ solution. The main technical challenge in this algorithm is solving these recursion relations. Obtaining the canonical momenta from a given solution and constructing the inhomogeneous term for the next order can also be tedious, but it's straightforward. As we will see momentarily, the solution of the recursion relations can be streamlined using the integration technique developed in \cite{Papadimitriou:2011qb}. Solving the HJ equation then becomes entirely algorithmic and it is ideally suited for implementation in a symbolic computation package such as {\tt xAct} \cite{xAct}.   

The recursion relations (\ref{kth-order-0}), (\ref{kth-order-1-space}) and (\ref{kth-order-1-time}) are identical in form to the equations appearing in the recursive solution of the HJ equation for relativistic backgrounds 
\cite{Papadimitriou:2011qb} and exactly the same techniques can be applied here. Indeed, many of the results in \cite{Papadimitriou:2011qb} are directly relevant. Firstly, note that the solutions of (\ref{kth-order-0}), (\ref{kth-order-1-space}) and (\ref{kth-order-1-time}) are qualitatively different depending on whether $u_0'+\frac{Z'}{Z}u_1$ is zero or not. Using (\ref{u-asymptotics}) we see that this quantity asymptotes to the constant parameter $\m$ and so there are three cases to examine: i) $\m\neq 0$, ii) $\m=0$ but $u_0'+\frac{Z'}{Z}u_1$ not identically zero, and iii)  $u_0'+\frac{Z'}{Z}u_1=0$, at least up to normalizable modes. We will consider two examples of case iii) in Section \ref{examples}. We will not discuss case ii) further here because it requires a specification the subleading terms in $u_1$ and $u_0$ that determine the asymptotic form of the scalar in this case. This can be easily done but would take us away from the generic case. In this section we will instead focus on case i), which is the generic situation.   

Provided the parameter $\m$ is not zero, all recursion relations (\ref{kth-order-0}), (\ref{kth-order-1-space}) and (\ref{kth-order-1-time}) admit a the homogeneous solution of the form
\be
\cf\sub{2k,2\ell}[\g]e^{-[(d+1-2k)+(z-1)(1-2\ell)]\ca(\f)},
\ee
where
\be
e^{\ca(\f)}=Z_\x^{-\frac{1}{2(\e-z)}}\sim e^{\f/\m},
\ee
and $\cf\sub{2k,2\ell}[\g]$ is a simultaneous eigenfunction of $\Hat\d^0$ and $\d_B^0$ with respective eigenvalues $d+1-2k$ and $1-2\ell$. Crucially, $\cf\sub{2k,2\ell}[\g]$ does not depend on $\f$. By construction, such an eigenfunction behaves asymptotically as 
\be
\cf\sub{2k,2\ell}[\g]\sim e^{[(d+1-2k)+(z-1)(1-2\ell)]r},
\ee  
which implies that the homogeneous solution is finite and so it corresponds to the usual renormalization scheme dependence.\footnote{The homogeneous solution is also related to the integration functions of the complete integral, and hence to the 1-point functions and the normalizable modes in the Fefferman-Graham expansions. This is discussed in more detail in Section \ref{ward}.} 
The inhomogeneous solutions of (\ref{kth-order-0}), (\ref{kth-order-1-space}) and (\ref{kth-order-1-time}) can be written formally in the form
\begin{align}
\label{inhomogeneous-sol}
\boxed{
\begin{aligned}
&\cl^0_{(2k,2\ell)}[\g,\f]= e^{-\cc_{k,\ell}\ca(\f)}\int^\f d\bar\f \ck(\bar\f) e^{\cc_{k,\ell}\ca(\bar\f)}\car^0_{(2k,2\ell)}[\g,\bar\f],\\
&\bs^i_j\cl_{(2k,2\ell)}^{1j}[\g(x),\f(x);x']=Z_\x^{\frac12} e^{-\cc_{k,\ell}\ca(\f)}\int^{\f(x)} d\bar\f \ck(\bar\f) e^{\cc_{k,\ell}\ca(\bar\f)} Z_\x^{-\frac12}\bs^i_j\car^{1j}_{(2k,2\ell)}[\g(x),\bar\f;x'],\\
&B_{oj}(x) \cl_{(2k,2\ell)}^{1j}[\g(x),\f(x);x']=\Om^{-1}e^{-\cc_{k,\ell}\ca(\f)}\int^{\f(x)} d\bar\f\ck(\bar\f) e^{\cc_{k,\ell}\ca(\bar\f)}\Om B_{oj}\Hat\car^{1j}_{(2k,2\ell)}[\g(x),\bar\f;x'],
\end{aligned}}
\end{align} 
where
\be
\ck(\f):=\frac{\a}{e^{\x\f}\left(u'_0+\frac{Z'}{Z} u_{1}\right)}\sim -\frac1\m,\qquad \Om\sim e^{-\D_-\f/\m}.
\ee
As in Eq.~(2.36)-(2.37) of \cite{Papadimitriou:2011qb}, 
the expressions (\ref{inhomogeneous-sol}) for the inhomogeneous solutions are formal since the source terms, such as $\car^0_{(2k,2\ell)}[\g,\bar\f]$,  generically contain derivatives of the scalar $\f$. In \cite{Papadimitriou:2011qb} these formal integrals were defined by systematically tabulating all possible derivative structures involving the scalar, up to four derivatives, and the corresponding integrals were evaluated generically. The results, adapted to the present problem, are summarized in Table \ref{integration}. As in \cite{Papadimitriou:2011qb} we have introduced the shorthand notation
\be
\label{shorthand}
\fint_{k,\ell,m}^\f \equiv (\mf A_{k,\ell}')^{m}e^{-\mf A_{k,\ell}}\int^{\f} d\bar\f \ck(\bar\f) e^{\mf A_{k,\ell}}(\mf A_{k,\ell}')^{-m},
\ee
where
\be\label{factor}
\mf A_{k,\ell}:=\left\{\begin{array}{l}
\cc_{k,\ell}\ca,\\
\cc_{k,\ell}\ca-\frac12\log Z_\x,\\
\cc_{k,\ell}\ca+\log\Om,
\end{array}\right.
\ee
depending on which integral in (\ref{inhomogeneous-sol}) one considers.
Using the map between integrands involving derivatives of the scalar and the corresponding integrals in Table 
\ref{integration}, any integral containing zero or two derivatives of the scalar can be directly evaluated.  
Most integrals containing four derivatives on the scalar can be evaluated directly using this table as well, but there are few cases which require an extension of the results in Table \ref{integration} because only certain tensor structures at the four-derivative level were considered in \cite{Papadimitriou:2011qb}. It is straightforward to generalize these results to any tensor structure with four derivatives on the scalar following the procedure in Appendix A of \cite{Papadimitriou:2011qb}. However, we will not carry out this generalization here as we will not needed it explicitly.       
\begin{table}
\[\begin{array}{|l|l|}
\hline\hline
\mf R_{(2k,2\ell)}[\g,\f]& \mf L_{(2k,2\ell)}[\g,\f]\\ \hline &\\

r_{1^m}(\f) t^{i_1i_2\ldots i_{m}}\pa_{i_1}\f\pa_{i_2}\vf\ldots\pa_{i_m}\f &
\fint_{k,\ell,m}^\f r_{1^m}(\bar\f)t^{i_1i_2\ldots i_{m}}\pa_{i_1}\f\pa_{i_2}\f\ldots\pa_{i_m}\f\\ &\\

r_2(\f)t^{ij} D_i D_j\f & \fint_{k,\ell,1}^\f r_{2}(\bar\f)t^{ij} D_i D_j\f\\ &
-\fint_{k,\ell,2}^\f \ck^{-1}\mf A_{k,\ell}' \pa_{\bar\f}^2\left(\frac{1}{\mf A_{k,\ell}'}\right)\fint_{k,\ell,1}^{\bar\f}r_{2}(\tilde\f)t^{ij}\pa_i\f\pa_j\f\\ &\\

\left(r_{1^22}(\f)t_1^{ijkl}+s_{1^22}(\f)t_2^{ijkl}\right)\pa_i\f\pa_j\f  D_k D_l\f &
\fint_{k,\ell,3}^\f s_{1^22}(\bar\f)t_2^{ijkl}\pa_i\f\pa_j\f  D_k D_l\f \\ & \\

\left(r_{2^2}(\f)t_1^{ijkl}+s_{2^2}(\f)t_2^{ijkl}\right) D_i D_j\f  D_k D_l\f &
\left(\fint_{k,\ell,2}^\f r_{2^2}(\bar\f)t_1^{ijkl}+\fint_{k,\ell,2}^\f s_{2^2}(\bar\f)t_2^{ijkl}\right) D_i D_j\f  D_k D_l\f \\&
-2\fint_{k,\ell,3}^\f \ck^{-1}\mf A_{k,\ell}'\pa_{\bar\f}^2\left(\frac{1}{\mf A_{k,\ell}'}\right)
\fint_{k,\ell,2}^{\bar\f} s_{2^2}(\tilde\f)t_2^{ijkl}\pa_i\f\pa_j\f  D_k D_l\f \\&\\ 
\hline\hline
\end{array}\]
\caption{General integration identities for integrands that contain up to four derivatives on the scalars that were derived in \cite{Papadimitriou:2011qb}. The shorthand notation $\fint_{k,\ell,m}^\f$ is defined in (\ref{shorthand}). $\mf R_{(2k,2\ell)}$ stands for any of the source terms on the RHS of (\ref{inhomogeneous-sol}), while $\mf L_{(2k,2\ell)}$ stands for any of the quantities on the LHS. The tensors $t^{i_1i_2\ldots i_{m}}$ and $t^{ij}$ are arbitrary totally symmetric tensors independent of $\f$, while 
$t_1^{ijkl}=\frac13\left(\g^{ik}\g^{jl}+\g^{il}\g^{jk}+\g^{ij}\g^{kl}\right)$,   
$t_2^{ijkl}=\frac13\left(\g^{ik}\g^{jl}+\g^{il}\g^{jk}-2\g^{ij}\g^{kl}\right)$. These formulas suffice for all terms appearing in $\car^0_{(2,0)}$ and $\car^0_{(2,2)}$, but only for terms in $\car^0_{(4,0)}$, $\car^0_{(4,2)}$ and $\car^0_{(4,4)}$ that are contracted with the particular tensors $t_1^{ijkl}$ and $t_2^{ijkl}$. Although these tensors cover the most general 4-derivative terms in the relativistic case \cite{Papadimitriou:2011qb}, this is not in general the case for the non-relativistic boundary conditions we impose here. However, the relevant integration formulas that generalize this table can be derived as in \cite{Papadimitriou:2011qb}. Moreover, as we will see in Section \ref{examples}, these formulas are not required in the case of exponential potentials, since the  integrals over the scalar can be evaluated in general independently of the tensor structure in that case. } 
\label{integration}
\end{table}

We can now summarize the complete recursion algorithm. We start by organizing the source terms (\ref{hsource}) into eigenfunctions $\mf R_{(2k,2\ell)}$ of the operator $\d_B$, utilizing the results in Table \ref{quadratic-momenta}. Taylor expanding these expressions in $B_i-B_{oi}$ one obtains the source terms at each order of the Taylor expansion, which are eigenfunctions of $\Hat\d^0$ and $\d^0_B$. These eigenfunctions are then written in the form  
\be\label{scalar-source-decomposition} 
\mf R_{(2k,2\ell)}[\g,\f]=\frac{1}{2\k^2}\sqrt{-\g} \sum_{I=1}^{N_{k,\ell}}c_{k,\ell}^I(\f)\ct_{k,\ell}^I, 
\ee
where the tensors $\ct_{k,\ell}^I$ contain only derivatives of the scalar $\f$, but are otherwise independent of $\f$.   
Using the identities in Table \ref{integration}, the integrals in (\ref{inhomogeneous-sol}) can be evaluated to obtain $\mf L_{(2k,2\ell)}$ in the form
\be \label{scalar-solution-decomposition}
\mf L_{(2k,2\ell)}[\g,\f]=\frac{1}{2\k^2}\sqrt{-\g} \sum_{I=1}^{N_{k,\ell}}p_{k,\ell}^I(\f)\ct_{k,\ell}^I. 
\ee
This determines the complete solution of the HJ equation at order $k$ up to linear order in $B_i-B_{oi}$. To obtain the solution at order $k+1$ we need to evaluate the momenta from the order $k$ solution and substitute them in the source term (\ref{hsource}) for the order $k+1$ equation. We then proceed as before. This procedure is repeated in order to obtain the solution of the HJ equation up to the finite term, where the recursion procedure breaks down. We will discuss when precisely this happens and the significance of the finite part in Section \ref{ward}.

\subsection{Solution at order $k=1$}

In order to illustrate the recursion algorithm we now construct the general solution at order $k=1$ and up to order $\co(B-B_o)$ in the Taylor expansion. The source term (\ref{hsource}) for $k=1$ and to lowest order in $B-B_o$ is 
\bea
\car^0_{(2)}&=&\frac{\sqrt{-\g}}{2\k^2}e^{d\x\f}\left(-R[\g]+\a_\x\pa^i\f\pa_i\f+Z_\x(\f)F^{ij}_oF_{oij}\right)
-\frac12\frac{\k^2}{\sqrt{-\g}}e^{-d\x\f}W^{-1}_\x(\f)\p_{\om(0)}^2\\
&=&\frac{\sqrt{-\g}}{2\k^2}e^{d\x\f}\left(\rule{0cm}{0.5cm}-R[\g]+\a_\x\pa^i\f\pa_i\f+Z_\x F^{ij}_oF_{oij}-4e^{-2d\x\f}W^{-1}_\x\left[D_i\left(e^{(d+1)\x\f}Y_{o}^{-1}u_1(\f)B_o^i\right)\right]^2\right).\NO
\eea
The first step in the algorithm is to decompose this into eigenfunctions of $\d^0_B$. The last term is an eigenfunction of $\d^0_B$ with eigenvalue $-1$ and hence it belongs to $\car^0_{(2,2)}$. This can be deduced by directly evaluating the action of $\d^0_B$ on this term, or by invoking the last entry in Table \ref{operator-momenta} and noticing that $\s^i_kB_l\p^{kl}_{(0,0)}=0$. The same result can also be read off the last entry in Table \ref{quadratic-momenta}. The other three terms are not eigenfunctions of $\d^0_B$, but they can be  decomposed into eigenfunctions of $\d^0_B$ using the projection operator $\bs^i_j$. For the scalar we have 
\be
\frac{\sqrt{-\g}}{2\k^2}e^{d\x\f}\a_\x\pa^i\f\pa_i\f=
\frac{\sqrt{-\g}}{2\k^2}e^{d\x\f}\a_\x\left(\bs^{ij}+Y_o ^{-1}B_o^iB_o^j\right)\pa_i\f\pa_j\f,
\ee
where both terms in this decomposition are eigenfunctions of $\d^0_B$ with respective eigenvalues $1$ and $-1$. The decomposition of $F^{ij}_oF_{oij}$  gives 
\be
\frac{\sqrt{-\g}}{2\k^2}e^{d\x\f}Z_\x(\f)F^{ij}_oF_{oij}=
\frac{\sqrt{-\g}}{2\k^2}e^{d\x\f}Z_\x(\f)
\left(\bs^{ij}\bs^{kl}F_{oik}F_{ojl}+2Y_o^{-1}B_o^iB_o^j\bs^{kl}F_{oik}F_{ojl}\right),
\ee
where the first term has $\d^0_B$ eigenvalue 3 and the second 1. However, there cannot be any eigenfunction of $\d^0_B$ with eigenvalue 3 when $k=1$ and therefore $\bs^{ij}\bs^{kl}F_{oik}F_{ojl}$ must vanish identically. Finally, the Ricci scalar can be decomposed into two eigenfunctions of $\d^0_B$ with eigenvalues 1 and $-1$, but the decomposition is less trivial. Namely the naive decomposition  
\be
R=\bs^{ij}R_{ij}-\bb n^i\bb n^jR_{ij},
\ee
is not correct in this case because these two terms are eigenfunctions of $\d^0_{B}$ only up to total derivatives. In particular, 
\be
\d^0_{B}\left(\sqrt{-\g}\bs^{ij}R_{ij}\right)=\sqrt{-\g}\left(\bs^{ij}R_{ij}-2 D_i(\bb n^i\bb K)\right),\,\,\,
\d^0_{B}\left(\sqrt{-\g}\bb n^i\bb n^jR_{ij}\right)=\sqrt{-\g}\left(-\bb n^i\bb n^jR_{ij}+2D_i\bb q^i\right).
\ee
However, $\sqrt{-\g}D_i\bb q^i$ and $\sqrt{-\g}D_i(\bb n^i\bb K)$ are eigenfunctions of $\d^0_{B}$ with respective eigenvalues $1$ and $-1$. It follows that the Ricci scalar can be decomposed in terms of four eigenfunctions of $\d^0_{B}$ as 
\be
\sqrt{-\g}R=\sqrt{-\g}(\bs^{ij}R_{ij}-D_i(\bb n^i\bb K))-\sqrt{-\g}D_i\bb q^i-\sqrt{-\g}(\bb n^i\bb n^jR_{ij}-D_i\bb q^i)+\sqrt{-\g}D_i(\bb n^i\bb K),
\ee
where the first two eigenfunctions have eigenvalue $1$ and the last two $-1$. Using the decomposition of the Ricci tensor in Table \ref{anisotropic-geometry} it is easy to see why these particular combinations arise. In terms of anisotropic geometric quantities these become
\be
\bs^{ij}R_{ij}-D_i(\bb n^i\bb K)=\bb R-D_k\bb q^k,\quad 
 \bb n^i\bb n^jR_{ij}-D_i\bb q^i=-\bb K^{kl}\bb K_{kl}-\bb n^kD_k\bb K,
\ee  
which makes it manifest that the eigenfunction with eigenvalue 1 contains only spatial derivatives, while the one with eigenvalue $-1$ contains only time derivatives.

Next we need to write these terms in the form (\ref{scalar-source-decomposition}) by making explicit all the dependence on the scalar field $\f$. Since 
\be
B_{oi}=\sqrt{-Y_o}\;\bb n_i, \quad Y_o= -\frac{z-1}{2\e}Z_\x^{-1}(\f),
\ee
we have 
\be
F_{oij}=-\frac{Y'_o}{2\sqrt{-Y_o}}\left(\bb n_j\pa_i\f -\bb n_i\pa_j\f \right)+\sqrt{-Y_o}\;\bb f_{ij},
\ee
where $\bb f_{ij}$ is defined in Table \ref{anisotropic-geometry} in Appendix \ref{identities}. Hence, 
\be
\bs^{ik}\bs^{jl}F_{okl}=\sqrt{-Y_o}\bs^{ik}\bs^{jl}\bb f_{kl}=\sqrt{-Y_o}\bs^{ik}\bs^{jl}(\bb q_i\bb n_j-\bb q_j\bb n_i)=0,
\ee
which confirms the conclusion we reached above that $\bs^{ij}\bs^{kl}F_{oik}F_{ojl}$ must vanish identically based on its eigenvalue under $\d_B^0$. Moreover, 
\be
Y_o^{-1}B_o^k\bs^{jl}F_{okl}=-\bb n^k \bs^{jl} \bb f_{kl}+\frac12 Y_o'\bb n^k\bb n_k\bs^{jl}\pa_l\f=
-\bb q^j-\frac12 Y_o^{-1}Y_o'\bs^{jl}\pa_l\f,
\ee
and so 
\be
2Z_\x(\f)Y_o^{-1}B_o^iB_o^j\bs^{kl}F_{oik}F_{ojl}=
-\frac{z-1}{\e}\left(\bb q^i\bb q_i+ Y_o^{-1}Y_o'\bb q^i\pa_i\f+\frac14 (Y_o^{-1}Y_o')^2\bs^{ij}\pa_i\f\pa_j\f\right).
\ee
Finally, 
\be
D_i\left(e^{(d+1)\x\f}Y_{o}^{-1}u_1B_o^i\right)=
-\frac{e^{(d+1)\x\f}u_1}{\sqrt{-Y_o}}\left(D_i\bb n^i
+\left((d+1)\x+\frac12 Z_\x^{-1}Z_\x'+ \frac{u_1'}{u_1}\right)\bb n^i\pa_i\f\right).
\ee
Collecting all results, the source term $\car^0_{(2)}$ can be decomposed in terms of a convenient basis of eigenfunctions as described in Table \ref{k=1-HJ-solution-0}, where we also introduce the linear operator
\be\label{cov-phi}
\cd_\f :=\pa_\f+(d+1)\x+\frac12\frac{Z'_\x}{Z_\x}.
\ee
The corresponding coefficients of the solutions $\cl^0_{(2,0)}$ and $\cl^0_{(2,2)}$ of the HJ equation, in the parameterization (\ref{scalar-solution-decomposition}), are then obtained using the integration formulas in Table \ref{integration}, which appear in the last column of Table \ref{k=1-HJ-solution-0}.
\begin{table}
\[\begin{array}{|c|c|c|c|c|}
\hline\hline
\multicolumn{3}{|c|}{k=1,\;\co(1)}& \car^0_{(2,2\ell)} & \cl^0_{(2,2\ell)} \\ \hline\hline 
\ell & I & \ct^I_{1,\ell} & c^I_{1,\ell} & p^I_{1,\ell}\\ \hline &&&&\\
0 & 1 & \bb R & -e^{d\x\f} & \fint_{1,0,0}c_{1,0}^1\\ &&&&\\
  & 2  & D_i\bb q^i & 2e^{d\x\f} & \fint_{1,0,0}c_{1,0}^2\\ &&&&\\
  & 3 & \bb q^i\bb q_i & -\frac{z-1}{\e}e^{d\x\f} & \fint_{1,0,0}c_{1,0}^3\\ &&&&\\
  & 4 & \bb q^i\pa_i\f & \frac{z-1}{\e}e^{d\x\f}\frac{Z'_\x}{Z_\x} & \fint_{1,0,1}c_{1,0}^4 \\ &&&&\\
  & 5 & \bs^{ij}\pa_i\f\pa_j\f & \left(\a_\x-\frac{z-1}{4\e}\left(\frac{Z'_\x}{Z_\x}\right)^2 \right)e^{d\x\f} & \fint_{1,0,2}c_{1,0}^5\\ &&&&\\
 \hline &&&&\\
1 & 1 & \bb K^{kl}\bb K_{kl} & -e^{d\x\f} & \fint_{1,1,0}c_{1,1}^1\\ &&&&\\
  & 2  & \bb n^iD_i\bb K & -2e^{d\x\f} & \fint_{1,1,0}c_{1,1}^2 \\ &&&&\\
  & 3 & \bb K^2 & -e^{d\x\f}\left(1 +\frac{8\e}{z-1}e^{2\x\f}\frac{Z_\x}{W_\x} u_1^2\right) & \fint_{1,1,0}c_{1,1}^3\\ &&&&\\
  & 4 & \bb K \bb n^j \pa_j\f & -\frac{16\e}{z-1}e^{(d+2)\x\f}\frac{Z_\x}{W_\x} u_1\cd_\f u_1 & \fint_{1,1,1}c_{1,1}^4\\ &&&&\\
  & 5 & (\bb n^i \pa_i\f)^2 & -e^{d\x\f}\left(\a_\x+ \frac{8\e}{z-1}e^{2\x\f}\frac{Z_\x}{W_\x} (\cd_\f u_1)^2\right) & \fint_{1,1,2}c_{1,1}^5\\ 
  &&&&\\
 \hline\hline
\end{array}\]
\caption{General solution of the first recursion relation in (\ref{inhomogeneous-sol}) at order $k=1$. The second column from the right describes the source of the inhomogeneous equation in the form (\ref{scalar-source-decomposition}), while the last column gives the solution $\cl^0_{(2,0)}$ and $\cl^0_{(2,2)}$ in the parameterization (\ref{scalar-solution-decomposition}). The shorthand notation used in the last column is defined in (\ref{shorthand}).}
\label{k=1-HJ-solution-0}
\end{table}

Similarly we find that the $\co(B-B_o)$ source terms for $k=1$ are
\bea\label{order-1-sources}
\car^{1i}_{(2,0)}[\g(x),\f(x);x']&=&\frac{\sqrt{-\g}}{2\k^2}e^{d\x\f}Z_\x
8\sqrt{-Y_o}\bb n^{[i}\bb q^{j]}D_j^x\d^{(d+1)}(x-x'),\NO\\\NO\\
\car^{1i}_{(2,2)}[\g(x),\f(x);x']&=&-\frac{\sqrt{-\g}}{2\k^2}e^{-d\x\f}W_\x^{-1}8D_k^x\left(e^{(d+1)\x\f}Y_o^{-1}u_1B_o^k\right)\times\NO\\
&&D_j^x\left(e^{(d+1)\x\f}Y_o^{-1}(u_1\g^{ij}-4u_2\bb n^i\bb n^j)\d^{(d+1)}(x-x')\right).
\eea
Decomposing these in spatial and time components leads to the expressions presented in Table \ref{k=1-HJ-solution-1}. In each case, the corresponding solutions of (\ref{inhomogeneous-sol}), obtained using Table \ref{integration}, are listed in the last column. One must remember, however, that (\ref{order-1-sources}) do not provide the full source for $B_{oj}\cl^{1j}_{(2,2\ell)}$ given in (\ref{t-source}). In particular, the full source for $B_{oj}\cl^{1j}_{(2,2\ell)}$ contains terms involving the momenta obtained from the $\co(1)$ solution in the Taylor expansion.  
\begin{table}
\vskip1in	
\[\begin{array}{|c|c|c|c|c|}
\hline\hline
\multicolumn{3}{|c|}{k=1,\;\co(B-B_o),\; \mbox{space}}& \bs^i_j\car^{1j}_{(2,2\ell)} & \bs^i_j\cl^{1j}_{(2,2\ell)} \\ \hline\hline 
\ell & I & \ct^I_{1,\ell} & c^I_{1,\ell} & p^I_{1,\ell}\\ \hline &&&&\\
0 & 1 & \bb q^i\bb n^kD_k^x\d(x-x') & -4e^{d\x\f}Z_\x\sqrt{-Y_o} & \fint_{1,0,0}c_{1,0}^1\\ &&&&\\
 \hline &&&&\\
1 & 1 & \bb K\bb D^i_x\d(x-x') & -\frac{16\e}{z-1}\frac{e^{(d+2)\x\f}}{\sqrt{-Y_o}}\frac{Z_\x}{W_\x}u_1^2 & \fint_{1,1,0}c_{1,1}^1\\ &&&&\\
  & 2  & \bb K\bb q^i\d(x-x') & \frac{64\e}{z-1}\frac{e^{(d+2)\x\f}}{\sqrt{-Y_o}}\frac{Z_\x}{W_\x}u_1u_2 & \fint_{1,1,0}c_{1,1}^2 \\ &&&&\\
  & 3 & \bb n^k\pa_k\f\bb D^i_x\d(x-x') & -\frac{16\e}{z-1}\frac{e^{(d+2)\x\f}}{\sqrt{-Y_o}}\frac{Z_\x}{W_\x}u_1\cd_\f u_1 & \fint_{1,1,1}c_{1,1}^3\\ &&&&\\
  & 4 & \bb K\bb D^i_x\f\d(x-x') & -\frac{16\e}{z-1}\frac{e^{(d+2)\x\f}}{\sqrt{-Y_o}}\frac{Z_\x}{W_\x}u_1\cd_\f u_1 & \fint_{1,1,1}c_{1,1}^4\\ &&&&\\
  & 5 & \bb q^i\bb n^k\pa_k\f\d(x-x') & \frac{64\e}{z-1}\frac{e^{(d+2)\x\f}}{\sqrt{-Y_o}}\frac{Z_\x}{W_\x}u_2\cd_\f u_1 & \fint_{1,1,1}c_{1,1}^5\\ 
  &&&&\\
  & 6 & \bb D^i_x\f\bb n^k\pa_k\f\d(x-x') & -\frac{16\e}{z-1}\frac{e^{(d+2)\x\f}}{\sqrt{-Y_o}}\frac{Z_\x}{W_\x}(\cd_\f u_1)^2 & \fint_{1,1,2}c_{1,1}^6\\ 
    &&&&\\
 \hline\hline
 \multicolumn{3}{|c|}{k=1,\;\co(B-B_o),\; \mbox{time}}& B_{oj}\car^{1j}_{(2,2\ell)} & B_{oj}\cl^{1j}_{(2,2\ell)} \\ \hline\hline 
 \ell & I & \ct^I_{1,\ell} & c^I_{1,\ell} & p^I_{1,\ell}\\ \hline &&&&\\
 0 & 1 & \bb q^kD_k^x\d(x-x') & 4e^{d\x\f}Z_\x Y_o & \fint_{1,0,0}c_{1,0}^1\\ &&&&\\
 \hline &&&&\\
 1 & 1 & \bb K\bb n^j D_j^x\d(x-x') & -\frac{16\e}{z-1}e^{(d+2)\x\f}\frac{Z_\x}{W_\x}u_1(u_1+4u_2) & \fint_{1,1,0}c_{1,1}^1\\ &&&&\\
 & 2  & \bb K^2\d(x-x') & -\frac{64\e}{z-1}e^{(d+2)\x\f}\frac{Z_\x}{W_\x}u_1u_2 & \fint_{1,1,0}c_{1,1}^2 \\ &&&&\\
 & 3 & \bb n^k\pa_k\f\;\bb n^j D_j^x\d(x-x') & -\frac{16\e}{z-1}e^{(d+2)\x\f}\frac{Z_\x}{W_\x}(u_1+4u_2)\cd_\f u_1 & \fint_{1,1,1}c_{1,1}^3\\ &&&&\\
 & 4 & \bb K\bb n^k \pa_k\f\d(x-x') & -\frac{16\e}{z-1}e^{(d+2)\x\f}\frac{Z_\x}{W_\x} \left(4u_2\cd_\f u_1+u_1\cd_\f(u_1+4u_2)\right) & \fint_{1,1,1}c_{1,1}^4\\ &&&&\\
 & 5 & (\bb n^k\pa_k\f)^2\d(x-x') & -\frac{16\e}{z-1}e^{(d+2)\x\f}\frac{Z_\x}{W_\x}\cd_\f u_1\cd_\f(u_1+4u_2) & \fint_{1,1,2}c_{1,1}^5\\ 
 &&&&\\
 \hline\hline
\end{array}\]
\caption{General solution of the second and third recursion relations in (\ref{inhomogeneous-sol}) at order $k=1$. The second column from the right describes the sources  $\bs^i_j\car^{1j}_{(2,2\ell)}$ and $B_{oj}\car^{1j}_{(2,2\ell)}$ of the inhomogeneous equations in the form (\ref{scalar-source-decomposition}), while the last column gives the components $\bs^i_j\cl^{1j}_{(2,2\ell)}$ and $B_{oj}\cl^{1j}_{(2,2\ell)}$ of the solution in the parameterization (\ref{scalar-solution-decomposition}). The shorthand notation used in the last column is defined in (\ref{shorthand}). The results in this table can be extended to the full source $B_{oj}\Hat\car^{1j}_{(2,2\ell)}$ in (\ref{t-source}) once the canonical momenta at order $\co(1)$ in the Taylor expansion are evaluated.}
\label{k=1-HJ-solution-1}
\end{table}

\begin{flushleft}
{\bf Computation of momenta at order $k=1$}
\end{flushleft}

The general solution of the recursion relations (\ref{inhomogeneous-sol}) at order $k=1$ and to the first two orders in the $B_i-B_{oi}$ expansion is given in Tables \ref{k=1-HJ-solution-0} and \ref{k=1-HJ-solution-1}. In order to proceed to the next order in $k$, we need to compute all the canonical momenta from the solution at order $k=1$ by evaluating the corresponding functional derivatives. The identities (\ref{zero-order-momenta}) imply that the momenta obtained from both the $\co(1)$ and $\co(B-B_o)$ solutions of the HJ equation will contribute to the $\co(1)$ momenta. Similarly, the $\co(B-B_o)$ momenta will get contributions from both the $\co(B-B_o)$ and $\co(B-B_o)^2$ parts of the HJ solution. Since we have only computed the solution of the HJ equation up to $\co(B-B_o)$, we can only determine the $\co(1)$ momenta here. 

It is useful to write these momenta entirely in terms of quantities that directly pertain to the geometry of the spatial surfaces and their embedding in the constant radial slices $\S_r$, rather than covariant variables with respect to  $\S_r$ diffeomorphisms, since these variables are best suited to facilitate the decomposition of the inhomogeneous term $\car^0_{(2k)}$ at the next order in $k$ into eigenfunctions of $\d_B^0$. All these quantities and their geometric meaning is defined in Appendix \ref{identities}, where various useful identities are presented as well. In terms of the anisotropic variables the momenta following from the $\co(1)$ solution in Table \ref{k=1-HJ-solution-0} are 
\bea\label{k=1-l=0-metric-momenta}
\left.\p^{0ij}_{(2,0)}\right|_{p_{1,0}^1}&=&\frac{1}{2\k^2}\sqrt{-\g}\left(p_{1,0}^1(\f)\left(-\bb R^{ij}+\bb D^{(i}\bb q^{j)}+\bb q^i\bb q^j+\frac12\bs^{ij}\left(\bb R-2\bb D_k\bb q^k-2\bb q^k\bb q_k\right)-\frac12 \bb n^i\bb n^j\bb R\right)\right.\NO\\
&&\left.+p_{1,0}'^1(\f)\left(\bb D^{(i}\bb D^{j)}\f+2\bb q^{(i}\bb D^{j)}\f-\bs^{ij}\left(\bb D^2\f+2\bb q^kD_k\f\right)\right)+p_{1,0}''^1(\f)\left(\bb D^{(i}\f\bb D^{j)}\f-\bs^{ij}\bb D_k\f \bb D^k\f\right)\right),\NO\\
\left.\p^{0ij}_{(2,0)}\right|_{p_{1,0}^2}&=&\frac{1}{2\k^2}\sqrt{-\g}\left(-p_{1,0}'^2(\f)\left(\frac12\bs^{ij}\bb q^kD_k\f-\bb q^{(i}\bb D^{j)}\f+\frac12\bb n^i\bb n^j\bb D^2\f\right)-\frac12p_{1,0}''^2(\f)\bb n^i\bb n^j\bb D_k\f\;\bb D^k\f\right),\NO\\
\left.\p^{0ij}_{(2,0)}\right|_{p_{1,0}^3}&=&\frac{1}{2\k^2}\sqrt{-\g}\left(p_{1,0}^3(\f)\left(\frac12\g^{ij}\bb q^k\bb q_k-\bb q^i\bb q^j+\bb n^i\bb n^j(\bb D_k\bb q^k+\bb q^k\bb q_k)\right)+p_{1,0}'^3(\f)\bb n^i\bb n^j\bb q^kD_k\f \right),\NO\\
\left.\p^{0ij}_{(2,0)}\right|_{p_{1,0}^4}&=&\frac{1}{2\k^2}\sqrt{-\g}\left(p_{1,0}^4(\f)\left(\frac12\bs^{ij}\bb q^kD_k\f-\bb q^{(i}\bb D^{j)}\f+\frac12\bb n^i\bb n^j\bb D^2\f\right)+\frac12p_{1,0}'^4(\f)\bb n^i\bb n^j\bb D_k\f\;\bb D^k\f\right),\NO\\
\left.\p^{0ij}_{(2,0)}\right|_{p_{1,0}^5}&=&\frac{1}{2\k^2}\sqrt{-\g}p_{1,0}^5(\f)\left(\frac12\g^{ij}\bb D_k\f\;\bb D^k\f-\bb D^i\f\;\bb D^j\f\right),\\\NO\\ 
\left.\p^{0ij}_{(2,2)}\right|_{p_{1,1}^1}&=&\frac{1}{2\k^2}\sqrt{-\g}\left(p_{1,1}^1(\f)\left(\rule{0cm}{0.5cm}-\bs^i_p\bs^j_q\bb n^kD_k\bb K^{pq}-\bb K\bb K^{ij}+2\bb n^{(i}\bb D_k\bb K^{kj)}+\frac12(\bs^{ij}+\bb n^i\bb n^j)\bb K^{kl}\bb K_{kl}\right)\right.\NO\\
&&\left.+p_{1,1}'^1(\f)\left(2\bb n^{(i}\bb K^{j)k}D_k\f-\bb K^{ij}\bb n^kD_k\f\right)\rule{0cm}{0.5cm}\right),\NO\\
\left.\p^{0ij}_{(2,2)}\right|_{p_{1,1}^2}&=&\frac{1}{2\k^2}\sqrt{-\g}\left(p_{1,1}^2(\f)\left(\bs^{ij}\bb n^kD_k\bb K-2\bb n^{(i}\bb D^{j)}\bb K+\frac12\g^{ij}\bb K^2\right)-p_{1,1}'^2(\f)\left(\rule{0cm}{0,5cm}\bb K\bb n^{(i}\bb D^{j)}\f\right.\right.\NO\\
&&\left.\left.+\bb n^{(i}\bb D^{j)}(\bb n^kD_k\f)-\frac12\bs^{ij}\left(2\bb K\bb n^kD_k\f+\bb n^kD_k(\bb n^lD_l\f)\right)+\frac12\bb n^i\bb n^j\bb K\bb n^kD_k\f\right)\right.\NO\\
&&\left.-p_{1,1}''^2(\f)\left(\bb n^{(i}\bb D^{j)}\f\; \bb n^kD_k\f-\frac12\bs^{ij}(\bb n^kD_k\f)^2\right)\right),\NO\\
\left.\p^{0ij}_{(2,2)}\right|_{p_{1,1}^3}&=&\frac{1}{2\k^2}\sqrt{-\g}\left(p_{1,1}^3(\f)\left(2\bb n^{(i}\bb D^{j)}\bb K-\bs^{ij}\bb n^kD_k\bb K-\frac12\g^{ij}\bb K^2\right)+p_{1,1}'^3(\f)\bb K\left(2\bb n^{(i}\bb D^{j)}\f-\bs^{ij}\bb n^kD_k\f\right)\right),\NO\\
\left.\p^{0ij}_{(2,2)}\right|_{p_{1,1}^4}&=&\frac{1}{2\k^2}\sqrt{-\g}\left(p_{1,1}^4(\f)\left(\bb n^{(i}\bb D^{j)}(\bb n^kD_k\f)-\frac12\bs^{ij}\bb n^kD_k(\bb n^lD_l\f)-\bb K\left(\bb n^{(i}\bb D^{j)}\f-\frac12\bb n^i\bb n^j\bb n^kD_k\f\right)\right)\right.\NO\\
&&\left.+p_{1,1}'^4(\f)\bb n^kD_k\f\left(\bb n^{(i}\bb D^{j)}\f-\frac12\bs^{ij}\bb n^kD_k\f\right)\right),\NO\\
\left.\p^{0ij}_{(2,2)}\right|_{p_{1,1}^5}&=&\frac{1}{2\k^2}\sqrt{-\g}p_{1,1}^5(\f)\left(\frac12\left(\bs^{ij}+\bb n^i\bb n^j\right)\bb n^kD_k\f-2\bb n^{(i}\bb D^{j)}\f\right)\bb n^lD_l\f.\label{k=1-l=1-metric-momenta}\\\NO\\
\left.\p^{0}_{\f(2,0)}\right|_{p_{1,0}^1}&=&\frac{1}{2\k^2}\sqrt{-\g}p_{1,0}'^1(\f)\bb R,\NO\\
\left.\p^{0}_{\f(2,0)}\right|_{p_{1,0}^2}&=&\frac{1}{2\k^2}\sqrt{-\g}p_{1,0}'^2(\f)(\bb D_i\bb q^i+\bb q^i\bb q_i),\NO\\
\left.\p^{0}_{\f(2,0)}\right|_{p_{1,0}^3}&=&\frac{1}{2\k^2}\sqrt{-\g}p_{1,0}'^3(\f)\bb q^i\bb q_i,\NO\\
\left.\p^{0}_{\f(2,0)}\right|_{p_{1,0}^4}&=&-\frac{1}{2\k^2}\sqrt{-\g}p_{1,0}^4(\f)(\bb D_i\bb q^i+\bb q^i\bb q_i),\NO\\
\left.\p^{0}_{\f(2,0)}\right|_{p_{1,0}^5}&=&-\frac{1}{2\k^2}\sqrt{-\g}\left(p_{1,0}'^5(\f)\bb D^k\f\bb D_k\f+2p_{1,0}^5(\f)\left(\bb D^2\f+\bb q^kD_k\f\right)\right),\label{k=1-l=0-scalar-momenta}\\
\left.\p^{0}_{\f(2,2)}\right|_{p_{1,1}^1}&=&\frac{1}{2\k^2}\sqrt{-\g}p_{1,1}'^1(\f)\bb K^{kl}\bb K_{kl},\NO\\
\left.\p^{0}_{\f(2,2)}\right|_{p_{1,1}^2}&=&\frac{1}{2\k^2}\sqrt{-\g}p_{1,1}'^2(\f)\bb n^kD_k\bb K,\NO\\
\left.\p^{0}_{\f(2,2)}\right|_{p_{1,1}^3}&=&\frac{1}{2\k^2}\sqrt{-\g}p_{1,1}'^3(\f)\bb K^2,\NO\\
\left.\p^{0}_{\f(2,2)}\right|_{p_{1,1}^4}&=&-\frac{1}{2\k^2}\sqrt{-\g}p_{1,1}^4(\f)\left(\bb K^2+\bb n^kD_k\bb K\right),\NO\\
\left.\p^{0}_{\f(2,2)}\right|_{p_{1,1}^5}&=&-\frac{1}{2\k^2}\sqrt{-\g}\left(p_{1,1}'^5(\f)(\bb n^iD_i\f)^2+2p_{1,1}^5(\f)\left(\bb K\bb n^kD_k\f+\bb n^kD_k(\bb n^lD_l\f)\right)\right).\label{k=1-l=1-scalar-momenta}
\eea
The coefficients $p_{k,\ell}^I$ appearing in these expressions are given in the last column of Table \ref{k=1-HJ-solution-0}. Finally, the vector momenta do not require functional differentiation since they are given directly by the solution of the last two equations in (\ref{inhomogeneous-sol}). Namely, from (\ref{vector-momentum}) we have
\be
\p^{0i}_{(2k,2\ell)}=
\int d^{d+1}x'\bs^i_j(x)\cl^{1j}_{(2k,2\ell)}[\g(x'),\f(x');x]+Y_o^{-1}B_o^i\int d^{d+1}x'B_{oj}(x)\cl^{1j}_{(2k,2\ell)}[\g(x'),\f(x');x].
\ee


\subsection{Solution at order $k=2$}

Given the $\co(1)$ momenta we obtained in the previous subsection we can now evaluate the $\co(1)$ source term at order $k=2$.  At $k=2$ the source (\ref{hsource}) is 
\be
\car_{(4)}=-\frac{\k^2}{\sqrt{-\g}}e^{-d\x\f}\left(2\p^{ij}_{(2)}\p_{(2)ij}-\frac2d\p\sub{2}^2
+\frac14Z^{-1}_\x\p^{i}_{(2)}\p_{(2)i}+\frac{1}{2\a}\left(\p_\f\sub{2}-2\x\p_{(2)}\right)^2+W^{-1}_\x\p_\om\sub{0}\p_\om\sub{2}\right).
\ee
Table \ref{quadratic-momenta} allows us to decompose this into eigenfunctions of $\d_B$ as 
\be
\car_{(4)}=\car_{(4,0)}+\car_{(4,2)}+\car_{(4,4)},
\ee
where 
\bea
&&\car_{(4,0)}=-\frac{\k^2}{\sqrt{-\g}}e^{-d\x\f}\left\{\frac{1}{2\a}\left(\p_\f\sub{2,0}-2\x\p_{(2,0)}\right)^2-\frac2d\left(\p_{(2,0)}\right)^2+ 2\s_{ik}\s_{jl}\p^{ij}_{(2,0)}\p^{kl}_{(2,0)}\right.\NO\\
&&\left.+2\left(Y^{-1}B_{i}B_{j}\p^{ij}_{(2,0)}\right)^2+8Y^{-1}\s_{ij}B_{k}B_{l}\p^{ik}_{(2,0)}\p^{jl}_{(2,2)}-2W^{-1}_\x\p_\om\sub{0}D_i\left(Y^{-1}\s^i_kB_{l}\p_{(2,0)}^{kl}\right)\right.\NO\\
&&\left.+\frac14Z^{-1}_\x\left(Y^{-1}\left(B_{i}\p_{(2,0)}^{i}\right)^2-4Y^{-1}\cp_{(2,0)i}B_{j}\p_{(2,0)}^{ij}+8Y^{-2}\s_{ij}B_{k}B_{l}\p^{ik}_{(2,0)}\p^{jl}_{(2,2)}\right)\right\},\NO\\\NO\\
&&\car_{(4,2)}=-\frac{\k^2}{\sqrt{-\g}}e^{-d\x\f}\left\{\frac{1}{\a}\left(\p_\f\sub{2,0}-2\x\p_{(2,0)}\right)\left(\p_\f\sub{2,2}-2\x\p_{(2,2)}\right)-\frac4d \p_{(2,0)}\p_{(2,2)}\right.\NO\\
&&\left.+ 4\s_{ik}\s_{jl}\p^{ij}_{(2,0)}\p^{kl}_{(2,2)}+4\left(Y^{-1}B_{i}B_{j}\p^{ij}_{(2,0)}\right)\left(Y^{-1}B_{k}B_{l}\p^{kl}_{(2,2)}\right)+4Y^{-1}\s_{ij}B_{k}B_{l}\p^{ik}_{(2,2)}\p^{jl}_{(2,2)}\right.\NO\\
&&\left.+\frac14Z^{-1}_\x\left(2Y^{-1}B_{i}B_{j}\p_{(2,0)}^{i}\p_{(2,2)}^{j}-4Y^{-1}\cp_{(2,0)i}B_{j}\p_{(2,2)}^{ij}-4Y^{-1}\cp_{(2,2)i}B_{j}\p_{(2,0)}^{ij}+\cp_{(2,0)i}\cp_{(2,0)}^{i}\right.\right.\NO\\
&&\left.\left.+4Y^{-2}\s_{ij}B_{k}B_{l}\p^{ik}_{(2,2)}\p^{jl}_{(2,2)}\right)+W^{-1}_\x\p_\om\sub{0}D_i\left(\cp^{i}_{(2,0)}+Y^{-1}B^iB_{j}\p_{(2,0)}^{j}-2Y^{-1}\s^i_kB_{l}\p_{(2,2)}^{kl}\right)\rule{0cm}{0.5cm}\right\},\NO\\\NO\\
&&\car_{(4,4)}=-\frac{\k^2}{\sqrt{-\g}}e^{-d\x\f}\left\{\frac{1}{2\a}\left(\p_\f\sub{2,2}-2\x\p_{(2,2)}\right)^2-\frac2d\left(\p_{(2,2)}\right)^2+2\s_{ik}\s_{jl}\p^{ij}_{(2,2)}\p^{kl}_{(2,2)}\right.\NO\\
&&\left.+2\left(Y^{-1}B_{i}B_{j}\p^{ij}_{(2,2)}\right)^2+\frac14Z^{-1}_\x\left(Y^{-1}\left(B_{i}\p_{(2,2)}^{i}\right)^2-4Y^{-1}\cp_{(2,2)i}B_{j}\p_{(2,2)}^{ij}+2\cp_{(2,0)i}\cp_{(2,2)}^{i}\right)\right.\NO\\
&&\left.+W^{-1}_\x\p_\om\sub{0}D_i\left(\cp^{i}_{(2,2)}+Y^{-1}B^iB_{j}\p_{(2,2)}^{j}\right)\rule{0cm}{0.5cm}\right\}.
\eea
These expressions can be simplified by noticing that, based on the eigenvalues in Table \ref{quadratic-momenta}, the following quantities must vanish:
\be
\s_{ij}B_{k}B_{l}\p^{ik}_{(2,0)}\p^{jl}_{(2,0)}=0,\quad \cp_{(2,2)i}\cp_{(2,2)}^{i}=0.
\ee
Since $\s_{ij}$ is asymptotically positive definite it follows that 
\be 
\s^i_kB_{l}\p^{kl}_{(2,0)}=0,\quad \cp_{(2,2)}^{i}=0. 
\ee
These identities have derived abstractly using the eigenvalues of the derivative and gradation operators, but can be checked explicitly. The first of these identities is is easily seen to hold for the momenta (\ref{k=1-l=0-metric-momenta}). The second identity is less obvious at this point, but can be checked in the examples in Section \ref{examples}. 

Finally, using these identities, as well as  (\ref{zero-order-momenta}) in order to properly isolate
the $\co(1)$ part of $\car_{(4,0)}$, $\car_{(4,2)}$ and $\car_{(4,4)}$, we can write the inhomogeneous terms at order $k=2$ in the simpler form
\bea\label{k=2-sources}
\car^0_{(4,0)}= -\frac{\k^2}{\sqrt{-\g}}e^{-d\x\f}&&\left( 2\p^{0ij}_{(2,0)}\p^{0}_{(2,0)ij}-\frac2d\left(\p^0_{(2,0)}\right)^2+\frac{1}{2\a}\left(\F_{(2,0)}^0\right)^2+\frac{1}{d\z}\left(Q^0_{(2,0)}\right)^2-\frac{\z}{d}\left(P^0_{(2,0)}\right)^2\right),\NO\\\NO\\
\car^0_{(4,2)}= -\frac{\k^2}{\sqrt{-\g}}e^{-d\x\f}&&\left( 4\p^{0ij}_{(2,0)}\p^{0}_{ij(2,2)}-\frac4d \p^0_{(2,0)}\p^0_{(2,2)}+\frac{1}{\a}\F_{(2,0)}^0\F_{(2,2)}^0+\frac{2}{d\z}Q^0_{(2,0)}Q^0_{(2,2)}-\frac{2\z}{d}P^0_{(2,0)}P^0_{(2,2)}\right.\NO\\
&&\left.
+\frac{\e}{2(z-1)}S_{(2,2)}^{0i}S_{(2,2)i}^{0}-4\bs_{ij}\bb n_k\bb n_l\p^{0ik}_{(2,2)}\p^{0jl}_{(2,2)}\right.\NO\\
&&\left.+\frac{1}{\sqrt{-Y_o}}W^{-1}_\x\p_\om^0\sub{0}D_i\left(S_{(2,2)}^{0i}-\bb n^i\left(P_{(2,0)}^0+\frac{1}{\z(\f)}Q_{(2,0)}^0\right)\right)\rule{0cm}{0.5cm}\right),\NO\\\NO\\
\car^0_{(4,4)}= -\frac{\k^2}{\sqrt{-\g}}e^{-d\x\f}&&\left(2\bs_{ik}\bs_{jl}\p^{0ij}_{(2,2)}\p^{0kl}_{(2,2)}+2\left(\bb n_i\bb n_j\p^{0ij}_{(2,2)}\right)^2-\frac2d\left(\p^0_{(2,2)}\right)^2+\frac{1}{2\a}\left(\F_{(2,2)}^0\right)^2-\frac{\z}{d}\left(P^0_{(2,2)}\right)^2\right.\NO\\
&&\left.+\frac{1}{d\z}\left(Q^0_{(2,2)}\right)^2-\frac{1}{\sqrt{-Y_o}}W^{-1}_\x\p_\om^0\sub{0}D_i\left(\bb n^i\left(P_{(2,2)}^0+\frac{1}{\z(\f)}Q_{(2,2)}^0\right)\right)\rule{0cm}{0.5cm}\right),
\eea
where we have defined 
\be\label{zeta}
\z(\f):=\frac d2\left(\frac{\e}{z-1}-\frac{d-1}{d}-\frac{1}{4\a}\left(\frac{Z'}{Z}\right)^2\right),
\ee
and
\bea\label{FPQ}
&&\F_{(2k,2\ell)}^0:=\p^0_\f\sub{2k,2\ell}-2\x\p^0_{(2k,2\ell)},\NO\\
&&Q_{(2k,2\ell)}^0:=\p^0_{(2k,2\ell)}+d\bb n^i\bb n^j\p^{0ij}_{(2k,2\ell)}+\frac{d}{4\a}\frac{Z'}{Z}\F_{(2k,2\ell)}^0,\NO\\
&&P_{(2k,2\ell)}^0:=B_{ok}\p_{(2k,2\ell)}^{0k}-\frac{1}{\z(\f)}Q_{(2k,2\ell)}^0,\NO\\
&&S_{(2k,2\ell)}^{0i}:=\bs^i_j\left(2\bb n_k\p_{(2k,2\ell)}^{0jk}+\sqrt{-Y_o}\p_{(2k,2\ell-2)}^{0j}\right).
\eea
Moreover, the inhomogeneous term (\ref{t-source}) can be written as 
\bea\label{t-source-simple}
&&B_{oj}\Hat\car^{1j}_{(2k,2\ell)}[\g(x),\f(x);x']:=B_{oj}\car^{1j}_{(2k,2\ell)}[\g(x),\f(x);x']\\
&&-2e^{\x\f}\left[\frac1\a\left(u_1'+\frac12 \frac{Z'}{Z}u_1\right)\F_{(2k,2\ell)}^0+2u_1\bb n_k\bb n_l\p_{(2k,2\ell)}^{0kl}+\frac2d(u_1+4u_2)Q_{(2k,2\ell)}^0\right]\d^{(d+1)}(x-x').\NO
\eea
Inserting the expressions for the canonical momenta from the order $k=1$ solution in these inhomogeneous terms one can use Table \ref{integration} in order to obtain the corresponding solutions $\cl^0_{(4,0)}$, $\cl^0_{(4,2)}$ and $\cl^0_{(4,4)}$ of the recursion relations  (\ref{kth-order-0}).

\section{Asymptotic expansions, Ward identities \& the holographic dictionary}
\label{ward}

So far we have concentrated on the algorithm for obtaining the general asymptotic solution of the radial Hamilton-Jacobi equation with Lifshitz or hyperscaling violating Lifshitz boundary conditions. The purpose of the current section is to point out certain generic features of this solution and to explain its relevance in the context of holography. 

\subsection{General structure of the solution, boundary counterterms \& renormalized action}

In the previous sections we have shown that this solution takes the form of a graded covariant expansion in simultaneous eigenfunctions of the operators $\Hat\d$ and $\d_B$, where each term in this expansion is a functional Taylor expansion in $B_i-B_{oi}$. Schematically, 
\be\label{formal-sol}
\cs=\sum_{k=0}^{k_{max}}\sum_{\ell=0}^k\left(\cs^0_{(2k,2\ell)}+\int (B-B_{o})\cs^1_{(2k,2\ell)}+\int \int(B-B_{o})^2\cs^2_{(2k,2\ell)}+\cdots\right).
\ee
By construction, each term in this expansion has definite asymptotic behavior, which is counted by the dilatation operator, $\d_D$, defined via the leading asymptotic behavior of the operator $\pa_r$ \cite{Papadimitriou:2004ap}. In order to determine the form of the dilatation operator we need to identify which field components are allowed to have independent sources by the boundary conditions, as well as their asymptotic behavior. As we have seen in the Section \ref{algorithm}, Lifshitz boundary conditions are equivalent to the covariant constraint (\ref{Lifshitz-constraint}) and so the covariant fields permitted to have independent sources are the metric $\g_{ij}$, the scalar $\f$, and the time component of $B_i-B_{oi}$. More concretely, decomposing $B_i-B_{oi}$ in timelike and spacelike components using the projection operator 
\be
\bs^i_j=\d^i_j-Y_o^{-1}B_o^iB_{oj},
\ee
we get
\be
B_i-B_{oi}=\bs^j_i (B_j-B_{oj})+Y_o^{-1}B_{oi}B_o^j(B_j-B_{oj})=\bs^j_i B_j+Y_o^{-1}B_{oi}B_o^j(B_j-B_{oj}).
\ee
However, (\ref{Lifshitz-constraint}) implies that the source of $\bs^j_i B_j$ must vanish for Lifshitz boundary conditions and therefore, since $B_{oi}$ is a function of $\g_{ij}$ and $\f$, the only independent source in $B_i-B_{oi}$ is contained in the scalar field
\be\label{psi}\boxed{
\j:=Y_o^{-1}B_o^j(B_j-B_{oj}).}
\ee
It follows that the dilatation operator can be identified with the asymptotic form of the operator  
\be
\pa_r=\int d^{d+1}x\left(\dot\g_{ij}\frac{\d}{\d\g_{ij}}+\dot\j\frac{\d}{\d \j}+\dot\f\frac{\d}{\d\f}\right).
\ee
The leading asymptotic form of $\g_{ij}$ and $\f$  can be obtained immediately from (\ref{leading-flow-eqs}) and (\ref{U-asymptotics}), namely
\be
\dot\g_{ij}\sim 2\g_{ij}+2(z-1)Y_o^{-1}B_{oi}B_{oj},\quad \dot\f\sim \m.
\ee
The leading asymptotic behavior of $\j$ can be inferred from that of $Y-Y_o$ in (\ref{supI-flow-eqs}), but it is instructive to derive it from first principles in the present more general setting. From (\ref{b-derivative-gamma}) and (\ref{b-derivative-phi}) we obtain  
\be
\dot B_{oi} =\frac12\left(Y_o^{-1}B_o^k B_o^l\dot\g_{kl}-\frac{Z'_\x(\f)}{Z_\x(\f)}\dot\f\right)B_{oi}.
\ee
Combining this with (\ref{leading-flow-eqs}) (ignoring transverse derivatives for now) yields 
\bea
\dot B_i-\dot B_{oi}&=&-e^{-d\x\f}Z_\x^{-1}U_YB_i-2e^{-d\x\f}\left(U_YY_o^{-1}B_o^kB_o^lB_kB_l+\left(\frac{\a_\x}{2d\a}U+\frac{\x}{2\a}U_X-\frac{\a_\x+d^2\x^2}{d\a}YU_Y\right)\right.\NO\\
&&\left.+\frac{1}{4\a}\frac{Z_\x'}{Z_\x}\left(U_X-(d+1)\x U+2\x YU_Y\right)\right)B_{oi}\NO\\
&=&-\frac14 e^{\x\f}\left(\frac{2}{\a}\left(u_0'+\frac{Z'}{Z}u_1\right)\frac{Z'}{Z}-\frac{8\e}{z-1}u_1+\frac{4}{d}\left(u_0+2(d-1)u_1\right)\right)B_{oi}+\frac{2\e}{z-1}e^{\x\f}u_1\bs^j_iB_j\NO\\
&&-2e^{\x\f}\left(\frac{1}{2\a}\left(u_1'+(u_1+2u_2)\frac{Z'}{Z}\right)\frac{Z'}{Z}-\frac{(u_1+4u_2)}{d}+2(u_1+2u_2)-\frac{4\e u_2}{z-1}\right)\frac{B_o^j(B_j-B_{oj})B_{oi}}{Y_o}\NO\\
&&+\frac{2\e}{z-1}e^{\x\f}u_1Y_o^{-1}B_{oi}B_o^j(B_j-B_{oj})+\co(B-B_o)^2.
\eea
As expected, the two $\co(B-B_o)^0$ terms vanish. The first one is proportional to the constraint given in the third equation in (\ref{potentials}), which was imposed by the requirement that the Taylor expansion in $Y-Y_o$ -- and hence in $B_i-B_{oi}$ -- be well defined. The second vanishes because the Lifshitz condition 
(\ref{Lifshitz-constraint}) requires that there be no source for $\bs^j_iB_j$. The $\co(B-B_o)$ is proportional to the scalar $\j$ defined in (\ref{psi}). Noting that the terms inside the parenthesis at $\co(B-B_o)$ are identical to the numerator of the function $\Om$ defined in (\ref{Omega}), we obtain
\be\label{B-asymptotics}\boxed{
\dot B_i-\dot B_{oi}\sim( \e-\D_-)Y_o^{-1}B_o^j(B_j-B_{oj})B_{oi}+\co(B-B_o)^2=( \e-\D_-)\j B_{oi}+\co(B-B_o)^2.}
\ee  
Since $\dot B_{o}^i\sim(\e-2z)B_o^i$ and $Y-Y_o\sim 2B_o^i(B_i-B_{oi})$, this leads to 
\be
\dot Y-\dot Y_{o}\sim \left(2(\e-z)-\D_-\right)(Y-Y_o)+\co(Y-Y_o)^2,
\ee
in complete agreement with the result (\ref{supI-flow-eqs}) we obtained in Section \ref{bgrnds}. Moreover, (\ref{B-asymptotics}) implies that 
\be
\dot\j\sim-\D_-\j,
\ee
and therefore the dilatation operator takes the form
\be\label{dilatation}\boxed{
\pa_r \sim\d_D\equiv \Hat\d^0+(z-1)\d_B^0+\int d^{d+1}x\left(\m\frac{\d}{\d\f}-\D_-\j\frac{\d}{\d \j}\right).}
\ee

Several comments are in order here. Firstly, it is clear from this form of the dilatation operator that every term in the expansion (\ref{formal-sol}) has definite asymptotic behavior. Namely, 
\be\label{asymptotics}
\int\cdots\int(B-B_{o})^m\cs^m_{(2k,2\ell)}\sim e^{\left(\cc_{k,\ell}+d\m\x-m\D_-\right)r},
\ee 
where recall that
\be
\cc_{k,\ell}:=d+1-2k+(z-1)(1-2\ell).
\ee
Secondly, we can now state more precisely why the dilatation operator is in general not a suitable operator in whose eigenfunctions to expand the solution of the HJ equation in the presence of a scalar field $\f$. Namely, each term in (\ref{formal-sol}) is in general only an {\em asymptotic} eigenfunction of $\d/\d\f$. However, an expansion in simultaneous 
eigenfunctions of $\Hat\d^0$ and $\d_B^0$ allows us to determine the $\phi$-dependence in closed form. Finally, note that in the relativistic limit $z\to 1$
\be
\Hat\d^0+(z-1)\d_B^0-\D_-\int d^{d+1}x\j\frac{\d}{\d \j}\to \Hat\d^0,
\ee
which is the operator used in \cite{Papadimitriou:2011qb} for the corresponding relativistic problem.

The definite asymptotic form (\ref{asymptotics}) of each term in the expansion (\ref{formal-sol}) allows us to determine up to which order in $k$, $\ell$ and $m$ we need to go. The criterion is that we need to determine all the terms for which 
\be
\cc_{k,\ell}+d\m\x-m\D_-\geq 0.
\ee 
When this quantity is positive the corresponding term in (\ref{formal-sol}) clearly diverges in the UV and needs to be removed with a local counterterm. The terms for which the inequality is saturated (which can only happen for certain values of the parameters $z$, $\th=-d\m\x$ and $\D_-$) are also divergent, but only linearly in the radial UV cut-off $r_o$. This follows from the fact that a  term in the expansion (\ref{formal-sol}) corresponding to the integers $k$, $\ell$ and $m$ has a single factor of $\cc_{k,\ell}+d\m\x-m\D_-$ in the denominator. This can be seen directly from the recursion formulas (\ref{inhomogeneous-sol}). Terms corresponding to integers for which the above inequality is saturated (if there are any) consequently have poles.  
By the usual dimensional regularization trick \cite{Papadimitriou:2004ap} where the radial cut-off is defined via
\be\label{anomaly-condition}
\frac{1}{\cc_{k,\ell}+d\m\x-m\D_-}=\frac{1}{d+z-\th-2k-(z-1)\ell-m\D_-}=:\frac{1}{d-d_*}=:r_o,
\ee    
the pole is traded for explicit cut-off dependence. Such terms  normally give rise to conformal anomalies since the explicit cut-off dependence breaks the invariance of the corresponding term under radial translations. In the absence of a linear dilaton, i.e. when $\m=0$, this is the best one can do since there is no regularization scheme where full bulk diffeomorphism invariance is preserved. However, when $\m\neq 0$ the cut-off $r_o$ can be replaced with $\f/\m$, thus preserving complete diffeomorphism invariance \cite{Papadimitriou:2011qb}. The terms for which the above inequality is saturated, therefore, always require regularization but they only lead to conformal anomalies when $\m=0$. This makes sense from the dual field theory point of view: for $\m\neq 0$ the theory has a running coupling in the UV. 

Irrespectively of whether there are integers for which the inequality (\ref{anomaly-condition}) is saturated, there is always an independent solution of the HJ equation starting with dilatation weight zero and is therefore UV finite. Namely, the solution (\ref{formal-sol}) takes the form
\be\label{HJ-solution-generic}\boxed{
\cs=\sum_{k,\ell,m\;|\; \cc_{k,\ell}+d\m\x-m\D_-\geq 0}\int\cdots\int(B-B_{o})^m\cs^m_{(2k,2\ell)} + \Hat\cs_{reg} +\cdots.}
\ee
where $\Hat\cs_{reg}$ is the lowest order term of this new independent solution and the dots stand for terms of negative dilatation weight that vanish in the UV. $\Hat\cs_{reg}$ satisfies
\be
\d_D\Hat\cs_{reg} =0,
\ee
and can be parameterized as 
\be\label{Sreg}
\Hat\cs_{reg}=\int d^{d+1}x\left(\g_{ij}\Hat\p^{ij}+B_i\Hat \p^i+\f\Hat\p_\f\right), 
\ee
where the quantities $\Hat\p^{ij}$, $\Hat \p^i$ and $\Hat\p_\f$
correspond undetermined integration functions of the HJ equation, subject only to certain constraints that we will derive shortly. In particular they are not functions of the induced fields $\g_{ij}$, $B_i$ and $\f$. As we have discussed in Section \ref{bgrnds}, a solution of the HJ equation that contains as many integration `constants' as generalized coordinates is a complete integral of the HJ equation, meaning that it is a sufficiently general solution of the HJ equation to describe {\em all} solutions of the second order equations of motion. In particular, every solution of the second order equations corresponds to specific values for the integration constants $\Hat\p^{ij}$, $\Hat \p^i$ and $\Hat\p_\f$. On the space of solutions of the equations of motion that have arbitrary sources for the fields $\g_{ij}$, $B_i$ and $\f$ (as allowed by the boundary conditions) and satisfy a certain regularity condition in the IR the quantities $\Hat\p^{ij}$, $\Hat \p^i$ and $\Hat\p_\f$ become {\em non-local} functionals of the sources.

The significance of $\Hat\cs_{reg}$ stems from the fact that the solution, $\cs$, of the HJ equation is nothing but the on-shell action. More accurately, for every solution of the equations of motion, the corresponding on-shell action is exactly equal to a complete integral of the HJ equation, for a specific choice of the integration functions $\Hat\p^{ij}$, $\Hat \p^i$ and $\Hat\p_\f$. The AdS/CFT dictionary identifies the on-shell action, and hence the complete integral $\cs$, with the generating function of connected correlation functions. The on-shell action is UV divergent, but its identification with the asymptotic complete integral (\ref{HJ-solution-generic}) means that these UV divergences can removed by the local covariant counterterms defined by  
\be\label{counterterms}
\boxed{
	\cs_{ct}:=-\sum_{k,\ell,m\;|\; \cc_{k,\ell}+d\m\x-m\D_-\geq 0 }\int\cdots\int(B-B_{o})^m\cs^m_{(2k,2\ell)}.}
\ee
This means that $\Hat\cs_{reg}=\cs+\cs_{ct}$ is identified with the {\em regularized on-shell action}, and therefore (by the AdS/CFT dictionary) with the regularized generating function of connected correlation functions. The {\em renormalized} on-shell action, or generating function, is given by the limit 
\be
\cs_{ren}:=\lim_{r_o\to\infty}\Hat\cs_{reg}.
\ee

\subsection{Fefferman-Graham asymptotic expansions, sources and 1-point functions }

The one-to-one correspondence between solutions of the equations of motion and complete integrals of the form (\ref{HJ-solution-generic}) can be seen clearly by deriving the Fefferman-Graham expansions for the induced fields directly from the asymptotic solution (\ref{HJ-solution-generic}) of the HJ equation. Inserting the solution (\ref{HJ-solution-generic}) in the flow equations (\ref{flow-eqs}), namely 
\begin{align}
\begin{aligned}
&\dot\g_{ij}=-\frac{4\k^2}{\sqrt{-\g}}e^{-d\x\f}\left(\left(\g_{ik}\g_{jl}-\frac{\a_\x+d^2\x^2}{d\a}\g_{ij}\g_{kl}\right)\frac{\d}{\d\g_{kl}}-\frac{\x}{2\a}\g_{ij}\frac{\d}{\d\f}\right)\cs,\\
&\dot A_i=-\frac{\k^2}{2}\frac{1}{\sqrt{-\g}}e^{-d\x\f}Z_\x^{-1}(\f)
\g_{ij}\frac{\d}{\d A_j}\cs,\\
&\dot\f=-\frac{\k^2}{\a}\frac{1}{\sqrt{-\g}}e^{-d\x\f}
\left(\frac{\d}{\d\f}-2\x\g_{ij}\frac{\d}{\d\g_{ij}}\right)\cs,\\
&\dot\om=-\frac{\k^2}{\sqrt{-\g}}e^{-d\x\f}W^{-1}_\x(\f)\frac{\d}{\d\om}\cs.
\end{aligned} 
\end{align}
one can obtain the Fefferman-Graham expansions by integrating the flow equations order by order in the radial coordinate. This way of deriving the asymptotic expansions completely bypasses the second order equations of motion and requires no ansatz for the form of these expansions. Most of the work has already been done in obtaining the asymptotic solution (\ref{HJ-solution-generic}) of the HJ equation and the flow equations allow us to use this result to derive the asymptotic expansions much more efficiently. 

More importantly, the flow equations allow us to identify generically the complete set of modes parameterizing the symplectic space of asymptotic solutions, without deriving the full form of these solutions. We have already identified a set of integration constants that parameterize $\Hat\cs_{reg}$ in the asymptotic complete integral (\ref{HJ-solution-generic}) of the HJ equation. These integration constants enter in the flow equations as    
\begin{align}
\label{flow-eqs-vevs}
\begin{aligned}
&\dot\g_{ij}\sim-\frac{4\k^2}{\sqrt{-\g}}e^{-d\x\f}\left(\Hat\p_{ij}-\frac{\a_\x+d^2\x^2}{d\a}\g_{ij}\Hat\p-\frac{\x}{2\a}\g_{ij}\Hat\f\right),\\
&\dot A_i\sim-\frac{\k^2}{2}\frac{1}{\sqrt{-\g}}e^{-d\x\f}Z_\x^{-1}(\f)
\Hat\p_i,\\
&\dot\f\sim-\frac{\k^2}{\a}\frac{1}{\sqrt{-\g}}e^{-d\x\f}
\left(\Hat\f-2\x\Hat\p\right),\\
&\dot\om\sim-\frac{\k^2}{\sqrt{-\g}}e^{-d\x\f}W^{-1}_\x(\f)D_i\Hat\p^i,
\end{aligned} 
\end{align}
and they will therefore lead to integration constants in the Fefferman-Graham expansions. To determine the radial dependence of these modes we need the other set of modes parameterizing the asymptotic expansions which corresponds to the integration constants of the flow equations themselves. To leading order the flow equations (\ref{flow-eqs}) reduce those given in
(\ref{leading-flow-eqs}). As we have already determined in Section \ref{algorithm} and in this section for $\j$ in (\ref{psi}), the requirement of asymptotically locally Lifshitz 
boundary conditions together with the leading form of the flow equations determine that the full set of integration constants of the flow equations and the corresponding radial dependence are as follows: 
\begin{align}
\label{sources}
\boxed{
\begin{aligned}
&n\sim e^{zr}n\sub{0}(x),\quad
n_a\sim e^{2r}n\sub{0}_a(x),\quad
\s_{ab}\sim e^{2r}\s\sub{0}_{ab}(x),\\
&\om\sim \om\sub{0}(x),\quad
\f\sim \m r+\f\sub{0}(x), \quad \j\sim e^{-\D_-r}\j_{-}(x),
\end{aligned} }
\end{align}
where $n\sub{0}(x)$, $n\sub{0}_a(x)$, $\s\sub{0}_{ab}(x)$, $\om\sub{0}(x)$, $\f\sub{0}(x)$ and $\j_{-}(x)$ are arbitrary functions of the transverse coordinates, and the given asymptotic form of $\f$ is valid for $\m\neq 0$. For $\m=0$ the asymptotic form of $\f$ depends on the subleading terms in the potentials that define the bulk theory. Note that the asymptotic behavior of the gauge field $A_i$ is completely determined in terms of these fields and does not contain any additional source allowed by the asymptotic Lifshitz condition (\ref{Lifshitz-constraint}), namely\footnote{This seems to contradict some of the findings of \cite{Christensen:2013lma,Christensen:2013rfa}. We thank Jelle Hartong and Niels Obers for pointing this out to us. However, as our analysis shows, we believe there is no additional boundary vector source, at least in the metric formalism. Our findings are in agreement with earlier literature using either the vielbein \cite{Ross:2011gu} or the metric \cite{Ross:2009ar,Mann:2011hg,Baggio:2011cp} formulations.}  
\be
A_i\sim \sqrt{\frac{z-1}{2\e Z_o}}\;n\sub{0}(x)e^{(\e-z)\f\sub{0}(x)/\m}e^{\e r}\d_{it}\left(1+e^{-\D_-r}\j_-(x)\right)+\pa_i\om\sub{0}(x).
\ee
The source $\om\sub{0}(x)$, therefore, corresponds to a pure gauge transformation.

The radial dependence of the sources (\ref{sources}) allows us to determine the radial dependence of the modes $\Hat\p^{ij}$, $\Hat \p^i$ and $\Hat\p_\f$ parameterizing $\Hat\cs_{reg}$. Since the only fields with independent sources are those in (\ref{sources}), (\ref{b-derivative-phi}) and (\ref{b-derivative-gamma}) imply that
\be
\d B_i\sim \d B_{oi} +\d\j B_{oi}
\sim\left(\frac12Y_o^{-1}B_o^kB_o^l\d\g_{kl}+(\n+\x)\d\f +\d\j \right)B_{oi},
\ee
so that
\be
\d\g_{ij}\Hat\p^{ij}+\d B_i\Hat\p^i+\d\f\Hat\p_\f\sim \d\g_{ij}\left(\Hat\p^{ij}+\frac12Y_o^{-1}B_o^iB_o^jB_{ok}\Hat\p^k\right)+\d\j B_{oi}\Hat\p^i+\d\f\left(\Hat\p_\f+(\n+\x)B_{oi}\Hat\p^i\right).
\ee
This motivates us to define the following quantities: 
\bea\label{vevs-1}
\Hat\ct^{ij}&:=&-\frac{e^{-d\x\f}}{\sqrt{-\g}}\left(2\Hat\p^{ij}+Y_o^{-1}B_o^iB_o^jB_{ok}\Hat\p^k\right)=-\frac{e^{-d\x\f}}{\sqrt{-\g}}\left(2\Hat\p^{ij}-\bb n^i\bb n^jB_{ok}\Hat\p^k\right),\NO\\
\Hat\co_\f &:=&\frac{e^{-d\x\f}}{\sqrt{-\g}}\left(\Hat\p_\f+(\n+\x)B_{oi}\Hat\p^i\right),\NO\\
\Hat\co_\j&:=&\frac{e^{-d\x\f}}{\sqrt{-\g}}B_{oi}\Hat\p^i,\NO\\
\Hat\ce^i&:= & \frac{e^{-d\x\f}}{\sqrt{-\g}}\sqrt{-Y_o}\bs^i_j\Hat\p^j.
\eea
Note that the quantity $\bs^i_j\Hat\p^j$ couples to variations of $B_i$ orthogonal to $B_{oi}$ and hence it corresponds to the 1-point function of an irrelevant operator. Although Lifshitz boundary conditions do not allow for a source of this operator it can have a non-zero expectation value. In terms of these variables the general variation of $\Hat\cs_{reg}$ with respect to the sources becomes  
\be\label{variation}
\d\g_{ij}\Hat\p^{ij}+\d B_i\Hat\p^i+\d\f\Hat\p_\f\sim \sqrt{-\g}e^{d\x\f}\left(-\frac12\d\g_{ij}\Hat\ct^{ij}+\d\j\Hat\co_\j+\d\f\Hat\co_\f\right),
\ee
where
\be
\d\g_{ij}\Hat\ct^{ij}=-2n\d n\Hat\ct^{tt}+2\d n_a(\Hat\ct^{ta}+n^a\Hat\ct^{tt})+\d\s_{ab}(\Hat\ct^{ab}-n^an^b\Hat\ct^{tt}).
\ee
The integration functions defined in (\ref{vevs-1}) are the symplectic conjugate variables to the sources (\ref{sources}) (except for $\Hat\ce^i$ whose source is set to zero) and, therefore, they are identified via the holographic dictionary with the renormalized 1-point functions of the dual operators.  The asymptotic form of these 1-point functions follows from the asymptotic form of the sources (\ref{sources}), together with the fact that $\Hat\cs_{reg}$ has dilatation weight zero. Namely, 
\be\label{vevs}\begin{array}{|c|l|l|}
\hline \hline 
&\hspace{0.7in}\mbox{1-point function}& \hspace{0.2in}\mbox{source}\quad\\
\hline&&\\
\quad\mbox{spatial stress tensor}\quad & \quad\Hat\P^i_j:=\bs^i_k\bs_{jl}\ct^{kl}\sim e^{-(d+z+d\m\x)r}\P^i_j(x)\quad & \quad \s_{(0)ab}\\&&\\
\mbox{momentum density} & \quad\Hat\cp^i:=-\bs^i_k\bb n_l\ct^{kl}\sim e^{-(d+2+d\m\x)r}\cp^i(x)\quad& \quad n_{(0)a}\\&&\\
\mbox{energy density} & \quad\Hat\ce:=-\bb n_k\bb n_l\ct^{kl}\sim e^{-(d+z+d\m\x)r}\ce(x)\quad& \quad n_{(0)}\\&&\\
\mbox{energy flux} & \quad\Hat\ce^i\sim e^{-(d+2z+d\m\x)r}\ce^i(x)\quad  &\quad 0\\&&\\
\mbox{dilaton}&\quad\Hat\co_\f\sim  e^{-(d+z+d\m\x)r}\co_\f(x)\quad & \quad\f_{(0)}\\&&\\
\mbox{composite scalar}&\quad\Hat\co_\j\sim  e^{-\D_+r}\co_\j(x)\quad & \quad \j_-\\&&\\
\hline\hline
\end{array}
\ee
As we shall confirm shortly by deriving the Ward identities these modes satisfy, this is precisely the spectrum of the energy-momentum complex \cite{Ross:2011gu}, plus the two additional scalar operators $\co_\f(x)$ and $\co_\j(x)$. Note that the asymptotic form of the momentum density and the energy flux differ by a factor of $e^{-r}$ relative to the operators defined in \cite{Ross:2011gu}, which reflects the fact that the indices of the corresponding operators in that reference are frame indices and not spacetime indices. The operators in \cite{Ross:2011gu} can be obtained by contracting our $\Hat\cp^i$ and $\Hat\ce^i$ with a spatial vielbein. However, the operators that enter the covariant Ward identities are $\Hat\cp^i$ and $\Hat\ce^i$ and not the ones with frame indices. Inverting the relations (\ref{vevs-1}) and inserting the asymptotic behaviors (\ref{vevs}) in (\ref{flow-eqs-vevs}) we obtain the dependence of the asymptotic expansions on the normalizable modes.

\subsection{Holographic Ward identities}

The holographic Ward identities follow directly from the first class constraints (\ref{constraints}). The Hamiltonian constraint leads to the trace Ward identity, while the momentum and gauge constraints imply the anisotropic diffeomorphism Ward identities. However, the trace Ward identity can be derived much more easily from the invariance of the HJ solution under radial translations.  
\begin{flushleft}
{\em Diffeomorphism Ward identity}
\end{flushleft}
Combining the momentum and gauge constraints in (\ref{constraints}) and applying them to $\Hat\cs_{reg}$ gives
\be
-2D_j\Hat\p^{ji}+F^i{}_j\Hat\p^j+\Hat\p_\f\pa^i\f-B^iD_j\Hat\p^j=0.
\ee
The leading asymptotic form of the vector field, $B_i\sim (1+\j)B_{oi}$, implies that
\be
F_{ij}\sim (1+\j)F_{oij}+\pa_i\j B_{oj}-\pa_j\j B_{oi}\sim F_{oij}+\pa_i\j B_{oj}-\pa_j\j B_{oi},
\ee
where we have assumed that $\D_->0$ in the second step. The above constraint then takes the form
\be
-D_j\left(2\Hat\p^{ji}+B_o^iY_o^{-1}B_{o}^jB_{ok}\Hat\p^k\right)+\left(\Hat\p^\f+(\n+\x)B_{ok}\Hat\p^k\right)\pa^i\f+(B_{oj}\Hat\p^j)\bb D^i\j-B_{o}^i\bb D_j(\bs^j_k\Hat\p^k)=0.
\ee
Using the variables introduced in (\ref{vevs-1}) and (\ref{vevs}) we obtain the constraint
\be 
D_j(e^{d\x\f}\Hat\ct^{ji})+e^{d\x\f}\left(\Hat\co_\f\pa^i\f+\Hat\co_\j \bb D^i\j\right)-B_o^i\bb D_j\left(\frac{e^{d\x\f}}{\sqrt{-Y_o}}\Hat\ce^j\right)=0. 
\ee
Different components of this equation behave differently asymptotically. Isolating components with the same scaling behavior using the projection operator $\bs^i_j$ we arrive at the three anisotropic Ward identities with arbitrary sources
 \begin{align}
 \label{diff-ward}
 \boxed{
 \begin{aligned}
 &\bb D_j\Hat\P^i_i+\bb q_j\Hat\P^j_i+\bb n^jD_j\Hat\cp_i+\bb K\Hat\cp_i+\bb K^i_i\Hat\cp_j+\bb n_i\bb q_j\Hat\cp^j-\Hat\ce \bb q_i+\Hat\co_\f\bb D_i\f+\Hat\co_\j\bb D_i\j=0,\\
 &\bb n^iD_i\Hat\ce+\bb K\Hat\ce-\bb K^i_j\Hat\P^j_i+\bb D_i\Hat\ce^i+\Hat\co_\f\bb n^iD_i\f=0,\\
 &\bb D_i\Hat\cp^i+2\bb q_i\Hat\cp^i=0.
 \end{aligned}}
 \end{align}
When all sources are set to their background value for flat space these identities reduce to the Ward identities for the energy-momentum complex discussed in \cite{Ross:2011gu}, plus conservation of the momentum density.  
\begin{flushleft}
	{\em Trace Ward identity}
\end{flushleft} 
The trace Ward identity can be derived by considering the transformation of $\Hat\cs_{reg}$ under an infinitesimal {\em local} radial translation $r_o\to r_o+\d\s(x)$, which induces an anisotropic Weyl transformation on the boundary. Such a tranformation in general gives 
\bea
\d_\s\Hat\cs_{reg}&\sim& \sqrt{-\g}e^{d\x\f}\left(n\d_\s n\Hat\ct^{tt}-\d_\s n_a\Hat\ct^{ta}-\frac12\d_\s\s_{ab}\Hat\ct^{ab}+\d_\s\j\Hat\co_\j+\d_\s\f\Hat\co_\f\right)\NO\\
&\sim& \sqrt{-\g}e^{d\x\f}\left(z n^2\Hat\ct^{tt}-2 n_a\Hat\ct^{ta}-\s_{ab}\Hat\ct^{ab}-\D_-\j\Hat\co_\j+\m\Hat\co_\f\right)\d\s.
\eea
If there is no explicit dependence on the radial cut-off in the counterterms, this variation must vanish identically. If, however, there is an explicit dependence on the radial cut-off, then the counterterms are not invariant and hence there is an additional contribution from the coefficients of the radial cut-off in the counterterms, i.e. the conformal anomaly. In particular, 
\begin{align}\label{trace-ward}
\boxed{
\begin{aligned}	
&z\Hat\ce+\Hat\P^i_i+\D_-\j\Hat\co_\j-\m\Hat\co_\f=0,& \m\neq 0,\\
&z\Hat\ce+\Hat\P^i_i+\D_-\j\Hat\co_\j=\ca,& \m=0,
\end{aligned}}
\end{align}
where the conformal anomaly is given by 
\be\label{anomaly}
	r_o \ca:=-\frac{e^{-d\x\f}}{\sqrt{-\g}}\sum_{k,\ell,m\;|\; \cc_{k,\ell}+d\m\x-m\D_- = 0 }\int\cdots\int(B-B_{o})^m\cs^m_{(2k,2\ell)}. 
\ee
As we pointed out earlier, there in no conformal anomaly when $\m\neq 0$ since in that case there is a regularization scheme that does not break radial translations.

\newpage

\section{Examples}
\label{examples}

In order to appreciate how the algorithm for solving the HJ equation recursively works in practice it is instructive to work through a few examples.

\subsection{Einstein-Proca theory}
\label{EP}

Our first example is the Einstein-Proca theory, which corresponds to setting
\bea 
&&u'_0+\frac{Z'}{Z}u_1=0,\quad \m=0,\quad \e=z, \quad \x=0, \quad \frac{W_o}{Z_o}=2dz,\NO\\
&&u_0=d+z-1, \quad u_1=\frac{z-1}{2},\quad u_2=\frac{(z-1)}{8(d+z-1)}\left((2d-1)(z-1)-d\D_-\right), 
\eea
where $\D_-$ is given in (\ref{dimensions-simple})
and the scalar field is constant, at least up to normalizable modes. This example is particularly interesting since it corresponds to the theory discussed in most of the literature on Lifshitz holography \cite{Koroteev:2007yp,Kachru:2008yh,Taylor:2008tg,Ross:2009ar,Horava:2009vy,Ross:2011gu,Mann:2011hg,Baggio:2011cp,Baggio:2011ha,Griffin:2011xs,Griffin:2012qx}. The linear equations (\ref{kth-order-0}), (\ref{kth-order-1-space}) and (\ref{kth-order-1-time}) 
in this case reduce to the {\em algebraic} equations  
\begin{align}
\label{mu=0-recursions} 
\begin{aligned}
-\cc_{k,\ell}\cl^0_{(2k,2\ell)}&=\car^0_{(2k,2\ell)},\\
-\cc_{k,\ell}\bs^i_j\cl^{1j}_{(2k,2\ell)}&=\bs^i_j\car^{1j}_{(2k,2\ell)},\\
\left(\D_--\cc_{k,\ell}\right)B_{oj}\cl^{1j}_{(2k,2\ell)}&=B_{oj}\Hat\car^{1j}_{(2k,2\ell)}.
\end{aligned} 
\end{align}
Moreover, the source term (\ref{t-source-simple}) for the third recursion relation now reads
\bea 
B_{oj}\Hat\car^{1j}_{(2k,2\ell)}&=&B_{oj}\car^{1j}_{(2k,2\ell)}-2(z-1)\bb n_k\bb n_l\p_{(2k,2\ell)}^{0kl}\d^{(d+1)}(x-x')\NO\\
&&-\frac{2(z-1)\left[(d+z-1)+(2d-1)(z-1)-d\D_-\right]}{d(d+z-1)}Q^0_{(2k,2\ell)}\d^{(d+1)}(x-x'),
\eea
and from (\ref{FPQ}) we obtain
\be
Q^0_{(2k,2\ell)}:=\p^0_{(2k,2\ell)}+d\bb n_k\bb n_l\p_{(2k,2\ell)}^{0kl},\qquad
P^0_{(2k,2\ell)}:=B_{oj}\p^{0j}_{(2k,2\ell)}-\frac{2(z-1)}{(d+z-1)}Q^0_{(2k,2\ell)}.
\ee
The solution of the recursion relations (\ref{mu=0-recursions}) takes the form
\bea\label{recursive-solution-proca}
\cl^0_{(2k,2\ell)}&=&-\frac{1}{\cc_{k,\ell}}\car^0_{(2k,2\ell)},\NO\\
\bs^i_j\p^{0j}_{(2k,2\ell)}(x')&=&-\frac{1}{\cc_{k,\ell}}\int d^{d+1}x\bs^i_j\car^{1j}_{(2k,2\ell)}[\g(x);x'],\NO\\
P^0_{(2k,2\ell)}(x')&=&\frac{1}{(\D_--\cc_{k,\ell})}\left(\int d^{d+1}xB_{oj}\car^{1j}_{(2k,2\ell)}[\g(x);x']\right.\NO\\
&&\left.\hspace{0in}-2(z-1)\bb n_k\bb n_l\p_{(2k,2\ell)}^{0kl}-\frac{2(z-1)\left(2z-1-\cc_{k,\ell}\right)}{(d+z-1)}Q^0_{(2k,2\ell)}\right),
\eea
where $\car^0_{(2k,2\ell)}$, $\bs^i_j\car^{1j}_{(2k,2\ell)}$ and $B_{oj}\car^{1j}_{(2k,2\ell)}$ for $k=1$ are given in Table \ref{k=1-HJ-solution-proca} and we have used (\ref{vector-momentum}) in the last two equations.
\begin{table}\vskip2.cm
\[\begin{array}{|c|c|c|c|c|}
\hline\hline
\multicolumn{3}{|c|}{k=1}& \car^0_{(2,2\ell)} & \cl^0_{(2,2\ell)} \\ \hline\hline 
\ell & I & \ct^I_{1,\ell} & c^I_{1,\ell} & p^I_{1,\ell}\\ \hline &&&&\\
0 & 1 & \bb R & -1 & \frac{1}{d+z-2}\\ &&&&\\
  & 2 & D_k\bb q^k & 2 & \frac{-2}{d+z-2}\\ &&&&\\
  & 3 & \bb q^i\bb q_i & -\frac{z-1}{z} & \frac{(z-1)}{z(d+z-2)}\\ &&&&\\
 \hline &&&&\\
1 & 1 & \bb K^{kl}\bb K_{kl} & -1 & \frac{1}{d-z}\\ &&&&\\
  & 2 & \bb n^kD_k\bb K & -2 & \frac{2}{d-z}\\ &&&&\\
  & 3 & \bb K^2 & -\frac{d+z-1}{d} & \frac{(d+z-1)}{d(d-z)}\\ &&&&\\
 \hline\hline
 \multicolumn{3}{|c|}{}& \bs^i_j\car^{1j}_{(2,2\ell)} & \bs^i_j\cl^{1j}_{(2,2\ell)}\\ \hline\hline 
 \ell & I & \ct^I_{1,\ell} & c^I_{1,\ell} & p^I_{1,\ell}\\ \hline &&&&\\
 0 & 1 & \bb q^i\bb n^kD_k^x\d(x-x') & -4Z_o\sqrt{-Y_o} & \frac{4Z_o\sqrt{-Y_o}}{d+z-2}\\ &&&&\\
 \hline &&&&\\
 1 & 1 & \bb K\bb D^i_x\d(x-x') & -\frac{1}{\sqrt{-Y_o}}\frac{4}{d}u_1 & \frac{4u_1}{d(d-z)\sqrt{-Y_o}}\\ &&&&\\
 & 2  & \bb K\bb q^i\d(x-x') & \frac{1}{\sqrt{-Y_o}}\frac{16}{d}u_2 & -\frac{16u_2}{d(d-z)\sqrt{-Y_o}} \\ &&&&\\
 \hline\hline
 \multicolumn{3}{|c|}{}& B_{oj}\car^{1j}_{(2,2\ell)} & B_{oj}\cl^{1j}_{(2,2\ell)} \\ \hline\hline 
 \ell & I & \ct^I_{1,\ell} & c^I_{1,\ell} & p^I_{1,\ell}\\ \hline &&&&\\
 0 & 1 & \bb q^kD_k^x\d(x-x') & 4Z_o Y_o & \frac{4Z_oY_o}{\D_--d-z+2}\\ &&&&\\
 \hline &&&&\\
 1 & 1 & \bb K\bb n^jD_j^x\d(x-x') & -\frac4d(u_1+4u_2) & -\frac{4(u_1+4u_2)}{d(\D_--d+z)}\\ &&&&\\
 & 2  & \bb K^2\d(x-x') & -\frac{16}{d}u_2 & -\frac{16u_2}{d(\D_--d+z)} \\ &&&&\\
 \hline\hline
\end{array}\]
\caption{General solution of the recursion relations (\ref{kth-order-0}) and (\ref{kth-order-1-time-degenerate}) at order $k=1$ for the Einstein-Proca theory. The second column from the right describes the sources $\car^0_{(2k,2\ell)}$, $\bs^i_j\car^{1j}_{(2k,2\ell)}$ and $B_{oj}\car^{1j}_{(2k,2\ell)}$ of these inhomogeneous equations in the form (\ref{scalar-source-decomposition}), while the last column gives the corresponding solutions $\cl^0_{(2k,2\ell)}$, $\bs^i_j\cl^{1j}_{(2k,2\ell)}$ and $B_{oj}\cl^{1j}_{(2k,2\ell)}$ in the parameterization (\ref{scalar-solution-decomposition}). For $B_{oj}\cl^{1j}_{(2k,2\ell)}$ only the terms corresponding to the source $B_{oj}\car^{1j}_{(2k,2\ell)}$ -- not the full $B_{oj}\Hat\car^{1j}_{(2k,2\ell)}$-- are listed here.  }
\label{k=1-HJ-solution-proca}
\end{table}

\begin{flushleft}
{\bf Solution at order $k=1$}
\end{flushleft}  

The solution (\ref{recursive-solution-proca}) at order $k=1$ is given in the last column of Table \ref{k=1-HJ-solution-proca} using the parameterization (\ref{scalar-solution-decomposition}). In particular, we can read off the solution for $\cl^0_{(2,2\ell)}$:
\bea\label{k=1-sol-proca}
\cl^0_{(2,0)}&=&\frac{1}{2\k^2}\frac{1}{d+z-2}\sqrt{-\g}\left(\bb R-2D_k\bb q^k+\frac{z-1}{z}\bb q^k\bb q_k\right),\NO\\
\cl^0_{(2,2)}&=&\frac{1}{2\k^2}\frac{1}{d-z}\sqrt{-\g}\left(\bb K^{kl}\bb K_{kl}+2\bb n^kD_k\bb K+\frac{d+z-1}{d}\bb K^2\right),
\eea
as well as for $\bs^i_j\p^{0j}_{(2,2\ell)}$:
\bea
\bs^i_j\p^{0j}_{(2,0)}&=&-\frac{1}{2\k^2}\frac{2(z-1)}{z(d+z-2)}\frac{1}{\sqrt{-Y_o}}\sqrt{-\g}\left(\bb n^kD_k\bb q^i+\bb K\bb q^i\right),\NO\\
\bs^i_j\p^{0j}_{(2,2)}&=&-\frac{1}{2\k^2}\frac{2}{d(d-z)}\frac{1}{\sqrt{-Y_o}}\sqrt{-\g}\left((z-1)\bb D^i\bb K+8u_2\bb K\bb q^i\right).
\eea
Differentiating the expressions in (\ref{k=1-sol-proca}) with respect to the metric $\g_{ij}$ leads to the momenta (\ref{k=1-l=0-metric-momenta}) and (\ref{k=1-l=1-metric-momenta}), which now take the form
\bea
\left.\p^{0ij}_{(2,0)}\right|_{p_{1,0}^1}&=&\frac{1}{2\k^2}\frac{1}{d+z-2}\sqrt{-\g}\left(-\bb R^{ij}+\bb D^{(i}\bb q^{j)}+\bb q^i\bb q^j+\frac12\bs^{ij}\left(\bb R-2\bb D_k\bb q^k-2\bb q^k\bb q_k\right)-\frac12 \bb n^i\bb n^j\bb R\right),\NO\\
\left.\p^{0ij}_{(2,0)}\right|_{p_{1,0}^2}&=&0,\NO\\
\left.\p^{0ij}_{(2,0)}\right|_{p_{1,0}^3}&=&\frac{1}{2\k^2}\frac{(z-1)}{z(d+z-2)}\sqrt{-\g}\left(\frac12\g^{ij}\bb q^k\bb q_k-\bb q^i\bb q^j+\bb n^i\bb n^j(\bb D_k\bb q^k+\bb q^k\bb q_k)\right),\label{k=1-l=0-metric-momenta-proca}\\\NO\\
\left.\p^{0ij}_{(2,2)}\right|_{p_{1,1}^1}&=&\frac{1}{2\k^2}\frac{1}{d-z}\sqrt{-\g}\left(\rule{0cm}{0.5cm}-\bs^i_p\bs^j_q\bb n^kD_k\bb K^{pq}-\bb K\bb K^{ij}+2\bb n^{(i}\bb D_k\bb K^{kj)}+\frac12(\bs^{ij}+\bb n^i\bb n^j)\bb K^{kl}\bb K_{kl}\right),\NO\\
\left.\p^{0ij}_{(2,2)}\right|_{p_{1,1}^2}&=&\frac{1}{2\k^2}\frac{2}{d-z}\sqrt{-\g}\left(\bs^{ij}\bb n^kD_k\bb K-2\bb n^{(i}\bb D^{j)}\bb K+\frac12\g^{ij}\bb K^2\right),\NO\\
\left.\p^{0ij}_{(2,2)}\right|_{p_{1,1}^3}&=&\frac{1}{2\k^2}\frac{(d+z-1)}{d(d-z)}\sqrt{-\g}\left(2\bb n^{(i}\bb D^{j)}\bb K-\bs^{ij}\bb n^kD_k\bb K-\frac12\g^{ij}\bb K^2\right).\label{k=1-l=1-metric-momenta-proca}
\eea
From these we obtain
\bea
&&\p^0_{(2,0)}=\frac{1}{2\k^2}\frac{1}{d+z-2}\sqrt{-\g}\left(\frac{d-1}{2}\left(\bb R+\frac{z-1}{z}\bb q_k\bb q^k-2D_k\bb q^k\right)-\frac{z-1}{z}D_k\bb q^k\right),\NO\\
&&\bb n_k\bb n_l\p^{0kl}_{(2,0)}=\frac{1}{2\k^2}\frac{1}{d+z-2}\sqrt{-\g}\left(-\frac12\bb R+\frac{z-1}{z}\left(D_k\bb q^k-\frac12 \bb q^k\bb q_k\right)\right),\NO\\
&&Q^{0}_{(2,0)}=\frac{1}{2\k^2}\frac{1}{d+z-2}\sqrt{-\g}\left(-\frac12\right)\left(\bb R+\frac{z-1}{z}\bb q^k\bb q_k+\frac{2(d-1)}{z}D_k\bb q^k\right),\\
\NO\\\NO\\
&&\p^0_{(2,2)}=\frac{1}{2\k^2}\frac{1}{d-z}\sqrt{-\g}\left(\frac{d-z}{2}\left(2\bb n^kD_k \bb K+\bb K^2+\bb K_{kl}\bb K^{kl}\right)+\frac{z-1}{2}\left(\bb K_{kl}\bb K^{kl}-\frac1d\bb K^2\right)\right),\NO\\ 
&&\bb n_k\bb n_l\p^{0kl}_{(2,2)}=\frac{1}{2\k^2}\frac{1}{d-z}\sqrt{-\g}\frac12\left(\bb K_{kl}\bb K^{kl}-\frac{d-z+1}{d}\bb K^2\right),\NO\\
&&Q^{0}_{(2,2)}=\frac{1}{2\k^2}\frac{1}{d-z}\sqrt{-\g}\left((d-z)\bb n^kD_k\bb K+\frac{2d-1}{2}\bb K_{kl}\bb K^{kl}-\frac{d+z-1}{2d}\bb K^2\right),\NO\\
&&S^{0i}_{(2,2)}=-\frac{1}{2\k^2}\sqrt{-\g}\left(\frac{2}{d-z}\left(\bb D_k\bb K^{ki}-\frac{d-z+1}{d}\bb D^i\bb K\right)+\frac{2(z-1)}{z(d+z-2)}\left(\bb n^kD_k\bb q^i+\bb K\bb q^i\right)\right).
\eea
Moreover, from Table \ref{k=1-HJ-solution-proca} we also obtain the source terms
\bea
\int d^{d+1}xB_{oj}\car^{1j}_{(2,0)}&=&\frac{1}{2\k^2}\sqrt{-\g}\frac{2(z-1)}{z}\bb D_k\bb q^k,\NO\\
\int d^{d+1}xB_{oj}\car^{1j}_{(2,2)}&=&\frac{1}{2\k^2}\frac{\sqrt{-\g}}{d}\left[2(z-1)\left(\bb n^k D_k+\bb K\right)+16 u_2\bb n^kD_k\right]\bb K.
\eea
Using these expressions we obtain the solution to the remaining third recursion relation in (\ref{mu=0-recursions}): 
\bea
P^{0}_{(2,0)}&:=&B_{oj}\p^{0j}_{(2,0)}-\frac{2(z-1)}{(d+z-1)}Q^0_{(2,0)}=-\frac{4z(z-1)}{(d+z-1)[\D_--(d+z-2)]}\left(Q^{0}_{(2,0)}+\frac{d+z-1}{2z^2}\frac{\sqrt{-\g}}{2\k^2}\bb q_k\bb q^k\right),\NO\\
P^{0}_{(2,2)}&:=&B_{oj}\p^{0j}_{(2,2)}-\frac{2(z-1)}{(d+z-1)}Q^0_{(2,2)}\NO\\
&=&\frac{1}{2\k^2}\sqrt{-\g}\frac{2 (z-1)}{(d-z) (d+z-1) (\D_{-}-d+z)}\left((d-z)(d-z-\D_{-})\bb n^{k}D_{k}\bb K\rule{0cm}{0.5cm}\right.\NO\\
&&\left.
+(d+z-1)\bb K^2-(d-z)\bb K_{kl}\bb K^{kl}+(d-3z+1)\left(\frac{z-1}{d}\bb K^2+d\bb K_{kl}\bb K^{kl}\right)\right).
\eea

\begin{flushleft}
{\bf Solution at order $k=2$}
\end{flushleft}  

The solutions to the three recursion relations (\ref{mu=0-recursions}) we obtained above determine the solution of the HJ equation up to and including order $k=1$ and $\co(B-B_o)$. With these results we can now determine the solution of the HJ equation at order $k=2$ but only to order $\co(1)$ in the Taylor expansion in $B-B_o$, which corresponds to the solution of only the first recursion relation in (\ref{mu=0-recursions}) for $k=2$. To obtain the solution of this recursion relation at $k=2$, the only non-trivial computation remaining is evaluating the inhomogeneous terms (\ref{k=2-sources}), which now read (dropping total derivatives in the last two)
\bea\label{k=2-sources-proca}
\car^0_{(4,0)}= -\frac{\k^2}{\sqrt{-\g}}&&\left(  2\p^{0ij}_{(2,0)}\p^{0}_{(2,0)ij}-\frac2d\left(\p^0_{(2,0)}\right)^2+\frac{2(z-1)}{d(d+z-1)}\left(Q^0_{(2,0)}\right)^2-\frac{(d+z-1)}{2d(z-1)} \left(P^0_{(2,0)}\right)^2\right),\NO\\\NO\\
\car^0_{(4,2)}= -\frac{\k^2}{\sqrt{-\g}}&&\left( 4\p^{0ij}_{(2,0)}\p^{0}_{ij(2,2)}-\frac4d \p^0_{(2,0)}\p^0_{(2,2)}+\frac{4(z-1)}{d(d+z-1)}Q^0_{(2,0)}Q^0_{(2,2)}-\frac{(d+z-1)}{d(z-1)} P^0_{(2,0)}P^0_{(2,2)}\right.\NO\\
&&\left.
+\frac{\sqrt{-\g}}{d\k^2} \left(S_{(2,2)}^{0i}-\bb n^i\left(P^0_{(2,0)}+\frac{2(z-1)}{(d+z-1)}Q^0_{(2,0)}\right)\right)D_i\bb K\right.\NO\\
&&\left.+\frac{z}{2(z-1)}S_{(2,2)}^{0i}S_{(2,2)i}^{0}-4\bs_{ij}\bb n_k\bb n_l\p^{0ik}_{(2,2)}\p^{0jl}_{(2,2)}\rule{0cm}{0.5cm}\right),\NO\\\NO\\
\car^0_{(4,4)}= -\frac{\k^2}{\sqrt{-\g}}&&\left(2\bs_{ik}\bs_{jl}\p^{0ij}_{(2,2)}\p^{0kl}_{(2,2)}+2\left(\bb n_i\bb n_j\p^{0ij}_{(2,2)}\right)^2-\frac2d\left(\p^0_{(2,2)}\right)^2+\frac{2(z-1)}{d(d+z-1)}\left(Q^0_{(2,2)}\right)^2\right.\NO\\
&&\left.-\frac{(d+z-1)}{2d(z-1)}\left(P^0_{(2,2)}\right)^2-\frac{\sqrt{-\g}}{d\k^2}  \left(P^0_{(2,2)}+\frac{2(z-1)}{(d+z-1)}Q^0_{(2,2)}\right)\bb n^kD_k\bb K\rule{0cm}{0.5cm}\right).
\eea
Using the results for the $k=1$ solution we obtained above we can write these source terms explicitly:
\bea
\car^0_{(4,0)}&=&-\frac{\sqrt{-\g}}{2\k^2}\frac{1}{(d+z-2)^2}\left\{
\left(\bb R^{ij}-\bb D^{(i}\bb q^{j)}-\frac1z \bb q^i\bb q^j\right)^2\right.\NO\\
&&\left.+\frac d4\left(\bb R-2D_k\bb q^k+\frac{z-1}{z}\bb q^k\bb q_{k}\right)^2+\frac14\left(\bb R+\frac{z-1}{z}(\bb q^k\bb q_{k}-2D_k\bb q^k)\right)^2\right.\NO\\
&&\left.-\left(\bb R-2D_k\bb q^k+\frac{z-1}{z}\bb q^k\bb q_{k}\right)\left(\bb R-D_k\bb q^k+\frac{z-1}{z}\bb q^k\bb q_{k}\right)\right.\NO\\
&&\left.-\frac1d\left(\frac{d-1}{2}\left(\bb R-2D_k\bb q^k+\frac{z-1}{z}\bb q^k\bb q_{k}\right)-\frac{z-1}{z}D_k\bb q^k\right)^2 \right.\NO\\
&&\left.+\frac{(z-1)}{4d(d+z-1)}\left(1-\frac{4z^2}{\left(\D_--(d+z-2)\right)^2}\right)\left(\bb R+\frac{z-1}{z}\bb q^k\bb q_k+\frac{2(d-1)}{z}D_k\bb q^k\right)^2\right.\\
&&\left.+\frac{2(z-1)(d+z-2)}{d\left(\D_--(d+z-2)\right)^2}\left(\bb R+\frac{2z(z-1)-(d+z-1)(d+z-2)}{2z^2}\bb q^k\bb q_k+\frac{2(d-1)}{z}D_k\bb q^k\right)\bb q_i\bb q^i\right\},\NO
\eea
\bea
\car^0_{(4,2)}&=&-\frac{\sqrt{-\g}}{2\k^2}\frac{1}{(d-z)(d+z-2)}\left\{2\left(\bb R^{ij}-\bb D^{(i}\bb q^{j)}-\frac1z\bb q^i\bb q^j\right)\left(\bs_{ip}\bs_{jq}\bb n^kD_k\bb K^{pq}+\bb K\bb K_{ij}\right)\right.\NO\\
&&\left.-\left(\bb R+\frac{z-1}{z}\bb q^i\bb q_i-D_i\bb q^i\right)\left(\bb K^{kl}\bb K_{kl}+\frac{d+z-1}{d}(2\bb n^kD_k\bb K+\bb K^2)\right)\right.\NO\\
&&\left.+\frac d2\left(\bb R+\frac{z-1}{z}\bb q^i\bb q_i-2D_i\bb q^i\right)\left(\bb K^{kl}\bb K_{kl}+\frac{d+z-1}{d}(2\bb n^kD_k\bb K+\bb K^2)-\frac2d\left(\bb n^kD_k\bb K+\bb K^2\right)\right)\right.\NO\\
&&\left.-\frac12\left(\bb R+\frac{z-1}{z}\left(\bb q^i\bb q_i-2D_i\bb q^i\right)\right)\left(\bb K^{kl}\bb K_{kl}-\frac{d+z-1}{d}\bb K^2\right)\right.\NO\\
&&\left.-\frac2d\left(\frac{d-1}{2}\left(\bb R+\frac{z-1}{z}\bb q_k\bb q^k-2D_k\bb q^k\right)-\frac{z-1}{z}D_k\bb q^k\right)\times\right.\NO\\
&&\left.\left(\frac{d-z}{2}\left(2\bb n^kD_k \bb K+\bb K^2+\bb K_{kl}\bb K^{kl}\right)+\frac{z-1}{2}\left(\bb K_{kl}\bb K^{kl}-\frac1d\bb K^2\right)\right)-\frac{(z-1)}{d(d+z-1)}\times\right.\NO\\
&&\left.\left(\bb R+\frac{z-1}{z}\bb q^k\bb q_k+\frac{2(d-1)}{z}D_k\bb q^k\right)\left((d-z)\bb n^kD_k\bb K+\frac{2d-1}{2}\bb K_{kl}\bb K^{kl}-\frac{d+z-1}{2d}\bb K^2\right)\right.\NO\\
&&\left.-\frac{2z(z-1)}{d(d+z-1)(\D_--d+z)(\D_--(d+z-2))}\left(\bb R-\frac{(d-1)(d+2z-2)}{z^2}\bb q^k\bb q_k+\frac{2(d-1)}{z}D_k\bb q^k\right)\right.\NO\\
&&\left.\times\left((d-z)(d-z-\D_{-})\bb n^{k}D_{k}\bb K
+(d+z-1)\bb K^2-(d-z)\bb K_{kl}\bb K^{kl}\rule{0cm}{0.5cm}\right.\right.\NO\\
&&\left.\left.+(d-3z+1)\left(\frac{z-1}{d}\bb K^2+d\bb K_{kl}\bb K^{kl}\right)\right)-\frac{2(d+z-2)}{(d-z)}\left(\bb D_k\bb K^{ki}-\frac{d-z+1}{d}\bb D^i\bb K\right)^2\right.\NO\\
&&\left.+\frac{z(d-z)(d+z-2)}{4(z-1)}\left(\frac{2}{d-z}\left(\bb D_k\bb K^{ki}-\frac{d-z+1}{d}\bb D^i\bb K\right)+\frac{2(z-1)}{z(d+z-2)}\left(\bb n^kD_k\bb q^i+\bb K\bb q^i\right)\right)^2\right.\NO\\
&&\left.-\frac{(d-z)(d+z-2)}{d}\left(\frac{2}{d-z}\left(\bb D_k\bb K^{ki}-\frac{d-z+1}{d}\bb D^i\bb K\right)+\frac{2(z-1)}{z(d+z-2)}\left(\bb n^kD_k\bb q^i+\bb K\bb q^i\right)\right)D_i\bb K\right.\NO\\
&&\left.+\frac{(z-1)(d-z)}{d(d+z-1)}\left(\bb R+\frac{z-1}{z}\bb q^k\bb q_k+\frac{2(d-1)}{z}D_k\bb q^k\right)\bb n^iD_i\bb K
\right.\\
&&\left.-\frac{2z(z-1)(d-z)}{d(d+z-1)(\D_--(d+z-2))}\left(\bb R-\frac{(d-1)(d+2z-2)}{z^2}\bb q^k\bb q_k+\frac{2(d-1)}{z}D_k\bb q^k\right)\bb n^iD_i\bb K\right\},\NO
\eea
\bea
\car^0_{(4,4)}&=&-\frac{\sqrt{-\g}}{2\k^2}\frac{1}{(d-z)^2}\left\{\left(\bs_{ip}\bs_{jq}\bb n^kD_k\bb K^{pq}+\bb K\bb K_{ij}\right)^2+\frac d4\left(\bb K^{kl}\bb K_{kl}+\frac{d+z-1}{d}(2\bb n^kD_k\bb K+\bb K^2)\right)^2\right.\NO\\
&&\left.-\left(\bb n^kD_k\bb K+\bb K^2\right)\left(\bb K^{kl}\bb K_{kl}+\frac{d+z-1}{d}(2\bb n^kD_k\bb K+\bb K^2)\right)+\frac14\left(\bb K_{kl}\bb K^{kl}-\frac{d-z+1}{d}\bb K^2\right)^2\right.\NO\\
&&\left.-\frac1d\left(\frac{d-z}{2}\left(2\bb n^kD_k \bb K+\bb K^2+\bb K_{kl}\bb K^{kl}\right)+\frac{z-1}{2}\left(\bb K_{kl}\bb K^{kl}-\frac1d\bb K^2\right)\right)^2\right.\NO\\
&&\left.+\frac{(z-1)}{d(d+z-1)}\left((d-z)\bb n^kD_k\bb K+\frac{2d-1}{2}\bb K_{kl}\bb K^{kl}-\frac{d+z-1}{2d}\bb K^2\right)^2\right.\NO\\
&&\left.-\frac{(z-1)}{d(d+z-1)(\D_--d+z) ^2}\left((d-z)(d-z-\D_{-})\bb n^{k}D_{k}\bb K\rule{0cm}{0.55cm}
+(d+z-1)\bb K^2-(d-z)\bb K_{kl}\bb K^{kl}\right.\right.\NO\\
&&\left.\left.+(d-3z+1)\left(\frac{z-1}{d}\bb K^2+d\bb K_{kl}\bb K^{kl}\right)\right)^2-\frac{2(z-1)(d-z)}{d(d+z-1)(\D_--d+z)}\times\right.\NO\\
&&\left.\left((d+z-1)\bb K^2-(d-z)\bb K_{kl}\bb K^{kl}+(d-3z+1)\left(\frac{z-1}{d}\bb K^2+d\bb K_{kl}\bb K^{kl}\right)\right)\bb n^iD_i\bb K\right.\NO\\
&&\left.-\frac{2(z-1)(d-z)}{d(d+z-1)}\left(\frac{2d-1}{2}\bb K_{kl}\bb K^{kl}-\frac{d+z-1}{2d}\bb K^2\right)\bb n^iD_i\bb K\right\}.
\eea
Correspondingly, the solution of the first recursion relation in (\ref{mu=0-recursions}) for $k=2$ is 
\be
\cl^0_{(4,0)}=-\frac{1}{d+z-4}\car^0_{(4,0)},\quad \cl^0_{(4,2)}=-\frac{1}{d-z-2}\car^0_{(4,2)},\quad \cl^0_{(4,4)}=-\frac{1}{d-3z}\car^0_{(4,4)}.
\ee

As an illustration let us consider the case $d=z=2$ which has been discussed before e.g. in \cite{Griffin:2011xs}. From (\ref{dimensions-simple}) follows that in this case $\D_-=0$ and hence $Y-Y_o\sim r$ as $r\to\infty$ and so we must set this mode to zero to ensure asymptotically locally Lif boundary conditions \cite{Ross:2009ar}. The zero order solution of the Taylor expansion in $B-B_o$ therefore gives the full solution in this case. The terms that contribute to the UV divergences, therefore are 
\be
\cs=\int d^{d+1}x\left(\cl^0_{(0)}+\cl^0_{(2,0)}+\cl^0_{(2,2)}+\cl^0_{(4,0)}\right),
\ee
where $\cl^0_{(0)}$ was given in (\ref{asymptotic-superpotential}). The terms $\cl^0_{(2,2)}$ and $\cl^0_{(4,0)}$ have poles at $d=z=2$ and therefore both contribute to the conformal anomaly. Setting $z=2$ and 
\be
\frac{1}{d-2}=r_o,
\ee 
where $r_o$ is the UV cut-off, these terms become 
\bea\label{horava}
\cl^0_{(0)}&=&\frac{\sqrt{-\g}}{2\k^2}6,\NO\\
\cl^0_{(2,0)}&=&\frac{\sqrt{-\g}}{2\k^2}\frac{1}{2}\left(\bb R-2D_k\bb q^k+\frac12\bb q^k\bb q_k\right)\simeq\frac{\sqrt{-\g}}{2\k^2}\frac{1}{2}\left(\bb R+\frac12\bb q^k\bb q_k\right),\NO\\
\cl^0_{(2,2)}&=&\frac{\sqrt{-\g}}{2\k^2}r_o\left(\bb K^{kl}\bb K_{kl}+2\bb n^kD_k\bb K+\frac32\bb K^2\right)\simeq\frac{\sqrt{-\g}}{2\k^2}r_o\left(\bb K^{kl}\bb K_{kl}-\frac12\bb K^2\right),\NO\\
\cl^0_{(4,0)}&=&\frac{\sqrt{-\g}}{2\k^2}\frac{r_o}{4}\left\{
\left(\bb R^{ij}-\bb D^{(i}\bb q^{j)}-\frac12 \bb q^i\bb q^j\right)^2+\frac 12\left(\bb R-2D_k\bb q^k+\frac12\bb q^k\bb q_{k}\right)^2+\frac14\left(\bb R+\frac12(\bb q^k\bb q_{k}-2D_k\bb q^k)\right)^2\right.\NO\\
&&\left.-\left(\bb R-2D_k\bb q^k+\frac12\bb q^k\bb q_{k}\right)\left(\bb R-D_k\bb q^k+\frac12\bb q^k\bb q_{k}\right)-\frac18\left(\bb R-3D_k\bb q^k+\frac12\bb q^k\bb q_{k}\right)^2 \right.\NO\\
&&\left.-\frac{1}{8}\left(\bb R+\frac12\bb q^k\bb q_k+D_k\bb q^k\right)^2+\frac12\left(\bb R-\frac14\bb q^k\bb q_k+D_k\bb q^k\right)\bb q_i\bb q^i\right\}\NO\\
&=&\frac{\sqrt{-\g}}{2\k^2}\frac{r_o}{4}\left\{\left(\bb D_i\bb q_j+\frac12\bb q_i\bb q_j-\frac12\bs_{ij}\left(\bb D_k\bb q^k+\frac12\bb q_k\bb q^k\right)\right)^2-\frac12\left(D_k\bb q^k-\frac12\bb q_k\bb q^k\right)^2+\frac12 \bb R\bb q_k\bb q^k\right\},
\eea
where $\simeq$ denotes equivalence up to total derivative terms and we have used the identities (see Table \ref{anisotropic-geometry}) $\bb D_{[i}\bb q_{j]}=0$, $D_k\bb q^k=\bb D_k\bb q^k+\bb q_k\bb q^k$, and $\bb R_{ij}=\frac12\bb R\bs_{ij}$ for $d=2$. Using the fact that up to total derivative terms 
\be
\bb D^i\bb q^j\bb D_i\bb q_j\simeq -\bb q^i\bb q^j\bb D_i\bb q_j-\frac12\bb R\bb q_k\bb q^k+\bb q^k\bb q_k\bb D_l\bb q^l+(\bb D_k\bb q^k)^2,
\ee
it is easy to check that $\cl^0_{(4,0)}$ vanishes identically in agreement with \cite{Griffin:2011xs}.

\FloatBarrier

\subsection{Exponential potentials with $\m=0$}

A second interesting example is a generalization of the Einstein-Proca theory discussed above obtained by relaxing the condition that the scalar be constant and that $\x=0$. In particular, the scalar is not necessarily constant in this case and the potentials defining the Lagrangian take the form  
\be 
V_{\xi}=V_o,\qquad  Z_{\xi}=Z_o e^{-2  (\xi +\n )\f},\qquad W_{\xi}=W=W_o e^{-2  (\xi +\n ) \f},
\ee
where 
\be
\m=0,\quad \e=z,\quad \n=-\frac{(d+z-1)\x}{z-1},\quad
W_o=2dzZ_o,\quad V_o=-\left(d(d+z)+z(z-1)\right).
\ee
The first three coefficients in the Taylor expansion of the superpotential correspondingly take the form
\bea
&&u_0(\f)=\left(d+z-1\right)e^{-\x\f},\quad
u_1(\f)=\frac12 (z-1)e^{-\x\f}, \NO\\
&&u_2(\f)=\frac{(z-1)^2\left((2d-1)(z-1)+\frac{d(2d+z-1)(d+z-1)}{(z-1)}\frac{\x^2}{\a}-d\D_-\right)}{8(d+z-1)\left(z-1-d(d+z-1)\frac{\x^2}{\a}\right)}e^{-\x\f},
\eea 
where $\D_-$ now must be evaluated using the general expression (\ref{dimensions}) instead of (\ref{dimensions-simple}). Note that as for the Einstein-Proca theory 
\be
u_0'+\frac{Z'}{Z}u_1=0,
\ee
and therefore the recursion relations that determine the HJ solution are still algebraic and in fact identical to those of the Einstein-Proca theory given in (\ref{mu=0-recursions}).  

The source term (\ref{t-source-simple}) of the third recursion relation now takes the form
\bea\label{t-source-simple-exp}
&&B_{oj}\Hat\car^{1j}_{(2k,2\ell)}[\g(x),\f(x);x']-\frac1\z(\D_--\cc_{k,\ell})Q_{(2k,2\ell)}^0\d^{(d+1)}(x-x')=\NO\\
&&B_{oj}\car^{1j}_{(2k,2\ell)}[\g(x),\f(x);x']-2(z-1)\bb n_k\bb n_l\p_{(2k,2\ell)}^{0kl}\d^{(d+1)}(x-x')-\frac{d\x}{\a}\F_{(2k,2\ell)}^0\d^{(d+1)}(x-x')\NO\\
&&-\frac{2(z-1)}{(d+z-1)\left(1+\frac{d\n\x}{\a}\right)}\left(2z-1-\frac{d\n\x}{\a}-\cc_{k,\ell}\right)Q_{(2k,2\ell)}^0\d^{(d+1)}(x-x'),
\eea
where $\bar u_2:= e^{\x\f}u_2$, and from (\ref{FPQ}) we obtain
\bea\label{FPQ-exp}
&&\F_{(2k,2\ell)}^0:=\p^0_\f\sub{2k,2\ell}-2\x\p^0_{(2k,2\ell)},\NO\\
&&Q_{(2k,2\ell)}^0:=\p^0_{(2k,2\ell)}+d\bb n^i\bb n^j\p^{0ij}_{(2k,2\ell)}-\frac{d\n}{2\a}\F_{(2k,2\ell)}^0,\NO\\
&&P_{(2k,2\ell)}^0:=B_{ok}\p_{(2k,2\ell)}^{0k}-\frac{2(z-1)}{(d+z-1)\left(1+\frac{d\n\x}{\a}\right)}Q_{(2k,2\ell)}^0.
\eea
The solutions to the three recursion relations (\ref{mu=0-recursions}) can therefore be written in the form
\bea\label{recursive-solution-exp}
\cl^0_{(2k,2\ell)}&=&-\frac{1}{\cc_{k,\ell}}\car^0_{(2k,2\ell)},\NO\\
\bs^i_j\p^{0j}_{(2k,2\ell)}(x')&=&-\frac{1}{\cc_{k,\ell}}\int d^{d+1}x\bs^i_j\car^{1j}_{(2k,2\ell)}[\g(x),\f(x);x'],\NO\\
P_{(2k,2\ell)}^0(x')&=&\frac{1}{\D_--\cc_{k,\ell}}\left(\rule{0cm}{.7cm}\int d^{d+1}xB_{oj}\car^{1j}_{(2k,2\ell)}[\g(x),\f(x);x']-2(z-1)\bb n_k\bb n_l\p_{(2k,2\ell)}^{0kl}\right.\NO\\
&&\left.-\frac{d\x}{\a}\F_{(2k,2\ell)}^0-\frac{2(z-1)}{(d+z-1)\left(1+\frac{d\n\x}{\a}\right)}\left(2z-1-\frac{d\n\x}{\a}-\cc_{k,\ell}\right)Q_{(2k,2\ell)}^0\right),
\eea
and again we have used (\ref{vector-momentum}). Note that in the limit $\x\to 0$ these expressions reduce to the corresponding ones in (\ref{recursive-solution-proca}) for the Einstein-Proca theory. The source terms $\car^0_{(2k,2\ell)}$, $\bs^i_j\car^{1j}_{(2k,2\ell)}$ and $B_{oj}\car^{1j}_{(2k,2\ell)}$ for $k=1$ are given in Tables \ref{k=1-HJ-solution-exp} and \ref{k=1-HJ-solution-1-exp}.
\begin{table}\vspace{2in}
\[\begin{array}{|c|c|c|c|c|}
\hline\hline
\ell & I & \ct^I_{1,\ell} & c^I_{1,\ell} & p^I_{1,\ell}\\ \hline &&&&\\
0 & 1 & \bb R & -e^{d\x\f} & \frac{1}{d+z-2}e^{d\x\f}\\ &&&&\\
  & 2 & D_k\bb q^k & 2e^{d\x\f} & \frac{-2}{d+z-2}e^{d\x\f}\\ &&&&\\
  & 3 & \bb q^i\bb q_i & -\frac{z-1}{z}e^{d\x\f} & \frac{(z-1)}{z(d+z-2)}e^{d\x\f}\\ &&&&\\
  & 4 & \bb q^i\pa_i\f & \frac{2d\x}{z}e^{d\x\f} & -\frac{2d\x}{z(d+z-2)}e^{d\x\f} \\ &&&&\\
  & 5 & \bs^{ij}\pa_i\f\pa_j\f & \left(\a_\x-\frac{d^2\x^2}{z(z-1)}\right)e^{d\x\f} & -\frac{1}{d+z-2}\left(\a_\x-\frac{d^2\x^2}{z(z-1)}\right)e^{d\x\f}\\ &&&&\\
 \hline &&&&\\
1 & 1 & \bb K^{kl}\bb K_{kl} & -e^{d\x\f} & \frac{1}{d-z}e^{d\x\f}\\ &&&&\\
  & 2 & \bb n^kD_k\bb K & -2e^{d\x\f} & \frac{2}{d-z}e^{d\x\f}\\ &&&&\\
  & 3 & \bb K^2 & -\frac{d+z-1}{d}e^{d\x\f} & \frac{(d+z-1)}{d(d-z)}e^{d\x\f}\\ &&&&\\
  & 4 & \bb K \bb n^j \pa_j\f & -2\x z e^{d\x\f} & \frac{2\x z}{d-z}e^{d\x\f}\\ &&&&\\
  & 5 & (\bb n^i \pa_i\f)^2 & -\left(\a_\x+ \frac{d\x^2z^2}{z-1}\right)e^{d\x\f} & \frac{1}{d-z}\left(\a_\x+ \frac{d\x^2z^2}{z-1}\right)e^{d\x\f}\\ 
  &&&&\\
 \hline\hline
\end{array}\]
\caption{General solution of the first recursion relation in (\ref{inhomogeneous-sol}) at order $k=1$ for exponential potentials and $\m=0$. The second column from the right describes the source of the inhomogeneous equation in the form (\ref{scalar-source-decomposition}), while the last column gives the solution $\cl^0_{(2,0)}$ and $\cl^0_{(2,2)}$ in the parameterization (\ref{scalar-solution-decomposition}).}
\label{k=1-HJ-solution-exp}\vspace{1.6in}
\end{table}
\begin{table}
	\vskip0.8in	
	\[\begin{array}{|c|c|c|c|c|}
	\hline\hline
	\multicolumn{3}{|c|}{k=1,\;\co(B-B_o),\; \mbox{space}}& \bs^i_j\car^{1j}_{(2,2\ell)} & \bs^i_j\cl^{1j}_{(2,2\ell)} \\ \hline\hline 
	\ell & I & \ct^I_{1,\ell} & c^I_{1,\ell} & p^I_{1,\ell}\\ \hline &&&&\\
	0 & 1 & \bb q^i\bb n^kD_k^x\d(x-x') & -4e^{d\x\f}Z_\x\sqrt{-Y_o} & \frac{4e^{d\x\f}Z_\x\sqrt{-Y_o}}{d+z-2}\\ &&&&\\
	\hline &&&&\\
	1 & 1 & \bb K\bb D^i_x\d(x-x') & -\frac{e^{d\x\f}}{\sqrt{-Y_o}}\frac{2(z-1)}{d} & \frac{e^{d\x\f}}{\sqrt{-Y_o}}\frac{2(z-1)}{d(d-z)}\\ &&&&\\
	& 2  & \bb K\bb q^i\d(x-x') & \frac{e^{d\x\f}}{\sqrt{-Y_o}}\frac{16\bar u_2}{d} & -\frac{e^{d\x\f}}{\sqrt{-Y_o}}\frac{16\bar u_2}{d(d-z)} \\ &&&&\\
	& 3 & \bb n^k\pa_k\f\bb D^i_x\d(x-x') & -\frac{e^{d\x\f}}{\sqrt{-Y_o}}2\x z& \frac{e^{d\x\f}}{\sqrt{-Y_o}}\frac{2\x z}{(d-z)}\\ &&&&\\
	& 4 & \bb K\bb D^i_x\f\d(x-x') & -\frac{e^{d\x\f}}{\sqrt{-Y_o}} 2\x z& \frac{e^{d\x\f}}{\sqrt{-Y_o}}\frac{2\x z}{(d-z)}\\ &&&&\\
	& 5 & \bb q^i\bb n^k\pa_k\f\d(x-x') & \frac{e^{d\x\f}}{\sqrt{-Y_o}}\frac{16\bar u_2\x z}{z-1} & -\frac{e^{d\x\f}}{\sqrt{-Y_o}}\frac{16\bar u_2\x z}{(z-1)(d-z)}\\ 
	&&&&\\
	& 6 & \bb D^i_x\f\bb n^k\pa_k\f\d(x-x') & -\frac{e^{d\x\f}}{\sqrt{-Y_o}}\frac{2d\x^2z^2}{z-1} & \frac{e^{d\x\f}}{\sqrt{-Y_o}}\frac{2d\x^2z^2}{(z-1)(d-z)}\\ 
	&&&&\\
	\hline\hline
	\multicolumn{3}{|c|}{k=1,\;\co(B-B_o),\; \mbox{time}}& B_{oj}\car^{1j}_{(2,2\ell)} & B_{oj}\cl^{1j}_{(2,2\ell)} \\ \hline\hline 
	\ell & I & \ct^I_{1,\ell} & c^I_{1,\ell} & p^I_{1,\ell}\\ \hline &&&&\\
	0 & 1 & \bb q^kD_k^x\d(x-x') & 4e^{d\x\f}Z_\x Y_o & -\frac{4e^{d\x\f}Z_\x Y_o}{d+z-2-\D_-}\\ &&&&\\
	\hline &&&&\\
	1 & 1 & \bb K\bb n^j D_j^x\d(x-x') & -e^{d\x\f}\frac{2(z-1+8\bar u_2)}{d} & e^{d\x\f}\frac{2(z-1+8\bar u_2)}{d(d-z-\D_-)}\\ &&&&\\
	& 2  & \bb K^2\d(x-x') & -e^{d\x\f}\frac{16\bar u_2}{d} & e^{d\x\f}\frac{16\bar u_2}{d(d-z-\D_-)}\\ &&&&\\
	& 3 & \bb n^k\pa_k\f\;\bb n^j D_j^x\d(x-x') & -e^{d\x\f}\frac{2\x z(z-1+8\bar u_2)}{z-1} & e^{d\x\f}\frac{2\x z(z-1+8\bar u_2)}{(z-1)(d-z-\D_-)}\\ &&&&\\
	& 4 & \bb K\bb n^k \pa_k\f\d(x-x') & -e^{d\x\f}\frac{2\x z(z-1+16\bar u_2)}{z-1} & e^{d\x\f}\frac{2\x z(z-1+16\bar u_2)}{(z-1)(d-z-\D_-)}\\ &&&&\\
	& 5 & (\bb n^k\pa_k\f)^2\d(x-x') & -e^{d\x\f}\frac{2d\x^2z^2(z-1+8\bar u_2)}{(z-1)^2} & e^{d\x\f}\frac{2d\x^2z^2(z-1+8\bar u_2)}{(z-1)^2(d-z-\D_-)}\\ 
	&&&&\\
	\hline\hline
	\end{array}\]
	\caption{General solution of the second and third recursion relations in (\ref{inhomogeneous-sol}) at order $k=1$ for exponential potentials and $\m=0$. The second column from the right describes the sources  $\bs^i_j\car^{1j}_{(2,2\ell)}$ and $B_{oj}\car^{1j}_{(2,2\ell)}$ of the inhomogeneous equations in the form (\ref{scalar-source-decomposition}), while the last column gives the components $\bs^i_j\cl^{1j}_{(2,2\ell)}$ and $B_{oj}\cl^{1j}_{(2,2\ell)}$ of the solution in the parameterization (\ref{scalar-solution-decomposition}). 
	Moreover, we have defined $\bar u_2:= e^{\x\f}u_2$.  	
	The results in this table can be extended to the full source $B_{oj}\Hat\car^{1j}_{(2,2\ell)}$ in (\ref{t-source}) once the canonical momenta at order $\co(1)$ in the Taylor expansion are evaluated.}
	\label{k=1-HJ-solution-1-exp}
\end{table}

\begin{flushleft}
{\bf Solution at order $k=1$}
\end{flushleft} 

The solution (\ref{recursive-solution-exp}) at order $k=1$ can be read off the last column of Tables \ref{k=1-HJ-solution-exp} and \ref{k=1-HJ-solution-1-exp}. Namely, from Table \ref{k=1-HJ-solution-exp} we see that the solution for $\cl^0_{(2,2\ell)}$ is:
\bea\label{k=1-sol-exp}
\cl^0_{(2,0)}&=&\frac{1}{2\k^2}\frac{1}{d+z-2}\sqrt{-\g}e^{d\x\f}\left(\bb R-2D_k\bb q^k+\frac{z-1}{z}\bb q^k\bb q_k-\frac{2d\x}{z}\bb q^i\pa_i\f-\left(\a_\x-\frac{d^2\x^2}{z(z-1)}\right)\bs^{ij}\pa_i\f\pa_j\f\right),\NO\\
\cl^0_{(2,2)}&=&\frac{1}{2\k^2}\frac{1}{d-z}\sqrt{-\g}e^{d\x\f}\left(\bb K^{kl}\bb K_{kl}+2\bb n^kD_k\bb K+\frac{d+z-1}{d}\bb K^2+2\x z \bb K\bb n^k\pa_k\f+\left(\a_\x+\frac{d\x^2z^2}{z-1}\right)(\bb n^k\pa_k\f)^2\right).\NO\\
\eea
Moreover, Table \ref{k=1-HJ-solution-1-exp} gives for $\bs^i_j\p^{0j}_{(2,2\ell)}$:
\bea
\bs^i_j\p^{0j}_{(2,0)}&=&-\frac{1}{2\k^2}\frac{2(z-1)}{z(d+z-2)}\frac{1}{\sqrt{-Y_o}}\sqrt{-\g}e^{d\x\f}\left(\bb n^kD_k\bb q^i+\bb K\bb q^i+\frac{d\x z}{z-1}\bb q^i\bb n^k\pa_k\f\right),\NO\\
\bs^i_j\p^{0j}_{(2,2)}&=&-\frac{1}{2\k^2}\frac{2}{d(d-z)}\frac{1}{\sqrt{-Y_o}}\sqrt{-\g}e^{d\x\f}\left((z-1)\bb D^i+8\bar u_2\bb q^i\right)\left(\bb K+\frac{d\x z}{z-1}\bb n^k\pa_k\f\right).
\eea
\FloatBarrier
The momenta following from (\ref{k=1-sol-exp}) are given by (\ref{k=1-l=0-metric-momenta}), (\ref{k=1-l=1-metric-momenta}), (\ref{k=1-l=0-scalar-momenta}) and (\ref{k=1-l=1-scalar-momenta}), which become
\bea\label{k=1-l=0-metric-momenta-exp}
\left.\p^{0ij}_{(2,0)}\right|_{p_{1,0}^1}&=&\frac{1}{2\k^2}\sqrt{-\g}p_{1,0}^1(\f)\left(-\bb R^{ij}+\bb D^{(i}\bb q^{j)}+\bb q^i\bb q^j+\frac12\bs^{ij}\left(\bb R-2\bb D_k\bb q^k-2\bb q^k\bb q_k\right)-\frac12 \bb n^i\bb n^j\bb R\right.\NO\\
&&\left.+d\x\left(\bb D^{(i}\bb D^{j)}\f+2\bb q^{(i}\bb D^{j)}\f-\bs^{ij}\left(\bb D^2\f+2\bb q^kD_k\f\right)\right)+d^2\x^2\left(\bb D^{(i}\f\bb D^{j)}\f-\bs^{ij}\bb D_k\f \bb D^k\f\right)\right),\NO\\
\left.\p^{0ij}_{(2,0)}\right|_{p_{1,0}^2}&=&\frac{1}{2\k^2}\sqrt{-\g}p_{1,0}^2(\f)\left(-d\x\left(\frac12\bs^{ij}\bb q^kD_k\f-\bb q^{(i}\bb D^{j)}\f+\frac12\bb n^i\bb n^j\bb D^2\f\right)-\frac12d^2\x^2\bb n^i\bb n^j\bb D_k\f\;\bb D^k\f\right),\NO\\
\left.\p^{0ij}_{(2,0)}\right|_{p_{1,0}^3}&=&\frac{1}{2\k^2}\sqrt{-\g}p_{1,0}^3(\f)\left(\frac12\g^{ij}\bb q^k\bb q_k-\bb q^i\bb q^j+\bb n^i\bb n^j(\bb D_k\bb q^k+\bb q^k\bb q_k)+d\x\bb n^i\bb n^j\bb q^kD_k\f \right),\NO\\
\left.\p^{0ij}_{(2,0)}\right|_{p_{1,0}^4}&=&\frac{1}{2\k^2}\sqrt{-\g}p_{1,0}^4(\f)\left(\frac12\bs^{ij}\bb q^kD_k\f-\bb q^{(i}\bb D^{j)}\f+\frac12\bb n^i\bb n^j\bb D^2\f+\frac12d\x\bb n^i\bb n^j\bb D_k\f\;\bb D^k\f\right),\NO\\
\left.\p^{0ij}_{(2,0)}\right|_{p_{1,0}^5}&=&\frac{1}{2\k^2}\sqrt{-\g}p_{1,0}^5(\f)\left(\frac12\g^{ij}\bb D_k\f\;\bb D^k\f-\bb D^i\f\;\bb D^j\f\right), \\
\left.\p^{0ij}_{(2,2)}\right|_{p_{1,1}^1}&=&\frac{1}{2\k^2}\sqrt{-\g}p_{1,1}^1(\f)\left(-\bs^i_p\bs^j_q\bb n^kD_k\bb K^{pq}-\bb K\bb K^{ij}+2\bb n^{(i}\bb D_k\bb K^{kj)}+\frac12(\bs^{ij}+\bb n^i\bb n^j)\bb K^{kl}\bb K_{kl}\right.\NO\\
&&\left.+d\x\left(2\bb n^{(i}\bb K^{j)k}D_k\f-\bb K^{ij}\bb n^kD_k\f\right)\right),\NO\\
\left.\p^{0ij}_{(2,2)}\right|_{p_{1,1}^2}&=&\frac{1}{2\k^2}\sqrt{-\g}p_{1,1}^2(\f)\left(\bs^{ij}\bb n^kD_k\bb K-2\bb n^{(i}\bb D^{j)}\bb K+\frac12\g^{ij}\bb K^2-d\x\left(\rule{0cm}{0,5cm}\bb K\bb n^{(i}\bb D^{j)}\f+\bb n^{(i}\bb D^{j)}(\bb n^kD_k\f)\right.\right.\NO\\
&&\left.\left.-\frac12\bs^{ij}\left(2\bb K\bb n^kD_k\f+\bb n^kD_k(\bb n^lD_l\f)\right)+\frac12\bb n^i\bb n^j\bb K\bb n^kD_k\f\right)\right.\NO\\
&&\left.-d^2\x^2\left(\bb n^{(i}\bb D^{j)}\f\; \bb n^kD_k\f-\frac12\bs^{ij}(\bb n^kD_k\f)^2\right)\right),\NO\\
\left.\p^{0ij}_{(2,2)}\right|_{p_{1,1}^3}&=&\frac{1}{2\k^2}\sqrt{-\g}p_{1,1}^3(\f)\left(2\bb n^{(i}\bb D^{j)}\bb K-\bs^{ij}\bb n^kD_k\bb K-\frac12\g^{ij}\bb K^2+d\x\bb K\left(2\bb n^{(i}\bb D^{j)}\f-\bs^{ij}\bb n^kD_k\f\right)\right),\NO\\
\left.\p^{0ij}_{(2,2)}\right|_{p_{1,1}^4}&=&\frac{1}{2\k^2}\sqrt{-\g}p_{1,1}^4(\f)\left(\bb n^{(i}\bb D^{j)}(\bb n^kD_k\f)-\frac12\bs^{ij}\bb n^kD_k(\bb n^lD_l\f)-\bb K\left(\bb n^{(i}\bb D^{j)}\f-\frac12\bb n^i\bb n^j\bb n^kD_k\f\right)\right.\NO\\
&&\left.+d\x\bb n^kD_k\f\left(\bb n^{(i}\bb D^{j)}\f-\frac12\bs^{ij}\bb n^kD_k\f\right)\right),\NO\\
\left.\p^{0ij}_{(2,2)}\right|_{p_{1,1}^5}&=&\frac{1}{2\k^2}\sqrt{-\g}p_{1,1}^5(\f)\left(\frac12\left(\bs^{ij}+\bb n^i\bb n^j\right)\bb n^kD_k\f-2\bb n^{(i}\bb D^{j)}\f\right)\bb n^lD_l\f,\label{k=1-l=1-metric-momenta-exp}\\
\left.\p^{0}_{\f(2,0)}\right|_{p_{1,0}^1}&=&\frac{1}{2\k^2}\sqrt{-\g}p_{1,0}^1(\f)d\x \bb R,\NO\\
\left.\p^{0}_{\f(2,0)}\right|_{p_{1,0}^2}&=&\frac{1}{2\k^2}\sqrt{-\g}p_{1,0}^2(\f)d\x(\bb D_i\bb q^i+\bb q^i\bb q_i),\NO\\
\left.\p^{0}_{\f(2,0)}\right|_{p_{1,0}^3}&=&\frac{1}{2\k^2}\sqrt{-\g}p_{1,0}^3(\f)d\x\bb q^i\bb q_i,\NO\\
\left.\p^{0}_{\f(2,0)}\right|_{p_{1,0}^4}&=&-\frac{1}{2\k^2}\sqrt{-\g}p_{1,0}^4(\f)(\bb D_i\bb q^i+\bb q^i\bb q_i),\NO\\
\left.\p^{0}_{\f(2,0)}\right|_{p_{1,0}^5}&=&-\frac{1}{2\k^2}\sqrt{-\g}p_{1,0}^5(\f)\left(d\x\bb D^k\f\bb D_k\f+2\left(\bb D^2\f+\bb q^kD_k\f\right)\right),\label{k=1-l=0-scalar-momenta-exp}\\
\left.\p^{0}_{\f(2,2)}\right|_{p_{1,1}^1}&=&\frac{1}{2\k^2}\sqrt{-\g}p_{1,1}^1(\f)d\x\bb K^{kl}\bb K_{kl},\NO\\
\left.\p^{0}_{\f(2,2)}\right|_{p_{1,1}^2}&=&\frac{1}{2\k^2}\sqrt{-\g}p_{1,1}^2(\f)d\x\bb n^kD_k\bb K,\NO\\
\left.\p^{0}_{\f(2,2)}\right|_{p_{1,1}^3}&=&\frac{1}{2\k^2}\sqrt{-\g}p_{1,1}^3(\f)d\x\bb K^2,\NO\\
\left.\p^{0}_{\f(2,2)}\right|_{p_{1,1}^4}&=&-\frac{1}{2\k^2}\sqrt{-\g}p_{1,1}^4(\f)\left(\bb K^2+\bb n^kD_k\bb K\right),\NO\\
\left.\p^{0}_{\f(2,2)}\right|_{p_{1,1}^5}&=&-\frac{1}{2\k^2}\sqrt{-\g}p_{1,1}^5(\f)\left(d\x(\bb n^iD_i\f)^2+2\left(\bb K\bb n^kD_k\f+\bb n^kD_k(\bb n^lD_l\f)\right)\right),\label{k=1-l=1-scalar-momenta-exp}
\eea
where $p^I_{k,\ell}$ in these expressions are listed in the last column of Table \ref{k=1-HJ-solution-exp}. Using these momenta we evaluate
\bea
\p^0_{(2,0)}&=&\frac{1}{2\k^2}\frac{\sqrt{-\g}e^{d\x\f}}{(d+z-2)}\left[\frac{d-1}{2}\left(\bb R+\frac{z-1}{z}\bb q_k\bb q^k-2D_k\bb q^k\right)-\frac{z-1}{z}D_k\bb q^k\right.\NO\\
&&\left.-d\x\left(d-\frac1z\right)\bb D^2\f-d\x\left(d+\frac{d+z-3}{z}\right)\bb q^kD_k\f\right.\NO\\
&&\left.-\left(\frac{d-1}{2}\left(\a_\x-\frac{d^2\x^2}{z(z-1)}\right)+d^2\x ^2\left(d-\frac1z\right)\right)\bb D^k\f\bb D_k\f\right],\NO\\
\bb n_k\bb n_l\p^{0kl}_{(2,0)}&=&\frac{1}{2\k^2}\frac{\sqrt{-\g}e^{d\x\f}}{(d+z-2)}\left[-\frac12\left(\bb R+\frac{z-1}{z}\left( \bb q^k\bb q_k-2D_k\bb q^k\right)\right)+\frac{d\x(z-1)}{z}\bb D^2\f\right.\NO\\
&&\left.+\frac{d\x(z-1)}{z}\bb q^kD_k\f+\left(\frac12\left(\a_\x-\frac{d^2\x^2}{z(z-1)}\right)+\frac{d^2\x^2(z-1)}{z}\right)\bb D^k\f\bb D_k\f\right],\NO\\
\F^0_{(2,0)}&=&\frac{1}{2\k^2}\frac{\sqrt{-\g}e^{d\x\f}}{(d+z-2)}\left[\x\left(\bb R+\frac{z-1}{z}\bb q_k\bb q^k-2D_k\bb q^k\right)+2\left(\a_\x+d^2\x^2-\frac{d(d+z-1)\x^2}{z(z-1)}\right)\bb D^2\f\right.\NO\\
&&\left.+\frac{2(d+z-1)\x}{z}D_k\bb q^k+2\left(\a_\x+d^2\x^2+\frac{d\x^2}{z(z-1)}\left((d+z-1)(z-2)-(z-1)\right)\right)\bb q^kD_k\f\right.\NO\\
&&\left.+\x\left((2d-1)\left(\a_\x-\frac{d^2\x^2}{z(z-1)}\right)+2d^2\x^2\left(d-\frac1z\right)\right)\bb D^k\f\bb D_k\f\right],\NO\\
Q^0_{(2,0)}&=&\frac{1}{2\k^2}\frac{\sqrt{-\g}e^{d\x\f}}{(d+z-2)}\left(1+\frac{d\x\n}{\a}\right)\left[-\frac12\left(\bb R+\frac{z-1}{z}\bb q_k\bb q^k-2D_k\bb q^k\right)-\frac{d+z-1}{z}D_k\bb q^k\right.\NO\\
&&\left.+d\x\left(1+\frac{d+z-1}{z(z-1)}\right)\bb D^2\f+d\x\left(\frac{d+z-1}{z(z-1)}-\frac{d-2}{z}\right)\bb q^kD_k\f\right.\NO\\
&&\left.+\left(\frac12\left(\a_\x+2d^2\x^2-\frac{d^2\x^2}{z(z-1)}\right)+\frac{d^2\x^2(d+z-1)}{z(z-1)}\right)\bb D^k\f\bb D_k\f\right],\\\NO\\\NO\\
\p^0_{(2,2)}&=&\frac{1}{2\k^2}\frac{\sqrt{-\g}e^{d\x\f}}{(d-z)}\left[\frac{d-1}{2}\left(2\bb n^kD_k \bb K+\bb K^2+\bb K_{kl}\bb K^{kl}\right)-(z-1)\left(\bb n^kD_k \bb K+\frac{(d+1)}{2d}\bb K^2\right)\right.\NO\\
&&\left.+d\x\left(d-z+1-\frac zd\right)\bb K\bb n^kD_k\f+d\x(d-z)\bb n^k\pa_k(\bb n^l\pa_l\f)\right.\NO\\
&&\left.+\left(\frac{d-1}{2}\left(\a_\x+\frac{dz^2\x^2}{z-1}\right)+d^2\x^2(d-z)\right)(\bb n^k\pa_k\f)^2 \right],\NO\\ \NO\\
\bb n_k\bb n_l\p^{0kl}_{(2,2)}&=&\frac{\sqrt{-\g}e^{d\x\f}}{2\k^2}\frac{1}{(d-z)}\frac12\left[\bb K_{kl}\bb K^{kl}-\frac{d-z+1}{d}\bb K^2-2\x(d-z)\bb K\bb n^k\pa_k\f+\left(\a_\x+\frac{dz^2\x^2}{z-1}\right)(\bb n^k\pa_k\f)^2\right],\NO\\ \NO\\
\F^0_{(2,2)}&=&\frac{1}{2\k^2}\frac{ \sqrt{-\g}e^{d\x\f}}{(d-z)}\left[\x\left(2\bb n^kD_k \bb K+\bb K^2+\bb K_{kl}\bb K^{kl}\right)+\frac\x d(z-1-2d)\bb K^2\right.\NO\\
&&\left.
-2\x\bb n^kD_k\bb K
-2\left(\a+d(z-1)\x^2+\frac{\x^2z(d-z+1)}{z-1}\right)\bb n^k\pa_k(\bb n^l\pa_l\f)
\right.\NO\\
&&\left.-2\left(\a+\frac{\x^2z(d-z+1)}{z-1}\right)\bb K\bb n^k\pa_k\f-\x\left((2d-1)\left(\a+\frac{d\x^2}{z-1}\right)+d(d-z)\x^2\right)(\bb n^k\pa_k\f)^2\right],\NO\\
Q^0_{(2,2)}&=&\frac{1}{2\k^2}\frac{ \sqrt{-\g}e^{d\x\f}}{(d-z)}\left[\left(d-\frac12\left(1+\frac{d\x\n}{\a}\right)\right)\left(2\bb n^kD_k \bb K+\bb K^2+\bb K_{kl}\bb K^{kl}\right)\right.\NO\\
&&\left.+\left(\left(1+\frac{d\x\n}{\a}\right)-(d+z)\right)\bb n^kD_k \bb K-\left(d+1+\left(\frac{z-1}{2d}-1\right)\left(1+\frac{d\x\n}{\a}\right)\right)\bb K^2\right.\NO\\
&&\left.+\left(\frac{d\n}{\a}\left(\a+\frac{\x^2z(d-z+1)}{z-1}\right)+(d-z)\x\right)\bb K\bb n^k\pa_k\f\right.\NO\\
&&\left.+\left(\frac{d\n}{\a}\left(\a+d(z-1)\x^2+\frac{\x^2z(d-z+1)}{z-1}\right)+d\x(d-z)\right)\bb n^k\pa_k(\bb n^l\pa_l\f)\right.\NO\\
&&\left. +\frac12\left(1+\frac{d\x\n}{\a}\right)\left((2d-1)\left(\a+\frac{d\x^2}{z-1}\right)+d(d-z)\x^2\right)(\bb n^k\pa_k\f)^2\right].
\eea
Moreover, from Table \ref{k=1-HJ-solution-1-exp} we obtain 
\bea
\int d^{d+1}xB_{oj}\car^{1j}_{(2,0)}&=&\frac{\sqrt{-\g}e^{d\x\f}}{2\k^2}\frac{2(z-1)}{z}\left(\bb D_k\bb q^k+d\x\bb q^kD_k\f\right),\\
\int d^{d+1}xB_{oj}\car^{1j}_{(2,2)}&=&\frac{\sqrt{-\g}e^{d\x\f}}{2\k^2}\left[\frac{2(z-1+8\bar u_2)}{d}\left(\bb n^k \pa_k+\bb K-\frac{d\x}{z-1}\bb n^k\pa_k\f\right)-\frac{16\bar u_2}{d}\bb K\right]\left(\bb K+\frac{dz\x}{z-1}\bb n^l\pa_l\f\right).\NO
\eea
Using these expressions in the last equation in (\ref{recursive-solution-exp}) one obtains $P_{(2,0)}^0$ and $P_{(2,2)}^0$, thus completing the solution of the recursion problem at $k=1$ and $\co(B-B_o)$. We will not write explicitly the expressions for $P_{(2,0)}^0$ and $P_{(2,2)}^0$ here since they are rather lengthy and they can easily be evaluated using Mathematica from the expressions we give above.  

As for the Einstein-Proca theory in the previous example, the solutions to the recursion relations (\ref{mu=0-recursions}) we obtained above determine the solution of the HJ equation up to and including order $k=1$ and $\co(B-B_o)$. These suffice in order to determine the solution of the HJ equation at order $k=2$ but only to order $\co(1)$ in the Taylor expansion in $B-B_o$, corresponding to the solution of only the first recursion relation in (\ref{mu=0-recursions}) for $k=2$. Again we will not write these solutions explicitly since they are too lengthy. But they can be evaluated straightforwardly with Mathematica by inserting the $k=1$ results above into (\ref{k=2-sources}).

This example can be compared directly with the model discussed in 
\cite{Chemissany:2012du}, which corresponds to the following values of our parameters: 
\bea
&& d=\e=z=2, \quad \x=\frac12,\quad \m=0,\quad \n=-\frac32,\quad \s=1, \quad \a=1,\NO\\
&& Z_o=\frac14,\quad W_o=2,\quad V=-10e^{-\f},\quad V_o=-10.
\eea
Moreover, the two scalars in \cite{Chemissany:2012du} are related to the scalar $\f$ here as
\be
\F_{\rm there}=\f,\quad \f_{\rm there}=-\log(k/2)+\f,
\ee
with $\f\to 0$ in the UV. Dropping terms with derivatives on the scalar $\f$ in this case we get the same result for $\cl^0_{(0)}$, $\cl^0_{(2,0)}$ and $\cl^0_{(2,2)}$ as in (\ref{horava}), but for $\cl^0_{(4,0)}$ we now get
\bea
\cl^0_{(4,0)}&=&\frac{\sqrt{-\g}}{2\k^2}e^{\f}\frac{r_o}{4}\left\{
\left(\bb R^{ij}-\bb D^{(i}\bb q^{j)}-\frac12 \bb q^i\bb q^j\right)^2+\frac 12\left(\bb R-2D_k\bb q^k+\frac12\bb q^k\bb q_{k}\right)^2+\frac14\left(\bb R+\frac12(\bb q^k\bb q_{k}-2D_k\bb q^k)\right)^2\right.\NO\\
&&\left.-\left(\bb R-2D_k\bb q^k+\frac12\bb q^k\bb q_{k}\right)\left(\bb R-D_k\bb q^k+\frac12\bb q^k\bb q_{k}\right)-\frac18\left(\bb R-3D_k\bb q^k+\frac12\bb q^k\bb q_{k}\right)^2 \right.\NO\\
&&\left.+\frac{1}{24}\left(\bb R+\frac12\bb q^k\bb q_k+D_k\bb q^k\right)^2+\frac{1}{12}\left(\bb R-\bb q^k\bb q_k+D_k\bb q^k\right)^2\right\}\NO\\
&\simeq& -\frac{\sqrt{-\g}}{2\k^2} \frac{e^{\f}}{16}\left(\bb R-\bb q^k\bb q_k+D_k\bb q^k\right)^2,
\eea
where again $\simeq$ denotes equality up to total derivative terms. This quantity is the only non-trivial conformal invariant with four spatial derivatives in $d=2$ and for $z=2$ \cite{Griffin:2011xs}. Note that this model is related to the Einstein-Proca theory of the previous example only by a change of frame since $\x=1/2$ here. So the effect of going from the Einstein frame (where no purely spatial anomaly is generated) to a non-Einstein frame is to generate a non-zero coefficient for this conformal invariant in the anomaly. However, the expression for the anomaly given in \cite{Chemissany:2012du} does not agree with our result. Namely, in our notation the purely spatial part of the expression in \cite{Chemissany:2012du} is 
\bea
\ca^{(4)}&\sim& \frac{e^\f}{8}\sqrt{-\g}\left(\left(\bb R_{ij}-\bb D_i\bb q_j-\frac12\bb q_i\bb q_j\right)^2-\frac12(D_k\bb q^k)^2+\frac12(\bb q_k\bb q^k)^2-\frac13\left(\bb R+\frac12\bb q^k\bb q_k-2D_k\bb q^k\right)^2\right)\NO\\
&\simeq& \frac{e^\f}{8}\sqrt{-\g}\left(\frac12(\bb q_k\bb q^k)^2-(D_k\bb q^k)^2\right),
\eea
which is in fact not a conformal invariant. We have traced the discrepancy to the fact that the $\co(B-B_o)$ contribution to the 2-derivative momenta has not be taken into account in \cite{Chemissany:2012du}.

\FloatBarrier

\subsection{Exponential potentials with $\m\neq 0$}

As a final example we consider a model with exponential potentials 
\be 
V_{\xi}=V_o,\qquad  Z_{\xi}=Z_o e^{-2  (\xi +\n )\f},\qquad W_{\xi}=W=W_o e^{-2  (\xi +\n ) \f},
\ee
corresponding to the first three superpotential coefficients ($\D_-$ is again given by (\ref{dimensions}))
\bea
&&u_0(\f)=\left(z-1+d(1+\m\x)\right)e^{-\x\f},\quad
u_1(\f)=\frac12 (z-1)e^{-\x\f}, \NO\\
&&u_2(\f)=\frac{1}{8\left(\frac{\n^2}{\a}+\frac{d-1}{d}-\frac{\e}{z-1}\right)}\left(\D_--(z-1)\left((2\n+\x)\frac{\n}{\a}+\frac{2d-1}{d}\right)\right)e^{-\x\f},
\eea 
but without any restriction on the parameters that define the boundary conditions. In particular, the crucial difference in this example relative to the previous two is that $\m\neq 0$ and so the recursion relations (\ref{kth-order-0}), (\ref{kth-order-1-space}) and (\ref{kth-order-1-time}) are no longer algebraic. However, there is still some simplification due to the fact that the potentials are exactly -- not merely asymptotically -- exponentials. 

The inhomogeneous solutions (\ref{inhomogeneous-sol}) become
\begin{align}
	\label{inhomogeneous-sol-exp-mu}
	\boxed{
		\begin{aligned}
			&\cl^0_{(2k,2\ell)}[\g,\f]= -\frac{1}{\m}e^{-\cc_{k,\ell}\f/\m}\int^\f d\bar\f  e^{\cc_{k,\ell}\bar\f/\m}\car^0_{(2k,2\ell)}[\g,\bar\f],\\
			&\bs^i_j\cl_{(2k,2\ell)}^{1j}[\g(x),\f(x);x']=-\frac{1}{\m} e^{-(\cc_{k,\ell}+\e-z)\f/\m}\int^{\f(x)} d\bar\f  e^{(\cc_{k,\ell}+\e-z)\bar\f/\m} \bs^i_j\car^{1j}_{(2k,2\ell)}[\g(x),\bar\f;x'],\\
			&B_{oj}(x) \cl_{(2k,2\ell)}^{1j}[\g(x),\f(x);x']=-\frac{1}{\m} e^{-(\cc_{k,\ell}-\D_-)\f/\m}\int^{\f(x)} d\bar\f e^{(\cc_{k,\ell}-\D_-)\bar\f/\m} B_{oj}\Hat\car^{1j}_{(2k,2\ell)}[\g(x),\bar\f;x'],
		\end{aligned}}
	\end{align} 
where we have used the fact that for the present example 
\be\label{exp-mu-simplification}
\ca(\f)=\f/\m, \quad \ck(\f)=-\frac{1}{\m}, \quad \Om\propto e^{-\D_-\f/\m},\quad Z_{\x}\propto e^{-2(\e-z)\f/\m},
\ee
which implies that $\mf A_{k,\ell}$ defined in (\ref{factor}) is always linear in $\phi$ and so $\mf A_{k,\ell}''=0$. This implies that the integrals defined in (\ref{shorthand}) simplify as 
\be
\fint_{k,\ell,m} =\fint_{k,\ell,0}\;,\quad \forall m.
\ee
Using this, together with $\mf A_{k,\ell}''=0$, we see that the integrals in Table (\ref{integration}) reduce in this case to ordinary integrals over the exponential coefficients of any tensor structure involving derivatives on the scalar. In fact, since the overall exponential function of the scalar in the source terms in (\ref{inhomogeneous-sol}) is easily determined to be
\be
\car^0_{(2k,2\ell)}\sim e^{d\x\f},\quad \bs^i_j\car^{1j}_{(2k,2\ell)}\sim e^{(d\m\x+z-\e)\f/\m},\quad B_{oj}\Hat\car^{1j}_{(2k,2\ell)}\sim e^{d\x\f}, 
\ee 
we can perform the integrals over the scalar fields generically without any reference to the explicit form of these source terms.   

The source term (\ref{t-source-simple}) of the third recursion relation in (\ref{inhomogeneous-sol}) can be written as
\bea\label{t-source-simple-exp-mu}
&&B_{oj}\Hat\car^{1j}_{(2k,2\ell)}[\g(x),\f(x);x']-\frac1\z(\D_--\cc_{k,\ell}-d\m\x)Q_{(2k,2\ell)}^0\d^{(d+1)}(x-x')=\NO\\
&&B_{oj}\car^{1j}_{(2k,2\ell)}[\g(x),\f(x);x']-2(z-1)\bb n_k\bb n_l\p_{(2k,2\ell)}^{0kl}\d^{(d+1)}(x-x')+\frac1\a(z-1)(\n+\x)\F_{(2k,2\ell)}^0\d^{(d+1)}(x-x')\NO\\
&&+\frac1\z\left(\cc_{k,\ell}+d\m\x-\left(\e+z-1+\frac1\a \n(\n+\x)(z-1)\right)\right)Q_{(2k,2\ell)}^0\d^{(d+1)}(x-x'),
\eea
while from (\ref{FPQ}) we get
\be\label{FPQ-exp-mu}
Q_{(2k,2\ell)}^0:=\p^0_{(2k,2\ell)}+d\bb n^i\bb n^j\p^{0ij}_{(2k,2\ell)}-\frac{d\n}{2\a}\F_{(2k,2\ell)}^0,\quad P_{(2k,2\ell)}^0:=B_{ok}\p_{(2k,2\ell)}^{0k}-\frac{1}{\z}Q_{(2k,2\ell)}^0,
\ee
where $\z$ defined in (\ref{zeta}) now becomes 
\be
\z=\frac d2\left(\frac{\e}{z-1}-\frac{d-1}{d}-\frac{\n^2}{\a}\right).
\ee
Performing the integrations over the scalar field in (\ref{inhomogeneous-sol}) we arrive at the solutions
\bea
\label{recursive-solution-exp-mu}
\cl^0_{(2k,2\ell)}&=&-\frac{1}{\cc_{k,\ell}+d\m\x}\car^0_{(2k,2\ell)},\NO\\
\bs^i_j\p^{0j}_{(2k,2\ell)}(x')&=&-\frac{1}{\cc_{k,\ell}+d\m\x}\int d^{d+1}x\bs^i_j\car^{1j}_{(2k,2\ell)}[\g(x),\f(x);x'],\NO\\
P_{(2k,2\ell)}^0(x')&=&\frac{1}{\D_--\cc_{k,\ell}-d\m\x}\left(\int d^{d+1}xB_{oj}\car^{1j}_{(2k,2\ell)}[\g(x),\f(x);x']-2(z-1)\bb n_k\bb n_l\p_{(2k,2\ell)}^{0kl}\right.\\
&&\left.+\frac1\a(z-1)(\n+\x)\F_{(2k,2\ell)}^0+\frac1\z\left(\cc_{k,\ell}+d\m\x-\left(\e+z-1+\frac1\a \n(\n+\x)(z-1)\right)\right)Q_{(2k,2\ell)}^0\right),\NO
\eea
where again we have used (\ref{vector-momentum}). In the limit $\m\to 0$ these expressions reduce to the corresponding ones in (\ref{recursive-solution-exp}) of the previous example. The source terms $\car^0_{(2k,2\ell)}$, $\bs^i_j\car^{1j}_{(2k,2\ell)}$ and $B_{oj}\car^{1j}_{(2k,2\ell)}$ for $k=1$ are given in Tables \ref{k=1-HJ-solution-exp-mu} and \ref{k=1-HJ-solution-1-exp-mu}. Note that since the hyperscaling parameter $\th$ in the Einstein frame is given by the combination $-d\m\x$, we see that the denominators in these recursion relations are shifted by $\th$ relative to the previous examples. 
\begin{table}
\[\begin{array}{|c|c|c|c|c|}
\hline\hline
\ell & I & \ct^I_{1,\ell} & c^I_{1,\ell} & p^I_{1,\ell}\\ \hline &&&&\\
0 & 1 & \bb R & -e^{d\x\f} & \frac{1}{d+z-2+d\m\x}e^{d\x\f}\\ &&&&\\
  & 2 & D_k\bb q^k & 2e^{d\x\f} & \frac{-2}{d+z-2+d\m\x}e^{d\x\f}\\ &&&&\\
  & 3 & \bb q^i\bb q_i & -\frac{z-1}{\e}e^{d\x\f} & \frac{(z-1)}{\e(d+z-2+d\m\x)}e^{d\x\f}\\ &&&&\\
  & 4 & \bb q^i\pa_i\f & -\frac{2(z-1)(\e-z)}{\e\m}e^{d\x\f} & \frac{2(z-1)(\e-z)}{\e\m(d+z-2+d\m\x)}e^{d\x\f} \\ &&&&\\
  & 5 & \bs^{ij}\pa_i\f\pa_j\f & \left(\a_\x-\frac{(z-1)(\e-z)^2}{\e\m^2} \right)e^{d\x\f} & \frac{(z-1)(\e-z)^2-\a_\x\m^2\e}{\e\m^2(d+z-2+d\m\x)}e^{d\x\f}\\ &&&&\\
 \hline &&&&\\
1 & 1 & \bb K^{kl}\bb K_{kl} & -e^{d\x\f} & \frac{1}{d-z+d\m\x}e^{d\x\f}\\ &&&&\\
  & 2 & \bb n^kD_k\bb K & -2e^{d\x\f} & \frac{2}{d-z+d\m\x}e^{d\x\f}\\ &&&&\\
  & 3 & \bb K^2 & -\frac{(d+2z+d\m\x-1-\e)}{d+d\m\x+z-\e}e^{d\x\f} & \frac{(d+2z+d\m\x-1-\e)}{(d+d\m\x+z-\e)(d-z+d\m\x)}e^{d\x\f}\\ &&&&\\
  & 4 & \bb K \bb n^j \pa_j\f & -\frac{2(z-1)(d\m\x+z-\e)}{\m(d+d\m\x+z-\e)}e^{d\x\f} & \frac{2(z-1)(d\m\x+z-\e)}{\m(d+d\m\x+z-\e)(d-z+d\m\x)}e^{d\x\f}\\ &&&&\\
  & 5 & (\bb n^i \pa_i\f)^2 & -\left(\a_\x+ \frac{(z-1)(d\m\x+z-\e)^2}{\m^2(d+d\m\x+z-\e)}\right)e^{d\x\f} & \frac{\a_\x\m^2(d+d\m\x+z-\e)+(z-1)(d\m\x+z-\e)^2}{\m^2(d+d\m\x+z-\e)(d-z+d\m\x)}e^{d\x\f}\\ 
  &&&&\\
 \hline\hline
\end{array}\]
\caption{General solution of the first recursion relation in (\ref{inhomogeneous-sol}) at order $k=1$ for exponential potentials and $\m\neq 0$. The second column from the right describes the source of the inhomogeneous equation in the form (\ref{scalar-source-decomposition}), while the last column gives the solution $\cl^0_{(2,0)}$ and $\cl^0_{(2,2)}$ in the parameterization (\ref{scalar-solution-decomposition}).}
\label{k=1-HJ-solution-exp-mu}
\end{table}
\begin{table}
	\vskip0.7in	
	\[\begin{array}{|c|c|c|c|c|}
	\hline\hline
	\multicolumn{3}{|c|}{k=1,\;\co(B-B_o),\; \mbox{space}}& \bs^i_j\car^{1j}_{(2,2\ell)} & \bs^i_j\cl^{1j}_{(2,2\ell)} \\ \hline\hline 
	\ell & I & \ct^I_{1,\ell} & c^I_{1,\ell} & p^I_{1,\ell}\\ \hline &&&&\\
	0 & 1 & \bb q^i\bb n^kD_k^x\d(x-x') & -4e^{d\x\f}Z_\x\sqrt{-Y_o} & \frac{4e^{d\x\f}Z_\x\sqrt{-Y_o}}{d+z+d\m\x-2}\\ &&&&\\
	\hline &&&&\\
	1 & 1 & \bb K\bb D^i_x\d(x-x') & -\frac{e^{d\x\f}}{\sqrt{-Y_o}}\frac{2(z-1)}{(d+z+d\m\x-\e)} & \frac{e^{d\x\f}}{\sqrt{-Y_o}}\frac{2(z-1)}{(d+z+d\m\x-\e)(d-z+d\m\x)}\\ &&&&\\
	& 2  & \bb K\bb q^i\d(x-x') & \frac{e^{d\x\f}}{\sqrt{-Y_o}}\frac{16\bar u_2}{(d+z+d\m\x-\e)} & -\frac{e^{d\x\f}}{\sqrt{-Y_o}}\frac{16\bar u_2}{(d+z+d\m\x-\e)(d-z+d\m\x)} \\ &&&&\\
	& 3 & \bb n^k\pa_k\f\bb D^i_x\d(x-x') & -\frac{e^{d\x\f}}{\sqrt{-Y_o}}\frac{2(z-1)((d-1)\x-\n)}{(d+z+d\m\x-\e)} & \frac{e^{d\x\f}}{\sqrt{-Y_o}}\frac{2(z-1)((d-1)\x-\n)}{(d+z+d\m\x-\e)(d-z+d\m\x)}\\ &&&&\\
	& 4 & \bb K\bb D^i_x\f\d(x-x') & -\frac{e^{d\x\f}}{\sqrt{-Y_o}}\frac{2(z-1)((d-1)\x-\n)}{(d+z+d\m\x-\e)} & \frac{e^{d\x\f}}{\sqrt{-Y_o}}\frac{2(z-1)((d-1)\x-\n)}{(d+z+d\m\x-\e)(d-z+d\m\x)}\\ &&&&\\
	& 5 & \bb q^i\bb n^k\pa_k\f\d(x-x') & \frac{e^{d\x\f}}{\sqrt{-Y_o}}\frac{16\bar u_2((d-1)\x-\n)}{(d+z+d\m\x-\e)} & -\frac{e^{d\x\f}}{\sqrt{-Y_o}}\frac{16\bar u_2((d-1)\x-\n)}{(d+z+d\m\x-\e)(d-z+d\m\x)}\\ 
	&&&&\\
	& 6 & \bb D^i_x\f\bb n^k\pa_k\f\d(x-x') & -\frac{e^{d\x\f}}{\sqrt{-Y_o}}\frac{2(z-1)((d-1)\x-\n)^2}{(d+z+d\m\x-\e)} & \frac{e^{d\x\f}}{\sqrt{-Y_o}}\frac{2(z-1)((d-1)\x-\n)^2}{(d+z+d\m\x-\e)(d-z+d\m\x)}\\ 
	&&&&\\
	\hline\hline
	\multicolumn{3}{|c|}{k=1,\;\co(B-B_o),\; \mbox{time}}& B_{oj}\car^{1j}_{(2,2\ell)} & B_{oj}\cl^{1j}_{(2,2\ell)} \\ \hline\hline 
	\ell & I & \ct^I_{1,\ell} & c^I_{1,\ell} & p^I_{1,\ell}\\ \hline &&&&\\
	0 & 1 & \bb q^kD_k^x\d(x-x') & 4e^{d\x\f}Z_\x Y_o & -\frac{4e^{d\x\f}Z_\x Y_o}{d+z+d\m\x-2-\D_-}\\ &&&&\\
	\hline &&&&\\
	1 & 1 & \bb K\bb n^j D_j^x\d(x-x') & -e^{d\x\f}\frac{2(z-1+8\bar u_2)}{(d+z+d\m\x-\e)} & e^{d\x\f}\frac{2(z-1+8\bar u_2)}{(d+z+d\m\x-\e)(d-z+d\m\x-\D_-)}\\ &&&&\\
	& 2  & \bb K^2\d(x-x') & -e^{d\x\f}\frac{16\bar u_2}{(d+z+d\m\x-\e)} & e^{d\x\f}\frac{16\bar u_2}{(d+z+d\m\x-\e)(d-z+d\m\x-\D_-)}\\ &&&&\\
	& 3 & \bb n^k\pa_k\f\;\bb n^j D_j^x\d(x-x') & -e^{d\x\f}\frac{2(z-1+8\bar u_2)((d-1)\x-\n)}{(d+z+d\m\x-\e)} & e^{d\x\f}\frac{2(z-1+8\bar u_2)((d-1)\x-\n)}{(d+z+d\m\x-\e)(d-z+d\m\x-\D_-)}\\ &&&&\\
	& 4 & \bb K\bb n^k \pa_k\f\d(x-x') & -e^{d\x\f}\frac{2(z-1+16\bar u_2)((d-1)\x-\n)}{(d+z+d\m\x-\e)} & e^{d\x\f}\frac{2(z-1+16\bar u_2)((d-1)\x-\n)}{(d+z+d\m\x-\e)(d-z+d\m\x-\D_-)}\\ &&&&\\
	& 5 & (\bb n^k\pa_k\f)^2\d(x-x') & -e^{d\x\f}\frac{2(z-1+8\bar u_2)((d-1)\x-\n)^2}{(d+z+d\m\x-\e)} & e^{d\x\f}\frac{2(z-1+8\bar u_2)((d-1)\x-\n)^2}{(d+z+d\m\x-\e)(d-z+d\m\x-\D_-)}\\ 
	&&&&\\
	\hline\hline
	\end{array}\]
	\caption{General solution of the second and third recursion relations in (\ref{inhomogeneous-sol}) at order $k=1$ for exponential potentials and $\m\neq 0$. The second column from the right describes the source terms $\bs^i_j\car^{1j}_{(2,2\ell)}$ and $B_{oj}\car^{1j}_{(2,2\ell)}$ of the inhomogeneous equations in the form (\ref{scalar-source-decomposition}), while the last column gives the components $\bs^i_j\cl^{1j}_{(2,2\ell)}$ and $B_{oj}\cl^{1j}_{(2,2\ell)}$ of the solution in the parameterization (\ref{scalar-solution-decomposition}). The constant $\bar u_2\equiv e^{\x\f} u_2$ has been introduced to simplify the expressions. The results in this table can be extended to the full source $B_{oj}\Hat\car^{1j}_{(2,2\ell)}$ in (\ref{t-source}) once the canonical momenta at order $\co(1)$ in the Taylor expansion are evaluated.}
	\label{k=1-HJ-solution-1-exp-mu}
\end{table}

\vspace{1.in}
\begin{flushleft}
	{\bf Solution at order $k=1$}
\end{flushleft} 

At order $k=1$ the solution (\ref{recursive-solution-exp-mu}) can be read off the last column of Tables \ref{k=1-HJ-solution-exp-mu} and \ref{k=1-HJ-solution-1-exp-mu}. From Table \ref{k=1-HJ-solution-exp-mu} we see that the solution for $\cl^0_{(2,2\ell)}$ is:
\bea\label{k=1-sol-exp-mu}
\cl^0_{(2,0)}&=&\frac{1}{2\k^2}\frac{1}{d+z-2+d\m\x}\sqrt{-\g}e^{d\x\f}\left(\bb R-2D_k\bb q^k+\frac{z-1}{\e}\bb q^k\bb q_k+\frac{2(z-1)(\e-z)}{\e\m}\bb q^i\pa_i\f\right.\NO\\
&&\left.+\frac{(z-1)(\e-z)^2-\a_\x\m^2\e}{\e\m^2}\bs^{ij}\pa_i\f\pa_j\f\right),\NO\\
\cl^0_{(2,2)}&=&\frac{1}{2\k^2}\frac{1}{d-z+d\m\x}\sqrt{-\g}e^{d\x\f}\left(\bb K^{kl}\bb K_{kl}+2\bb n^kD_k\bb K+\frac{d+2z+d\m\x-1-\e}{d+d\m\x+z-\e}\bb K^2\right.\\
&&\left.+\frac{2(z-1)(d\m\x+z-\e)}{\m(d+d\m\x+z-\e)} \bb K\bb n^k\pa_k\f+\frac{\a_\x\m^2(d+d\m\x+z-\e)+(z-1)(d\m\x+z-\e)^2}{\m^2(d+d\m\x+z-\e)}(\bb n^k\pa_k\f)^2\right).\NO
\eea
Moreover, from Table \ref{k=1-HJ-solution-1-exp-mu} we obtain:
\bea
\bs^i_j\p^{0j}_{(2,0)}&=&-\frac{1}{2\k^2}\frac{2(z-1)}{\e(d+z+d\m\x-2)}\frac{\sqrt{-\g}e^{d\x\f}}{\sqrt{-Y_o}}\left(\bb n^kD_k\bb q^i+\bb K\bb q^i+\frac{d\m\x+z-\e}{\m}\bb q^i\bb n^k\pa_k\f\right),\\
\bs^i_j\p^{0j}_{(2,2)}&=&-\frac{1}{2\k^2}\frac{2(z-1)}{(d+z+d\m\x-\e)(d-z+d\m\x)}\frac{\sqrt{-\g}e^{d\x\f}}{\sqrt{-Y_o}}\left(\bb D^i+\frac{8\bar u_2}{z-1}\bb q^i\right)\left(\bb K+\frac{d\m\x+z-\e}{\m}\bb n^k\pa_k\f\right),\NO
\eea
where $\bar u_2 \equiv e^{\x\f} u_2$. The metric and scalar momenta corresponding to the solution (\ref{k=1-sol-exp-mu}) are again given by the simplified formulas (\ref{k=1-l=0-metric-momenta}), (\ref{k=1-l=1-metric-momenta}),  (\ref{k=1-l=0-scalar-momenta}) and (\ref{k=1-l=1-scalar-momenta}), but with the coefficients $p^I_{k,\ell}$ listed in Table \ref{k=1-HJ-solution-exp-mu} now. From these canonical momenta we obtain
\bea
\p^0_{(2,0)}&=&\frac{1}{2\k^2}\frac{\sqrt{-\g}e^{d\x\f}}{(d+d\m\x+z-2)}\left[\frac{d-1}{2}\left(\bb R+\frac{z-1}{\e}\bb q_k\bb q^k-2D_k\bb q^k\right)-\frac{z-1}{\e}D_k\bb q^k\right.\NO\\
&&\left.-\left(d^2\x+\frac{(z-1)(\n+\x)}{\e}\right)\bb D^2\f+\left(\frac{(z-1)((d-2)(\n+\x)-d\x)}{\e}-d^2\x\right)\bb q^kD_k\f\right.\NO\\
&&\left.-\left(\frac{d-1}{2}\left(\a_\x-\frac{(z-1)(\n+\x)^2}{\e}\right)+d\x\left(d^2\x+\frac{(z-1)(\n+\x)}{\e}\right)\right)\bb D^k\f\bb D_k\f\right],\NO\\
\bb n_k\bb n_l\p^{0kl}_{(2,0)}&=&\frac{1}{2\k^2}\frac{\sqrt{-\g}e^{d\x\f}}{(d+d\m\x+z-2)}\left[-\frac12\left(\bb R+\frac{z-1}{\e}\left( \bb q^k\bb q_k-2D_k\bb q^k\right)\right)+\left(d\x+\frac{(z-1)(\n+\x)}{\e}\right)\bb D^2\f\right.\NO\\
&&\left.+\frac{d\x(z-1)}{\e}\bb q^kD_k\f+\left(\frac12\left(\a_\x-\frac{(z-1)(\n+\x)^2}{\e}\right)+d\x\left(d\x+\frac{(z-1)(\n+\x)}{\e}\right)\right)\bb D^k\f\bb D_k\f\right],\NO\\
\F^0_{(2,0)}&=&\frac{1}{2\k^2}\frac{\sqrt{-\g}e^{d\x\f}}{(d+d\m\x+z-2)}\left[\x\left(\bb R+\frac{z-1}{\e}\bb q_k\bb q^k-2D_k\bb q^k\right)+2\left(\a_\x+d^2\x^2-\frac{(z-1)(\n+\x)\n}{\e}\right)\bb D^2\f\right.\NO\\
&&\left.-\frac{2(z-1)\n}{\e}D_k\bb q^k+2\left(\a_\x+d^2\x^2-\frac{z-1}{\e}\left((\n+\x)^2+(d-2)(\n+\x)\x-d\x^2\right)\right)\bb q^kD_k\f\right.\NO\\
&&\left.+\x\left((2d-1)\left(\a_\x-\frac{(z-1)(\n+\x)^2}{\e}\right)+2d\x\left(d^2\x+\frac{(z-1)(\n+\x)}{\e}\right)\right)\bb D^k\f\bb D_k\f\right],\NO\\
Q^0_{(2,0)}&=&\frac{1}{2\k^2}\frac{\sqrt{-\g}e^{d\x\f}}{(d+d\m\x+z-2)}\left[-\frac12\left(1+\frac{d\x\n}{\a}\right)\left(\bb R+\frac{z-1}{\e}\bb q_k\bb q^k-2D_k\bb q^k\right)-\frac{2(z-1)\z}{\e}D_k\bb q^k\right.\NO\\
&&\left.+\left(d\x\left(1+\frac{d\x\n}{\a}\right)-\frac{2(z-1)(\n+\x)\z}{\e}\right)\bb D^2\f+\left(\left(1+\frac{d\n\x}{\a}\right)\times\rule{0cm}{0.7cm}\right.\right.\NO\\
&&\left.\left.\left(d\x+\frac{(z-1)\left[(d-1)(\n+\x)-d\x\right]}{\e}\right)-d\left(d\x+\frac{(z-1)\left[\left(\frac{2\z}{d}+1\right)(\n+\x)-d\x\right]}{\e}\right)\right)\bb q^kD_k\f\right.\NO\\
&&\left.+\left(\frac12\left(1+\frac{d\n\x}{\a}\right)\left(\a_\x+2d^2\x^2-\frac{(z-1)(\n+\x)^2}{\e}\right)-\frac{2d\x(z-1)(\n+\x)\z}{\e}\right)\bb D^k\f\bb D_k\f\right],
\eea
and
\bea
\p^0_{(2,2)}&=&\frac{1}{2\k^2}\frac{\sqrt{-\g}e^{d\x\f}}{(d+d\m\x-z)}\left[\frac{d-1}{2}\left(2\bb n^kD_k \bb K+\bb K^2+\bb K_{kl}\bb K^{kl}\right)-\frac{d(z-1)}{d+d\m\x+z-\e}\left(\bb n^k\pa_k \bb K+\frac{(d+1)}{2d}\bb K^2\right)\right.\NO\\
&&\left.+\left(d^2\x+\frac{(z-1)[\n+\x-d(d+1)\x]}{d+d\m\x+z-\e}\right)\bb K\bb n^kD_k\f+\left(d^2\x+\frac{d(z-1)(\n+\x-d\x)}{d+d\m\x+z-\e}\right)\bb n^k\pa_k(\bb n^l\pa_l\f)\right.\NO\\
&&\left.+\left(\frac{d-1}{2}\left(\a_\x+\frac{(z-1)(\n+\x-d\x)^2}{d+d\m\x+z-\e}\right)+\frac{d^2\x(z-1)(\n+\x-d\x)}{d+d\m\x+z-\e}+d^3\x^2\right)(\bb n^k\pa_k\f)^2 \right],\NO\\ \NO\\
\bb n_k\bb n_l\p^{0kl}_{(2,2)}&=&\frac{\sqrt{-\g}e^{d\x\f}}{2\k^2}\frac{1}{(d+d\m\x-z)}\left[\frac12\bb K_{kl}\bb K^{kl}-\frac12\left(1-\frac{z-1}{d+d\m\x+z-\e}\right)\bb K^2\right.\NO\\
&&\left.-\left(d\x+\frac{(z-1)(\n+\x-d\x)}{d+d\m\x+z-\e}\right)\bb K\bb n^k\pa_k\f+\frac12\left(\a_\x+\frac{(z-1)(\n+\x-d\x)^2}{d+d\m\x+z-\e}\right)(\bb n^k\pa_k\f)^2\right],\NO\\ \NO\\
\F^0_{(2,2)}&=&\frac{1}{2\k^2}\frac{ \sqrt{-\g}e^{d\x\f}}{(d+d\m\x-z)}\left[\x\left(2\bb n^kD_k \bb K+\bb K^2+\bb K_{kl}\bb K^{kl}\right)+\frac{(z-1)\left[2(\n+\x)+\x\right]}{d+d\m\x+z-\e}\bb K^2\right.\NO\\
&&\left.
+\frac{2(z-1)(\n+\x)}{d+d\m\x+z-\e}\bb n^kD_k\bb K
-2\left(\a_\x+d^2\x^2+\frac{(z-1)(\n+\x-d\x)(\n+2\x-d\x)}{d+d\m\x+z-\e}\right)\bb n^k\pa_k(\bb n^l\pa_l\f)
\right.\NO\\
&&\left.-2\left(\a_\x+d^2\x^2+\frac{(z-1)\left[(\n+\x-d\x)(\n+2\x-d\x)-d^2\x^2\right]}{d+d\m\x+z-\e}\right)\bb K\bb n^k\pa_k\f\right.\NO\\
&&\left.-\left((2d-1)\x\left(\a_\x+d^2\x^2+\frac{(z-1)(\n+\x-d\x)(\n+\x)}{d+d\m\x+z-\e}\right)\right.\right.\NO\\
&&\left.\left.+\frac{d\x^2(z-1)(\n+\x-d\x)}{d+d\m\x+z-\e}+d^2\x^3\right)(\bb n^k\pa_k\f)^2\right],\NO\\
Q^0_{(2,2)}&=&\frac{1}{2\k^2}\frac{ \sqrt{-\g}e^{d\x\f}}{(d+d\m\x-z)}\left[\left(d-\frac12\left(1+\frac{d\x\n}{\a}\right)\right)\left(2\bb n^kD_k \bb K+\bb K^2+\bb K_{kl}\bb K^{kl}\right)\right.\NO\\
&&\left.-d\left(1+\frac{(z-1)\left(1+\frac{(\n+\x)\n}{\a}\right)}{d+d\m\x+z-\e}\right)\bb n^kD_k \bb K-\left(d+\frac{(z-1)\left[\frac12(1+\frac{d\n\x}{\a})+\frac{d(\n+\x)\n}{\a}\right]}{d+d\m\x+z-\e}\right)\bb K^2\right.\NO\\
&&\left.+\left(\frac{d\n}{\a}\left(\a_\x+d^2\x^2+\frac{(z-1)\left[(\n+\x-d\x)(\n+2\x-d\x)-d^2\x^2\right]}{d+d\m\x+z-\e}\right)\right.\right.\NO\\
&&\left.\left.\rule{0cm}{0.6cm}-\frac{(z-1)[(d-1)(\n+\x)+d\x]}{d+d\m\x+z-\e}\right)\bb K\bb n^k\pa_k\f+\left(d^2\x+\frac{d(z-1)(\n+\x-d\x)}{d+d\m\x+z-\e}\right.\right.\NO\\
&&\left.\left.+\frac{d\n}{\a}\left(\a_\x+d^2\x^2+\frac{(z-1)(\n+\x-d\x)(\n+2\x-d\x)}{d+d\m\x+z-\e}\right)\right)\bb n^k\pa_k(\bb n^l\pa_l\f)\right.\NO\\
&&\left. +\left(1+\frac{d\x\n}{\a}\right)\left(\frac{2d-1}{2}\left(\a_\x+d^2\x^2+\frac{(z-1)(\n+\x-d\x)(\n+\x)}{d+d\m\x+z-\e}\right)\right.\right.\NO\\
&&\left.\left.+\frac{d\x}{2}\left(d\x+\frac{(z-1)(\n+\x-d\x)}{d+d\m\x+z-\e}\right)\right)(\bb n^k\pa_k\f)^2\right].
\eea
Moreover, from Table \ref{k=1-HJ-solution-1-exp-mu} we get
\bea
\int d^{d+1}xB_{oj}\car^{1j}_{(2,0)}&=&\frac{1}{2\k^2}\sqrt{-\g}e^{d\x\f}\frac{2(z-1)}{\e}\left(\bb D_k\bb q^k+d\x\bb q^kD_k\f\right),\NO\\
\int d^{d+1}xB_{oj}\car^{1j}_{(2,2)}&=&\frac{1}{2\k^2}\frac{\sqrt{-\g}e^{d\x\f}}{d+d\m\x+z-\e}\left[2(z-1+8\bar u_2)\left(\bb n^k D_k+\bb K+(\n+\x)\bb n^kD_k\f\right)-16\bar u_2\bb K\right]\times\NO\\
&&\left(\bb K+\frac{d\m\x+z-\e}{\m}\bb n^lD_l\f\right).
\eea
These expressions allow us to write explicitly the solutions $P_{(2,0)}^0$ and $P_{(2,2)}^0$ of the third recursion relation in (\ref{recursive-solution-exp-mu}), but as in the previous example, the resulting expressions are forbiddingly lengthy to be presented explicitly here. However, the results we have presented allow one to evaluate $P_{(2,0)}^0$ and $P_{(2,2)}^0$ easily by evaluating the last expression in (\ref{recursive-solution-exp-mu}) using Mathematica. The same holds for the solution at $k=2$ and $\co(B-B_o)^0$, which can be obtained by inserting the $k=1$ results in (\ref{k=2-sources}).

\FloatBarrier

\section{Concluding remarks}

In this paper we have developed a general algorithm for constructing the holographic dictionary for a large class of theories that admit asymptotically locally Lifshitz and hyperscaling violating Lifshitz boundary conditions with arbitrary dynamical exponents. This dictionary only exists for $\th\leq d+z$, $z\geq 1$, since there are no well defined asymptotic expansions for $\th> d+z$ and $z\leq 1$.     

The algorithm we developed relies entirely on the metric formulation of the dynamics and there is no need for the introduction of vielbeins at any point. The objective of the algorithm is the systematic construction of the most general asymptotic solution of the radial Hamilton-Jacobi equation subject to asymptotically locally Lifshitz and hyperscaling violating Lifshitz boundary conditions. This is achieved by expanding the solution of the Hamilton-Jacobi equation in simultaneous eigenfunctions of two commuting functional operators, which generalizes the standard expansion in eigenfunctions of the dilatation operator to non-relativistic and non-scale invariant boundary conditions.
The resulting recursive procedure does not require any ansatz and it is entirely algorithmic. In future work we hope we will be able to implement this algorithm in a symbolic computation package.
 
The entire holographic dictionary can be derived from this asymptotic solution of the Hamilton-Jacobi equation as is shown in Section \ref{ward}. In particular, the asymptotic Fefferman-Graham expansions, including the sources and 1-point functions, are derived directly from this asymptotic solution of the Hamilton-Jacobi equation, without any need for solving the second order equations of motion. In fact, the Hamilton-Jacobi equation leads to a much more efficient method for computing renormalized correlation functions as well \cite{Papadimitriou:2004rz,McFadden:2010vh,Papadimitriou:2013jca}. Our method provides a solid basis for computing correlation functions in asymptotically Lifshitz and hyperscaling violating Lifshitz backgrounds, and we intend to explore this direction in future work. Another potential application of the present work is in the holographic computation of entanglement entropy.    

Finally, we have shown that the unique non-trivial conformal invariant for $z=2$ in 2 dimensions with four spatial derivatives appears in the conformal anomaly of an Einstein-Proca theory, provided the latter is coupled with a dilaton and one moves away from the Einstein frame. To our knowledge, this is the first example where this term is actually generated, implying that the detailed balance condition does not hold in this case \cite{,Griffin:2011xs}. More generally, the algorithm presented here provides a systematic tool for generating non-relativistic conformal invariants for any dimension and any value of the dynamical exponent $z\geq 1$.

\section*{Acknowledgments}

We thank Blaise Gout\'eraux, Sean Hartnoll, Cynthia Keeler, Simon Ross and Kostas Skenderis for useful discussions and correspondence. IP would also like to thank the Asian Pacific Center for Theoretical Physics (APCTP) as well as the Centro de Ciencias de Benasque Pedro Pascual for the hospitality during the completion of this work. WC is supported by the SITP, Stanford and the Arab Fund for Economic and Social Development. The work of IP is funded by the Consejo Superior de Investigaciones Cient\'ificas and the European Social Fund under the contract JAEDOC068. This work has also been supported by the ESF Holograv Programme, the Spanish Ministry of Economy and Competitiveness under grant FPA2012-32828, Consolider-CPAN (CSD2007-00042), the Spanish MINECO's ``Centro de Excelencia Severo Ochoa'' Programme under grant SEV-2012-0249, as well as by the grant HEPHACOS-S2009/ESP1473 from the C.A. de Madrid.

\appendix

\renewcommand{\theequation}{\Alph{section}.\arabic{equation}}

\setcounter{section}{0}

\section*{Appendices}
\setcounter{equation}{0}

\section{Hamiltonian description of constrained systems}
\label{constrained-systems}

In this appendix we provide a Hamiltonian description for the effective point particle Lagrangian 
\bea\label{pp-lagnrangian-no-omega}
L_{eff}&=&\frac{1}{2\k^2}e^{f+dh+d\x\f}\left(\left(1+\frac{d^2\x^2}{\a_\x}\right)(\dot f+d\dot h)^2-(\dot f^2+d\dot h^2)-\a_\x\left(\dot \f-\frac{d\x}{\a_\x}(\dot f+d\dot h)\right)^2\right.\NO\\
&&\left.\rule{0cm}{0.6cm}\rule{2.8cm}{0cm}+2Z_\x(\f)e^{-2f}\dot a^2+W_\x(\f)e^{-2f}a^2-V_\x(\f)\right),
\eea
which is obtained from (\ref{pp-lagrangian}) by setting $\om=const.$, subject to the constraint 
(\ref{pp-constraint}). There are two equivalent ways to deal with such constrained dynamical systems and we discuss both.

\begin{flushleft}
{\bf i) Hamiltonian analysis after implementing the constraint}
\end{flushleft}

The most straightforward way to study the constrained system is to strongly implement the constraint from the onset. In the present case this leads to the reduced Lagrangian
\bea\label{pp-lagrangian-reduced}
\lbar L_{eff}&=&\frac{1}{2\k^2}e^{f+dh+d\x\f}\left(\left(1+\frac{d^2\x^2}{\a_\x}\right)(\dot f+d\dot h)^2-(\dot f^2+d\dot h^2)-\a_\x\left(\dot \f-\frac{d\x}{\a_\x}(\dot f+d\dot h)\right)^2\right.\NO\\
&&\left.\rule{0cm}{0.6cm}\rule{2.8cm}{0cm}+\frac{z-1}{\e}\left(\dot f-\frac12\frac{Z'_\x}{Z_\x}\dot \f\right)^2+\frac{z-1}{2\e}W_\x(\f)Z_\x^{-1}(\f)-V_\x(\f)\right),
\eea
The corresponding reduced canonical momenta are 
\bea\label{pp-momenta-reduced}
&&\lbar\p_f=\frac{1}{\k^2}e^{f+dh+d\x\f}\left(d\left(\dot h+\x\dot\f\right)+\frac{z-1}{\e}\left(\dot f-\frac12\frac{Z'_\x}{Z_\x}\dot \f\right)\right),\NO\\ &&\lbar\p_h=\frac{1}{\k^2}e^{f+dh+d\x\f}d\left(\dot f+(d-1)
\dot h+d\x\dot\f\right),\NO\\  &&\lbar\p_{\f}=-\frac{1}{\k^2}e^{f+dh+d\x\f}\left(\a_\x\left(\dot\f-\frac{d\x}{\a_\x}(\dot f+d\dot h)\right)+\frac{z-1}{2\e}\frac{Z'_\x}{Z_\x}\left(\dot f-\frac12\frac{Z'_\x}{Z_\x}\dot \f\right)\right),
\eea
and the Hamiltonian takes the form
\bea \label{pp-hamiltonian-reduced}
\lbar H_{eff}&=&\frac{\k^2}{2}e^{-f-dh-d\x\f}\left(\rule{0cm}{0.6cm}\frac{1}{d}\p_f\left(2\p_h-(d-1)\p_f\right)-\frac{1}{\a}\left(\p_\f-\x(\p_f+\p_h)\right)^2\right.\NO\\
&&\left.+\frac{z-1}{2\e}\left(\left(\frac1d-\frac{\x}{2\a}\frac{Z'}{Z}\right)\left(\p_h-(d-1)\p_f\right)+\frac{1}{2\a}\frac{Z'}{Z}\left(\p_\f-d\x\p_f\right)\right)^2\right)\NO\\
&&+\frac{1}{2\k^2}e^{f+dh+d\x\f}\left(-\frac{z-1}{2\e}W_\x(\f)Z_\x^{-1}(\f)+V_\x(\f)\right).
\eea

\begin{flushleft}
{\bf ii) Incorporating the constraint using Dirac's algorithm}
\end{flushleft}
  
The same Hamiltonian can be obtained following the Dirac algorithm for constrained systems \cite{H&T,Hong:2003vd,Rothe:2003qn,Rothe&Rothe}. In this procedure we start by adding the constraint (\ref{pp-constraint}) using a Lagrange multiplier, i.e. 
\be 
L_{eff}\rightarrow L_0=L_{eff}-\l \cg,
\ee
with
\be  \cg:=a-\sqrt{\frac{z-1}{2\e}}Z_{\xi}^{-1/2}(\f)e^{f}.
\ee 
The canonical momentum conjugate to the Lagrange multiplier $\l$ vanishes identically, which leads to the \emph{primary} constraint
\be 
\P=\frac{\pa L_o}{\pa \dot{\l}}=0.
\ee
The corresponding Hamiltonian is
\be 
H_o=H_{eff}+\l \cg,
\ee
where 
\bea \label{pp-hamiltonian-no-omega}
H_{eff}&=&\frac{\k^2}{2}e^{-f-dh-d\x\f}\left(\frac{1}{d}\p_f\left(2\p_h-(d-1)\p_f\right)-\frac{1}{\a}\left(\p_\f-\x(\p_f+\p_h)\right)^2+\frac12Z_\x^{-1}e^{2f}\p_a^2\right)\NO\\
&&+\frac{1}{2\k^2}e^{f+dh+d\x\f}\left(-W_\x(\f)e^{-2f}a^2+V_\x(\f)\right),
\eea
and we define the total Hamiltonian by adding the primary constraint as
\be 
H_{T}=H_o+u\P,
\ee
where $u$ is another Lagrange multiplier. Correspondingly, the extended Poisson bracket is defined as
\be
\{\ca,\cb\}_{ext}:=\frac{\pa\ca}{\pa \p_h}\frac{\pa\cb}{\pa h}+\frac{\pa\ca}{\pa \p_f}\frac{\pa\cb}{\pa f}+\frac{\pa\ca}{\pa \p_a}\frac{\pa\cb}{\pa a}+\frac{\pa\ca}{\pa \p_\f}\frac{\pa\cb}{\pa \f}+\frac{\pa\ca}{\pa \P}\frac{\pa\cb}{\pa \l}-\ca\leftrightarrow\cb, 
\ee
for any pair of local phase space functions $\ca,\cb$. With this definition of the Poisson bracket the radial derivative of a local phase space function is given by
\be
\dot\ca=\{H_T,\ca\}_{ext}.
\ee

Starting with the primary constraint $\P$, the next step is to generate all {\em secondary} constraints by Poisson commuting the constraints with the total Hamiltonian and with all previous constraints. At each step of the iteration process, if a given constraint Poisson commutes with all previous constraints but does not commute with $H_T$, then this Poisson bracket gives rise to a new constraint. If on the other hand we reach a point where a constraint does not commute with at least one of the previous constraints, then the iteration procedure stops and some of the Lagrange multipliers must be determined in terms of the phase space variables. 

In the current system, the Poisson bracket of the primary constraint with the total Hamiltonian generates the original constraint $\cg$,  
\bea  
-\{H_{T},\P \}_{ext}&=&\frac{\pa H_{T}}{\pa \l }=\cg,
\eea 
which now emerges as a secondary constraint. At the next step of the iterative process we compute the Poisson brackets
\be
\{\P,\cg\}_{ext}=0,
\ee
and 
\bea 
\dot{\cg}&=&\{H_{T},\cg\}_{ext}=\frac{\pa H_{eff}}{\pa \p_f}\frac{\pa\cg}{\pa f}+\frac{\pa H_{eff}}{\pa \p_a}\frac{\pa\cg}{\pa a}+\frac{\pa H_{eff}}{\pa \p_\f}\frac{\pa\cg}{\pa \f}\NO\\
&\approx&\frac{\k^2}{2}e^{-f-dh-d\x\f}\left(Z_\x^{-1}e^{2f}\p_a-2a\left(\frac1d\left(\p_h-(d-1)\p_f\right)+\frac{1}{2\a}\frac{Z'}{Z}\left(\p_\f-\x(\p_h+\p_f)\right)\right)\right)\NO\\
&=:&\frac{\k^2}{2}e^{-f-dh-d\x\f}\wt\cg,
\eea
where the $\approx$ sign here means equal up to previous constraints. $\wt \cg=0$ is therefore a second secondary constraint. However, evaluating its Poisson bracket with $\cg$ we find that it does not vanish, which means that there are no more secondary constraints and the iterative procedure stops at this level. Namely,    
\be
\{\cg,\wt\cg\}=2a\left(\frac{1}{4\a}\left(\frac{Z'}{Z}\right)^2+\frac{d-1}{d}-\frac{\e}{z-1}\right)\sim \left(\frac{\a_2 y_o+\a_1}{\a_1y_o}\right) a\neq 0,
\ee
where these constants are defined in subsection \ref{sup}. Note that $\a_2 y_o+\a_1$ is related to the coefficient of the derivative square term in (\ref{ansatzI-eq}) and the fact that it is non-zero  strongly influences the dynamics of the system. The same quantity appears in (\ref{zeta}). The fact that this Poisson bracket is non vanishing means that the Poisson bracket  
\be \label{thirdc} \dot{\wt{\cg}}=\{H_{T},\wt{\cg}\}_{ext}=\{H_{eff},\wt{\cg}\}+\l\{\cg,\wt{\cg}\},
\ee
can be set to zero by a choice of the Lagrange multiplier $\l$. Finally, solving the constraints $\cg$ and $\wt\cg$ for $a$ and $\pi_{a}$ and inserting them into (\ref{pp-hamiltonian}), one obtains precisely the Hamiltonian $\lbar H_{eff}$ in  (\ref{pp-hamiltonian-reduced}).

\begin{flushleft}  
{\bf Hamilton-Jacobi formulation of the constrained system} 
\end{flushleft}
 
We finally want to show that the Hamiltonian (\ref{pp-hamiltonian-reduced}) of the constrained system gives rise to a Hamilton-Jacobi equation that describes the Lifshitz solutions (\ref{Lif-sol}) in the standard way. Taking the potentials of the Lagrangian as in (\ref{exp-potentials}) the canonical momenta become 
\bea 
&&\lbar\p_f=\frac{1}{\k^2}e^{f+dh+d\x\f}\left(d+z+d\m\x-1\right),\NO\\ &&\lbar\p_h=\frac{1}{\k^2}e^{f+dh+d\x\f}d\left(d+z+d\m\x-1\right),\NO\\  &&\lbar\p_{\f}=\frac{1}{\k^2}e^{f+dh+d\x\f}d\x\left(d+z+d\m\x-1\right),
\eea
which can be written as gradients 
\be 
\lbar\pi_{f}=\frac{\pa \lbar\cs_{eff}}{\pa f},\qquad \lbar\pi_{h}=\frac{\pa \lbar\cs_{eff}}{\pa h}, \qquad \lbar\pi_{\f}=\frac{\pa \lbar\cs_{eff}}{\pa \f},
\ee
with the simple Hamilton-Jacobi function
\be 
\label{Seff}
\lbar\cs_{eff}=\frac{1}{\k^{2}}e^{f+d h+d\xi\phi} (d+z+d\m\x-1).
\ee
This clearly demonstrates that the Lifshitz solutions (\ref{Lif-sol}) are the trajectories of a constrained dynamical system.

\section{Remarks on functional operators}
\label{functional-operators}

Let $\vf(x)$ be a generic tensor field and consider the functional operator
\be
d_f(x):=f[\vf]\frac{\d}{\d \vf(x)}, 
\ee
where $f[\vf]$ is a local functional of $\vf(x)$. There is an integrated version of this operator, namely 
\be
\d_f:=\int d^{d+1}xd_f(x)=\int d^{d+1}x f[\vf]\frac{\d}{\d \vf(x)}.
\ee
Suppose that for a local functional $\vr[\vf]$ of $\vf(x)$,
\be
\d_f\int d^{d+1}x \vr[\vf(x)]=\l_f \int d^{d+1}x \vr[\vf(x)],
\ee
holds. It follows that 
\be
\int d^{d+1}x\left(\d_f \vr[\vf(x)]-\l_f \vr[\vf(x)]\right)=0,
\ee
and hence
\be 
\d_f \vr[\vf(x)]=\l_f \vr[\vf(x)]+\pa_iv^i_f, 
\ee
for some $v^i_f$. Therefore,
\be
\int d^{d+1}x'\left\{f[\vf(x')]\frac{\d}{\d \vf(x')} \vr[\vf(x)]-\d^{(d+1)}(x-x')\left(\l_f g[\vf(x)]+\pa_iv^i_f(x)\right)\right\}=0,
\ee
or
\be 
f[\vf(x')]\frac{\d}{\d \vf(x')} \vr[\vf(x)]=\d^{(d+1)}(x-x')\left(\l_f \vr[\vf(x)]+\pa_iv^i_f(x)\right)
+\pa'_i\left(\d^{(d+1)}(x-x')u_f^i(x')\right), 
\ee
for some $u^i_f$. This in turn implies that
\be 
d_f(x')\int d^{d+1}x \vr[\vf(x)]=f[\vf(x')]\frac{\d}{\d \vf(x')} \int d^{d+1}x \vr[\vf(x)]=\l_f \vr[\vf(x')]+\pa'_i\left(v^i_f(x')+u_f^i(x')\right). 
\ee

\begin{lemma}
\label{functional-lemma-1}
For any $\vr[\vf]$ such that 
\be\label{eigenfunction-condition-lemma}
\d_\vf\int d^{d+1}x \vr[\vf(x)]=\l_\vf \int d^{d+1}x \vr[\vf(x)],
\ee
where 
\be
\d_\vf:=\int d^{d+1}x \vf(x)\frac{\d}{\d \vf(x)},
\ee
we have
\be
\d_\vf \vr[\vf(x)]=\l_\vf \vr[\vf(x)],
\ee
i.e. $v_\vf^i=0$. 
\end{lemma}
The proof is straightforward. Namely, the most general $\vr[\vf]$ that satisfies (\ref{eigenfunction-condition-lemma}) is a polynomial in $\vf$ and its derivatives. Acting explicitly with $\d_\vf$ on such a polynomial one arrives at the above result. $\square$

\section{Anisotropic geometry}
\label{identities}

In this appendix we collect a number of results on the anisotropic description of the dynamics in terms of the ADM \cite {ADM} variables introduced in (\ref{decomposition}). These variables are suitable for the uneven treatment of space and time required by Lifshitz or hvLf boundary conditions. Table \ref{anisotropic-geometry} is a compilation of the most relevant geometric identities for the ADM description of the dynamics. 
\begin{table}\vskip2.1cm
\begin{align}
\boxed{
\begin{aligned}
\hskip0.2in & \hskip5.2in\\
&\text{\em Unit normal:}\\
&\mathbb{n}_i=(n,0),\quad \bb n^i=(1/n,-n^a/n),\quad \bb n_i\bb n^i =-1\\\\
&\text{\em Induced metric:}\\
&\bs_{ij} := \g_{ij}+\bb n_i \bb n_j,\quad \bb n^i\bs_{ij}=\bs_{ij}\bb n^j=0\\\\
&\text{\em Covariant derivative:}\\
&\bb D_i T^{i_1\ldots i_m}{}_{j_1\ldots j_n}:=
\bs^{i_1}_{k_1}\ldots\bs^{i_m}_{k_m}\bs^{l_1}_{j_1}\ldots\bs^{l_n}_{j_n}\bs^{j}_i D_j T^{k_1\ldots k_m}{}_{l_i\ldots l_n},\quad \bb D_i\bs_{jk}=0,\quad \bb D_{[i}\bb n_{j]}=0\\
&\bb D_i(a^{i_1\ldots i_m}b^{i_1\ldots i_n})=a^{i_1\ldots i_m}\bb D_i(b^{i_1\ldots i_n})+b^{i_1\ldots i_n}\bb D_i(a^{i_1\ldots i_m})\quad \mbox{iff $a$ and $b$ are transverse}\\\\
&\text{\em Extrinsic curvature:}\\
&\bb K_{ij}:=\frac12 \pounds_n\bs_{ij}=\bb D_i \bb n_j=\bs^k_iD_k\bb n_j,\quad \bb K_{ij}=\bb K_{ji},\quad \bb K := \bb K^i_i=D_i\bb n^i,\quad \bb n^i\bb K_{ij}=0\\\\
&\text{\em Orthogonal transport and twist:}\\
&\bb q^i:=\bb n^kD_k\bb n^i,\quad  \bb f_{ij}:=\pa_i\bb n_j-\pa_j\bb n_i=D_i\bb n_j-D_j\bb n_i=\bb q_i\bb n_j-\bb q_j\bb n_i\\
&\bb n_i\bb q^i=0,\quad \bb q_j=\bb n^i\bb f_{ij},\quad \bb f^{ij}\bb f_{ij}=-2\bb q_i\bb q^i \\
&D_k\bb q^k=\bb D_k\bb q^k+\bb q_k\bb q^k,\quad \bb D_{[i}\bb q_{j]}=0\NO\\\\
&\text{\em Riemann tensor:}\\
&\bs^i_p\bs^q_j\bb n^k\bb n^l R^p{}_{kql}[\g]=-\bs^i_p\bs^q_j\bb n^k D_k\bb K^p_q-\bb K^i_k\bb K^k_j+\bb D^i\bb q_j+\bb q^i\bb q_j\\
&\bs^i_p\bs^q_j\bs^l_k\bb n^s R^p{}_{lqs}[\g]=\bb D^i \bb K_{kj}-\bb D_k\bb K^i_j\\
&\bs^i_p\bs^k_q\bs^s_j\bs^t_l R^{pq}{}_{st}[\g]=\bb R^{ik}{}_{jl}+\bb K^i_j\bb K^k_l-\bb K^i_l\bb K^k_j\\\\
&\text{\em Ricci tensor:}\\
&\bb n^i\bb n^jR_{ij}[\g]=\bb K^2-\bb K_{ij}\bb K^{ij}-D_i\left(\bb n^iD_k \bb n^k-\bb n^kD_k\bb n^i\right)\\
&\bs^i_j\bb n^kR_{ik}[\g]=\bb D_i\bb K^i_j-\bb D_j \bb K\\
&\bs^i_k\bs^l_jR^k_l[\g]=\bb R^i_j+\bb K^i_j\bb K+\bs^i_p\bs^q_j\bb n^k D_k\bb K^p_q-\bb D^i\bb q_j-\bb q^i\bb q_j\\\\
&\text{\em Ricci scalar:}\\
&R[\g]=\bb R[\s]-\bb K^2+\bb K_{ij}\bb K^{ij}+2D_i\left(\bb n^iD_k \bb n^k-\bb n^kD_k\bb n^i\right)\\\\
\end{aligned}  }
\end{align}
\caption{Geometric identities related to the ADM decomposition (\ref{decomposition}).} 
\label{anisotropic-geometry}
\end{table}
It should be stressed that the ADM decomposition (\ref{decomposition}) differs from the one in (\ref{ADM-metric}) in two crucial ways. Firstly, the slicing in (\ref{ADM-metric} is along a spacelike direction, while in (\ref{decomposition}), as in the usual ADM decomposition \cite{ADM}, it is along a timelike direction. This introduces some sign differences. More importantly, in (\ref{decomposition}) the lapse and shift functions, respectively $n$ and $n_a$, are dynamical since they are part of the induced metric $\g_{ij}$ and hence they cannot be gauge-fixed at will.    

Besides the standard geometric objects, such as the extrinsic curvature $\bb K_{ij}$, in Table \ref{anisotropic-geometry} we have introduced the `orthogonal transport' vector field $\bb q^i$ and the `twist field' $\bb f_{ij}$, both of which  measure the failure of $\bb n_i$ to be a geodesic vector field. In terms of components, the only non zero component of $\bb f_{ij}$ is $\bb f_{ta}=\pa_a n$,
while $\bb q_i$ takes the form
\be
\bb q_t=\frac{n^c}{n}\pa_cn,\quad \bb q_a=\frac1n\pa_a n,
\ee
and hence
\be
\bb q_i\bb q^i=\frac{1}{n^2}\bs^{ab}\pa_an\pa_b n.
\ee
From these relations follows that $\bb q_i=0 \Leftrightarrow \bb f_{ij}=0 \Leftrightarrow \pa_a n=0$.

Throughout this paper we use extensively the following identities expressing the asymptotic form $B_{oi}$ of the vector field and its derivatives in terms of geometric quantities 
\bea
&&B_{oi}=\sqrt{\frac{z-1}{2\e}}Z_\x^{-1/2}(\f)\bb n_i,\\
&&D_i B_{oj}=\sqrt{\frac{z-1}{2\e}}Z_\x^{-1/2}\left[-\frac12 \left(\frac{Z'_\x}{Z_\x}\right)\bb n_j\pa_i\f+D_i\bb n_j\right],\\
&&F_{oij}=\sqrt{\frac{z-1}{2\e}}Z_\x^{-1/2}\left[\frac12 \left(\frac{Z'_\x}{Z_\x}\right)(\bb n_i\pa_j\f-\bb n_j\pa_i\f)+\bb f_{ij}\right],\\
&&D_iB_o^i=\sqrt{\frac{z-1}{2\e}}Z_\x^{-1/2}\left[-\frac12 \left(\frac{Z'_\x}{Z_\x}\right)\bb n^i\pa_i\f+\bb K\right],
\eea
as well as the functional derivatives
\bea
&&\frac{\d B_{oi}(x')}{\d\f(x)}=-\frac12 \frac{Z'_\x(\f)}{Z_\x(\f)}B_{oi}\d^{(d+1)}(x-x'),\label{b-derivative-phi}\\
&&\frac{\d B_{oi}(x')}{\d\g_{kl}(x)}=-\frac{\e}{z-1}Z_\x B_o^kB_o^lB_{oi}\d^{(d+1)}(x-x'),\label{b-derivative-gamma}\\
&&\frac{\d \bb n_{i}(x')}{\d\g_{kl}(x)}=-\frac12 \bb n^k\bb n^l\bb n_{i}\d^{(d+1)}(x-x')\label{n-down-derivative-gamma},\\
&&\frac{\d \bb n^i(x')}{\d\g_{kl}(x)}=\left(\frac12 \bb n^k\bb n^l\bb n^i-\bs^{i(k}\bb n^{l)}\right)\d^{(d+1)}(x-x')\label{n-up-derivative-gamma},\\
&&\frac{\d \bs_{ij}(x')}{\d\g_{kl}(x)}=\left(\d^k_i\d^l_j-\bb n^k\bb n_i\bb n^l\bb n_j\right)\d^{(d+1)}(x-x')\label{sigma-down-derivative-gamma},\\
&&\frac{\d \bs^{ij}(x')}{\d\g_{kl}(x)}=-\bs^{ik}\bs^{jl}\d^{(d+1)}(x-x')\label{sigma-up-derivative-gamma}.
\eea

\FloatBarrier

\newpage


\bibliography{hvLf}
 \bibliographystyle{utphysmodb}
\end{document}